\documentclass[12pt,preprint]{aastex}

\usepackage{lscape}
\usepackage{rotating}
\usepackage{natbib}
\usepackage{lscape}
\usepackage{color}
\usepackage{graphicx}
\tighten

\newcommand{\bb}{\mbox{$\ |\ $}}  

\newcommand{\mearth}{\mbox{$M_\oplus$}}
\newcommand{\mjup}{\mbox{$M_{Jup}$}}
\newcommand{\msun}{\mbox{$M_\odot$}}

\newcommand{\mic}{\mbox{$\mu$m}}
\newcommand{\app}{\mbox{$\sim$}}

\newcommand{\pp}{\mbox{$\pm$}}

\newcommand{\dg}{\mbox{$^\circ$}}

\makeatletter
\newcommand{\Rmnum}[1]{\expandafter\@slowromancap\romannumeral #1@}
\makeatother

\begin{document}

\title {The Gemini NICI Planet-Finding Campaign: \\
The Frequency of Giant Planets Around Debris Disk Stars\footnotemark}

\author {Zahed Wahhaj\altaffilmark{2}, 
Michael C. Liu\altaffilmark{3},  
Eric L. Nielsen\altaffilmark{3},
Beth A. Biller\altaffilmark{4},
Thomas L. Hayward\altaffilmark{5},
Laird M. Close\altaffilmark{6},
Jared R. Males\altaffilmark{6},
Andrew Skemer\altaffilmark{6},
Christ Ftaclas\altaffilmark{3},
Mark Chun\altaffilmark{3},
Niranjan Thatte\altaffilmark{7},
Matthias Tecza\altaffilmark{7},
Evgenya L. Shkolnik\altaffilmark{8},
Marc Kuchner\altaffilmark{9},
I.\ Neill Reid\altaffilmark{10},
Elisabete M. de Gouveia Dal Pino\altaffilmark{11},
Silvia H. P. Alencar\altaffilmark{12},
Jane Gregorio-Hetem\altaffilmark{11},
Alan Boss\altaffilmark{13},
Douglas N. C. Lin\altaffilmark{14}
Douglas W. Toomey\altaffilmark{7}}

\slugcomment{Accepted to {\it Ap.J.} on June 24, 2013.}

\altaffiltext{2} {European Southern Observatory, Alonso de Cordova 3107,  Vitacura, Casilla 19001, Santiago, Chile}
\altaffiltext{3} {Institute for Astronomy, University of Hawaii, 2680 Woodlawn Drive, Honolulu, HI 96822, USA}
\altaffiltext{4} {Max-Planck-Institut für Astronomie, Königstuhl 17, D-69117 Heidelberg, Germany}
\altaffiltext{5} {Gemini Observatory, Southern Operations Center, c/o AURA, Casilla 603, La Serena, Chile}
\altaffiltext{6} {Steward Observatory, University of Arizona, 933 North Cherry Avenue, Tucson, AZ 85721, USA}
\altaffiltext{7} {Department of Astronomy, University of Oxford, DWB, Keble Road, Oxford OX1 3RH, U.K. }
\altaffiltext{8} {Lowell Observatory, 1400 West Mars Hill Road, Flagstaff, Arizona, 86001, USA}
\altaffiltext{9} {NASA Goddard Space Flight Center, Exoplanets and Stellar Astrophysics Laboratory, Greenbelt, MD 20771, USA}
\altaffiltext{10} {Space Telescope Science Institute, 3700 San Martin Drive, Baltimore, MD 21218, USA}
\altaffiltext{11} {Universidade de Sao Paulo, IAG/USP, Departamento de Astronomia, Rua do Matao, 1226, 05508-900, Sao Paulo, SP, Brazil}
\altaffiltext{12} {Departamento de Fisica - ICEx - UFMG, Av.\ Ant\^onio Carlos, 6627, 30270-901, Belo Horizonte, MG, Brazil}
\altaffiltext{13} {Department of Terrestrial Magnetism, Carnegie Institution of Washington, 5241 Broad Branch Road, NW, Washington, DC 20015, USA}
\altaffiltext{14} {Department of Astronomy and Astrophysics, University of California, Santa Cruz, CA, USA}
\altaffiltext{15} {Mauna Kea Infrared, LLC, 21 Pookela St., Hilo, HI 96720, USA}


\begin{abstract}

We have completed a high-contrast direct imaging survey for giant planets around 57 debris disk stars 
as part of the Gemini NICI Planet-Finding Campaign. We achieved median $H$-band contrasts of 
12.4 mag at $0.5''$ and 14.1 mag at $1"$ separation. 
Follow-up observations of the 66 candidates with projected separation $<$ 500~AU show that all of them are background objects. 
\textcolor{black}{To establish statistical constraints on the underlying giant planet population based on our imaging data, we have developed a new Bayesian formalism that incorporates
(1) non-detections,  (2) single-epoch candidates, (3) astrometric and (4) photometric
information, and (5) the possibility of multiple planets per star to constrain the planet
population. Our formalism allows us to include in our analysis the
previously known $\beta$~Pictoris and the HR~8799 planets. Our results show at 95\% confidence that  
$<$13\% of debris disk stars have a $\geq$5\mjup\ planet beyond 80~AU, and
$<$21\% of debris disk stars have a $\geq$3\mjup\  planet outside of
40~AU, based on hot-start evolutionary models.
We model the population of directly-imaged planets as $d^2
N/dM da \propto m^{\alpha}
a^{\beta}$, where $m$ is planet mass
and $a$ is orbital semi-major axis (with a maximum value of
$a_{max}$). We find that $\beta < -0.8$ and/or $\alpha >
1.7$. Likewise, we find that $\beta < -0.8$ and/or $a_{max} < 200$~AU. 
For the case where the planet frequency rises sharply with mass ($\alpha > 1.7$), 
this occurs because all the planets detected to date have masses above 5\mjup ,
but planets of lower mass could easily have been detected by our search.
If we ignore the $\beta$~Pic and HR~8799 planets (should they belong
to a rare and distinct group), we find that \textcolor{black}{$<$20\% of debris disk
stars have a $\geq$3\mjup\ planet beyond 10~AU,} and $\beta < -0.8$ and/or $\alpha <
-1.5$. Likewise, $\beta < -0.8$ and/or $a_{max} < 125$~AU. 
Our Bayesian constraints are not strong enough to reveal any dependence of the planet frequency on stellar host mass.}
Studies of transition disks have suggested that about 20\% of stars
are undergoing planet formation; our non-detections at large separations show that planets with orbital separation $>40$~AU and planet masses $>3$~\mjup\ 
do not carve the central holes in these disks.

  
\end{abstract}

\footnotetext[1]{Based on observations obtained at the Gemini
  Observatory, which is operated by the Association of Universities for
 Research in Astronomy, Inc., under a cooperative agreement with the
  NSF on behalf of the Gemini partnership: the National Science
  Foundation (United States), the Science and Technology Facilities
  Council (United Kingdom), the National Research Council (Canada),
  CONICYT (Chile), the Australian Research Council (Australia),
  Minist\'{e}rio da Ci\^{e}ncia e Tecnologia (Brazil) and Ministerio de
  Ciencia, Tecnolog\'{i}a e Innovaci\'{o}n Productiva (Argentina).}

\section {INTRODUCTION}
Debris disks are second-generation dust disks found mostly around relatively young (0.01--1~Gyr) stars.
They are made of collisional debris from planetesimals left over after the primordial dust 
disks have disappeared \citep[e.g.][]{2001ARA&A..39..549Z, 2008ARA&A..46..339W}. Observational studies of debris disks have been a major part of the effort 
to understand planet formation. Since the first image of the $\beta$ Pictoris disk \citep{1984Sci...226.1421S}, high-resolution 
imaging has revealed the morphology of many debris disks. The presence of holes and azimuthal asymmetries 
in the dust distribution can give clues to the presence of planetary bodies in the disks \citep[e.g.][]{2000ApJ...537L.147O, 2002MNRAS.334..589W, 2003ApJ...588.1110K}. 
Moreover, the debris disks around younger stars are brighter \citep[e.g.][]{2005ApJ...620.1010R,2006ApJ...653..675S}, 
a result which provides a constraint on planet formation models.
These links to planet formation make debris disks excellent targets for planet searches.

Radial velocity (RV) and transit searches for extrasolar planets have resulted in over 800 discoveries, providing a wealth of information 
on the frequency of planets with small orbital radii ($\leq$ 5~AU) in 
old systems (\app\ 3 Gyr) \citep[e.g.][]{2008PASP..120..531C,2012ApJS..201...15H}. Meanwhile, direct imaging searches 
have begun to set strong constraints on the population of planets at larger separations, targeting young systems since they have brighter planets.  
\citet{2007ApJS..173..143B} conducted a study of 54 young nearby 
stars of spectral types A--M, using the 6.5-m MMT and the 8-m VLT. Selecting somewhat older stars (median age$\sim$250 Myr), 
\citet{2007ApJ...670.1367L} obtained adaptive optics (AO) imaging of a sample of 85 nearby stars, using the ALTAIR system on the Gemini North telescope.
In a comprehensive statistical analysis of these samples, \citet{2010ApJ...717..878N} estimated that 95\% of stars have 
no planets at separations larger than 65~AU, at 95\% confidence. Later, \citet{2010ApJ...716.1551L} presented results on a sample of 58 stars,
including objects with known disks and known exoplanets, finding that
less than 20\% of such stars can have $>$40~\mjup\ companions between 10 and 50~AU. 
\citet{2011ApJ...736...89J} have imaged 18 stars of spectral type earlier than A0 and concluded that less than 32\% of such systems 
have planets on wide orbits (however, see discussion in Nielsen et
al.\ 2013). \textcolor{black}{Finally, \citet{2012arXiv1206.4048V} presented a direct-imaging survey which included 42 A~and F~stars
and estimated that 5.9--18\% of A and F stars have planets at
separations between 5--320~AU (though age estimates for many of their
targets were somewhat optimistic; see Nielsen et al.\ 2013.) }
 
An interesting result of direct imaging searches so far is that many of the systems 
with exoplanet discoveries also have debris disks, i.e. $\beta$ Pictoris, HR~8799 
and Fomalhaut \citep{2009A&A...493L..21L,2008Sci...322.1348M,2008Sci...322.1345K}. 
Even though the discovery rate for large-separation exoplanets in debris disk systems is low ($<$2\%; this work), 
the discovery rate in non-debris disk systems is even lower. In addition, \citet{2012arXiv1206.2370W} find that 
the  RV-discovered systems with only Saturn-mass planets have a
higher-than-expected debris disk fraction, 4 out of 6 (67\% ) 
compared to 4 of 11 (36\%) in the full sample of stars with RV planets
(debris disk fraction in RV Jupiter-mass systems is constrained to $<$20\%).  The correlation between higher disk fraction and lower-mass 
planets suggests that the formation mechanism for Saturn-only systems
results in large, stable debris disks which can produce dust for a
long time, though based on small number statistics. 

Since late 2008, the Gemini NICI Planet-Finding Campaign has been conducting a direct-imaging search for exo-planets 
around a large sample nearby young stars \citep{2010SPIE.7736E..53L}.  
In this paper, we present the NICI Campaign results for 57 debris disk stars, the largest and most sensitive 
direct-imaging search for planets around debris disk systems to date. 
Companion papers by Biller et al.\ (2013) and Nielsen et al.\ (2013) present survey results for the nearby 
moving group stars and young A \& F~stars, respectively.

\section{OBSERVATIONS}
\textcolor{black}{
A sample of 57 debris disk stars (Table~1) was observed as part of the NICI Campaign from 2008 to 2012 (see Table~2). 
The ages, distances and spectral types of the debris disk stars are
displayed in Figure~1 and their disk properties are shown in Table~3.  A little more than half the stars (29/57) 
are FGKM stars, while the rest (28/57) are B or A stars. The age estimates
for our B and A stars come from a new Bayesian method discussed in Nielsen et al.\ 2013. The ages for the rest come 
from the compilations of debris disks in \citet{2005ApJ...620.1010R}, \citet{2006ApJ...653..675S}, \citet{2006ApJ...644..525M} 
and \citet{2007ApJ...660.1556R}. The main methods for estimating the ages can be found in \citet{2004ARA&A..42..685Z}.} 

\textcolor{black}{
The ages of our target stars range from 5 to 1130 Myr with a median of
100 Myr and an RMS of 205 Myr.
The distribution is actually bimodal with 25 stars younger than 60
Myr with a median of 12 Myrs and RMS of 12 Myr, and 32 stars older
than 60 Myr with a median of 224 Myr and RMS of 220 Myr.
The two subsets represent moving group members and older debris disks, and 
thus our sample is about evenly distributed between the two subsets.
The distances to the target stars range from 3.2 to 112~pc.  The 15
stars that lie beyond 60~pc are B, A or F stars. }

The debris disk stars were observed using our standard Campaign observing modes described in detail in Wahhaj et al.\ (2013).
To summarize, we observed each target in two modes, Angular Spectral Differential Imaging (ASDI) and Angular 
Differential Imaging (ADI), in order to optimize sensitivity to both methane-bearing and non-methane-bearing companions. 
In ASDI mode, we imaged simultaneously in the off- (1.578~\mic ; $CH_4 S$~4\%) 
and on-methane (1.652~\mic ;  $CH_4 L$~4\%) bands using NICI's dual near-IR imaging cameras. 
In ADI mode, we only observed in the $H$-band using one camera. The ASDI mode is more sensitive 
to companions at separations less than \app\ 1.5$''$, while the ADI mode is more sensitive at larger separations. 
In both observing modes, the primary star was placed behind a partially transmissive 
focal plane mask with a half-power radius of 0.32$''$ and a full extent of 0.55$''$. 

When using the ADI technique, the telescope rotator is turned off, and the sky rotates with respect to the instrument detectors. 
This is done so that the instrument and the telescope optics stay aligned with each other and fixed with respect to the detector.
Their respective speckle patterns, caused by imperfections in the optics, are thus decoupled from any astronomical objects.
When reducing images, a reference PSF made by stacking the speckle-aligned images was subtracted from the individual images, so that some 
fraction of the speckle pattern was removed. At large separations from the target star ($\gtrsim$~1.5\arcsec),
our sensitivity is limited by throughput not by residual speckle structure. 
Thus in the ADI mode, all the light is sent to one camera in order to achieve maximum sensitivity.
We usually obtained 20 1-minute images using the standard $H$-band filter, which is about four times wider than the 4\% methane filters.

To search for close-in planets, we combined NICI's angular
and spectral difference imaging modes into a single unified sequence that we call 
``ASDI''. In this observing mode, a 50-50 beam splitter in NICI
divides the incoming light between the off-  and
on-methane filters which pass the light into the two
imaging cameras, henceforth designated the ``blue'' and ``red'' channels
respectively. The two cameras are read out simultaneously for each exposure and thus the 
corresponding images have nearly identical speckle patterns, which can be subtracted from 
each other, prior to ADI processing.
In the ASDI mode, we typically obtain 45 1-minute frames.

\section{DATA REDUCTION}

The Campaign data reduction procedures are described in detail in Wahhaj et al.\ (2013).
Briefly, in our standard ASDI reduction, we reduce the images in five steps:
(1) do basic reduction of flatfield, distortion and image orientation corrections;
(2) determine the centroid of the primary star and apply image filters (e.g.\ smoothing);
(3) subtract the red channel from the blue channel (SDI);
(4) subtract the median of the entire stack of images from the individual difference images (ADI); and
(5) de-rotate each individual image to a common sky orientation and then stack the images. 
The ADI reduction is the same as the ASDI reduction, except that since we are only dealing with the 
blue channel images, we do not perform the SDI reduction in step~3.

Flatfield images have been obtained during most NICI Campaign observing runs, and thus 
most datasets have corresponding flatfields obtained within a few days of their observing date.
The images are divided by the flatfield, which we estimate to have an uncertainty of 0.1\%.  
This was estimated from the fractional difference between separate flatfields made from two halves 
of a sequence of flatfield images. 

In NICI Campaign datasets, the primary star is usually unsaturated as it is imaged through a 
focal plane mask which is highly attenuating
\citep[$\Delta CH_4 $(4\%)$S= $6.39\pp0.03~mag,\\$\Delta CH_4 $(4\%)$L= $ 6.20\pp0.05~mag, $\Delta H= $5.94\pp0.05~mag;][]{2011ApJ...729..139W}. 
Thus the locations of the primary star in these images are accurately determined and used later for image registration. The centroiding accuracy for unsaturated peaks or peaks 
in the non-linear regime is 0.2~mas (Wahhaj et al.\ 2013) or 1\% of a NICI pixel. In ASDI datasets for bright stars ($H <$ 3.5 mag) and 
in most ADI datasets, the primary is saturated. In these images, the peak of the primary is still discernible 
as a negative image. \textcolor{black}{The pixel with the smallest value in the negative
image is taken to be the centroid.}. We have estimated that the centroiding accuracy of the saturated images is 9~mas by
comparing the estimated centroids in these images to those of the unsaturated short-exposure images 
obtained right before and after the long exposures.  

\section{RESULTS}
\label{section:results}

\subsection{Contrast Curves}
For each of the reduced data sets, the companion detection sensitivities are presented as  
95\% completeness contrast curves in delta magnitudes as a function of separation from the primary star.
Wahhaj et al.\ (2013) describe how the 95\% completeness limits are defined and measured: 
95\% of all objects in the field brighter than the indicated contrast level should be detected by our observations.
We have previously shown that the 95\% contrast curve in most cases agree well (at the \app\ 10--30\% level) with the nominal 5$\sigma$ 
contrast curves which are usually presented in direct imaging analyses.  We use the 95\% curves in this paper, 
because they are statistically more meaningful and also more reliable.

These completeness curves are calculated by using the unsaturated 
images of the primary, seen through the coronagraphic mask, as simulated companions. The simulated companions are embedded into the 
raw images and recovered in the final reduced images using fixed (automatic) detection criteria, as described in Wahhaj et al.\ (2013.)
\textcolor{black}{The companions are inserted into the images with separations such that they do not interfere with each other during ADI processing.}
For this paper, we will use the term {\it contrast curve} to mean {\it 95\% completeness contrast curve}.

In a few data sets, a bright stellar companion reduced the contrast achieved in some portion of the field. For 
simplicity, we consider these low contrast regions (worse than the median contrast in an annuli by $\geq$1~mag) 
as regions without data. For each angular separation we also report the 
coverage fraction, or the annular fraction with data (Table~\ref{tab:contrasts}). Because the primary was not usually placed in the center 
of the detector, the coverage fraction also falls off gradually at separations $\geq$~7$''$.

The median contrasts achieved for the debris disk sample at 0.5, 1.0, 2.0, 4.0$''$ were 
12.4, 14.1, 15.5 and 16.2 mag with standard deviations of 0.5, 0.82, 0.87 and 1.3 mag, respectively, 
across all contrast curves. For separations less than 1.5$''$, 
the contrast curves from the ASDI reduction are usually more sensitive that those from the ADI reduction. 
For larger separations, the ADI contrast curves are more sensitive. 
If the images are saturated for one reduction, we report
the contrast curve from the other reduction(s) for the relevant separations.
When estimating sensitivities to planets, we use all available contrast curves independently, as described in \S~\ref{sec:monte}. 
We report our final contrasts in Table~\ref{tab:contrasts} and
Figure~\ref{fig:contrasts}.  \textcolor{black}{The companion mass
  limits corresponding to these contrasts are presented in
  Table~\ref{tab:mass}, along with the mass limits at the debris disk
  edges}. The brightness to mass conversions are obtained using the hot-start models in \citet{2003IAUS..211...41B}.

\subsection{Follow-up of Candidate Companions}

Many candidate companions were detected in our observations.
The targets with candidates were observed again several months to years later to check if the candidates were 
comoving with the primary or were fixed in the sky as expected for background objects. Usually, a second-epoch observation was 
obtained only when the expected relative motion between the science target and any background star was 
more than about 3 NICI PSF widths ($150$~mas). The primary's motion was calculated from its 
proper motion and parallax, usually from the revised Hipparcos Catalog \citep{2007A&A...474..653V}. 
Multiple epochs of follow-up observations were obtained when the astrometric 
uncertainties were too high to determine if candidates were background from the second epoch data. 
The uncertainties in the separation and PA for each epoch are estimated to be 0.009$''$ and $0.2$\dg\ respectively when the primary is unsaturated,
and  0.018$''$ and $0.5$\dg\ when the primary is saturated (Wahhaj et al.\ 2013).

We detected a total of 78 planet candidates around 23 stars. Follow-up observations of 19 targets with 66 of the most promising 
candidates (projected separation $<$ 500~AU) show that all of them are background objects.
The astrometry and photometry for these objects are presented in
Table~\ref{tab:cands} and \ref{tab:cands2}.  
Figures 3--6 show the motion of 
all the candidates with respect to the primary. 
The candidates with large projected separation ($>$ 500 AU) were treated as lower priority follow-up targets and thus second-epoch observation of some of these 
have not been obtained.  The astrometry for the 12 candidates around 4 stars which were not followed 
up are presented in Table~\ref{tab:cands_nofollow}, along with their
projected physical separations in AU.
We discuss how our statistical formalism correctly treats these
single-epoch detections in \S 4.8.


\subsection{Contrast limit at the location of  Fomalhaut b}
\citet{2008Sci...322.1345K} presented the detection of a planet at 0.6
and 0.8~\mic\ around Fomalhaut \textcolor{black}{using the 
Hubble Space Telescope \citep[see also][]{2012ApJ...760L..32C,2013ApJ...769...42G}}.  
The planet was detected at a separation of 12.61$''$ and a PA of 316.86\dg\ in 2004 and a separation of 12.72$''$ and 317.49\dg\ in 2006. 
The planet was not detected in their $H$-band or $CH_4S$-band images obtained in 2005 with 3$\sigma$ detection limits of $H$=22.9~mag and $CH_4S$=20.6~mag. In this section, we present the detection limits achieved at the location of the planet in a NICI observation 
obtained on UT 2008 November 17.

In the 2008 NICI observations of Fomalhaut, we placed Fomalhaut in one corner of the NICI detector (only 3.4$''$ from either edge), 
so that the location of the planet would lie near the center of the 18.43$''$ wide detector. We imaged Fomalhaut through 
the 0.9$''$ coronagraphic mask in the $CH_4L$ 4\% filter with the blue camera only. We obtained 79 images, each with 66.9 seconds of exposure (1.47 hours total). We also obtained an unsaturated short exposure image 
with an exposure time of 0.38 seconds ($\times$ 10 coadds) in the $CH_4S$~1\% filter. 
The long exposure images are saturated out to a separation of \app\ 1.2$''$.
The total rotation obtained in the ADI data set was 6.33\dg , from which we estimate that the planet would move through \app\ 25 NICI PSF FWHMs
 relative to the detector. Thus, we expect almost no self-subtraction of the planet during the ADI image processing. 
Moreover, the planet would only change position by 
0.33 PSF cores between adjacent exposures, and thus smearing during an individual exposure on the detector was also negligible. 
The airmasses for the observation ranged from 1.04 to 1.33.

The location of the primary was measured from the saturated images, where the peak of the star can be recognized as a negative image as described in 
Wahhaj et al.\ (2013). The precision of these measurements has been estimated to be \app\ 9~mas. The images were reduced using the 
standard ADI Campaign pipeline (Wahhaj et al.\ 2013). No sources were detected in the reduced image near the locations of the \citet{2008Sci...322.1345K} detections or anywhere else in the image. 
 
To determine the contrast limit near the location of the \citet{2008Sci...322.1345K} planet, we needed to measure 
the opacity of the 0.9$''$ focal plane mask in the $CH_4S$~1\% filter. This was done by scanning a calibration light 
source internal to the AO system across the mask and measuring the change in its relative brightness.
As a result, the opacity at the center of the mask was measured to be 6.2\pp0.1~mag. 

To estimate the contrast at the location of the planet, we simulated planets by scaling the star spot in the short exposure image 
to different contrasts relative to the primary and inserted them into the raw images. The simulated planets 
were inserted at 25 locations in a sector of the image,
of radial extent 2$''$ and total PA extent 10\dg\  (\app\ 2.2$''$ arc length). The sector was centered 
near the location of the planet at a separation of 12.65$''$
and a PA of 317\dg . We determine the contrast at which all 25 planets are recovered according to the procedure described in Wahhaj et al.\ (2013). 
This procedure was repeated four times, changing the center of the insertion region by 0.1$''$ each time in the positive and negative separation and PA directions.
In this way, we determine the contrast at which $>$99\% (100 out of 100) planets are recovered near the location of Fomalhaut~b.
This contrast is 20.2~mag \textcolor{black}{($\Delta CH_4S$~1\%)}. In Figure~\ref{fig:fomal}, we show the location of \citet{2008Sci...322.1345K} detections 
and the simulated planets at the contrast limit of the reduced image. 

The Strehl ratio of the NICI images were degraded with increasing
angular separation from the primary due to anisoplanatism, 
but we do not have an independent estimate of the degradation for NICI observations.
In the absence of anisoplanatism data for Gemini-South, we simply follow \citet{2007ApJ...670.1367L} and assume that 
the reduction in Strehl for the ALTAIR AO system is $e^{ -(separation/12.5)^2}$, with the separation in arcseconds, 
or 1.1~mag at the location of the planet. We apply this correction 
to the contrast curves and assume an uncertainty of 0.3~mag.  For Fomalhaut, $V=  $ 1.16 (SIMBAD) 
and $V-H = $ 0.28 \citep[A4V star][]{1995ApJS..101..117K}. 
Assuming $H-CH_4S = 0$, we get $CH_4S =$ 0.88~mag for the primary star.
Adding the contrast limit and Fomalhaut's brightness, we estimate with 99\% confidence that 
there are no planets at the location of Fomalhaut~b with $CH_4S <$ 20.0\pp0.3~mag. 
We do not improve much on the $CH_4S >$ 20.6  contrast limit presented in \citet{2008Sci...322.1345K},
although we note that the contrast quoted therein was a nominal 3$\sigma$ limit, obtained 
without testing with simulated planets. In Figure~\ref{fig:fomal}, 
we plot the nominal 5$\sigma$ contrast curve for our 2008 Fomalhaut 
dataset scaled to match the contrast measurement at the location of the planet.  Inside of 1.8$''$ 
separation, we plot the contrast curve from a UT 2011 October 12 ASDI data set.

Using the \citet{2003A&A...402..701B} models, Fomalhaut's Hipparcos parallax of 129.81~mas and 
assuming an age of 450\pp40~Myr for Fomalhaut \citep{2012ApJ...754L..20M},
we estimate an upper limit of 12--13\mjup\ for the mass of the planet from our $CH_4S$ non-detection. 
This is well above the dynamical mass estimate of $< 3$\mjup\ presented in the 2008 discovery paper, 
and the upper limit of $< 2$\mjup\ from the 4.5~\mic\ and $J$-band non-detections \citep{2012ApJ...747..116J, 2012ApJ...760L..32C}. 
Thus, it is not surprising that we do not detect the planet.
 
\subsection{Especially Low Mass-limits Achieved within 15 AU}

\textcolor{black}{
Especially high-sensitivities to planets within 15~AU projected
separation were achieved for GJ 803 (4\mjup\  at 5~AU; 2.5~\mjup\ at 10~AU), 
and $\beta$ Pic (6.5~\mjup\ at 10~AU; $\beta$~Pic~b was detected
interior to this separation). High sensitivities were 
also achieved for TWA~7 (4~\mjup\ at 14~AU), HIP 25486 (5~\mjup\ at 13~AU), HD~17255A (6~\mjup\ at 15~AU),
Fomalhaut (10~\mjup\ at 15~AU) and $\epsilon$~Eridani (11~\mjup\ at 10~AU). Although these are 
higher sensitivities than achieved in previous surveys, we are not yet
sensitive to true Jupiter-analogs.} \textcolor{black}{However, such
sensitivities have already been obtained in the one special case of AP~Col, a very nearby (8.4~pc) young (12--50~Myr) M-dwarf. \citep{2012ApJ...754..127Q}.}

\subsection{Bayesian Inference on Planet Populations}
  
In this section, we constrain the population of extrasolar planets around nearby debris disk stars using Bayesian 
statistics. Bayes' equation tells us that
\begin{equation}
 P( model \bb data, I) = P( data \bb model, I)\ \frac {P( model \bb I)} {P( data \bb I)},
\end{equation}
where $P( data \bb model, I)$ is the probability of the $data$ given the $model$ 
and $I$ is any relevant prior information we have about our targets and our observations.
The $model$ is our hypothesis about the planet population expressed in terms of some interesting parameters, 
such as the average number of planets around AFGKM type stars.
The $data$ are relevant information from the observations which most strongly constraint the 
interesting parameters.

In our case, there are two interesting model parameters to constrain using our observations: 
(1) the frequency of planets (average number of planets around a sample star) 
and (2) the fraction of stars with at least one planet. 
For both parameters, we are considering stars with detected debris disks and planets in a certain mass and 
semi-major axis (SMA) range, as these are the fundamental constraints from our dataset.

\subsubsection{The Frequency of Giant Planets}
\label{sec:bayes_F}
We model the probability distribution of the number of planets in a single system as a Poisson distribution:
\begin{equation}
P(n \bb F) = \frac {1} {n!} {F^ne^{-F}} ,
\label{eq:pois_mult}
\end{equation}
where $n$ is the number of planets in the system and $F$ is the 
expectation value of the distribution i.e., the frequency of planets which is one of the 
model parameters of interest.

For the $j$th star in our sample of $N$ stars, we use $f_j$ to denote the fraction of planets that are detectable 
in a certain mass and SMA range, given that such planets are distributed according to some population model. 
These $f_j$ are calculated using the contrast limits for the $j$th star and 
Monte Carlo simulations as in \citet{2010ApJ...717..878N}. Thus $nf_j$ is the expection value
of the number of planets detected for a system with $n$ planets. Let $n_j$ be the number of planets actually detected
for that star. Then the likelihood for $n_j$ detections given that $nf_j$ is the expected value is given by another Poisson 
distribution:
\begin{equation}
     P(n_j \bb nf_j) = \frac {1} {n_j!} {(nf_j)^{n_j}e^{-nf_j}} .
\label{eq:pois_det}
\end{equation}
We would like to determine the model parameters, $F$ and $n$.
The contrast curve for the $j$th star and the data are $\{c_j\}$ and $\{n_j\}$, respectively.
However the contrast curves are only relevant to calculating the $f_j$'s and thus 
$P(n_j \bb n, c_j)$ is the same as $P(n_j \bb nf_j)$, the probability of $n_j$ detections given that $n$ planets exist and a fraction $f_j$ 
of a model population are above the contrast limits.

Thus, using Bayes's equation we get
\begin{equation}
  P(F , n \bb \{n_j\} , \{c_j\}, I) = \prod_{j=1}^{N} P(n_j \bb  n f_j)\ P( n \bb F).
\label{eq:double_pois}
\end{equation}
This expression is exact under certain assumptions: (1) the observation result from one star 
does not affect the information we have another star and (2)  the priors $P(n_j, f_j \bb I),  P(n_j \bb f_j , I)$, 
$P(f_j \bb I), P(f_j \bb F, n, I)$ are flat (=1) because we have no information about them. 
The Appendix provides a complete derivation of Eq.~\ref{eq:double_pois}.

We obtain the probability distribution for the frequency of planets, $F$, by summing over the parameter, $n$.
\begin{equation}
P(F  \bb \{n_j\} , \{c_j\}, I) = \prod_{j=1}^{N}\  \Bigg( \sum_{n} P(n_j \bb n f_j)\ P( n \bb F) \Bigg).
\label{eq:bayes_expan2}
\end{equation}

We could sum $n$ from zero to infinity to be
exact. \textcolor{black}{However, the calculations will be sufficiently precise when the upper bound is 
large compared to the maximum number of  actual detections in a single system.} So, we sum $n$ from zero to 20.

For any star with multiple planets, we have to use the above general expression. However 
for our stars, $n_j=0$ because we did not detect any planets, except
in the case of $\beta$ Pic~b where we count the known planet
(astrometry and photometry taken from Bonnefoy et al.\ 2013). 
When we have zero detections ($n_j=0$), the general expression given by Eq.~\ref{eq:double_pois} simplifies to
\begin{equation}
P(F , n \bb n_j =0, f_j, I) =  \sum_{n}  \frac {1} {n!} F^ne^{-(F+n f_j)}.
\end{equation}

\subsubsection{The Fraction of Stars with Planets}
\label{sec:fsp}
Another number of interest is the fraction of stars with planets, which can be derived from $F$, if we assume 
a Poission distribution for the multiplicity of planets as in Eq.~\ref{eq:pois_mult}.
The fraction of stars with planets, $F_{sp} = 1-F_{s0}$, where $F_{s0}$ is the number of stars with zero planets.
But $F_{s0} = P(F \bb n=0) =  e^{-F} \Rightarrow F_{sp} = 1-e^{-F}$. 
Thus, we can easily derive $P(F_{sp} \bb data, I)$ from $P(F \bb data, I)$, or the probability 
distribution for the number of stars with planets from that of the frequency of planets, $F$. 
In previous work, the fraction of stars with planets has been equated to $F$ \citep{2007ApJ...670.1367L,2012arXiv1206.4048V} (See Appendix for 
further discussion of previous work).
However, this is only true in the special case that all stars have at most one planet or when $F << 1$ (substituting in $F_{sp} = 1-e^{-F}$).

We will later employ a simpler statistical formulation in which we constrain only the average number of planets 
per star for the given sample. However, such an approach would not distinguish between different multiplicities of systems,
and we would not be able to estimate the fraction of stars with zero planets. In other words, this approach provides weaker 
constraints on the model population as we are throwing away multiplicity information. The simple formulation would be 
\begin{equation}
P(F|data) = P(n_d | n_m) = \frac{1}{n_d!}e^{-n_m}{n_m}^{n_d},
\end{equation}
where $n_m = \sum_{j=1}^N Ff_j = F\sum_{j=1}^Nf_j$  is the total number of planet detections predicted by the model. 
In other words, the number of detections predicted by the model 
is the detection probability given that exactly one planet exists, times the average planet multiplicity, summed over all stars.
Also, $n_d$ is the total number of planets detected in the sample.
 
This is in fact the likelihood used in \citet{2008ApJ...674..466N} and \citet{2010ApJ...717..878N} for the case of zero 
planet detections.
We show in \S~\ref{sec:stat_lims} that if the distribution of planet multiplicities is in fact Poissonian, then this likelihood can be 
used to obtain the correct result even if there are actual planet detections in a survey.

 
\subsection{Monte Carlo Simulations to Calculate Detection Probabilities}
\label{sec:monte}

In this section, we calculate the $f_j$ values (the planet detection probabilities given exactly one planet 
exists around each star) in our Bayesian formulation employing a Monte-Carlo technique \citep[see][]{2008ApJ...674..466N}.
For each star $j$, $f_j$ is the probability of detecting a planet within 
some chosen mass and SMA range, $m = [m_0,m_1]$ and $a=[a_0,a_1]$,
given that exactly one planet exists in that range and that no planets exist at larger $a$. For each star, 
we simulate $10^4$ planets with a distribution  given by
$$\frac{d^2 N} {dM da} = n_{11} m^{\alpha} a^{\beta}.$$
\textcolor{black}{
The $10^4$ planets are only assigned masses between 
0.5 and 13~\mjup\ and SMA between 0.5 and 1000~AU. 
Initially, we set $\alpha=-1.31$ and $\beta=-0.61$, using the
estimates from RV surveys \citep{2008PASP..120..531C}.} 

The quantity $n_{11}$ is the number of planets per star per unit mass range and SMA range.  
From the RV results, we calculate $n_{11}$ to be 0.044.
Thus once we have constraints on $\alpha$ and $\beta$, we will have estimates for the number of planets in any $m$ and $a$ range,
and also be able to compare our results with those from the RV surveys. 
If they are not consistent we can hypothesize that the planet formation behavior
changes beyond some physical regime (e.g. for large SMA).
We note here that the $f_j$ computations do not depend on $n_{11}$. 
  
Next, we transform the $m$ and $a$ values of the simulated planets to 
$\Delta H$ and $\rho$,  the flux ratios and 
their angular separations relative to their primary stars, respectively. The calculation from $a$ to $\rho$ is performed using 
the distance to the primary and samplings from prior distributions in orbital parameters and random viewing angles as described in \citet{2008ApJ...674..466N}. 
The calculation from $m$ to $\Delta H$ is performed using
\citet{2003IAUS..211...41B} models also as described in \citet{2008ApJ...674..466N}. 
The models use a ``hot-start'' initial condition for the evolution of planet fluxes with age.  
In Figure~\ref{fig:mass_lim}, we show the mass sensitivity limits achieved at 0.5$''$ and 3$''$ using the \citet{2003IAUS..211...41B} models.
It is evident that the contrast achieved at the two separations probe significantly different planet masses. 
\citet{2010ApJ...717..878N} have shown the consequences 
of using the core-accretion models of
\citet{2008ApJ...683.1104F}. These ``cold start'' models predict much lower near-infrared fluxes for planets at ages
below 100~Myrs, but agree with the ``hot start'' models at older ages. However, given the discoveries of the planet around $\beta$
Pictoris and the four planets around HR~8799,  it is evident that at least some planets 
have luminosities comparable to those produced by ``hot start'' models \citep{2010ApJ...710.1408F}.

Now we calculate what fraction of the planets have 
$\Delta H$ and $\rho$ above the NICI imaging contrast limits. For the $j$th star, this fraction is our desired $f_j$ value.
In Figure~\ref{fig:completeness}, we illustrate the computation of $f_j$,
the detection probability of a planet given that only one planet
exists for the case of 49 Cet. The detection probabilities for 
our survey stars ranged from ~1\% to 30\%. 

\subsection{Limits on the Average Number of Planets around Debris Disk Stars}
\label{sec:stat_lims}
When we have actual detections, the number of detections in each $\{\Delta H(\rho), \rho\}$ bin should be compared to 
the number of planets predicted by our number distribution model in those bins. Otherwise our model parameters, $\alpha$, $\beta$, etc. 
will not be optimally constrained as we will be ignoring information about the brightness and separation of the detected planets.
In the case of zero detected planets, however, the likelihood expression for the model parameters is exactly the same, whether we count
the predicted detections in each bin separately or count the detections in all the bins together. 

For our survey we only have one detection, 
the known planet around $\beta$ Pictoris. 
Thus, we start with the simplest statistical formulation 
for the average planet multiplicity, $F$. As we discussed earlier, in this Poissonian 
formulation we only compare the total number of planet detections expected to the total number actually detected.
Then from $F$, we can calculate $F_{sp}$ the fraction of stars with planets, as described in \S\ref{sec:fsp}. 

We compute the probability distributions for $F$ and $F_{sp}$ in the two
ways, \textcolor{black}{one using the full likelihood and the other 
using the simple likelihood. For both methods, we have set
$\alpha=-1.31$  and $\beta=-0.61$ based on \citet{2008PASP..120..531C}}. 
First, we consider the entire planet mass and SMA range from $0.5$ to $13$\mjup\ and $0.5$ to $1000$~AU,
respectively . From Figure~\ref{fig:limits_fp}, we can see that the 2$\sigma$ upper limit on the average number of planets 
around our debris disk sample is 0.42, for the chosen range of mass and SMA. Assuming a Poisson 
multiplicity distribution,  the upper limit on $F_{sp}$ is then 35\%. The lower limit of 2\% is 
entirely due to the $\beta$ Pictoris planet detection. 

As a check on the validity of both methods, Figure~\ref{fig:limits_fp} also compares the results obtained from the simple likelihood 
to the full likelihood derived in \S\ref{sec:bayes_F}. Interestingly,
the results from the two likelihoods are quite similar.
Although the simple likelihood cannot constrain the multiplicity distribution, if we assume a Poisson 
distribution, we obtain a good result. Of course, if the real planet multiplicity distribution is not Poissonian,
the simpler likelihood will incorrectly yield better constraints than the full likelihood.

As another check, we conducted a simulation with 
the number of planets and detections around each star following a Poisson distribution. 
To make matters simple, brightnesses and separations were not simulated, just the number of planets were. 
Thus, to each star in our survey, we assigned a number of planets drawn from a Poisson distribution with mean of 3.
The number of detections for the star is then just $f_j$ times the number of planets around the star.
Figure~\ref{fig:fake_limits_fp} shows that the results from the two
likelihoods are quite similar.
Again, although the simple likelihood cannot constrain the multiplicity distribution, if we assume a Poission 
distribution, we obtain a good result. 

Figure~\ref{fig:limits_mass_sep} shows the same upper limits computed as a function of planet mass 
and orbital semi-major axis. In this case, we ignored the power-law model for the distribution of 
the planets and simply computed the upper limit to the average number 
of planets in a given $\{m,a\}$ bin. Of course, the likelihood for $F$ in a particular bin cannot
be calculated ignoring the average multiplicity outside that bin, unless we stipulate that there 
are no planets outside. Since we are interested in the absolute upper limit to the number of planets 
in a bin, we do make that stipulation. 

Based on Figure~\ref{fig:limits_mass_sep}, at the 95\% confidence level we state that 
$<$10\% of debris disk stars have a $\geq$5\mjup\ planet at 60~AU, and
$<$30\% of debris disk stars have a $\geq$3.5\mjup\  planet at 
20~AU. Also, the average number of planets with masses above 3.5\mjup\ 
at 10~AU is less than 1, and at 25~AU it is less than 0.25. \textcolor{black}{In all
cases, only smaller fractions are permitted at larger semi-major axes
(up to 500~AU).}

\subsection{Limits on the Mass and Semi-Major Axis Distribution of Planets}


We now consider how our observations 
constrain $\alpha$, $\beta$ and the average planet-multiplicity, if we ignore the 
constraints from the RV studies 
both for these parameters and the average planet-multiplicity.
Moreover, we add an additional parameter,
the SMA cutoff, $a_{max}$, beyond which no planets are allowed.

The joint probability distribution for the four model parameters is given by 
$$ P(\alpha, \beta, a_{max}, F | data)  = P( data | \alpha, \beta, a_{max}, F) * P(\alpha, \beta, a_{max}, F) / P(data)$$
and again the ratio of priors, the last factor in above expression, is set to unity. 
The data are as usual the contrast curves and the detections. 
In calculating the likelihood, $P(\alpha, \beta, a_{max}, F | data)$,
the actual and model-predicted detections in each $\Delta H$ and
$\rho$ bin are compared separately. This requires that for each combination of $\alpha, \beta, a_{max}, F$, we calculate the $f_j$ for each 
$\Delta H$ and $\rho$ bin separately. 

  \textcolor{black}{ Here, single-epoch detections which could neither be confirmed nor rejected as
  planets are also taken into account. For example, suppose for
  a certain star, a \{$\Delta H$, $\rho$\} bin has $n_c$ candidate companions (from
  single-epoch detections) and $n_d$ confirmed planet
  detections. Suppose also that the planet population model predicts $n_m$
  planets in that bin. Then the probability of the available
  information being true given
  the model is  $\sum_{n_p=n_d}^{n_d+n_c} P(n_p | n_m)$, where $P$
  is the Poisson distribution. In other words, we sum the
  probabilities for the range of possible planets in that bin.} 

 
\textcolor{black}{
  To illustrate the value of including single epoch detections, we
  consider an observation that tells us that a range of planets are possible for a 
  certain star. This observation therefore contributes equal probabilities to all models which predict planets within that range. 
  Thus, the observation is a poor discriminant for those models. The probability
  contribution from each star is calculated separately and then
  multiplied (see Equation 5). Thus, when a star with many candidate
  planets is added to a survey, it does not weaken the final constraints on the planet population.
  It simply contributes less to the constraints than the other stars. The available information is used optimally and correctly, 
  and does not under- or over-constrain the final results.}

  \textcolor{black}{ We can easily appreciate the utility of this method if we consider the example of 
  a star with only very bright candidates detected at large physical separations. Suppose these
  candidates are too bright to be planets or are detected beyond
  500~AU from the primary. Then the planet population constraints on
  this system (inside 500 AU, if the detections were outside)
  are just as strong as they would be if there were no detections. 
  Another advantage of this method is that population constraints can be
  calculated in the middle of the survey, even before any planet
  candidates are re-observed to check for common proper-motion. }

The total likelihood is the product of the detection likelihoods calculated for the individual bins. 
We chose bin widths in $\Delta H$ of 2 mags, with bins from 5 to 23 mags.
We chose $\rho$ bins with inner edges from 0.36$''$ to 9.8$''$ (outer edges from 0.54$''$ to 14.8$''$).
The innermost bin is given a width of 0.18$''$ and each subsequent bin is larger by factor of 1.5.
Thus, we are using both astrometric and photometric information from
any detected planets and the detection  multiplicity information for
each system to calculate our constraints. 

To demonstrate the validity of our Bayesian approach, we simulate the planet detections 
from a fictitious planet population with $\alpha=-$1.1, $\beta=-$0.6, $a_{max}=$193~AU 
(corresponding to one $a_{max}$ bin), and average planet multiplicity
$F=$~3, \textcolor{black}{around our survey stars and assume the survey contrast curves. 
This fictitious population creates dozens of planet detections for our survey.} 
We compute the likelihoods for values of $\alpha$ and $\beta$ between $-2.5$ and $+2.5$, dividing the 
range into 50 equal-sized bins. For $a_{max}$, we use 50 values between 20 and 1000~AU, spaced 
logarithmically. For $F$, we use 100 values between 0.005 and 20, spaced logarithmically.
The resulting probability distributions are shown in Figure~\ref{fig:alpha_beta}.
The resulting 2$\sigma$ limits for the parameters are consistent with the simulation input values,
and show that our method is effective at narrowing the range of possible planet populations.
The 1$\sigma$ limits are also consistent with the simulation input values, yielding constraints:  
$\alpha=$[-1.35,-0.85], $\beta=$[-0.85,-0.2], $a_{max}=$[153,193]~AU and $F=$[1.6,3.2]. However, the limits
also show that even if we had dozens of detections, we would only be able to set moderately 
stringent constraints on the debris disk planet population, because of the sensitivity limits 
and the size of our sample of target stars. 

\textcolor{black}{We now have the tools to include in our analysis the  
HR~8799 bcd planet detections and their astrometry and photometry ($\rho=$ 1.72$''$, 0.96$''$, 0.61$''$
and $\Delta H=$12.6, 11.6, 11.6 mag respectively) from the \citet{2012arXiv1206.4048V}  
survey of A and F stars. We thus expand our sample to include the contrast curves from 
the six debris disk stars that were in the Vigan survey but not in ours: HR 8799, HIP 22192, HIP 10670, HIP 26309, HIP 69732, HIP 41152
(only HR~8799 had planet detections). We also include the \citet{2013arXiv1302.1160B} detection of $\beta$ Pictoris~b
($\rho \sim$0.5$''$, $\Delta H \sim$10 mag).}

\textcolor{black}{Figure~\ref{fig:alpha_beta2} shows the Bayesian constraints 
calculated for our augmented NICI debris disk survey.  Since the
2$\sigma$ limits show a correlation between $\alpha$ and $\beta$, and 
also $\beta$ and $a_{max}$, we quote the constraints as
follows. Either $\beta < -0.8$ or both $\alpha > 1.7$ and $a_{max} < 200$~AU.  
The planet frequency is forced to rise sharply with mass ($\alpha > 1.7$), 
because all our detected planets have masses above 5\mjup , even though
lower masses could easily have been detected at these
separations. Either the steep $\beta$ ($< -0.8$), or the small $a_{max}$
($<200$~AU) prevents planets from appearing at larger separations.
Additionally, we see that many planets (but $< 12$) are allowed around
most stars, since most of them are placed within 10~AU where the
survey sensitivity is poor (see Figure~\ref{fig:limits_mass_sep}.) 
However, as we see next, many fewer planets are allowed at larger SMA.} 

\textcolor{black}{
The upper limits on the number of planets for different planet mass and SMA ranges can now be
calculated from the 4-dimentional probability density function of
$\alpha$, $\beta$, $a_{max}$ and $F_p$. For
each mass and SMA bin (3-6~\mjup\ and 25--125~AU, for example),
we calculate the number of planets predicted by each model in the 4D
function. Thus in each bin, we have a range of predicted planet counts and
the associated probability for each total count. From this new
probability density function, we calculate 95\% upper limits on the
number of planets in each bin. The number of stars with at least one
planet is calculated from the planet multiplicity using their Poisson
relation as done in \S~\ref{sec:fsp}. The results are shown in
Tables~\ref{tab:planhostfrac} and \ref{tab:avgplanmult}. Notably, 
$<$21\% of stars are allowed to have a $\geq$3\mjup\ planet
outside of 40~AU, while $<$13\% of stars are allowed to 
have a $\geq$5\mjup\ planet outside of 80~AU at 95\% confidence level.}
\textcolor{black}{
Unlike Figure~\ref{fig:limits_mass_sep}, which shows the model-independent
upper limits to planet frequencies, Tables~\ref{tab:planhostfrac} and \ref{tab:avgplanmult} are consistent 
with the planet mass and SMA power-law models. 
Because of the HR~8799 planets, the mass distribution is forced to
rise at higher masses ($\alpha > 1.7$), and at small separations the upper
limits are nearly constant across the mass bins. 
There is of course a caveat to these results: they simply represent the best
models out of the ones we compared to the data, and it is quite possible that 
power-law distributions cannot describe the true planet population.
We should also remember that (1) our models do not produce companion
masses above 13~\mjup , and (2) only models with $\alpha$ and $\beta$ between -2.5 to
2.5, and $a_{max}$ between 20 and 1000~AU were compared to the data. 
} 

\textcolor{black}{
The $\beta$ Pic and HR~8799 planets could in principle belong to a small distinct sub-population
of massive planets separate from the true tail of the planet distribution
beyond 10~AU. Thus, we also calculate population constraints without
including these planets (Figure~\ref{fig:alpha_beta2_zd}). In this
case, the 2$\sigma$ constraints imply that either $\beta < -0.8$ or both $\alpha < -1.5$ and $a_{max} < 125$~AU. 
We also calculate the upper limits to the planet frequencies as a
function of mass and SMA (Tables~\ref{tab:planhostfrac_zd}  and \ref{tab:avgplanmult_zd}). Notably, 
$<$20\% of stars are allowed to have a $\geq$3\mjup\ planet
outside of 10~AU, while $<$13\% of stars are allowed to 
have a $\geq$5\mjup\ planet outside of 20~AU.}

\textcolor{black}{To investigate the dependence of planet multiplicity
  on stellar mass, we change the planet-population model so that the
  multiplicity is given by $F_P = (M_*)^{\gamma}$, where $M_*$ is the stellar mass. 
  We repeat the Bayesian analysis for the augmented NICI survey (with
  additions from the Vigan et al.\ survey).
  We find that the 1$\sigma$ limits on $\gamma$ are [0.3, 3.2], 
  which implies that an A5V star is predicted to 
  have 1.2 to 9.2 times more planets than a G2V star 
  (Figure~\ref{fig:stmass}). But given the 2$\sigma$ limits on
  $\gamma$  ([-0.5, 4.7]), it is still possible that the
  planet-frequency does not rise with stellar mass.}

\section{DISCUSSION}

Thus far, observations have only weakly linked debris disks and planet formation. 
Direct imaging surveys seem to suggest a relatively higher yield of
giant planets around A~stars with debris disks (i.e.\ $\beta$ Pic~b
and HR~8799~bcde). 
\textcolor{black}{These detections are consistent with the core-accretion process producing more planets 
around higher mass stars, which have more massive disks, and also with the extrapolations
from the RV planet population \citep[but see discussion in Nielsen et al.\ 2013]{2011ApJ...733..126C}.
Gravitational instability models also produce more planets around
higher mass stars and moreover are able to produce the HR~8799 planets \citep{2011ApJ...731...74B}
\citet{2009ApJ...702L.163N} found that disk instability models typically require disk masses of 0.03 to 1.3\msun\  
to produce substellar companions of mass 2--21\mjup\ at separations
beyond 60~AU, but \citet{2011ApJ...731...74B} has argued that these
calculations underestimate the efficiency of the process, because of
over-simplified cooling time assumptions.}

However, giant planets on very wide ($\geq$ 100~AU) orbits are generally rare, which also implies that planet formation by core accretion 
probably dominates over formation by disk instability \citep{2010ApJ...717..878N,2011ApJ...736...89J}.
From Figure~\ref{fig:limits_mass_sep} we see that such companions beyond 60~AU 
are absent for 90\% of debris disk stars, and thus they probably never experienced disk-instability.  

Until now, only a weak correlation has been found between debris disks and RV planets \citep{2007ApJ...658.1312M, 2009ApJ...705.1226B, 2012ApJ...752...53L, 2012arXiv1206.2370W}.
No correlation has been found yet between debris disks and stellar metallicity \citep{2006ApJ...652.1674B}, while there is a strong 
correlation between giant RV planets and metallicity \citep{2005ApJ...622.1102F}. However, this correlation weakens significantly for planets 
smaller than Neptune \citep{2012Natur.486..375B}. Thus it seems that high metallicity is not a requirement for debris disks or 
for small planets, although we do not know at this time whether the occurrence of the two are correlated. Not finding correlations 
with debris disks or stellar metallicities increases the probability
that small planets are abundant, since debris disks and
very high-metallicity stars are rare.

To properly interpret the constraints on the planet population around our target stars, we have to consider the selection effects
that went into creating our sample.  
The NICI Campaign stars were mostly selected by calculating the planet detection probabilities ($f_j$) for stars which were young or nearby or massive 
and choosing the highest values \citep{2010SPIE.7736E..53L}. This sample was then supplemented with interesting stars, 
such as stars with debris disks.

The sources for the debris disk targets were largely compilations of
\citet{2006ApJ...644..525M} and \citet{2007ApJ...660.1556R}, which were based on the 
IRAS all-sky survey and supplemented by the ISO survey. These compilations included stars of spectral types BAFGKM and systems with 
fractional disk luminosities ($L_d/L_*$) as low as $10^{-5}$. Also included in the Campaign were the most promising stars 
(based on planet detection probability) of those found to have excess in the A~star {\it Spitzer} surveys of \citet{2005ApJ...620.1010R} 
and \citet{2006ApJ...653..675S}  at 24~\mic\ and 70~\mic , respectively. These surveys
were sensitive to $L_d/L_*$ as low as $10^{-6}$, limited by calibration uncertainties 
at mid- to far-infrared wavelengths, which limit the excess disk emission that can be detected relative to the 
photospheric emission. Because debris disks are easier to detect around bright stars, our sample is biased towards A stars. 
{\it Spitzer} surveys have also been conducted around FGK stars \citep{2008ApJ...673L.181M}, but the 
most IR-luminous targets from these surveys were already included in the Campaign by drawing from the IRAS and ISO surveys mentioned 
above. Thus, our debris disk stars are the youngest and nearest of the known debris disks, which are mostly complete to $L_d/L_* =10^{-5}$. 

It is thought that almost no disk with $L_d/L_* < 10^{-2}$ is primordial, as the small dust grains responsible for the excess
luminosity are expected to be dispersed on time scales much shorter than the ages of the debris disk stars (Backman and Paresce 1993).  
A few debris disks are known to have small amounts of gas, e.g.\ $\beta$ Pic and 49~Cet \citep{2004A&A...413..681B,2005MNRAS.359..663D}.
However, it is believed that the gas probably has a non-primordial origin like planetesimal collisions or sublimation \citep{2007ApJ...660.1541C,2007A&A...466..201B,2007ApJ...666..466C}.
Thus, all the debris disks in our sample are very likely composed of second-generation dust created
by collisions between larger rocky bodies \citep{1993prpl.conf.1253B}.

There are three main explanations for how the detectable dust in debris disks are produced: (1) steady-state
collisions between $>$1-100~km size planetesimals \citep{2002MNRAS.334..589W, 2007MNRAS.380.1642Q} which gradually decrease over hundreds 
of Myrs \citep{2007ApJ...663..365W, 2003ApJ...598..626D}; (2)  chance, rare collisions between
\app\ 1500~km protoplanets \citep{2002MNRAS.334..589W}, which produce dust that 
is detectable for a few million years, and (3)  the delayed stirring of a planetesimal belt when 
a large object ($\geq$ 2000~km) is formed \citep{2004AJ....127..513K}. 
Other dust production mechanisms, such as sublimation of comets \citep{2005ApJ...626.1061B} or planet migration \citep{2005Natur.435..466G} 
also require already existing massive bodies.  
Thus, stars with debris disks are different from other young stars 
in the NICI Campaign, in that they very likely possess $>$1-100~km sized planetesimals which are the sources of their dust, 
and potentially protoplanets and planets which stir the smaller bodies.

The fraction of stars with inner dust disks decrease from $>$80\% to $<$5\% from 0.3 to 15~Myrs as seen in surveys of 
near- to mid-infrared studies of open clusters \citep{2007ApJ...662.1067H}. 
However, we know that 16\% of sun-like stars older than 1~Gyrs possess
debris disks \citep[see discussion in][]{2010ApJ...710L..26K, 2008ApJ...674.1086T}.
An even larger fraction possess close-in super earth planets \citep{2010Sci...330..653H, 2011ApJ...738...81W, 2011arXiv1109.2497M}.
For the rest of the stars, it is still possible that the dust has formed into pebbles, planetesimals or protoplanets, 
as yet not detectable, somewhere in the system. 

Observations of stars with transition disks, i.e.\ primordial disks
that have developed inner holes, provide additional statistics on 
the fraction of stars with planets.  All disks are thought to undergo an evolution from an accreting, 
massive disk phase to gas-poor, low fractional-luminosity debris disks.
The evolutionary change may not always be recognizable, but transition disks are thought to be objects from 
this period. \citet{2012ApJ...750..157C} found in a study of  
34 transition disks that roughly 18\% are in a planet-forming phase, 18\% are in a grain-growth phase (likely 
an earlier phase), and 64\% are in the debris disk or photo-evaporation phase (likely a later 
phase). The accreting disks that have rising spectral energy distributions in the mid to far-infrared and but low 
fluxes in the near to mid-infrared, indicating massive disks with inner holes are very likely to be undergoing 
planet formation. The sharp inner holes that are necessary to produce the observed SEDs cannot be produced 
by alternate explanations, i.e., photo-evaporation and grain growth.
\textcolor{black}{Indeed LkCa~15, a star with such a transition disk, was recently found to have a
planetary object in formation within its inner cavity \citep{2012ApJ...745....5K}.}
The results suggest that at least 18\% of stars form planets, while it is uncertain whether 
the stars in the other stages will ever undergo or have already undergone planet formation. However, these planets 
either have small orbital separations ($<40$~AU) or are too small
($<3$~\mjup ) to be detected by our debris disk survey.
  
Recently, using simulations that examined the survival of debris disks and terrestrial planets in 
systems with already existing giant planets (1~M$_{sat}$ to 3~\mjup ),
\citet{2011A&A...530A..62R,2012A&A...541A..11R} predicted that 
(1) debris disks should be anti-correlated with eccentric giant planets (usually in wide-orbits); 
(2) disks have a high probability (\app\ 95\%) of surviving in systems with low-mass giant planets ($\leq$ 1\mjup  ); 
and (3) massive outer disks tend to stabilize inner giant planets and also lead to long disk lifetimes.
Thus, the massive giant planets on wide orbits that the Campaign is sensitive to may be much less prevalent 
in our debris disk sample than in other non-debris Campaign stars.
At the same time, low-mass giant planets may be more prevalent in the debris disk sample.
The \citet{2012A&A...541A..11R} simulations also suggest that $\beta$ Pic and HR~8799 
were once accompanied by massive outer disks ($\sim$100~\mearth), since 
they both have very massive planets (5--10~\mjup ) which probably required a 
massive disk to stabilize them.

Twenty two of the 57 debris disks in our sample have resolved disks around them and most of these have asymmetries in 
them in the form of arcs, clumps, etc. Asymmetries are the strongest indicators of the influence of planetary mass objects \citep{2008ARA&A..46..339W}, although
the location of the unseen planets cannot be uniquely determined from them. This may be the most important distinction between 
debris disk stars and other groups of stars included in the Campaign.    

\section{CONCLUSIONS}
We have completed a direct imaging survey for giant planets around 57 debris disk stars 
as part of the Gemini NICI Planet-Finding Campaign. We achieved median contrasts at $H$-band of 12.4 mag at $0.5''$ and 14.1 mag at $1"$. 
We detected a total of 78 planet candidates around 23 stars. Follow-up observations of 19 targets with 66 of the most promising 
candidates (projected separation $<$ 500~AU), show that all of them are background objects.

We have developed a more general Bayesian formalism than previous studies, which allowed us to use 
(1) non-detections, (2) single-epoch detections, and (3) multiple
confirmed detections in a single system along with (4) their astrometric and
(5) photometric information to constrain the planet population. We demonstrated 
the validity of this approach by simulating an input planet population and recovering good estimates for 
the population parameters.
We also show that the statistical formulation used in 
\citet{2008ApJ...674..466N} and \citet{2010ApJ...717..878N} is consistent with our 
more general Bayesian formulation and we 
discuss the bounds of the applicability of the earlier method. 
We also discuss under what assumptions the method 
presented in \citet{2007ApJ...670.1367L} is consistent with 
ours.

In our new statistical formulation, we make a distinction 
between the fraction of stars with planets and the average 
planet multiplicity. We assume a Poisson distribution in 
planet multiplicity to represent the planet population model,
such that the two statistics are naturally related. However, 
it is also possible to study other population models 
within our Bayesian formulation. The most interesting aspect 
of our new formalism is that both astrometric and photometric 
information about detected planets can be used to constrain 
the planet population. Also multiple planet detections around 
a single star, such as in the HR~8799bcde system, 
can be incorporated into constraint calculations. 
Thus, the formulation can be naturally applied to the upcoming 
direct-imaging surveys, SPHERE and GPI, from which multiple 
planet detections are more feasible. 

\textcolor{black}{
We used our Bayesian method to analyze the statistical properties of
the underlying planet population, based on our contrast curves for all
targets (plus 6 extra stars from Vigan et al.\ 2012.)
For this total debris disk sample, we find at the 95\%
confidence level that 
$<$21\% of debris disk stars have a $\geq$3\mjup\ planet 
outside of 40~AU, and $<$13\% of stars have a $\geq$5\mjup\ planet 
beyond 80~AU. We also find that indeed multiple massive planets per system may still remain
undetected by direct-imaging surveys inside of 5~AU. 
The Bayesian constraints on the planet-mass power-law index ($\alpha$) 
and the SMA power-law index ($\beta$) show that either $\beta < -0.8$
or both $\alpha > 1.7$ and $a_{max} < 200$~AU,
where $a_{max}$ is the maximum allowed SMA.  
The planet frequency is forced to rise sharply with mass ($\alpha > 1.7$), 
because all our detected planets have masses above 5\mjup , even though
lower masses could easily have been detected at these separations. } 

\textcolor{black}{
Since, the $\beta$ Pic and HR~8799 planets may represent a distinct population
of massive planets separate from the true tail of the planet-distribution,
we also calculated population constraints without
including these planets. In this
analysis, our 2$\sigma$ constraints show that either $\beta < -0.8$ or
both $\alpha < -1.5$ and $a_{max} < 125$~AU. 
Also, we found a possible weak correlation between planet-frequency and stellar
mass but our 2$\sigma$ constraints are still consistent with no correlation.
We also estimate that $<$20\% of stars are allowed to have a $\geq$3\mjup\ planet
outside of 10~AU, while $<$13\% of stars are allowed to 
have a $\geq$5\mjup\ planet outside of 20~AU.
These constraints are stronger than what previous surveys have found because 
of the improved performance of NICI. 
}

We did not detect the Fomalhaut planet in our NICI observations of the star. 
With 99\% confidence that there are no planets with $CH_4S <$20.0\pp 0.3~mag 
near the location of the \citet{2008Sci...322.1345K} 
detection. The upper limit on the mass of the planet from the NICI observations
is 12--13~\mjup , assuming thermal emission and an age of 450\pp40~Myr for Fomalhaut \citep{2012ApJ...754L..20M}. 
Thus, it is not surprising that we do not detect the planet.

A study of transition disks (those with inner holes) by \citet{2012ApJ...750..157C} suggested 
that roughly 18\% are in a planet-forming phase,  18\% are in a grain-growth phase (likely 
an earlier phase), and 64\% are in the debris disk or photo-evaporation phase (likely a later 
phase). This suggests that at least 18\% of stars form planets, while it is uncertain whether 
the other stars will ever undergo or have already undergone planet formation. 
However, these planets either have small orbital separations ($<40$~AU) or are too 
small ($<3$~\mjup ) to be detected by our debris disk survey.
 
 
\acknowledgements

\textcolor{black}{
This work was supported in part by NSF grants AST-0713881 and
AST-0709484.  The Gemini Observatory is operated by the Association of Universities
for Research in Astronomy, Inc., under a cooperative agreement with
the NSF on behalf of the Gemini partnership: the National Science
Foundation (United States), the Science and Technology Facilities Council (United Kingdom),
the National Research Council (Canada), CONICYT (Chile), the Australian Research Council
(Australia), CNPq (Brazil), and CONICET (Argentina). 
Our research has employed the 2MASS data products; NASA's Astrophysical
Data System; the SIMBAD database operated at CDS, Strasbourg, France.
We thank Ruobing Dong and Lucas Cieza for discussions on the
importance of transition disks to planet formation. }

{\it Facilities:} Gemini-South (NICI).

\vfill
\eject

\appendix
\section{APPENDIX}

Here we present the full derivation of the probability distribution for $F$, the frequency of planets around single stars.
$$  P(F , n \bb \{n_j\} , \{f_j\}, I) = P(\{n_j\} , \{f_j\} \bb F , n , I) \frac {P(F , n \bb I)} {P(\{n_j\} , \{f_j\} \bb I)},$$
We set the denominator on the right side of the equation to 1, because we assume that there is no bias in our sampling 
of stars, other than that we have chosen young nearby stars with debris disks (we explicitly state that our conclusions are only valid for this sample).
Using the notation $D_j$  for $n_j , f_j$, and assuming that the $\{D_j\}$ are independent  
$$  P(\{D_j\} \bb F , n, I) = P(D_1 \bb D_{j=2 \cdot \cdot N}, F , n , I)  P(D_{j=2 \cdot \cdot N} \bb F , n , I) $$
$$   = \prod_{j=1}^N  P(D_j \bb F , n , I) = \prod_{j=1}^N  P(n_j , f_j \bb F , n , I) $$
$$   = \prod_{j=1}^N  P(n_j \bb f_j,  F , n , I) P(f_j \bb F , n , I) = \prod_{j=1}^N  P(n_j \bb f_j,  F , n , I),$$
since $P(f_j \bb F , n , I)=1$ as the $f_j$s only depend on the contrasts achieved and the power-law behavior of the population model and 
not on the overall normalization.
Also, we have
$$P(F , n \bb I) =  P(n \bb F, I) P(F \bb I) =  P(n \bb F, I),$$
since we assume that there is no prior information on the frequency of planets and thus set $P(F \bb I)=1$.
Thus, we have ( since $P(n_j \bb f_j,  F , n , I) = P(n_j \bb f_j, n , I))$
$$  P(F , n \bb \{n_j\} , \{f_j\}, I) = \prod_{j=1}^N  P(n_j \bb f_j, n , I) P(n \bb F, I).$$
Finally, summing over the nuissance parameter $n$,  we have the probability distribution for the frequency of planets as presented in \S~\ref{sec:bayes_F}:
$$P(F \bb \{n_j\} , \{f_j\}, I) = \prod_{j=1}^N  \Bigg( \sum_{n} P(n_j \bb f_j, n , I) P(n \bb F, I) \Bigg) . $$

\section{APPENDIX}
\label{sec:prev_work}
\citet{2007ApJ...670.1367L} used the following expression for the likelihood of $F_{sp}$ (in our notation):
\begin{equation}
P({d_j}| F_{sp}, I) = \prod_{j=1}^{N} (F_{sp} f_j)^{d_j} (1-F_{sp} f_j)^{1-d_j} ,
\end{equation}
where $d_j=0$ if no planets are detected around star $j$, and $d_j=1$ otherwise. The likelihood that
the star has one or more planets is written as $F_{sp} f_j$. In other words, in this model, every star has
exactly $F_{sp}$ planets within the chosen mass and SMA range, and thus it is not possible to accommodate stars with $0$ planets
and stars with $\geq 4$ planets (e.g. HR~8799) within the same population model.

The other limitation of this formulation is that all systems with non-zero detections are considered the same.
Thus, we are throwing away information and the constraints on the model are not optimum given the data. 

To examine the special case where our formulations agree, let us force a simpler model 
where a star can have exactly 1 or 0 planets \citep[similar to][]{2008ApJ...674..466N},
even though the model will not be able to describe real data sets with multiple planet discoveries.
Thus, $F=F_{sp} * 1 + (1-F_{sp}) * 0 = F_{sp}$. Also, let us set the probabilities to $P(1 \bb F)=F$
and $P(0 \bb F)=1-F$, instead of using the Poisson likelihoods.  Similarly, let us 
replace $P(1 \bb f_j)=f_j$ and  $P(0 \bb f_j)=1-f_j$. Then, using our Bayesian result from earlier and summing over $n=\{0,1\}$, we get
$$P(F\bb {n_j=0,1}, {f_j}, I) = \prod_{j=1}^N\Bigg(  P(n_j \bb 0)\ P( 0 \bb F) + P(n_j \bb 1 f_j)\ P( 1 \bb F) \Bigg) $$
$$= \prod_{j=1}^N \Bigg(  P(n_j \bb 0)\ (1-F) + P(n_j \bb f_j)\ F \Bigg) $$
$$= \prod_{j=1}^N {\Bigg(  P(1 \bb 0)\ (1-F) + P(1 \bb f_j)\ F \Bigg)}^{n_j}  {\Bigg(  P(0 \bb 0)\ (1-F) + P(0 \bb  f_j)\ F \Bigg)}^{1-n_j}$$
$$= \prod_{j=1}^N {\Bigg(  f_j F \Bigg)}^{n_j}  {\Bigg(  (1-F) + (1-f_j) F \Bigg)}^{1-n_j}$$
$$= \prod_{j=1}^N (f_j F)^{n_j}  (1-f_j F)^{1-n_j}.$$
This is the same expression as in \citet{2007ApJ...670.1367L}. Thus their approach agrees with ours 
for small $F$ and when a star can only have 1 or 0 planets.

\textcolor{black}{
For multiple planet systems, we can generalize the expression of
\citet{2007ApJ...670.1367L}~to
$$P(data | model, I)= P(\{d_{jk}\} | \{F_{nk}\} )= \prod_{j} \prod_{k}
\sum_{n}  F_{nk}^{d_{jk}}\ ,$$
where $F_{nk}$ is the probability of $k$ planet detections in systems with
$n$ planets,  and $d_{jk}$ is 1 if $k$ planets are detected around
star $j$ and 0 otherwise. Since the number of actual planets in a
system can vary, we sum over $n$ thus marginalizing over this parameter.
Now, we make the following expansion:
$$F_{nk}\ =\ (probability\ that\ n\ planets\ exist\ in\ system\ j) $$ 
$$\times\ (probability\ that\ k\ planets\ will\ be\ detected\ given\ the\ contrast\ limits)$$
$$= P(n \bb F)\ P(k_j \bb n f_j)$$
This expression is the same as our result in Eq.~\ref{eq:bayes_expan2}.}


\bibliographystyle{apj}
\bibliography{zrefs}
\vfill
\eject


\begin{deluxetable}{lrrrcrrrr}
\tabletypesize{\scriptsize}
\tablecaption{Debris Disk Stars Observed by the NICI Campaign.}
\tablewidth{0pt}
\tablehead{ \colhead{Target} &\colhead{R.A.} & \colhead{Dec.} & \colhead{Dist.\ (pc)} & \colhead{Sp.\ Type}&\colhead{Age$^*$\ (Myr)} & \colhead{$V$\ (mag)} & \colhead{$H$\ (mag)} & \colhead{$K$\ (mag)}}
\startdata
HR 9                 & 00:06:50 & $-$23:06:27 & 39.1    & F2 & 12$^{\ 1}$     & 6.20 & 5.33 & 5.24 \\
HIP 1481          & 00:18:26 & $-$63:28:39 & 41.0    & F9 & 30$^{\ 1}$     & 7.46 & 6.25 & 6.09 \\
49 Cet              & 01:34:37 & $-$15:40:34 & 61.0    & A1 & 40$^{\ 10}$     & 5.62 & 5.53 & 5.46 \\
HD 10472        & 01:40:24 & $-$60:59:56 & 67.0    & F2 & 30$^{\ 1}$      & 7.62 & 6.69 & 6.63 \\
HD 10939        & 01:46:06 & $-$53:31:19 & 57.0    & A1 & 346$^{\ 2}$   & 5.00 & 5.03 & 4.96 \\
HIP 10679        & 02:17:24 & +28:44:31    & 34.0    & G2 & 12$^{\ 1}$     & 7.80 & 6.36 & 6.26 \\
HD 15115        & 02:26:16 & +06:17:33    & 45.0    & F4 & 12$^{\ 1}$      & 6.79 & 5.81 & 5.77 \\
HD 17848        & 02:49:01 & $-$62:48:23 & 50.7    & A2 & 372$^{\ 2}$   & 5.30 & 5.16 & 4.97 \\
\enddata
\label{tab:star_props}
\tablecomments{
Age references$^*$: (1) \citet{2004ARA&A..42..685Z} (2) \citet{2007ApJ...660.1556R} 
(3) \citet{2011ApJ...732...61Z} 
(4) \citet{2010ApJ...717..878N}
(5) \citet{2006ApJ...644..525M} 
(5) \citet{2011ApJ...730L..29M}
(6) \citet{2005ApJ...620.1010R}
(7) \citet{2005ApJ...631.1170L}
(8) \citet{2012ApJ...754L..20M} 
(9) \citet{2006ApJS..166..351C}
(10) Nielsen et al.\ 2013}
\end{deluxetable}

\begin{deluxetable}{lccccc}
\tabletypesize{\scriptsize}
\tablecaption{Observation of Campaign Debris-Disk Targets}
\tablewidth{0pt}
\tablehead{ \colhead{Target}  &\colhead{Date} & \colhead{Obs Mode} & \colhead{Nunber of} & \colhead{Total}             & \colhead{Total} \\
                   \colhead{}           & \colhead{}        &\colhead{}                  & \colhead{Images}     & \colhead{Exp. Time (s)} & \colhead{Rotation (deg)} 
}
\startdata
HR 9 & 2009-12-05 & ADI &      20 &    1208 &     2.8 \\
HR 9 & 2009-12-05 & ASDI &      45 &    2736 &    22.8 \\
49 Cet & 2009-12-02 & ASDI &      45 &    2736 &    40.6 \\
49 Cet & 2009-12-02 & ADI &      20 &    1208 &    16.4 \\
HD 10939 & 2010-12-26 & ASDI &      44 &    2708 &     3.5 \\
HIP 10679 & 2011-10-16 & ASDI &      45 &    2701 &    13.3 \\
HD 15115 & 2009-12-04 & ADI &      20 &    1208 &     6.2 \\
HD 15115 & 2009-12-04 & ASDI &      43 &    2598 &    17.4 \\
\enddata
\label{tab:obs}
\end{deluxetable}

\begin{deluxetable}{llrcrcrrcr}
\tabletypesize{\scriptsize}
\tablecaption{Properties of Star and Disk}
\tablewidth{0pt}
\tablehead{ \colhead{Target} &\colhead{Assoc.} & \colhead{$L_d/L_*$} & \colhead{ref.} & \colhead{Disk Size} &\colhead{ref.} & \colhead{Disk Inc.} & \colhead{Disk PA} & \colhead{ref.}   \\
  \colhead{} &\colhead{} & \colhead{$\times 10^{-5}$} & \colhead{} & \colhead{AU} &\colhead{} & \colhead{degree} & \colhead{degree} & \colhead{}  }
\startdata
HR 9 & BPMG & 10 & 1 & 200 & 1 & - & - &   \\
HIP 1481 & THMG & 7.1 & 3 &  &  & - & - &   \\
49 Cet & - & 92 & 4 & 900 & 2 & 60 & 125 & 2  \\
HD 10472 & THMG & 67 & 1 & 700 & 1 & - & - &   \\
HD 10939 & - & 6.8 & 5 &  &  & - & - &   \\
HIP 10679 & BPMG & 80 & 1 & 200 & 1 & - & - &   \\
HD 15115 & BPMG & 49 & 4 & 872 & 2 & 90 & 278.5  & 2  \\
HD 17848 & - & 6.4 & 7 &  &  & - & - &   \\
\enddata
\label{tab:diskprops}
\tablecomments{The columns in order from left to right are: (1) target name, (2) membership in association, 
(3) fractional disk luminosity and (4) associated reference, (5) disk size in AU and (6) associated reference, 
(7) disk inclinication to the line of sight and (8) associated reference.
References: (1) \citet{2008ApJ...681.1484R} (2) Catalog of Resolved Debris Disks (http://circumstellardisks.org/).
(3) \citet{2011ApJ...732...61Z} (4) \citet{2006ApJ...644..525M} (5) \citet{2011ApJ...730L..29M}
(6) \citet{2006ApJ...653..675S} (7) \citet{2007ApJ...660.1556R} (8) \citet{2008ApJ...677..630H}
(9) \citet{2005ApJ...631.1170L} (10) \citet{2006ApJS..166..351C} (1i) \citet{2007ApJ...661..368W}
(1b) \citet{1996AJ....111..439H} (2b) \citet{2003AJ....126.2048G} (3b) \citet{2004ApJS..152..261W}
(4b) \citet{2006AJ....132..161G} (5b) \citet{2006MNRAS.372..163J} (6b) \citet{2008A&A...485..571J}
(7b) \citet{2010A&A...520A..79M}.}
\end{deluxetable}


\begin{landscape}
\begin{deluxetable}{lrrrrrrrrrrrrrrrrrr}
\tabletypesize{\scriptsize}
\tablecaption{95\% completeness $CH_4$ and $H$-band contrasts ($\Delta$mag)}
\tablewidth{0pt}
\tablehead{ \colhead{Target} &\colhead{0.36$''$} & \colhead{0.5$''$} & \colhead{0.75$''$} & \colhead{1$''$}&\colhead{1.5$''$} & \colhead{2$''$} & \colhead{3$''$} & \colhead{4$''$}&\colhead{5$''$}&\colhead{7$''$} & \colhead{Cov.} & \colhead{9$''$} & \colhead{Cov.}&\colhead{12$''$} & \colhead{Cov.} & \colhead{14.8$''$} & \colhead{Cov.}}
\startdata
HR 9, $CH_4$ & 11.0 & 12.6 & 13.6 & 14.1 & 14.8 & 15.0 & 15.1 & 15.0 & 15.1 & 14.8 & 0.98 & 14.5 & 0.74 & 13.7 & 0.35 & 12.9 & 0.07 \\
HR 9, $H$-band &  -  &  -  &  -  &  1.4 & 14.1 & 15.3 & 16.5 & 16.6 & 17.0 & 17.0 & 0.85 & 17.0 & 0.58 & 16.9 & 0.24 & 16.5 & 0.03 \\
HIP 1481, $CH_4$ & 10.7 & 11.9 & 13.2 & 13.6 & 14.0 & 14.2 & 14.1 & 14.1 & 14.0 & 13.9 & 0.99 & 13.4 & 0.74 & 12.4 & 0.31 &  -  &  -  \\
HIP 1481, $H$-band &  -  &  -  & 11.5 & 12.8 & 14.4 & 15.2 & 15.5 & 15.6 & 15.4 & 15.4 & 0.89 & 15.0 & 0.65 & 13.6 & 0.31 & 13.6 & 0.06 \\
49 Cet, $CH_4$ & 10.7 & 12.4 & 13.4 & 14.0 & 14.6 & 14.9 & 14.9 & 14.8 & 14.6 & 14.2 & 1.00 & 13.3 & 0.84 & 12.4 & 0.47 & 12.1 & 0.10 \\
49 Cet, $H$-band &  -  &  -  &  4.0 & 13.2 & 14.8 & 15.9 & 16.8 & 16.9 & 16.8 & 16.6 & 0.89 & 16.3 & 0.66 & 15.8 & 0.32 & 15.3 & 0.07 \\
HD 10939, $CH_4$ &  7.6 &  9.6 & 11.8 & 13.1 & 14.2 & 14.5 & 14.6 & 14.6 & 14.5 & 14.5 & 0.93 & 14.4 & 0.68 & 14.0 & 0.30 & 13.5 & 0.06 \\
HIP 10679, $CH_4$ & 11.0 & 12.3 & 13.3 & 13.9 & 14.4 & 14.6 & 14.7 & 14.8 & 14.6 & 14.5 & 0.93 & 14.2 & 0.68 & 13.7 & 0.28 & 13.0 & 0.05 \\
\enddata
\label{tab:contrasts}
\tablecomments{Contrasts achieved in the $CH_4$ and $H$-bands. Beyond 6.3$''$ only a fraction of each final image has data, so the coverage fraction is given in the column next to the contrast. All the contrasts are 95\% completeness contrasts, except at separations beyond 6.3$''$ where the nominal 1$\sigma$ contrast curve is used. A constant is added to the 1$\sigma$ curve so that both curves match at 6.3$''$.}
\end{deluxetable}

\end{landscape}

\begin{deluxetable}{lcccccccc}
\tabletypesize{\scriptsize}
\tablecaption{Lowest-Mass Companion Detectable at Different Projected Separations (\mjup).}
\tablewidth{0pt}
\tablehead{ \colhead{Target} &\colhead{0.36$''$} &\colhead{0.5$''$}
  &\colhead{0.75$''$} & \colhead{1$''$} & \colhead{2$''$} &
  \colhead{4$''$} & \colhead{8$''$} & \colhead{at disk edge}}
\startdata
      HR 9 &    7.4 &    5.7 &    5.0 &    4.6 &    3.6 &    2.4 &    2.2 &    2.7 \\
  HIP 1481 &   10.0 &    8.4 &    7.6 &    6.8 &    5.0 &    4.3 &    4.3 &      - \\
    49 Cet &   14.8 &   10.3 &    9.6 &    9.2 &    6.7 &    5.5 &    5.5 &    6.2 \\
  HD 10472 &      - &      - &    8.9 &    7.9 &    5.4 &    5.0 &    5.0 &    4.9 \\
  HD 10939 &      - &   60.7 &   35.9 &   27.8 &   24.7 &   24.7 &   24.6 &      - \\
 HIP 10679 &    6.0 &    5.0 &    4.0 &    3.6 &    3.2 &    3.0 &    3.0 &      - \\
  HD 15115 &    6.7 &    5.3 &    4.3 &    4.0 &    2.5 &    2.2 &    2.2 &    2.2* \\
  HD 17848 &   42.6 &   35.8 &   29.3 &   26.5 &   19.3 &   16.1 &   16.1 &      - \\
\enddata
\label{tab:mass}
\tablecomments{\textcolor{black}{The last column gives the lowest mass companion
  detectable at the outer radius of the debris disk, using the radius
  estimate from Table~\ref{tab:diskprops}. If the disk radius is too large
  (falls outside of the NICI field of view), we present the mass limit at the edge
  of the field (marked with * in table). If the disk radius falls inside the coronagraph, we do
  not present the mass limit.}}
\end{deluxetable}

\begin{deluxetable}{ccccccccccc}
\tabletypesize{\scriptsize}
\tablecaption{Properties of Candidate Companions\label{table_comp2a}}
\tablewidth{0pt}
\tablehead{
\colhead{Name} & \colhead{\#} & \colhead{Sep} & 
\colhead{PA} & 
\colhead{$\Delta H$} & \colhead{$\Delta \tau$} & 
\colhead{Epochs} & \colhead{$\chi^2_{\nu}$(BG)} & 
\colhead{$\chi^2_{\nu}$(CPM)} & \colhead{dof} & \colhead{Comp?} \\
\colhead{} & \colhead{} & \colhead{($"$)} & \colhead{(deg)} & 
\colhead{(mag)} & \colhead{(years)} & \colhead{} & \colhead{} & 
\colhead{} & \colhead{} & \colhead{}
}
\startdata
HD 19668 & 1 & 5.476 & 150.7 & 10.1 & 9.93 & 8 & 0.73 & 34.93 & 14 & BG  \\ 
HD 27290 & 1 & 7.745 & 244.0 & 10.6 & 1.05 & 2 & 1.64 & 106.91 & 2 & BG  \\ 
HD 31295 & 1 & 6.261 & 271.6 & 13.3 & 1.95 & 2 & 0.09 & 69.71 & 2 & BG  \\ 
HD 31295 & 2 & 8.805 & 316.7 & 14.5 & 1.95 & 2 & 0.02 & 187.48 & 2 & BG  \\ 
$\zeta$ Lep & 1 & 5.295 & 110.0 & 18.0 & 2.24 & 2 & 0.55 & 28.93 & 2 & BG  \\ 
HD 53143 & 1 & 4.479 & 286.2 & 10.3 & 0.98 & 2 & 2.62 & 163.81 & 2 & BG  \\ 
HD 54341 & 1 & 3.080 & 131.2 & 12.7 & 2.03 & 2 & 0.36 & 3.05 & 2 & BG  \\ 
HD 54341 & 2 & 4.372 & 137.3 & 11.8 & 2.03 & 2 & 0.98 & 2.56 & 2 & BG  \\
\enddata
\label{tab:cands}
\tablecomments{Summary of the candidate companions detected around 
each target star.  For each candidate we list the separation and 
position angle at the reference epoch, the contrast between 
candidate and host star, the time baseline for our astrometric 
data, the number of epochs, the reduced chi-square statistic for 
the companion to be background (BG) or common proper motion (CPM), 
the number of degrees of freedom, and the final determination of 
each companion: background or common proper motion.  Astrometry 
for individual epochs is in Table~\ref{tab:cands2}.
}
\end{deluxetable}

\begin{deluxetable}{cccccccccccccc}
\tabletypesize{\scriptsize}
\tablecaption{Astrometry of Candidate Companions\label{table_comp2b}}
\tablewidth{0pt}
\tablehead{
 & & & \multicolumn{4}{c}{Measured Position} & & 
\multicolumn{4}{c}{Background Position} & 
& \\
\cline{4-7}
\cline{9-12}
\colhead{Name} & \colhead{\#} & \colhead{Epoch} & \colhead{Sep} & 
\colhead{$\sigma_{Sep}$} & \colhead{PA} & \colhead{$\sigma_{PA}$} & 
\colhead{} & \colhead{Sep} & \colhead{$\sigma_{Sep}$} & 
\colhead{PA} & \colhead{$\sigma_{PA}$} & 
\colhead{Inst.} & \colhead{Comp?} \\
\colhead{} & \colhead{} & \colhead{} & \colhead{($"$)} & 
\colhead{} & \colhead{(deg)} & \colhead{} & \colhead{} & 
\colhead{($"$)} & \colhead{} & \colhead{(deg)} & \colhead{} & 
\colhead{} & \colhead{} \\
}
\startdata
HD 19668 & 1 & 2010.66 & 5.476 & 0.009 & 150.7 & 0.2 &   & \ldots & \ldots & \ldots & \ldots & N & BG \\
  &   & 2001.93 & 6.781 & 0.144 & 149.1 & 1.3 &   & 6.726 & 0.015 & 148.583 & 0.185 & S & BG \\
  &   & 2002.97 & 6.662 & 0.288 & 150.3 & 2.6 &   & 6.581 & 0.014 & 148.724 & 0.185 & S & BG \\
  &   & 2009.74 & 5.566 & 0.013 & 150.0 & 0.5 &   & 5.603 & 0.010 & 150.414 & 0.204 & V & BG \\
  &   & 2010.82 & 5.429 & 0.013 & 150.0 & 0.5 &   & 5.450 & 0.010 & 150.550 & 0.211 & V & BG \\
  &   & 2010.98 & 5.408 & 0.009 & 150.7 & 0.2 &   & 5.437 & 0.010 & 150.356 & 0.213 & N & BG \\
  &   & 2011.70 & 5.301 & 0.009 & 151.1 & 0.2 &   & 5.326 & 0.011 & 150.975 & 0.219 & N & BG \\
  &   & 2011.86 & 5.301 & 0.013 & 151.2 & 0.3 &   & 5.304 & 0.011 & 150.758 & 0.220 & L & BG \\
\enddata
\label{tab:cands2}
\tablecomments{Astrometry for each candidate companion detected around 
our target stars from NICI and archival observations.  At each 
epoch we give the measured separation, position angle, and 
uncertainties as well as the predicted separation and position 
angle for a background object based on the proper motion and 
parallax of the primary and the candidate position at the reference 
epoch, which is the first epoch listed for each candidate.  
Astrometry is taken from NICI (N), VLT NACO (V), 
Keck NIRC2 (K), VLT ISAAC (I), ESO 3.6m (E), and Gemini NIRI (G).
}
\end{deluxetable}

\begin{landscape}
\begin{deluxetable}{lrrrrrrrr}
\tabletypesize{\scriptsize}
\tablecaption{Single-epoch astrometry for sources with large physical separations.}
\tablewidth{0pt}
\tablehead{ \colhead{Target}  &\colhead{Obj. No.} &\colhead{Epoch} & \colhead{Sep. (AU)} & \colhead{Separation ($''$)} & \colhead{Sep. Unc. ($''$)} & \colhead{PA (deg)} & \colhead{PA Unc. (deg)} & \colhead{$\Delta H$ (mag)}}
\startdata
HD 131835 &        1 &       2009.28 &   594.8 &   5.360 &   0.009 &    20.1 &   0.2 &  14.8 \\
  &        2 &       2009.28 &   642.5 &   5.790 &   0.009 &    22.8 &   0.2 &  14.1 \\
  &        3 &       2009.28 &   683.6 &   6.160 &   0.009 &   -56.0 &   0.2 &  13.6 \\
  &        4 &       2009.28 &   894.5 &   8.060 &   0.009 &  -119.8 &   0.2 &  16.2 \\
  &        5 &       2009.28 &   994.2 &   8.960 &   0.009 &   100.4 &   0.2 &  16.4 \\
HD 157728 &        1 &       2011.38 &   491.2 &  11.424 &   0.009 &  -286.7 &   0.2 &  14.8 \\
HD 172555 &        1 &       2009.27 &   225.7 &   7.730 &   0.009 &   -41.1 &   0.2 &  15.5 \\
HD 182681 &        1 &       2010.66 &   313.5 &   4.544 &   0.009 &  -110.0 &   0.0 &  12.2 \\
  &        2 &       2010.66 &   365.1 &   5.292 &   0.009 &  -108.7 &   0.0 &  17.2 \\
  &        3 &       2010.66 &   593.1 &   8.595 &   0.009 &   -54.9 &   0.0 &  15.5 \\
  &        4 &       2010.66 &   612.1 &   8.871 &   0.009 &   -70.4 &   0.0 &  14.0 \\
  &        5 &       2010.66 &   659.2 &   9.554 &   0.009 &   -56.2 &   0.0 &  16.8 \\
\enddata
\label{tab:cands_nofollow}
\end{deluxetable}

\end{landscape}

\begin{deluxetable}{rccccccc}
\tabletypesize{\scriptsize}
\tablecaption{\textcolor{black}{Upper Limits on the Fraction of Stars
    with Planets Including $\beta$ Pic~b and HR~8799bcd.}}
\tablewidth{0pt}
\tablehead{ \colhead{} &\colhead{$>$0.5\mjup} &\colhead{$>$1\mjup}
  &\colhead{$>$3\mjup} &\colhead{$>$5\mjup} &\colhead{$>$7\mjup}
  &\colhead{$>$9\mjup} &\colhead{$>$11\mjup}}
\startdata
$>$0.5~AU   &      1.00  &   1.00 &    1.00       &  1.00        &1.00   &  1.00  &   0.99\\ 
$>$5~AU     &      0.60  &   0.55 &    0.50        &   0.47      &    0.43  &   0.35 &    0.23\\
$>$10~AU    &      0.48  &   0.41 &    0.35       &  0.33        &    0.29  &   0.23 &    0.15\\
$>$20~AU    &      0.39  &   0.35 &    0.29       &  0.26        &    0.23  &   0.19 &    0.11\\
$>$40~AU    &      0.25  &   0.25 & {\bf 0.21}  &  0.19          &  0.17    & 0.13   &  0.08\\
$>$80~AU    &      0.19  &   0.15 &    0.14       &  {\bf 0.13}  &   0.11   &  0.09  &   0.05\\
$>$160~AU   &      0.14  &   0.11 &    0.10      & 0.08          &    0.07  &   0.06 &    0.03\\

\enddata
\label{tab:planhostfrac}
\tablecomments{\textcolor{black}{These are the 95\% confidence-interval upper limits on
  the fraction of stars with $\geq$1 planet(s), when we include
  $\beta$ Pic~b and HR~8799bcd in the analysis.
  The limits are given for planet mass and
  semi-major axis greater than indicated in table and less than
  13\mjup\ and 1000~AU, respectively. They are
  calculated from the 4-dimensional probability density given as a 
  function of the mass, SMA, SMA-cutoff and planet multiplicity. 
  Thus unlike Figure~\ref{fig:limits_mass_sep}, these are
  model-dependent constraints. Planet detections and single-epoch
  detections are all taken into account in these limits.}}
\end{deluxetable}

\begin{deluxetable}{rccccccc}
\tabletypesize{\scriptsize}
\tablecaption{\textcolor{black}{Upper Limits on the Average Planet
    Multiplicity Including $\beta$ Pic~b and HR~8799bcd.}}
\tablewidth{0pt}
\tablehead{ \colhead{} &\colhead{$>$0.5\mjup} &\colhead{$>$1\mjup}
  &\colhead{$>$3\mjup} &\colhead{$>$5\mjup} &\colhead{$>$7\mjup}
  &\colhead{$>$9\mjup} &\colhead{$>$11\mjup}}
\startdata

$>$0.5~AU      & 11.33   &  11.10    & 10.95     &10.54   &9.62    &7.72    &4.68\\
$>$5~AU         &   0.91   &     0.80   &    0.69     &0.64    &0.56    &0.44    &0.26\\
$>$10~AU       &   0.65   &     0.53   &     0.44    &0.39    &0.34    &0.27    &0.16\\
$>$20~AU       &   0.49   &     0.43   &     0.34    &0.30    &0.26    &0.20    &0.12\\
$>$40~AU       &   0.29   &     0.28   &     0.24    &0.22    &0.18    &0.14    &0.08\\
$>$80~AU       &   0.20   &     0.16   &     0.15    &0.14    &0.12    &0.09    &0.05\\
$>$160~AU     &   0.15   &     0.12   &     0.10    &0.09    &0.08    &0.06    &0.03\\

\enddata
\label{tab:avgplanmult}
\tablecomments{\textcolor{black}{These are the 95\% confidence-interval upper limits on
  the average planet multiplicity, when we include
  $\beta$ Pic~b and HR~8799bcd in the analysis. Same as for
  Table~\ref{tab:planhostfrac}, the limits are given for planet mass and
  semi-major axis greater than indicated in table and less than
  13\mjup\ and 1000~AU, respectively.}}
\end{deluxetable}

\begin{deluxetable}{rccccccc}
\tabletypesize{\scriptsize}
\tablecaption{\textcolor{black}{Upper Limits on the Fraction of Stars
    with Planets Excluding $\beta$ Pic~b and HR~8799bcd.}}
\tablewidth{0pt}
\tablehead{ \colhead{} &\colhead{$>$0.5\mjup} &\colhead{$>$1\mjup}
  &\colhead{$>$3\mjup} &\colhead{$>$5\mjup} &\colhead{$>$7\mjup}
  &\colhead{$>$9\mjup} &\colhead{$>$11\mjup}}
\startdata
$>$0.5~AU         &    1.00   &    1.00   &    1.00   &    1.00   &    1.00   &    1.00   &    0.97\\
$>$5~AU            &    0.61   &    0.37   &    0.25   &    0.21   &    0.18   &    0.14   &    0.08\\
$>$10~AU          &    0.47   &    0.28   &     {\bf 0.20}   &    0.17   &    0.15   &    0.11   &    0.06\\
$>$20~AU          &    0.35   &    0.22   &    0.15   &     {\bf 0.13}   &    0.11   &    0.08   &    0.05\\
$>$40~AU          &    0.28   &    0.18   &    0.12   &    0.11   &    0.09   &    0.07   &    0.04\\
$>$80~AU          &    0.23   &    0.15   &    0.10   &    0.09   &    0.08   &    0.06   &    0.03\\
$>$160~AU        &    0.22   &    0.13   &    0.09   &    0.08   &    0.07   &    0.05   &    0.03\\
\enddata
\label{tab:planhostfrac_zd}
\tablecomments{\textcolor{black}{These are the 95\% confidence-interval upper limits on
  the fraction of stars with $\geq$1 planet(s), when we exclude
  $\beta$ Pic~b and HR~8799bcd in the analysis.
  The limits are given for planet mass and
  semi-major axis greater than indicated in table and less than
  13\mjup\ and 1000~AU, respectively. They are
  calculated from the 4-dimensional probability density given as a 
  function of the mass, SMA, SMA-cutoff and planet multiplicity. 
  Thus unlike Figure~\ref{fig:limits_mass_sep}, these are
  model-dependent constraints. Planet detections and single-epoch
  detections are all taken into account in these limits.}}
\end{deluxetable}

\begin{deluxetable}{rccccccc}
\tabletypesize{\scriptsize}
\tablecaption{\textcolor{black}{Upper Limits on the Average Planet
    Multiplicity Excluding $\beta$ Pic~b and HR~8799bcd.}}
\tablewidth{0pt}
\tablehead{ \colhead{} &\colhead{$>$0.5\mjup} &\colhead{$>$1\mjup}
  &\colhead{$>$3\mjup} &\colhead{$>$5\mjup} &\colhead{$>$7\mjup}
  &\colhead{$>$9\mjup} &\colhead{$>$11\mjup}}
\startdata
$>$0.5~AU    &    10.28   &   8.74   &   8.53    &  8.08    &  7.18& 5.77&    3.43\\
$>$5~AU       &    0.94     &  0.46    &   0.29    &   0.24   &    0.20&    0.15&    0.08\\
$>$10~AU     &     0.64    &   0.33   &    0.22   &    0.19  &     0.16&    0.11&    0.06\\
$>$20~AU     &     0.44    &   0.25   &    0.17   &   0.14  &     0.12&    0.09&    0.05\\
$>$40~AU     &     0.32    &   0.20   &    0.13   &    0.11  &     0.09&    0.07&    0.04\\
$>$80~AU     &     0.26    &   0.17   &    0.11   &    0.09  &     0.08&    0.06&    0.03\\
$>$160~AU   &      0.25   &    0.14  &     0.10  &     0.08 &      0.07&    0.06&    0.03\\

\enddata
\label{tab:avgplanmult_zd}
\tablecomments{\textcolor{black}{These are the 95\% confidence-interval upper limits on
  the average planet multiplicity, when we exclude
  $\beta$ Pic~b and HR~8799bcd in the analysis. Same as for
  Table~\ref{tab:planhostfrac}, the limits are given for planet mass and
  semi-major axis greater than indicated in table and less than
  13\mjup\ and 1000~AU, respectively.}}
\end{deluxetable}

\vfill
\eject
\vfill
\eject

\begin{figure}[H]
     \hbox {
        \includegraphics[height=8cm]{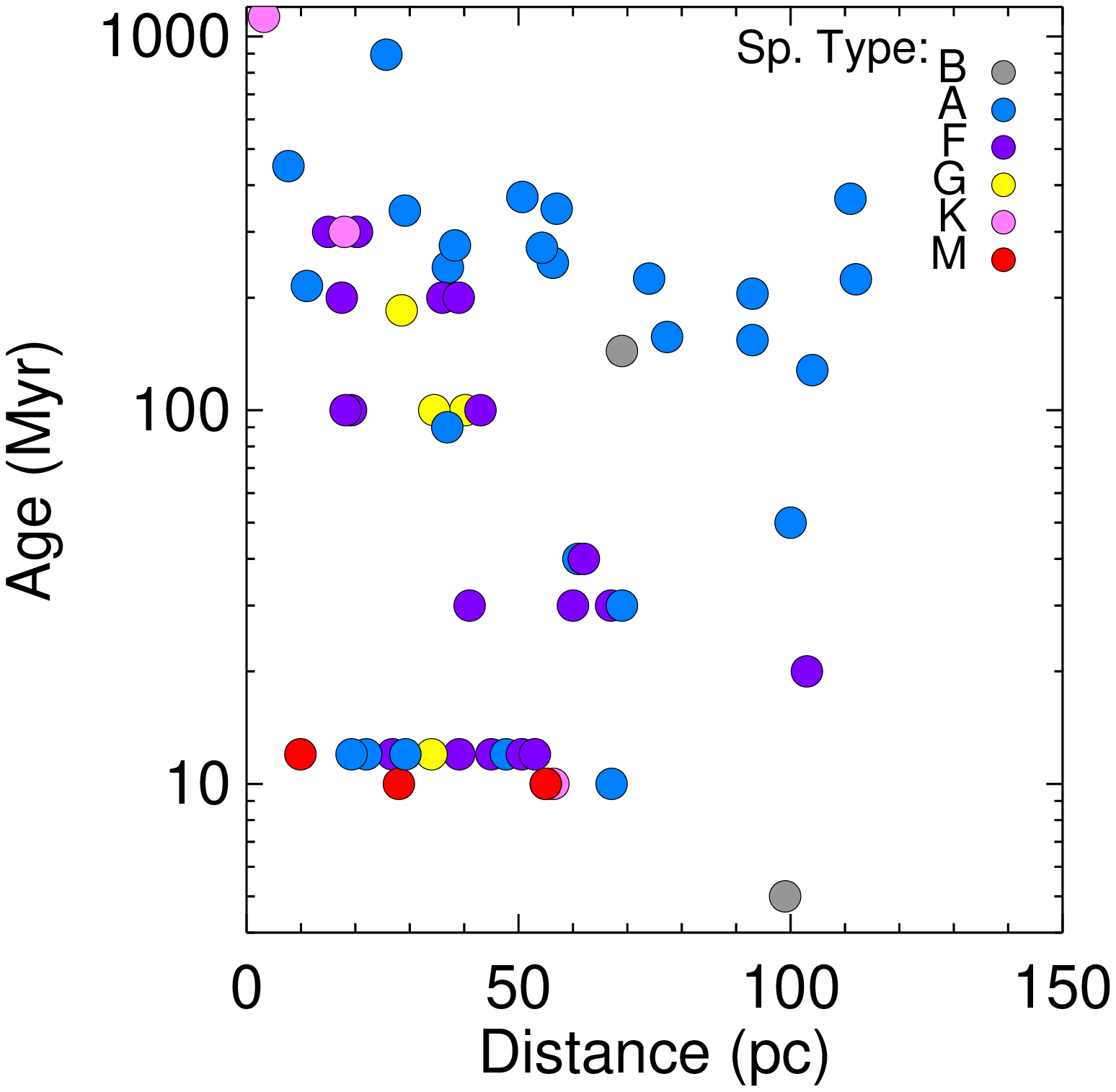}
        \includegraphics[height=8cm]{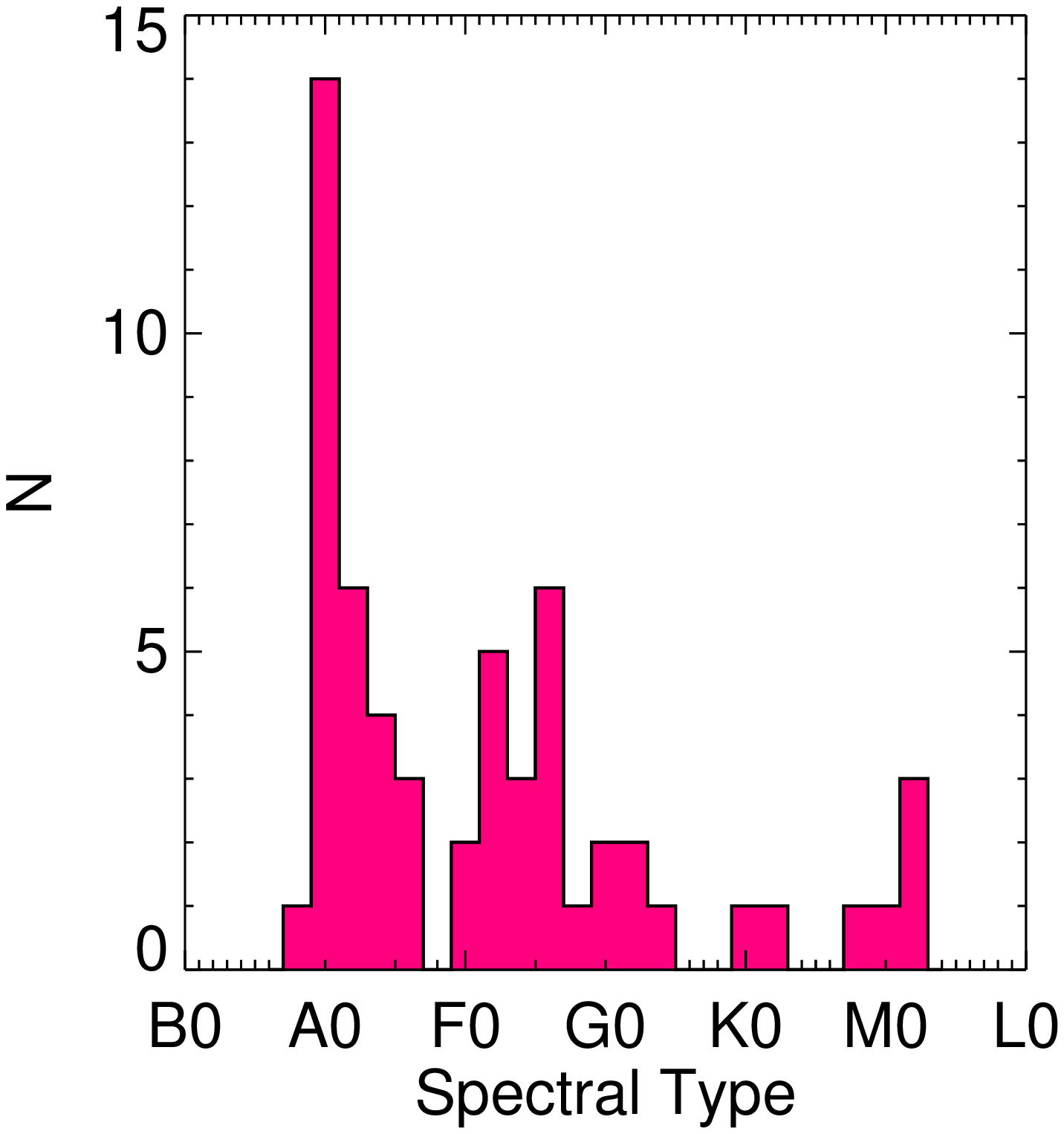}
      }  
  \caption{Ages, spectral types and distances of the NICI debris disk targets.}
  \label{fig:radec_dists}
\end{figure} 

\begin{figure}[H]
  \centerline{
  \vbox{
    \hbox {
      \includegraphics[width=6.3cm]{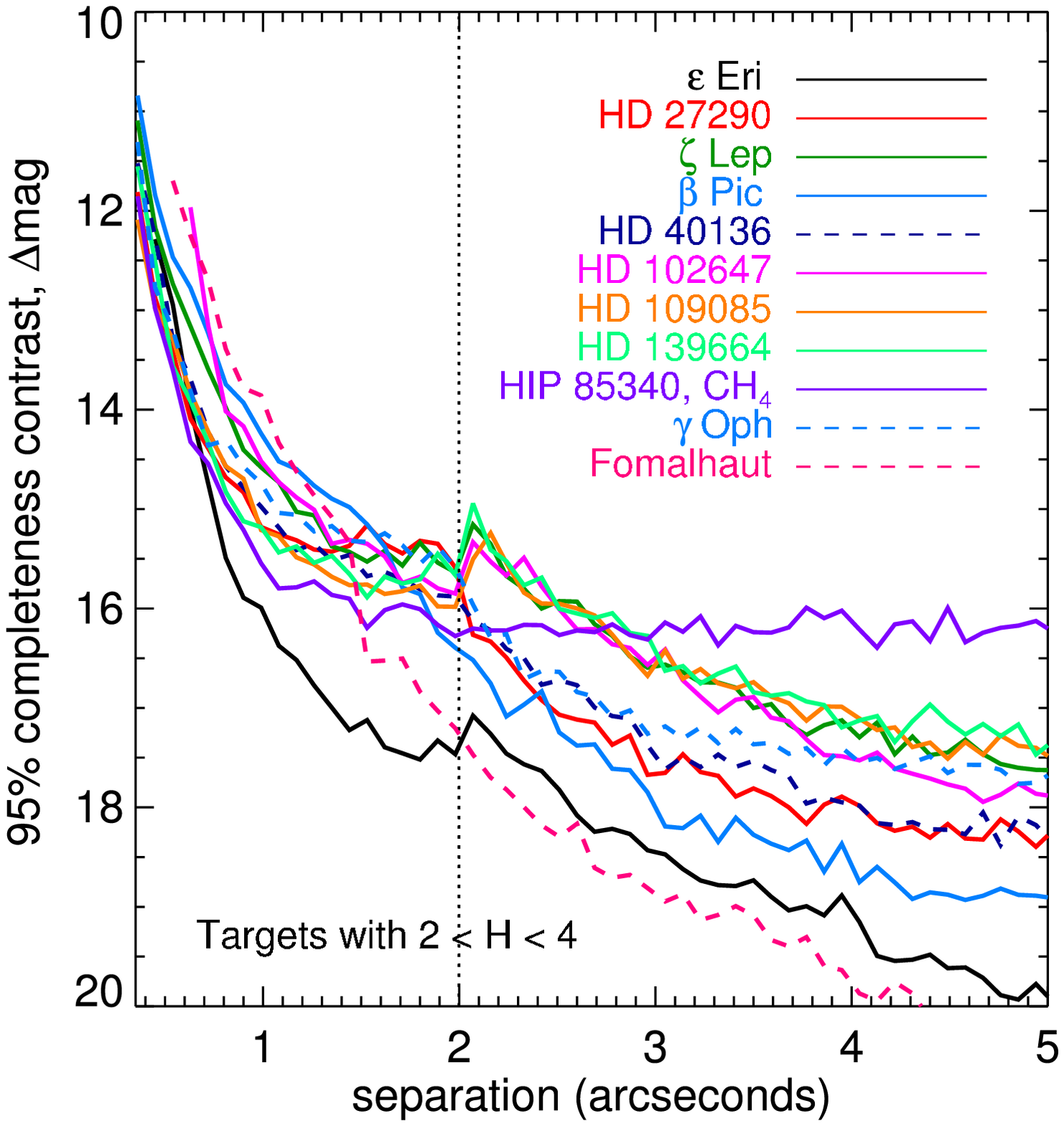}
      \includegraphics[width=6.3cm]{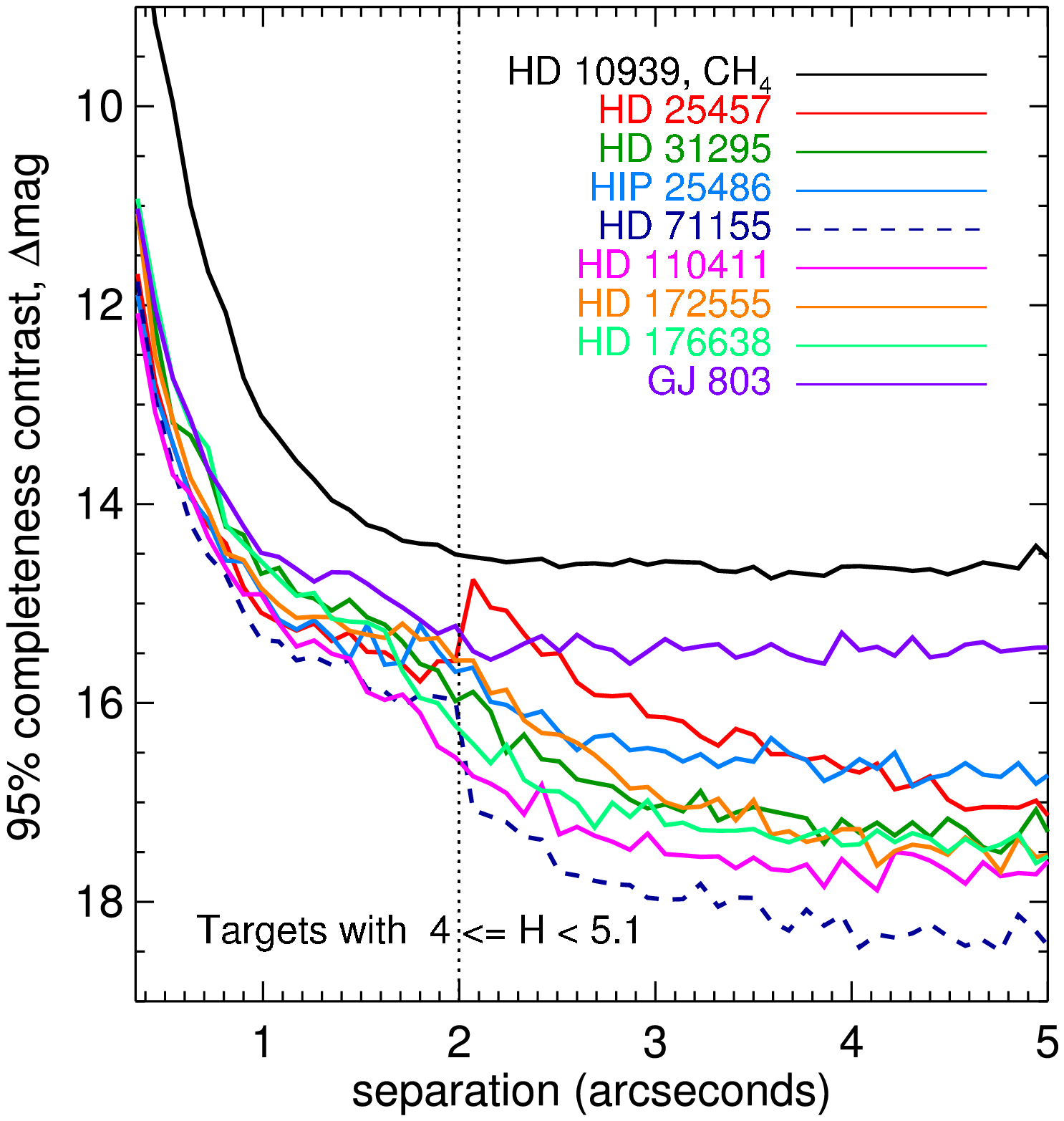}
    }  
    \hbox {
      \includegraphics[width=6.3cm]{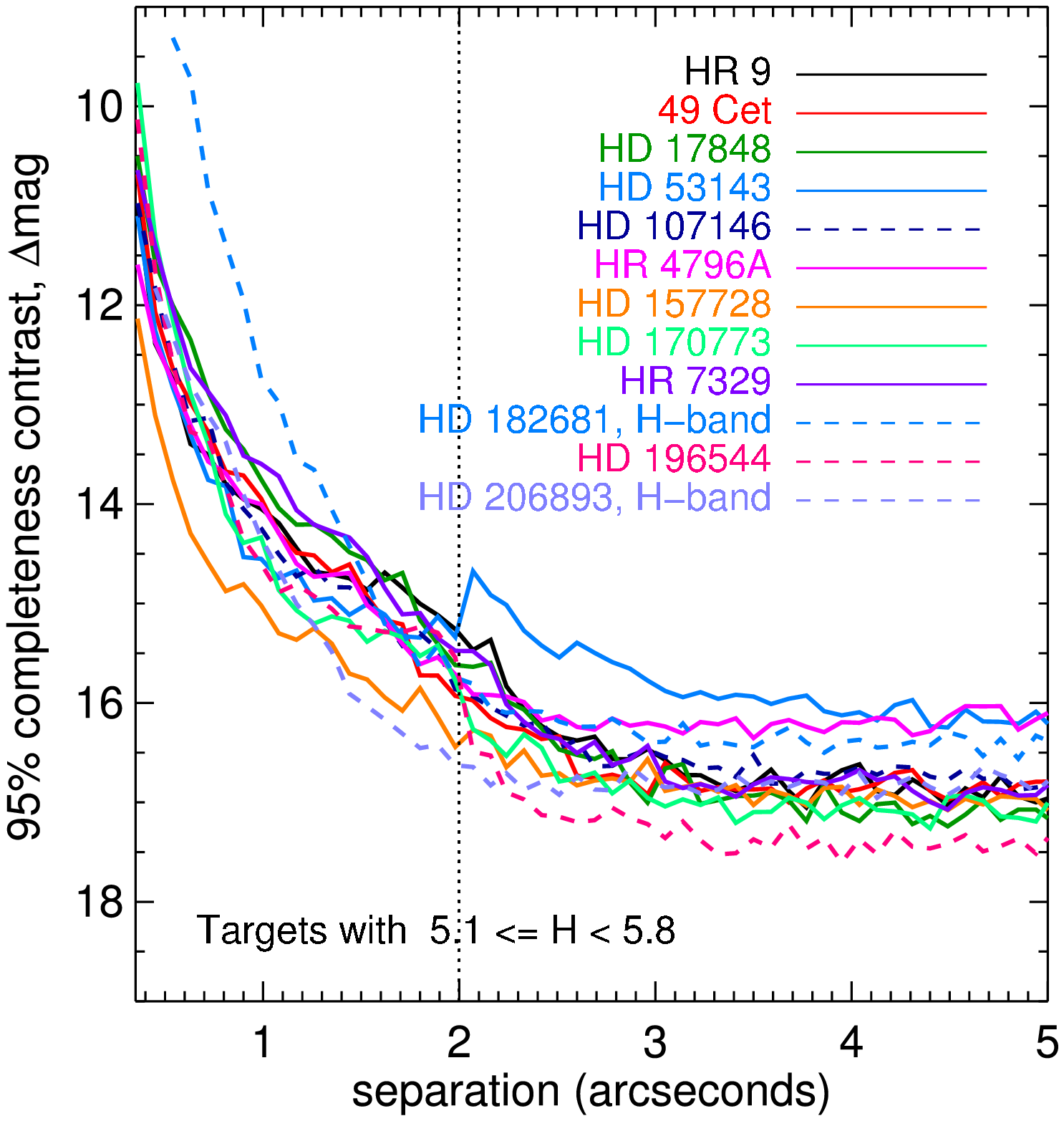}
      \includegraphics[width=6.3cm]{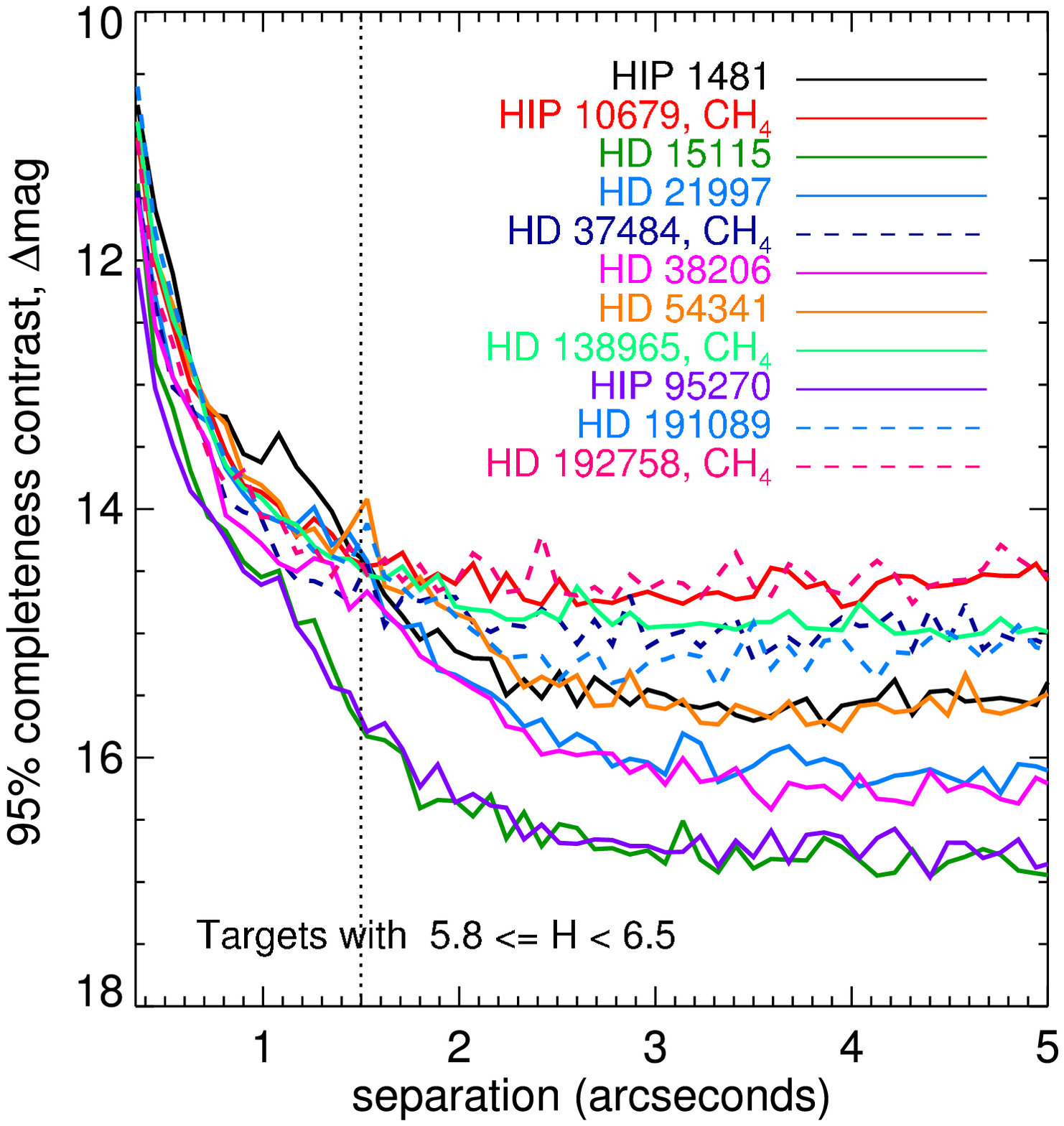}
    }  
    \hbox {
      \includegraphics[width=6.3cm]{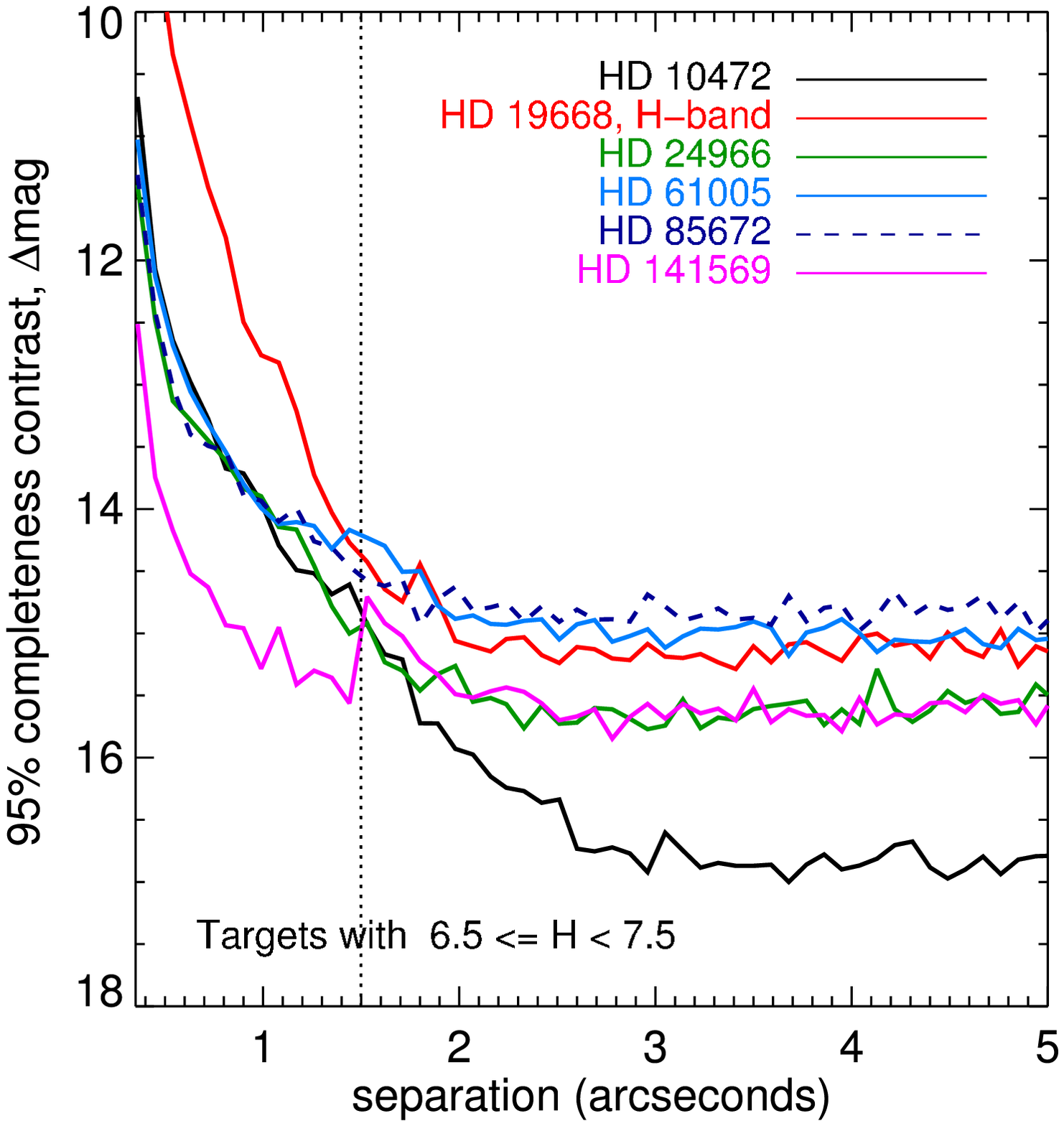}
      \includegraphics[width=6.3cm]{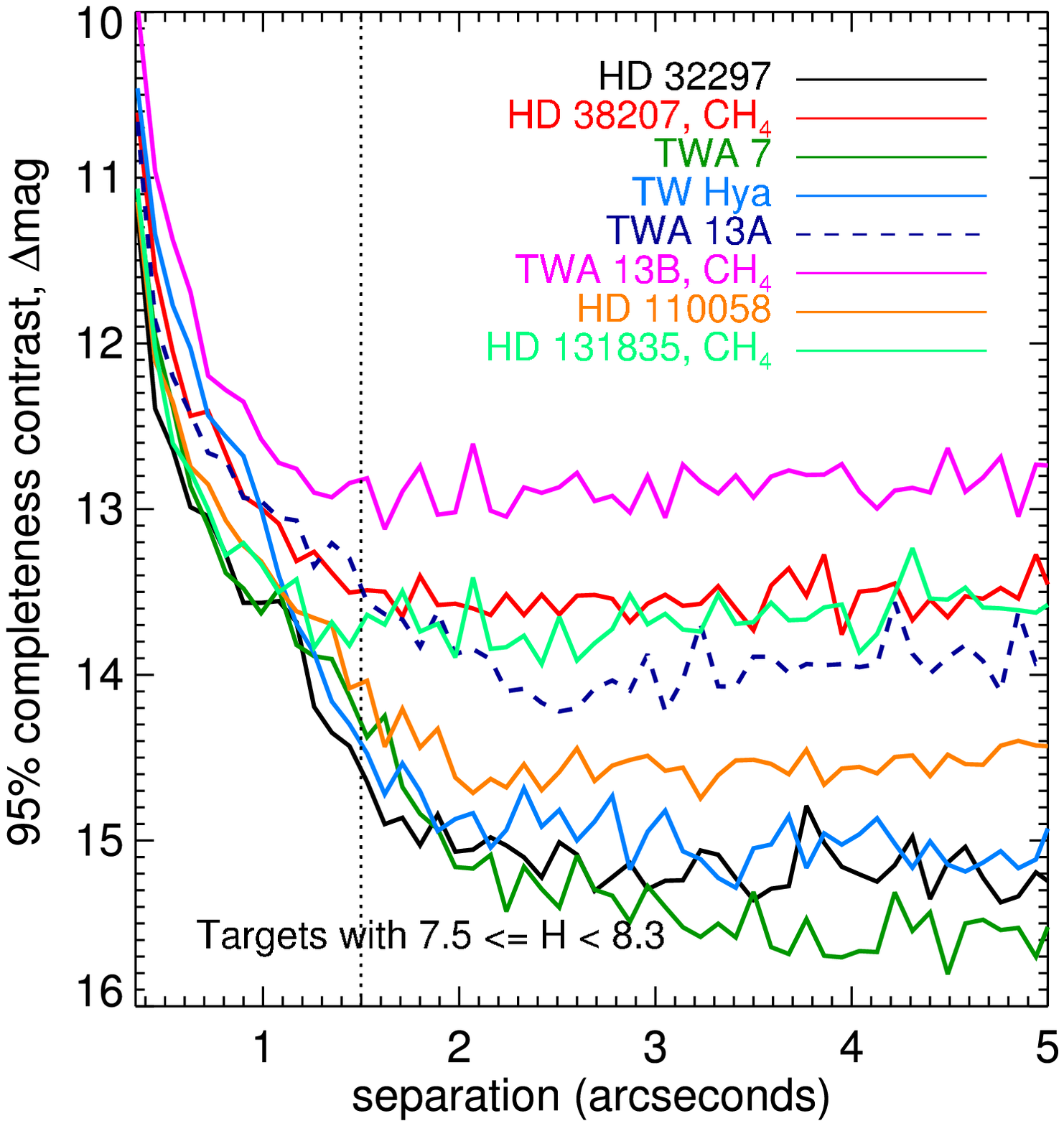}
    }  
  }}
  \caption{The contrast curves for all Campaign debris disk targets, categorized by $H$-band magnitude of the primary. 
    For separations less than \app1.5$''$, the $CH_4$ filter contrasts are
    usually better. For larger separations, the $H$-band contrasts are
    better. \textcolor{black}{In the figure above, beyond the
      dotted-line we show the $H$-band contrasts.}
    When only one filter is available, the star's name  in the legend is tagged with the filter name.}
  \label{fig:contrasts}
\end{figure} 

\begin{figure}
\centerline{
\includegraphics[width=2.0in]{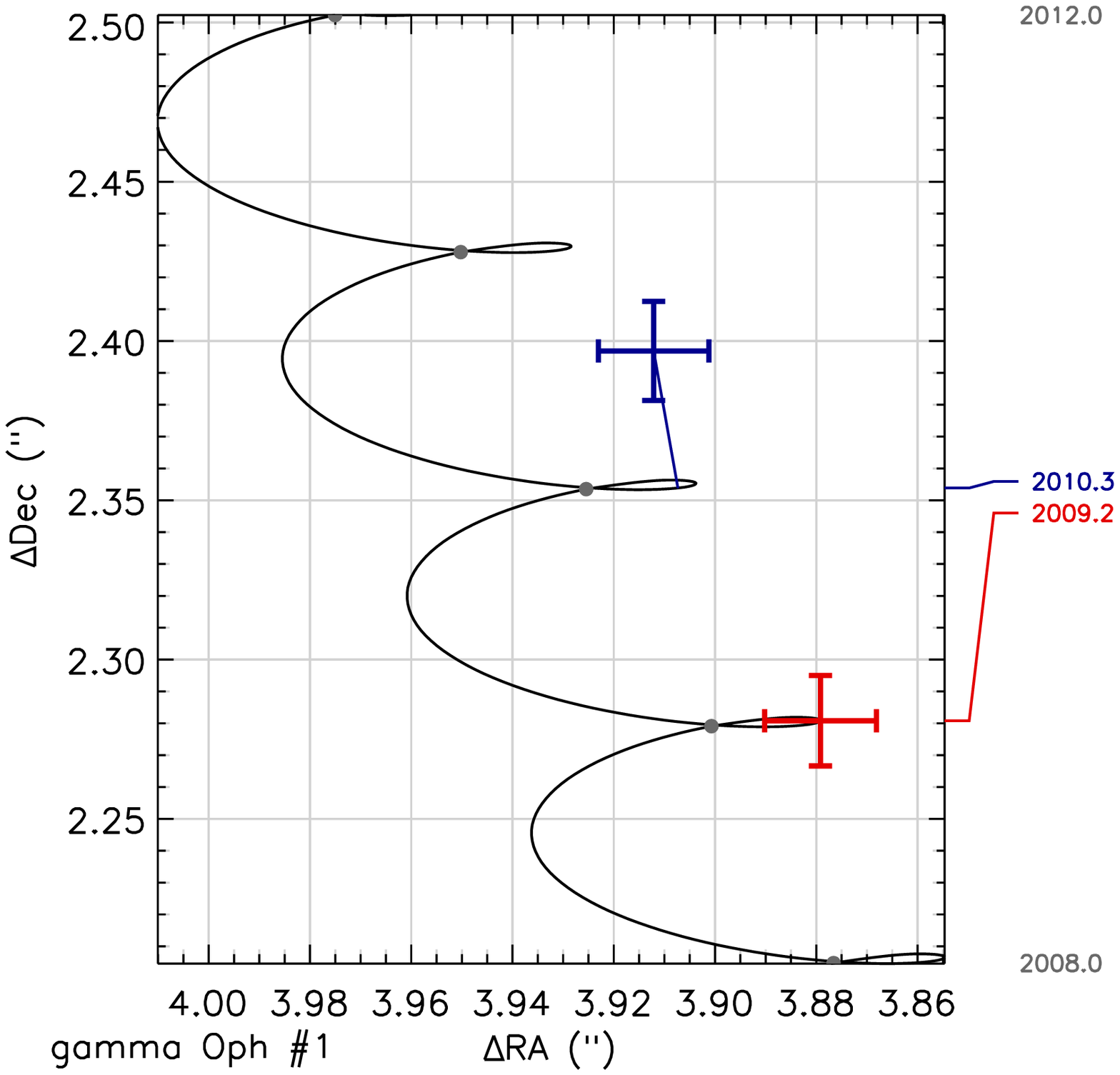}
\hskip -0.3in
\includegraphics[width=2.0in]{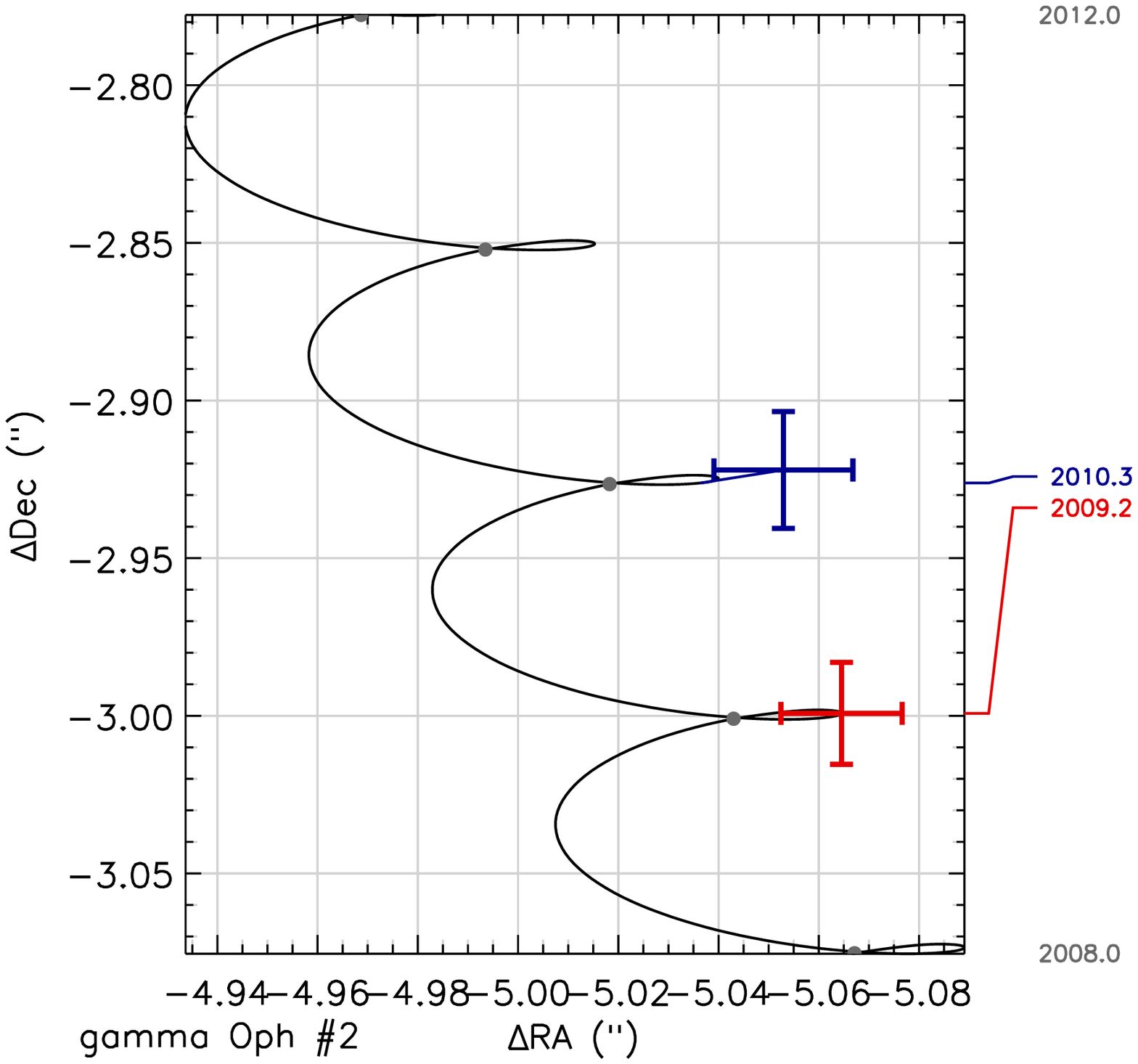}
\hskip -0.3in
\includegraphics[width=2.0in]{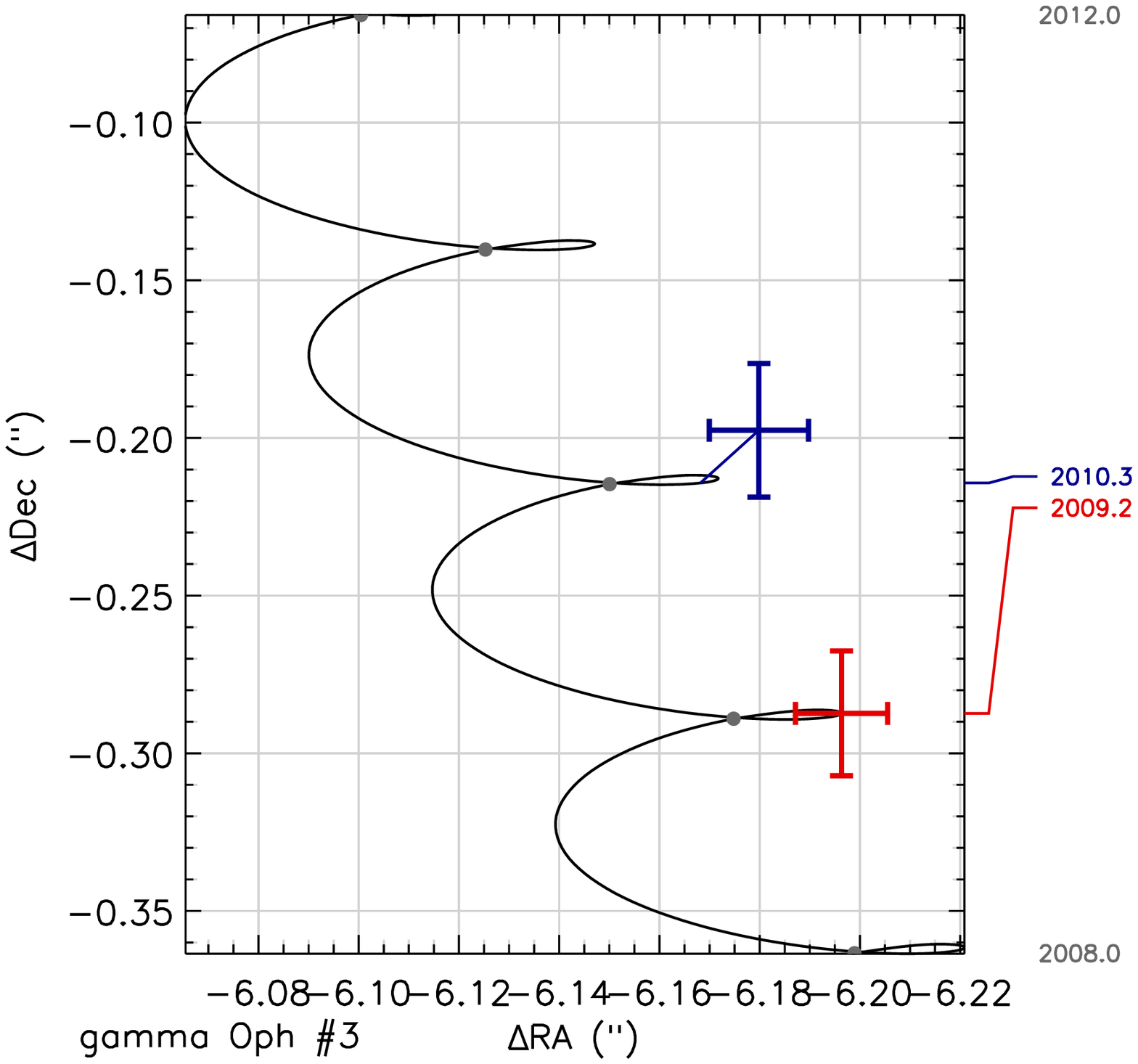}
\hskip -0.3in
\includegraphics[width=2.0in]{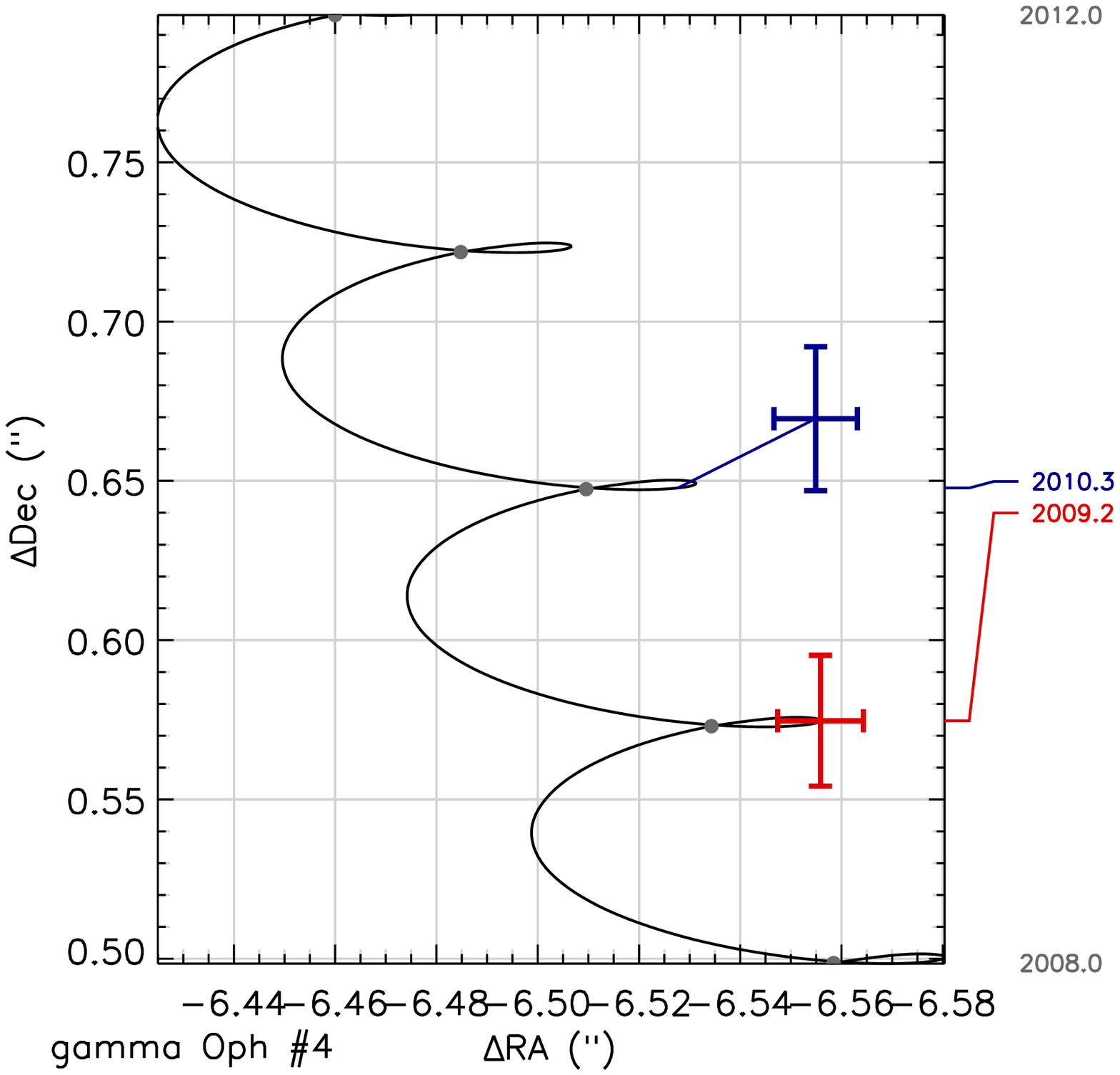}
}
\vskip -0.2in
\centerline{
\includegraphics[width=2.0in]{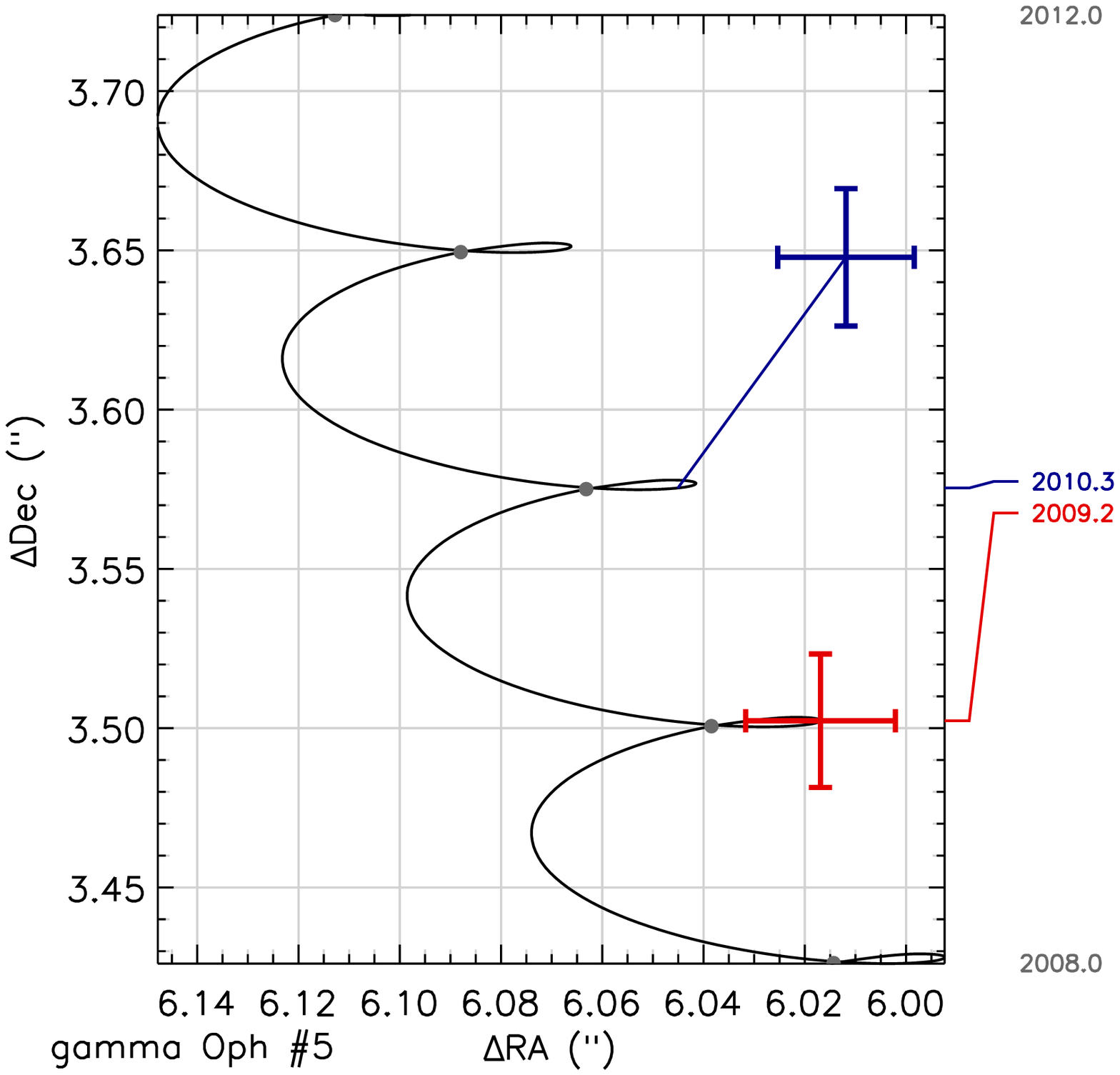}
\hskip -0.3in
\includegraphics[width=2.0in]{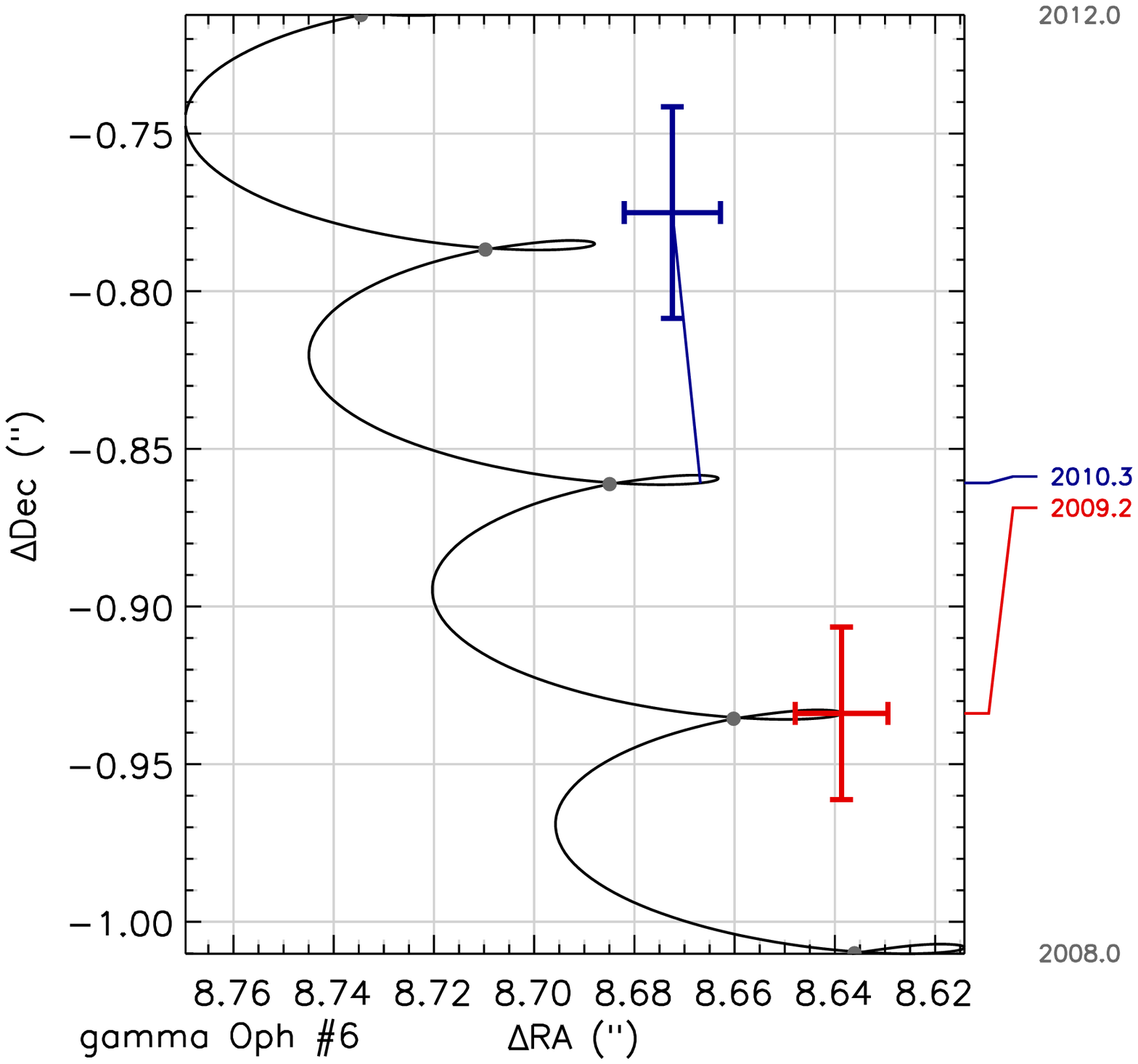}
\hskip -0.3in
\includegraphics[width=2.0in]{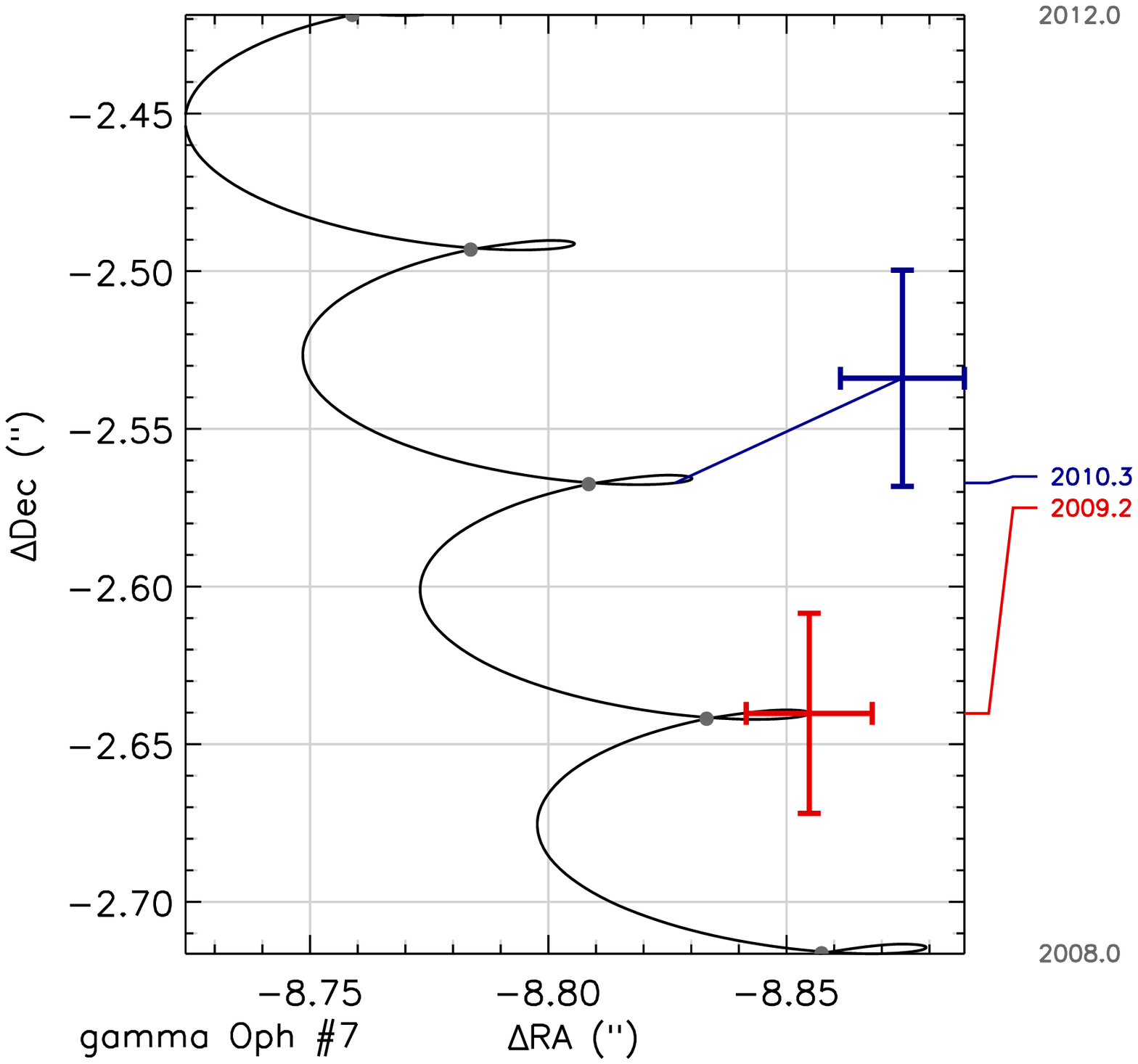}
\hskip -0.3in
\includegraphics[width=2.0in]{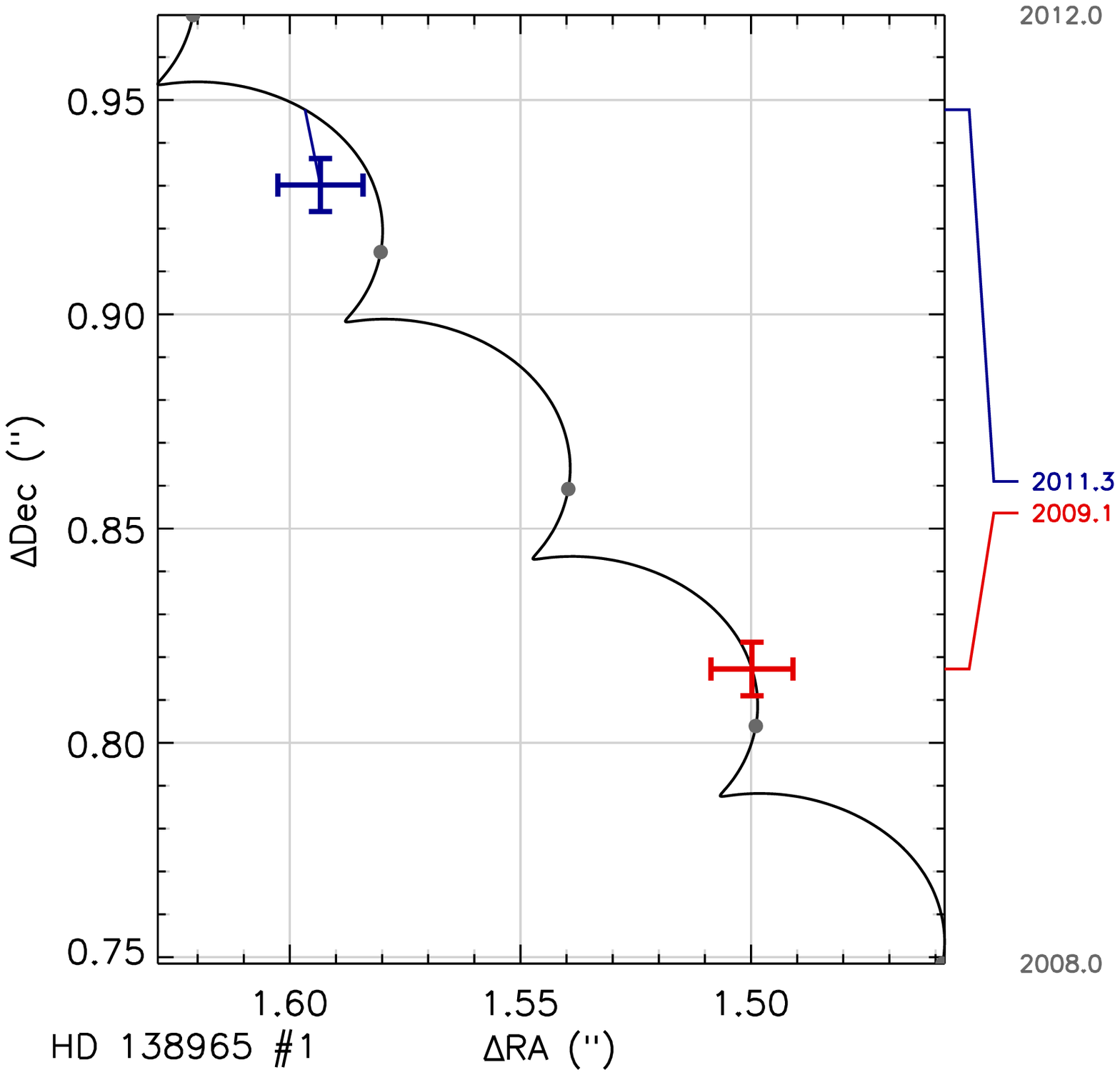}
}
\vskip -0.2in
\centerline{
\includegraphics[width=2.0in]{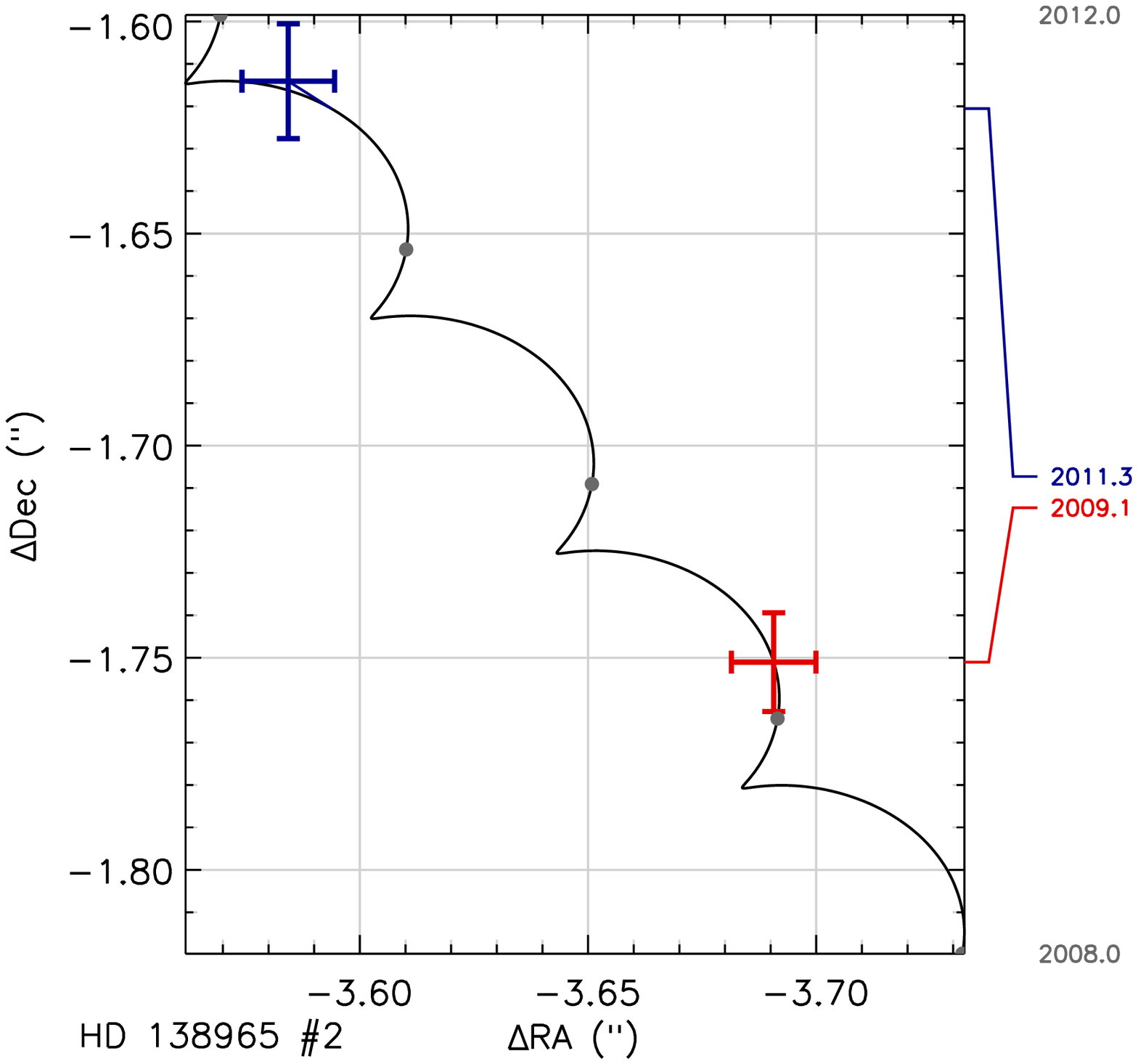}
\hskip -0.3in
\includegraphics[width=2.0in]{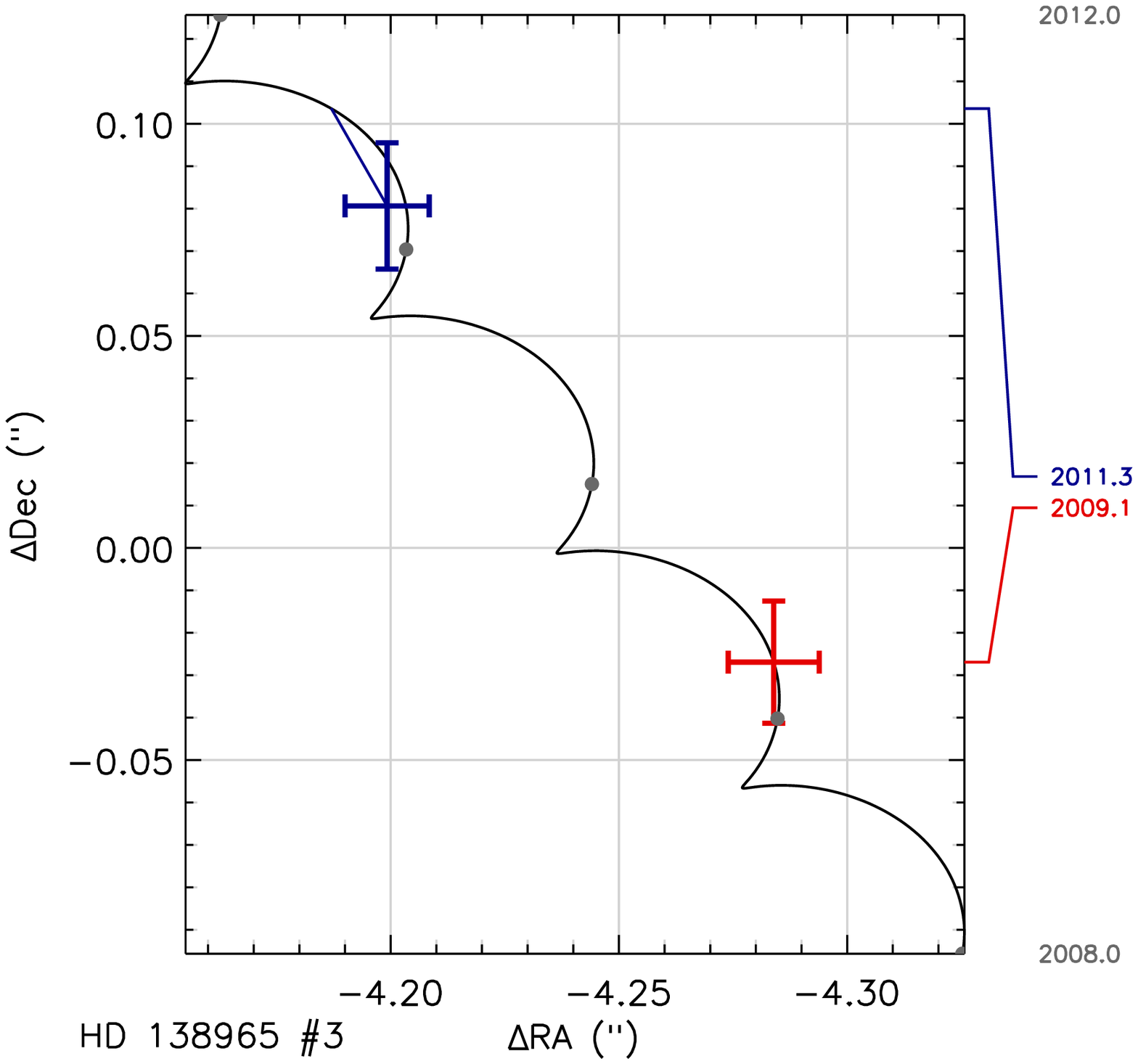}
\hskip -0.3in
\includegraphics[width=2.0in]{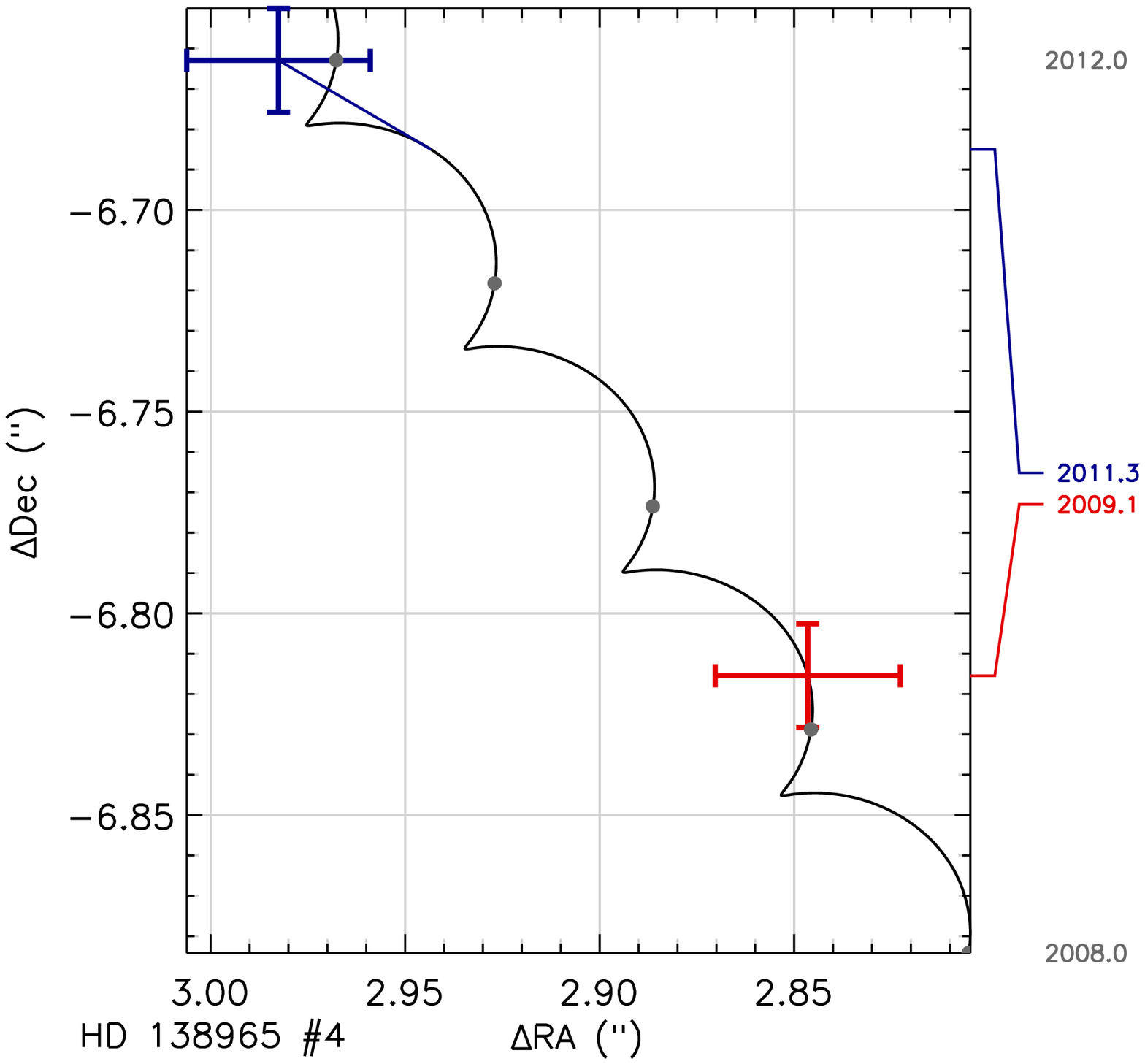}
\hskip -0.3in
\includegraphics[width=2.0in]{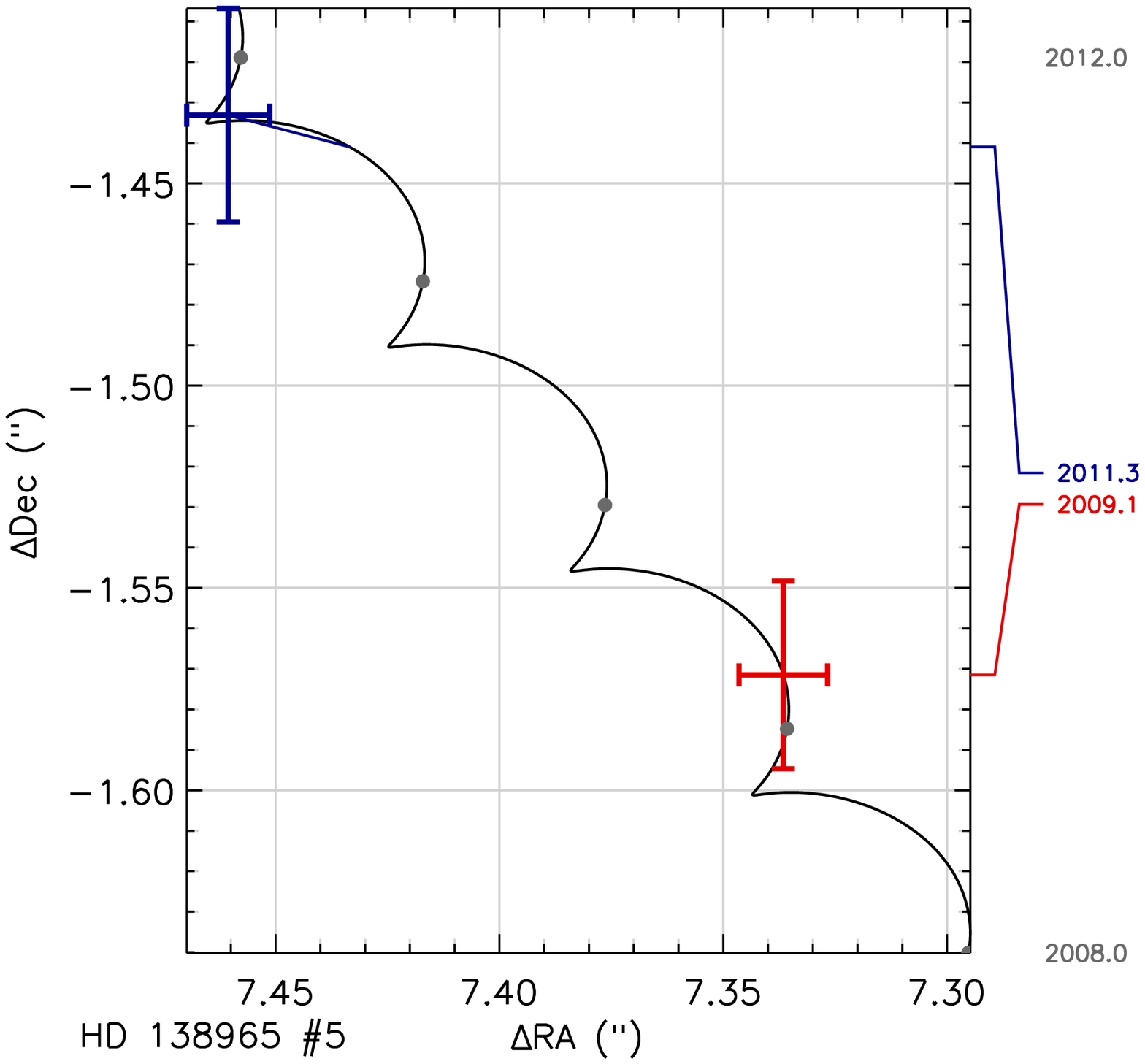}
}
\vskip -0.2in
\centerline{
\includegraphics[width=2.0in]{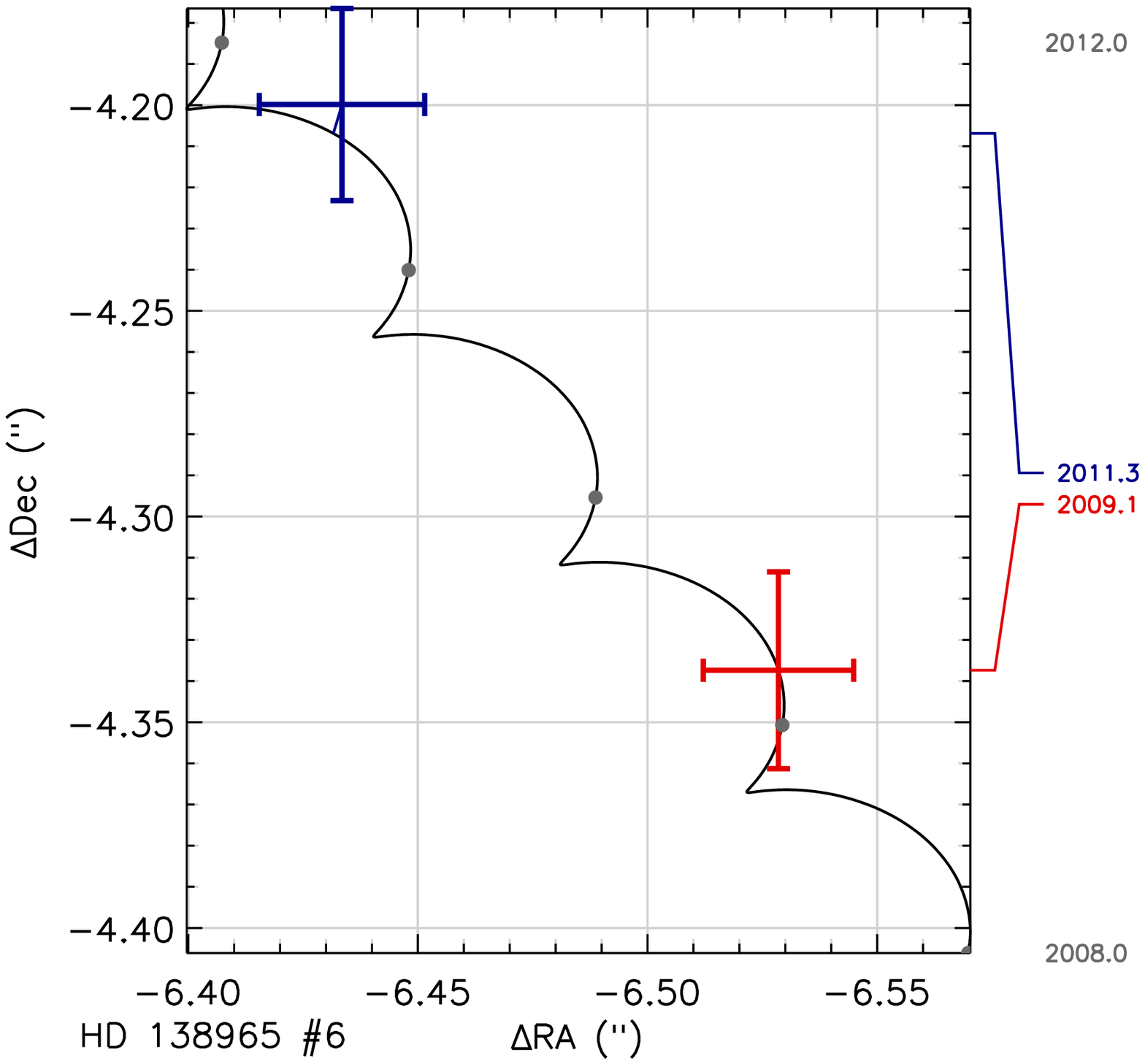}
\hskip -0.3in
\includegraphics[width=2.0in]{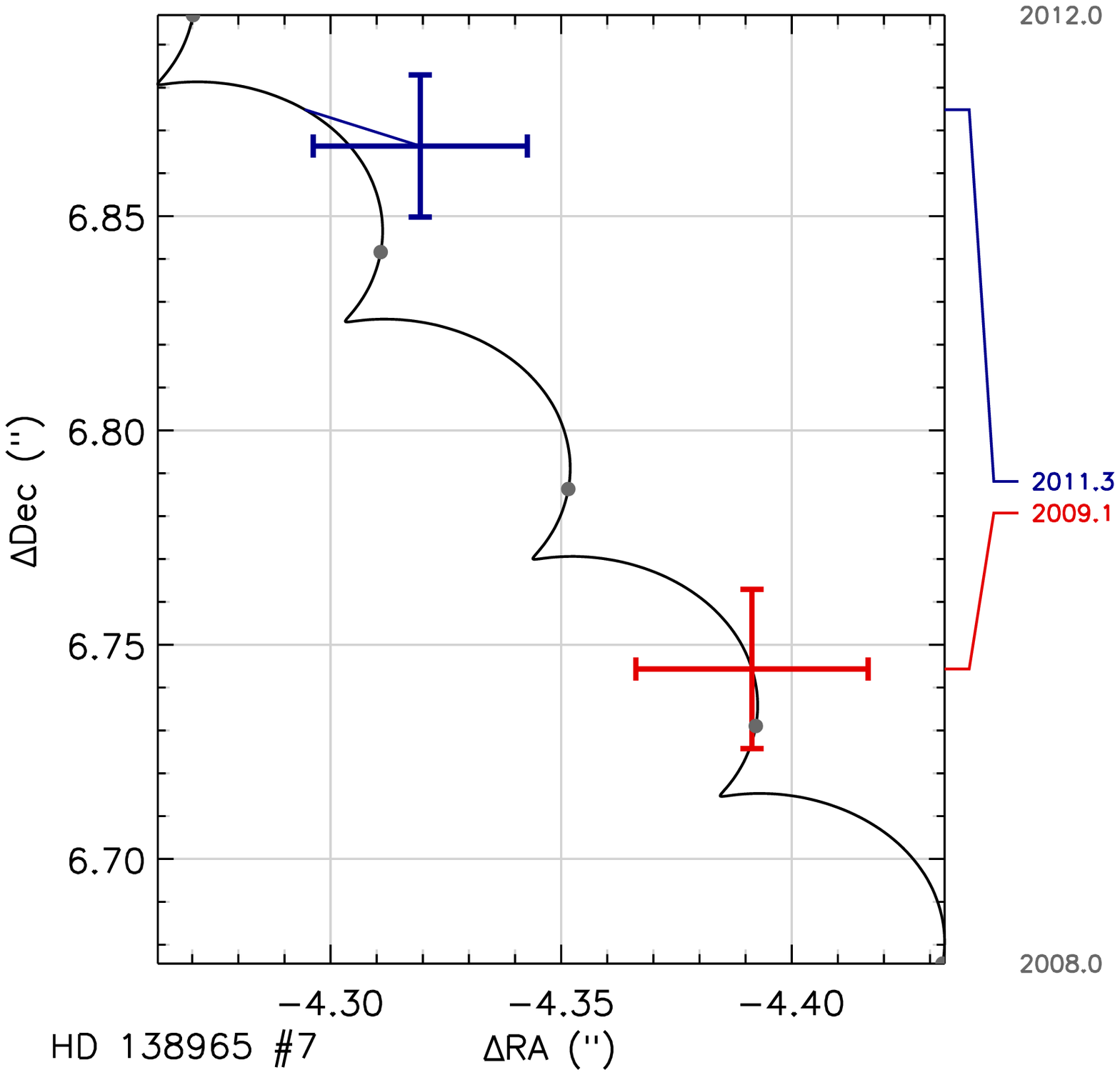}
\hskip -0.3in
\includegraphics[width=2.0in]{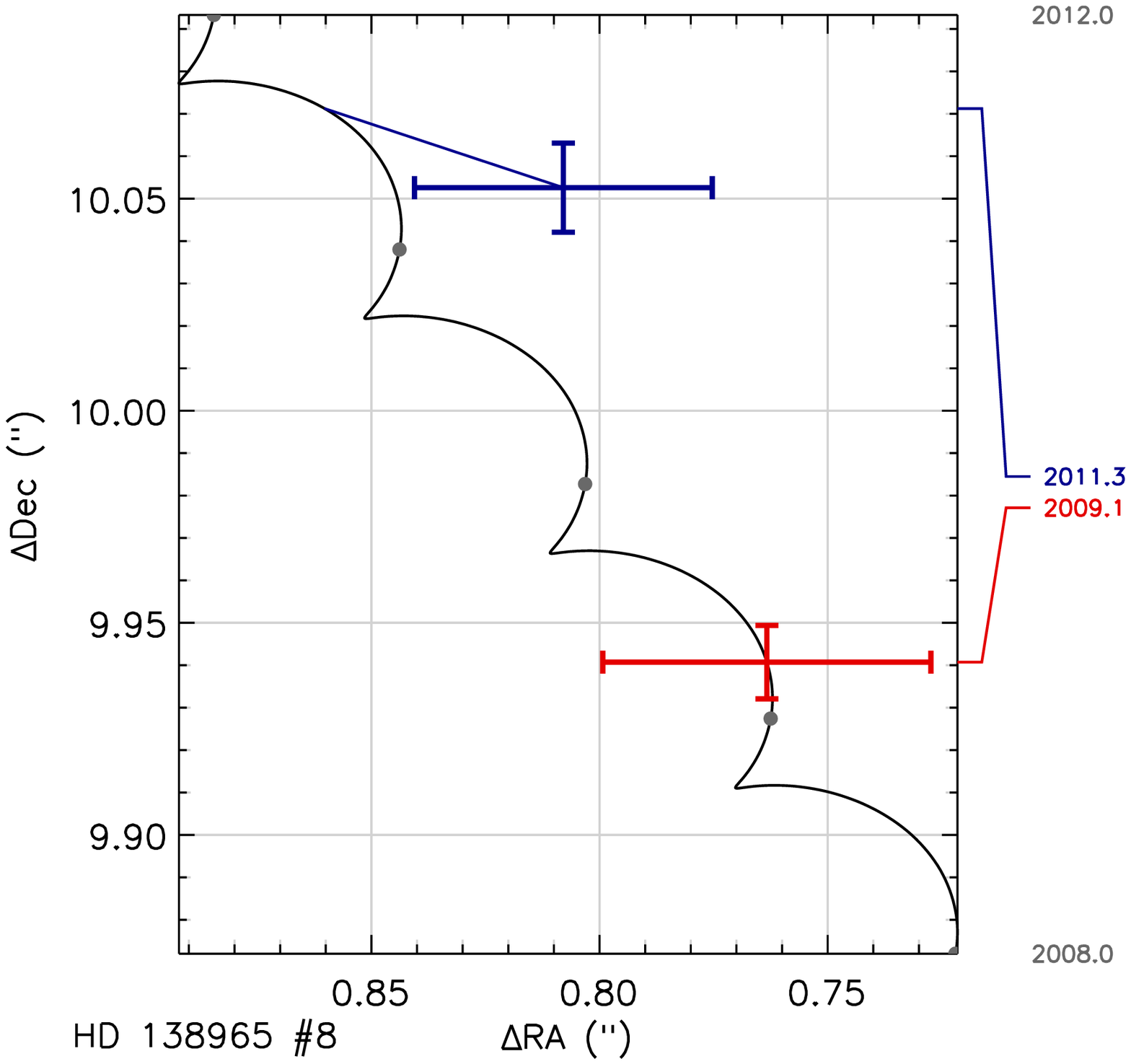}
\hskip -0.3in
\includegraphics[width=2.0in]{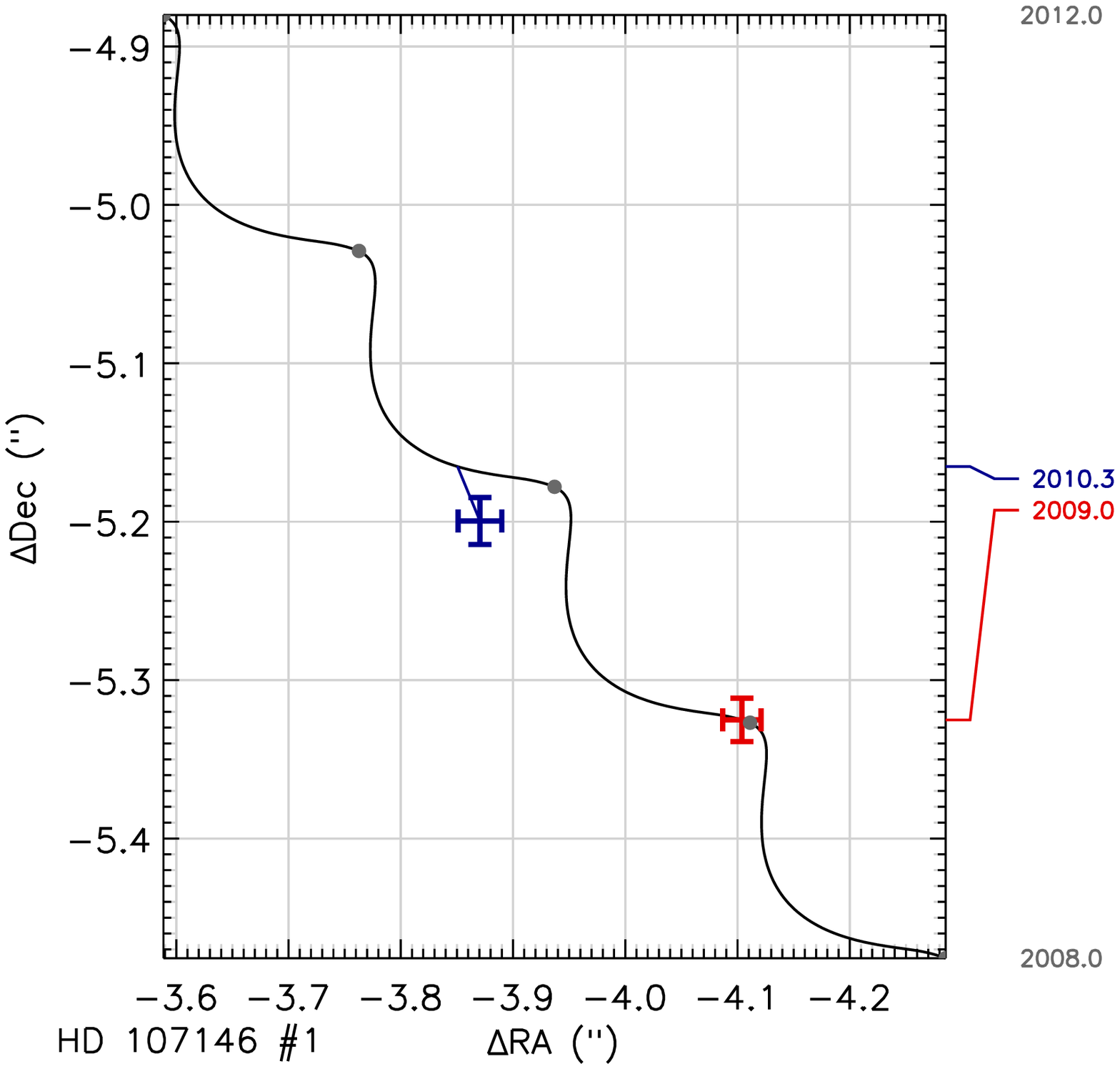}
}
\vskip -0.2in
\caption{The change in RA and DEC of candidate companions around our target stars compared with the expected change for a background object, given the parallax and proper motion of the primary star. The error bars represent the uncertainty in our astrometry.}
\label{fig:cand_motion}
\end{figure} 

\begin{figure}
\centerline{
\includegraphics[width=2.0in]{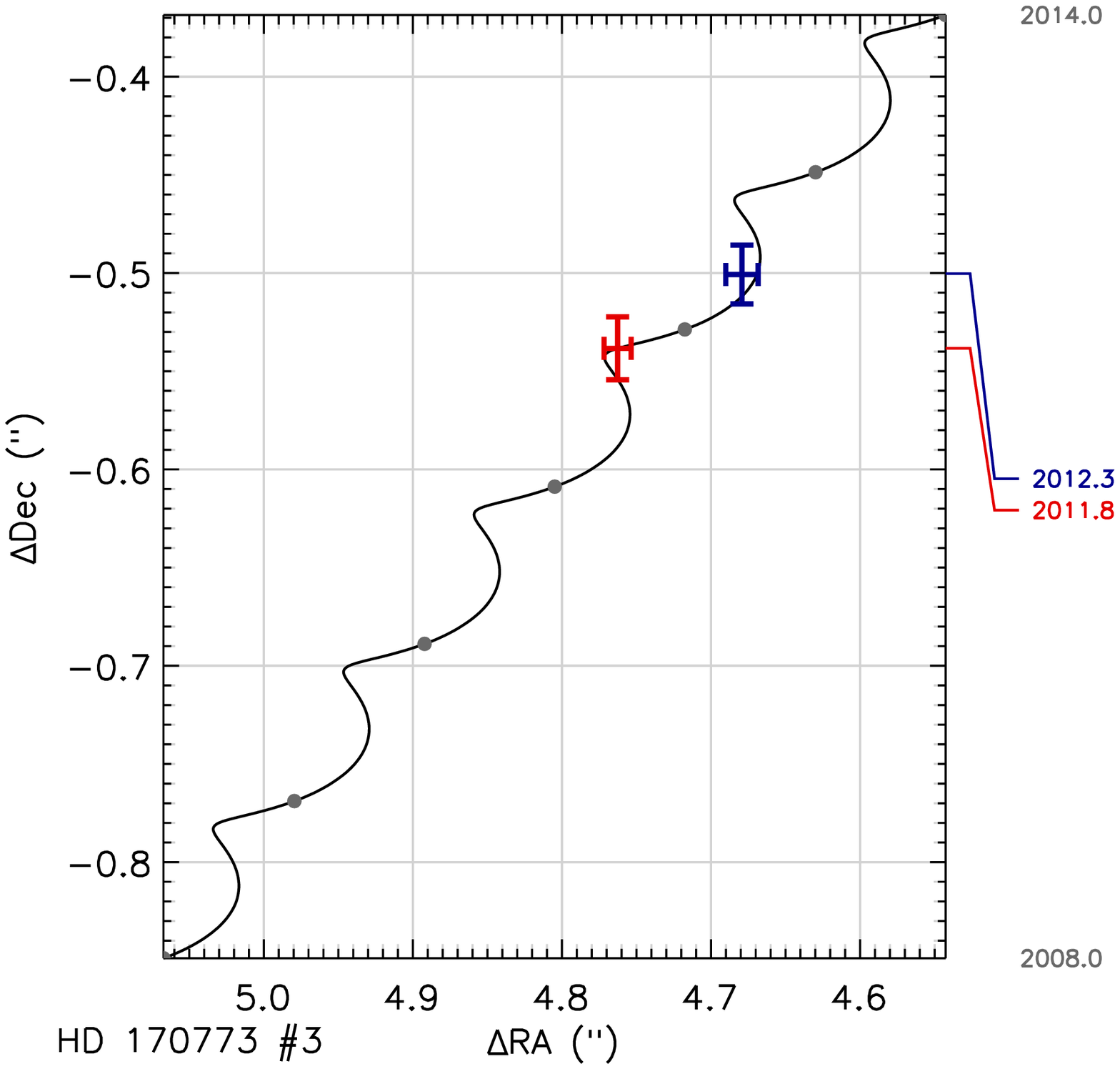}
\hskip -0.3in
\includegraphics[width=2.0in]{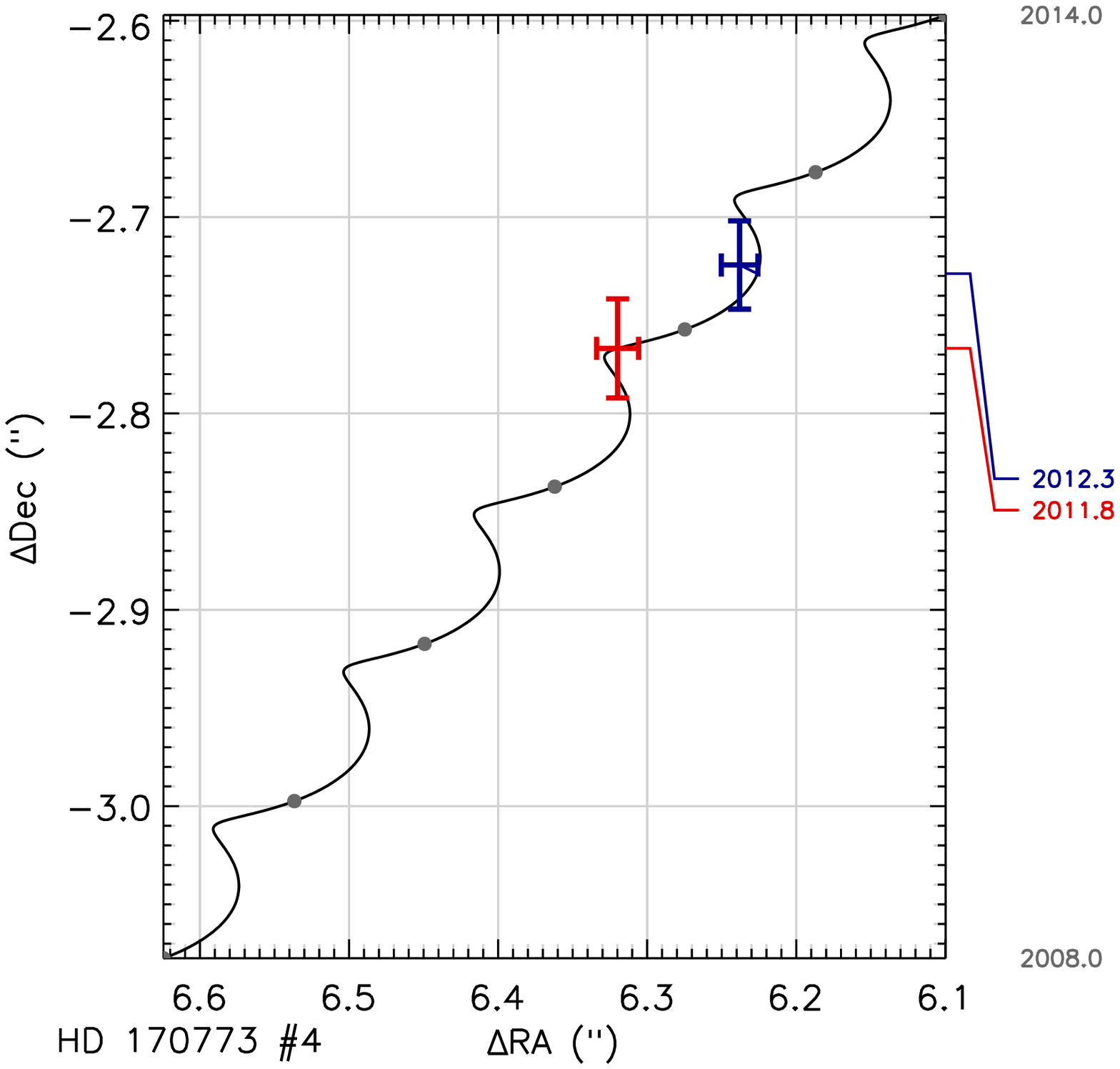}
\hskip -0.3in
\includegraphics[width=2.0in]{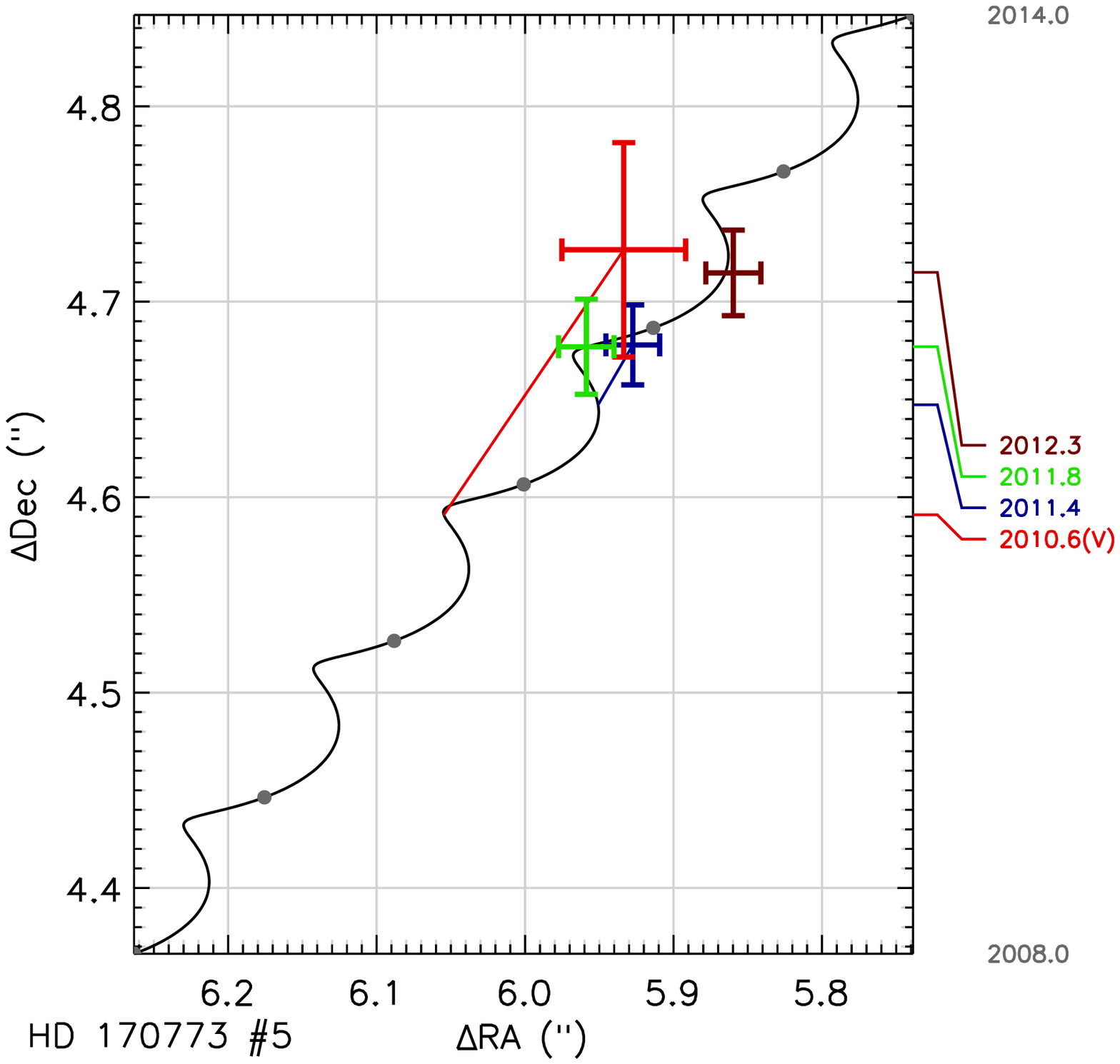}
\hskip -0.3in
\includegraphics[width=2.0in]{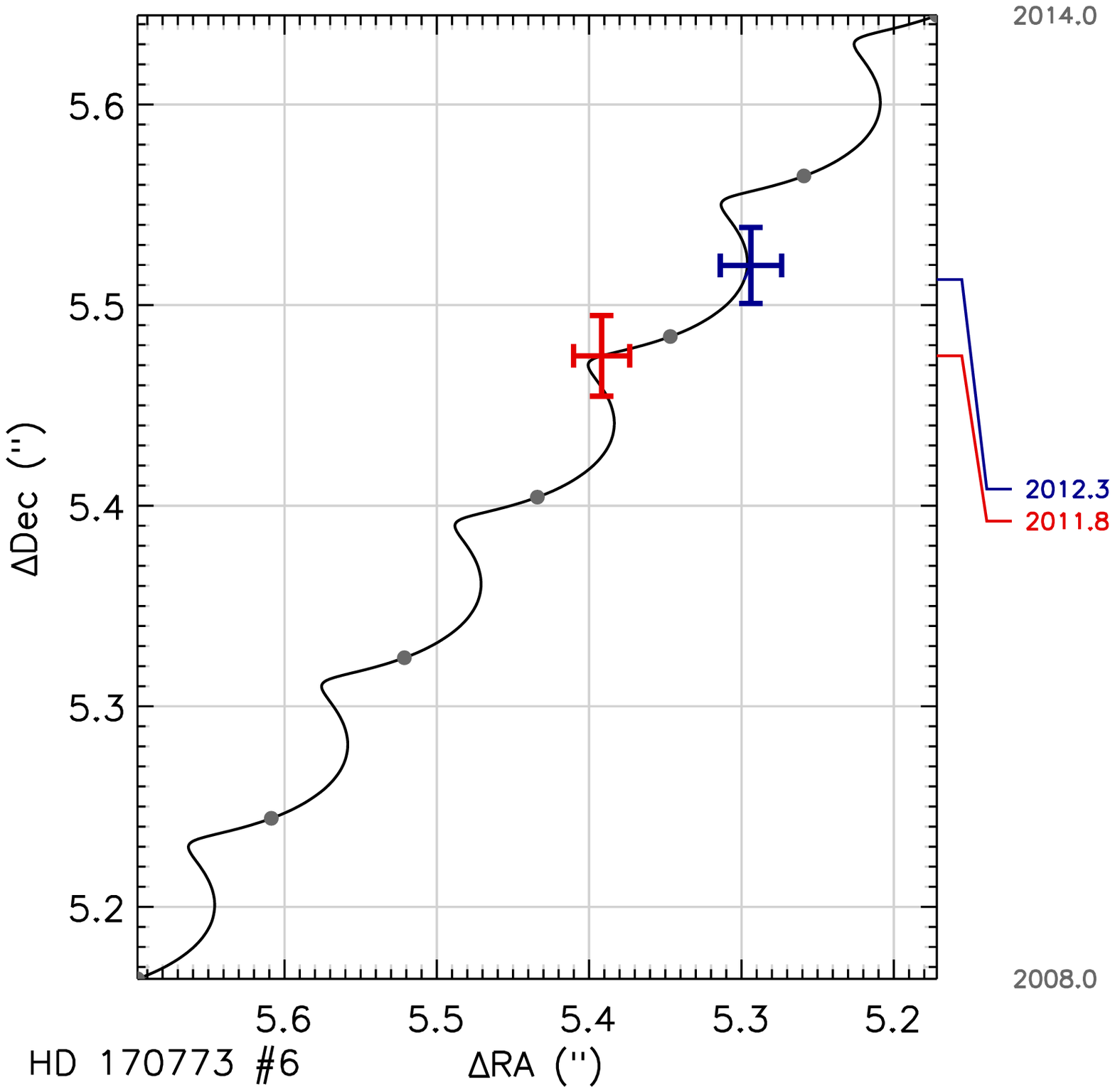}
}
\vskip -0.2in
\centerline{
\includegraphics[width=2.0in]{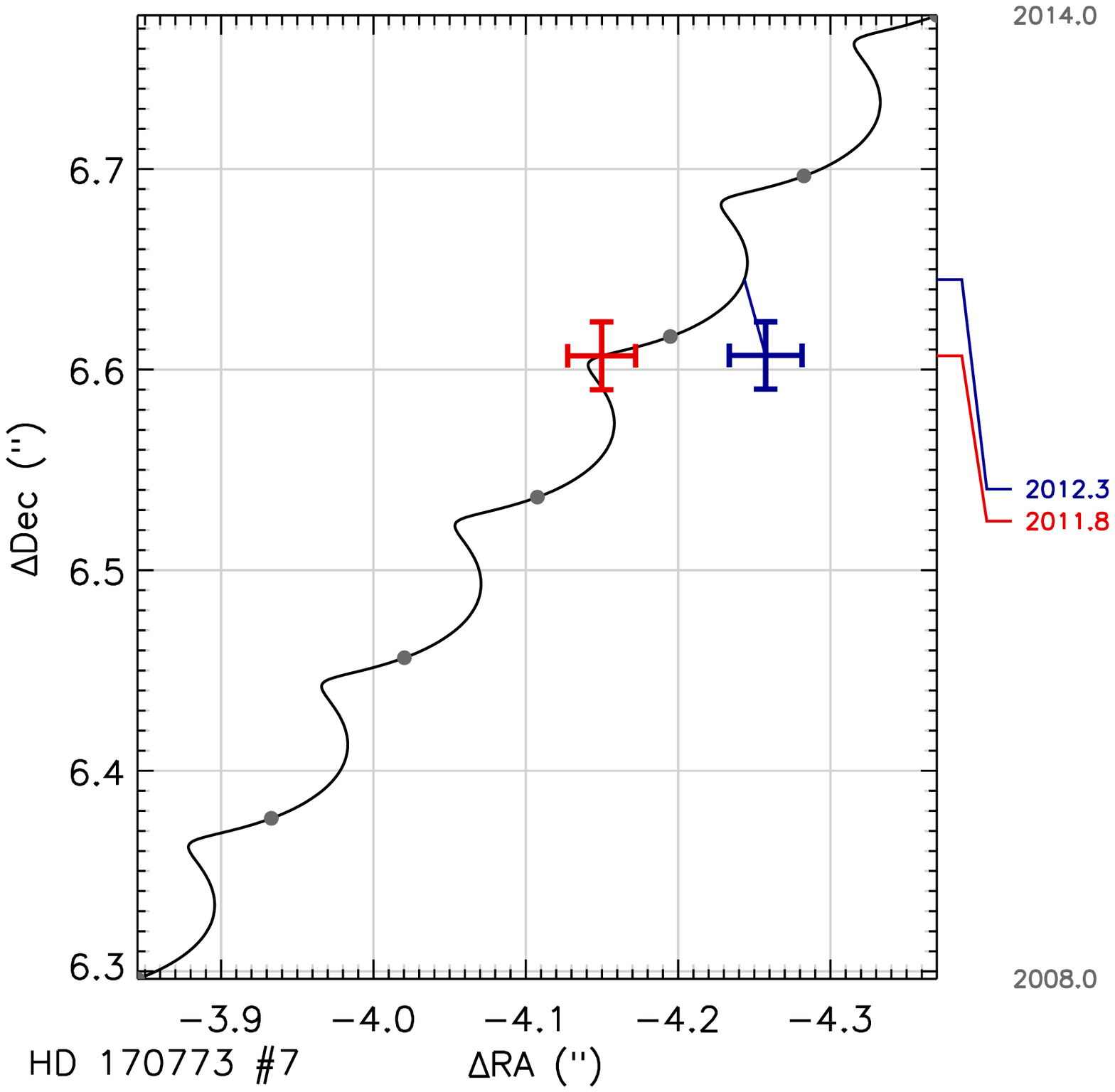}
\hskip -0.3in
\includegraphics[width=2.0in]{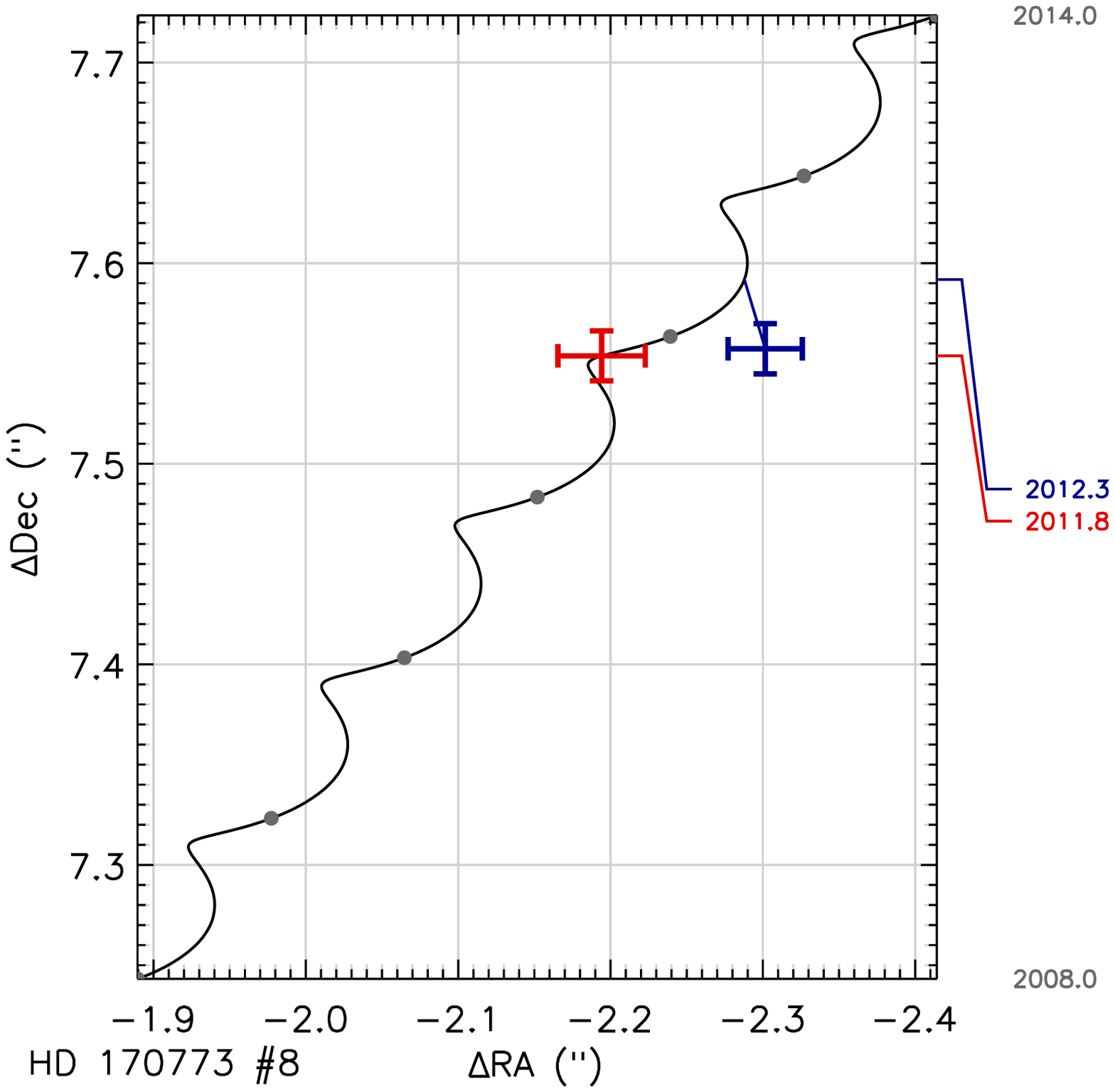}
\hskip -0.3in
\includegraphics[width=2.0in]{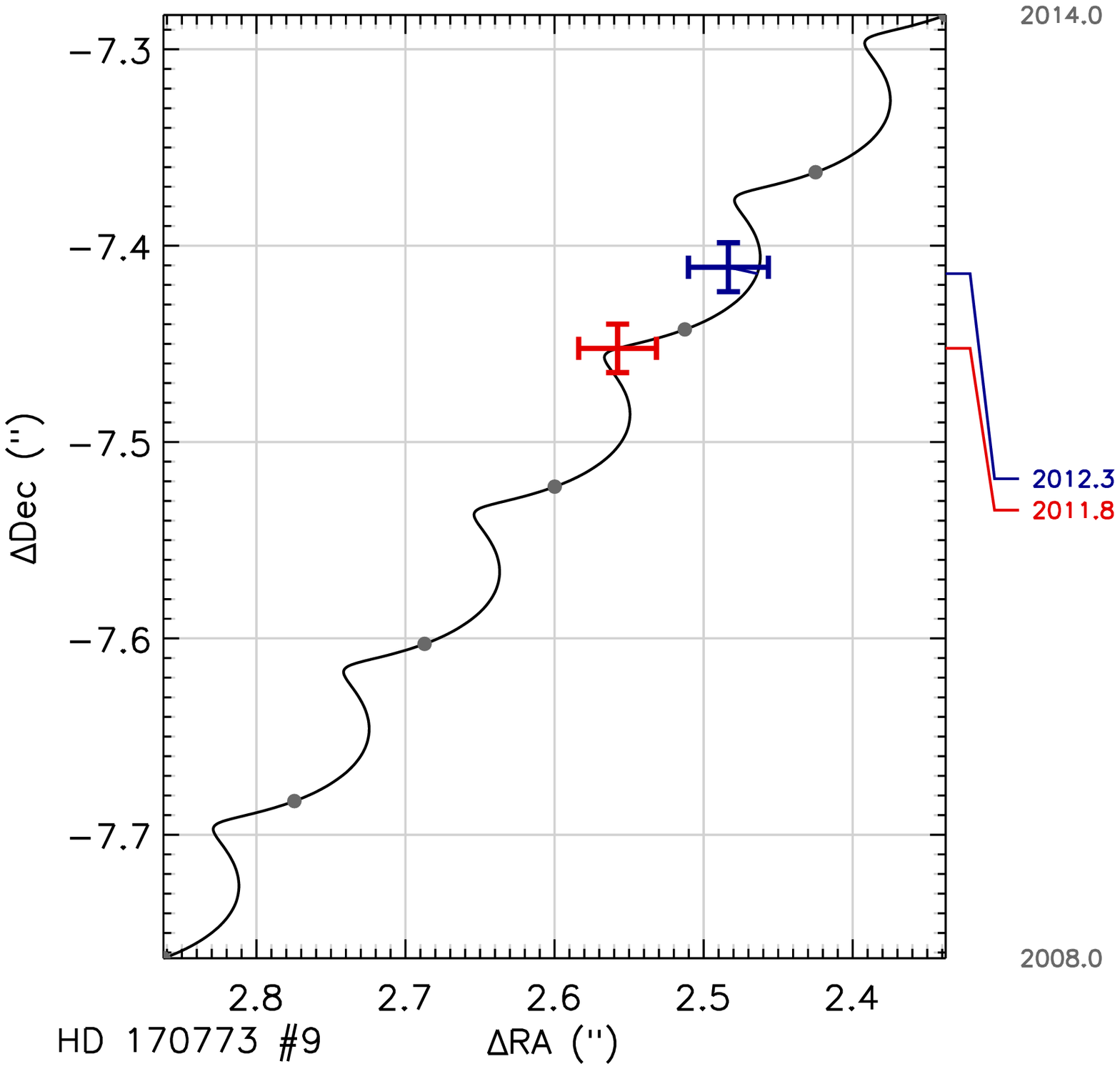}
\hskip -0.3in
\includegraphics[width=2.0in]{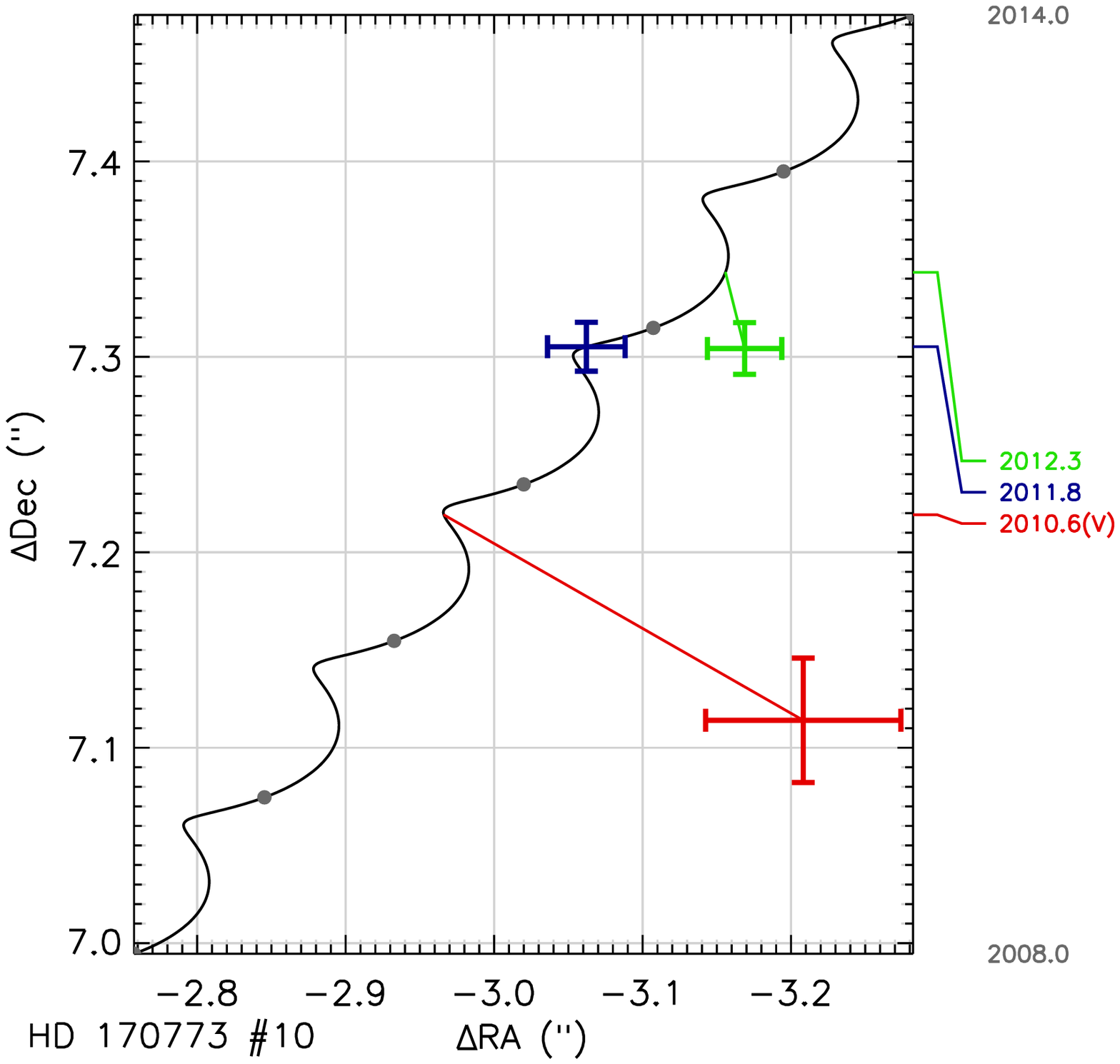}
}
\vskip -0.2in
\centerline{
\includegraphics[width=2.0in]{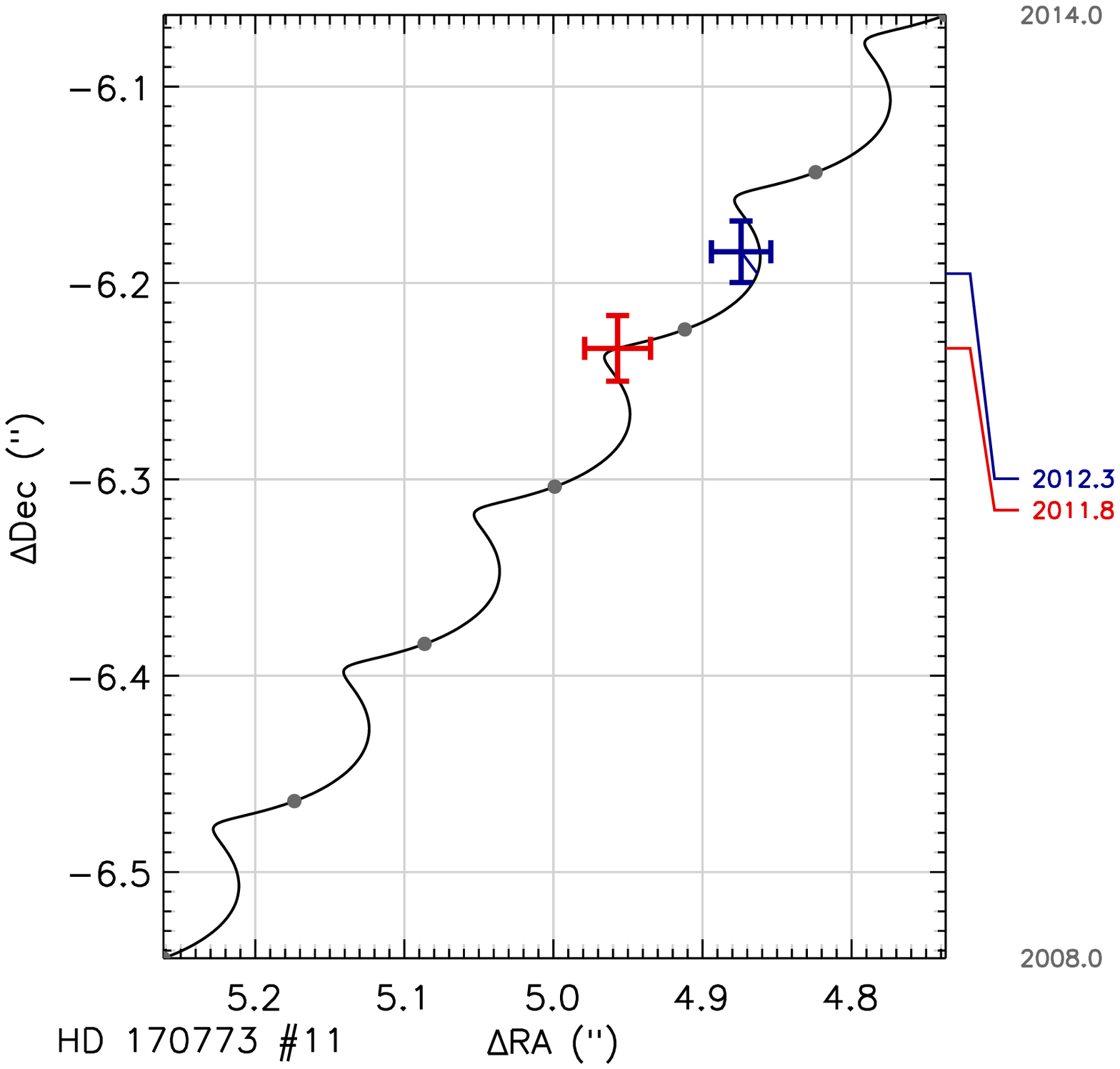}
\hskip -0.3in
\includegraphics[width=2.0in]{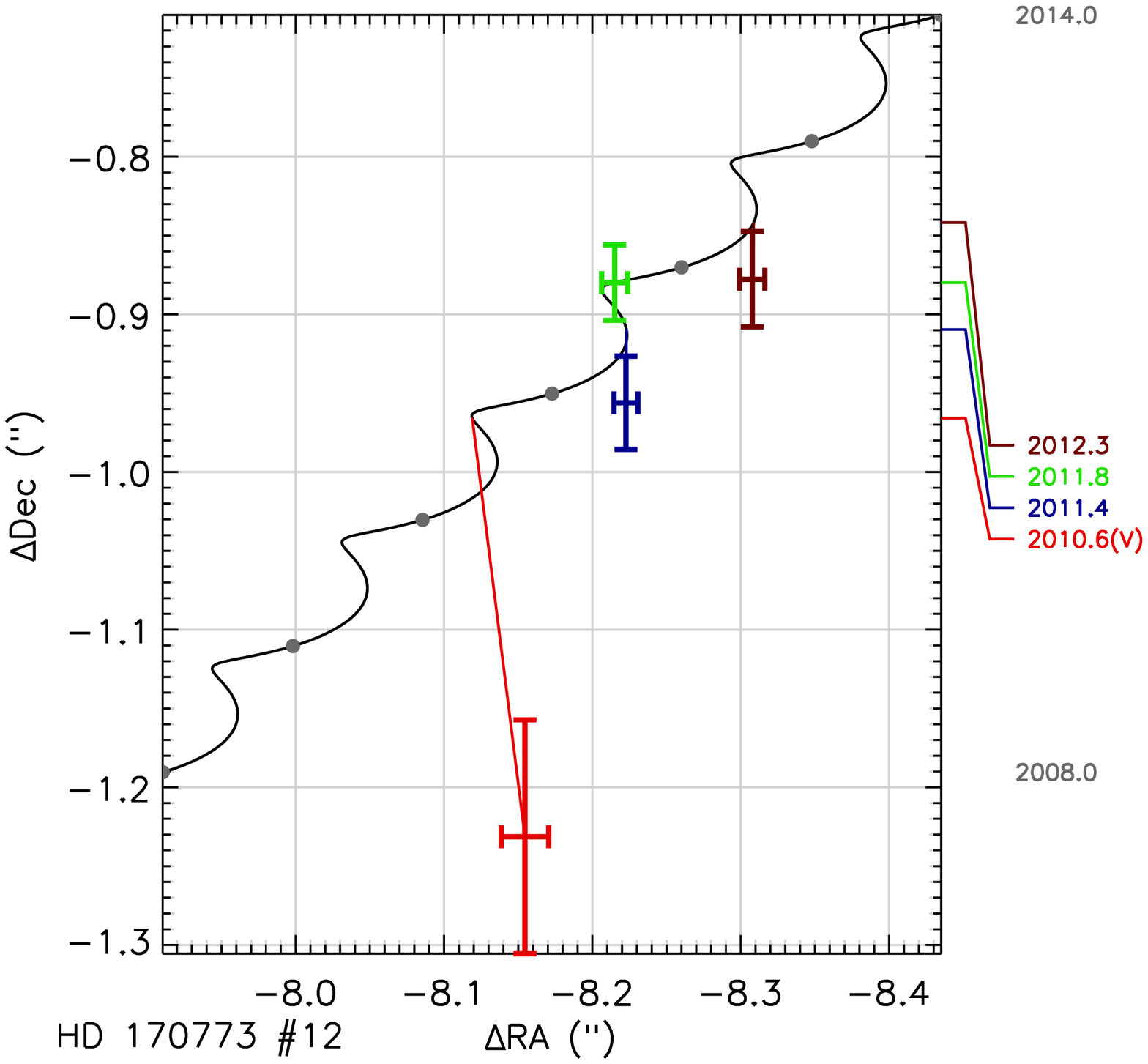}
\hskip -0.3in
\includegraphics[width=2.0in]{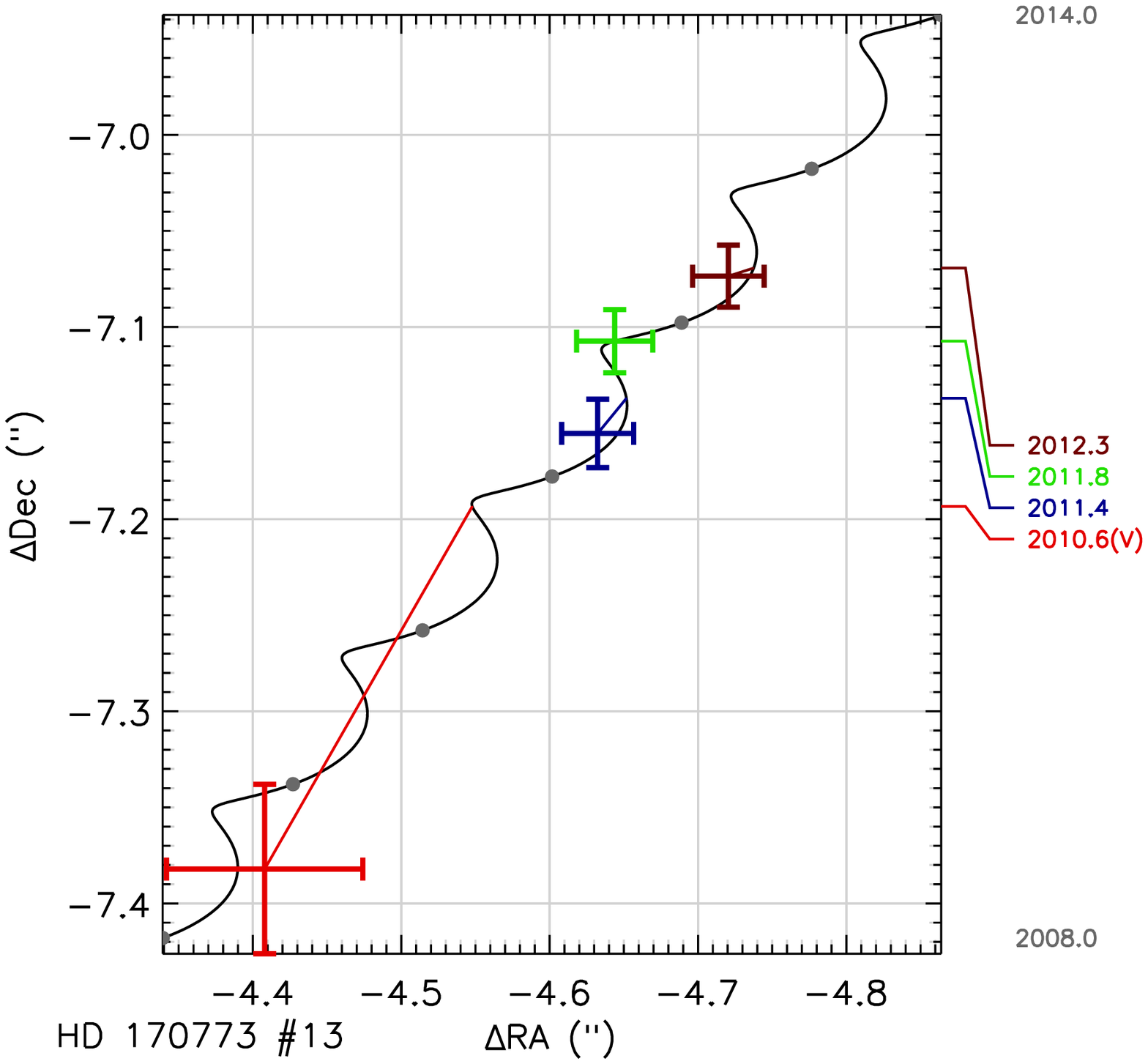}
\hskip -0.3in
\includegraphics[width=2.0in]{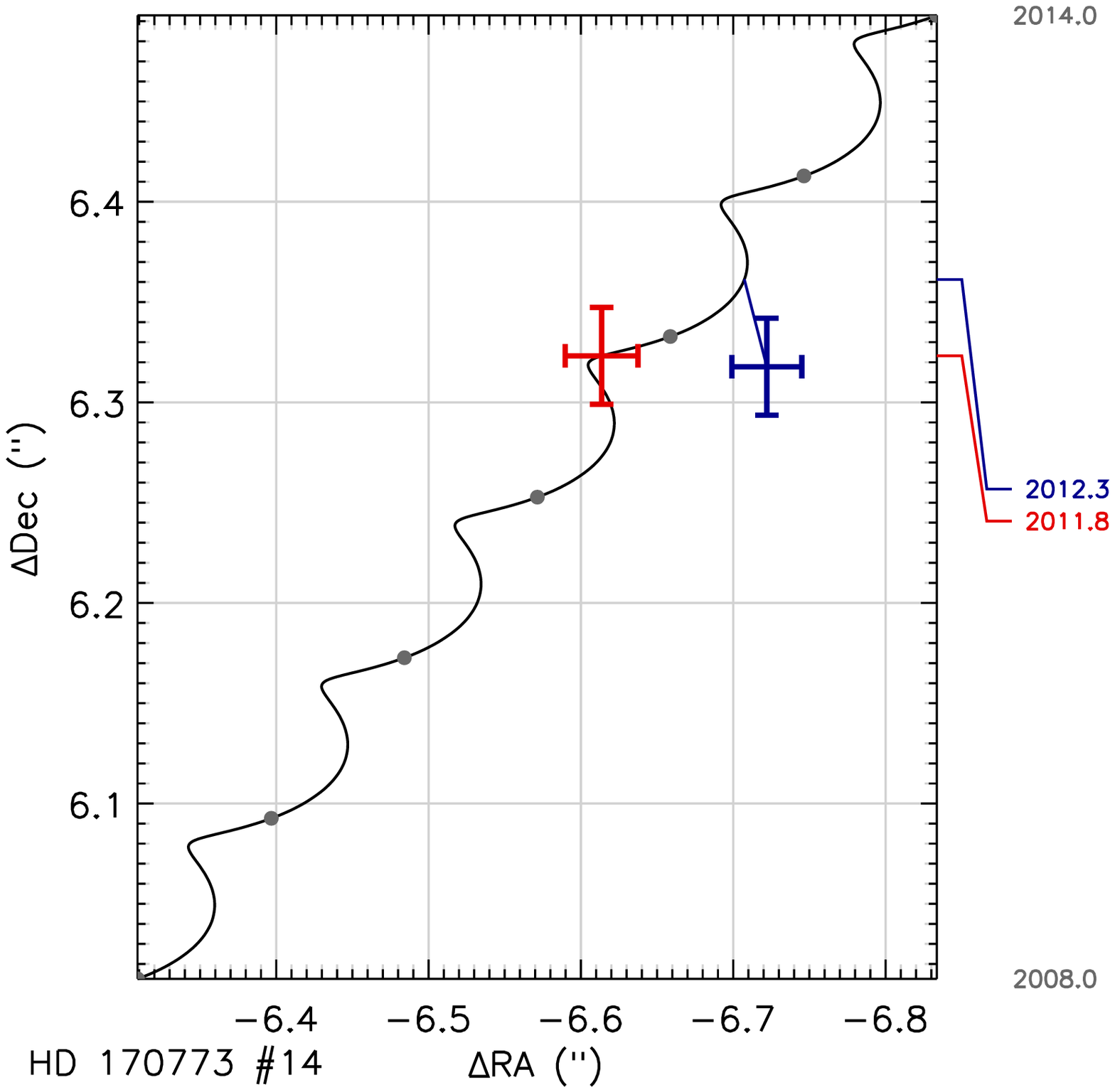}
}
\vskip -0.2in
\centerline{
\includegraphics[width=2.0in]{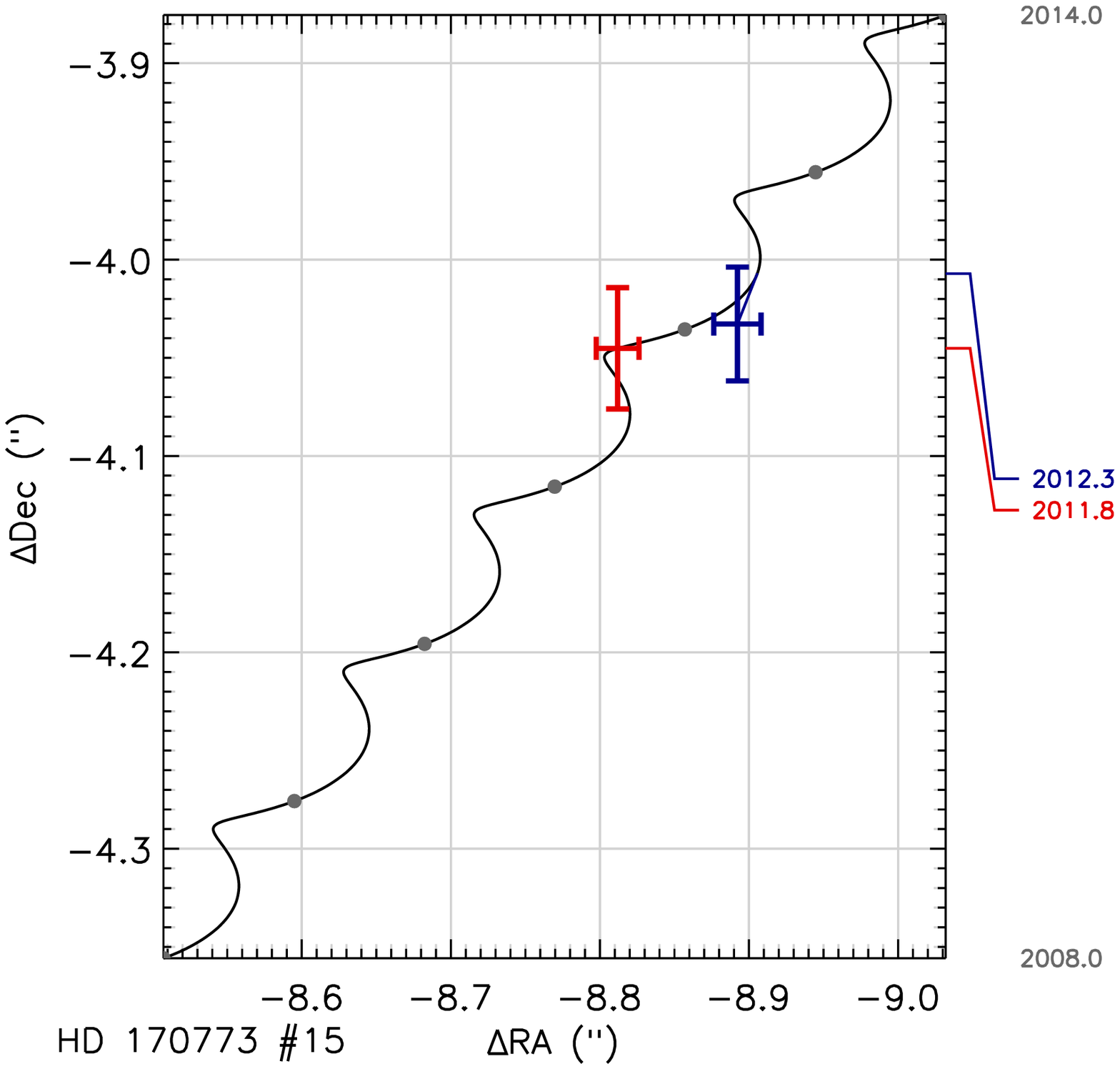}
\hskip -0.3in
\includegraphics[width=2.0in]{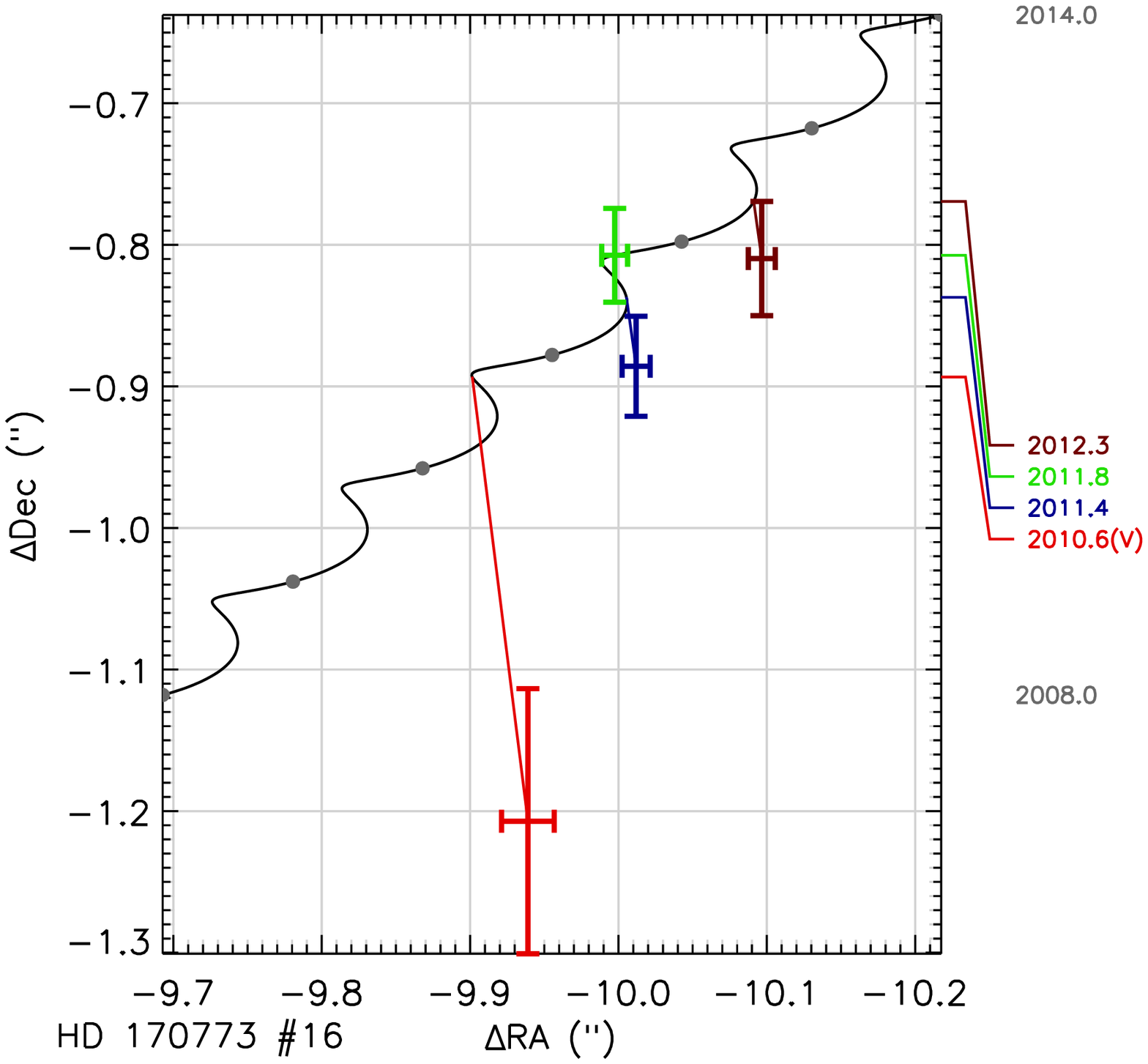}
\hskip -0.3in
\includegraphics[width=2.0in]{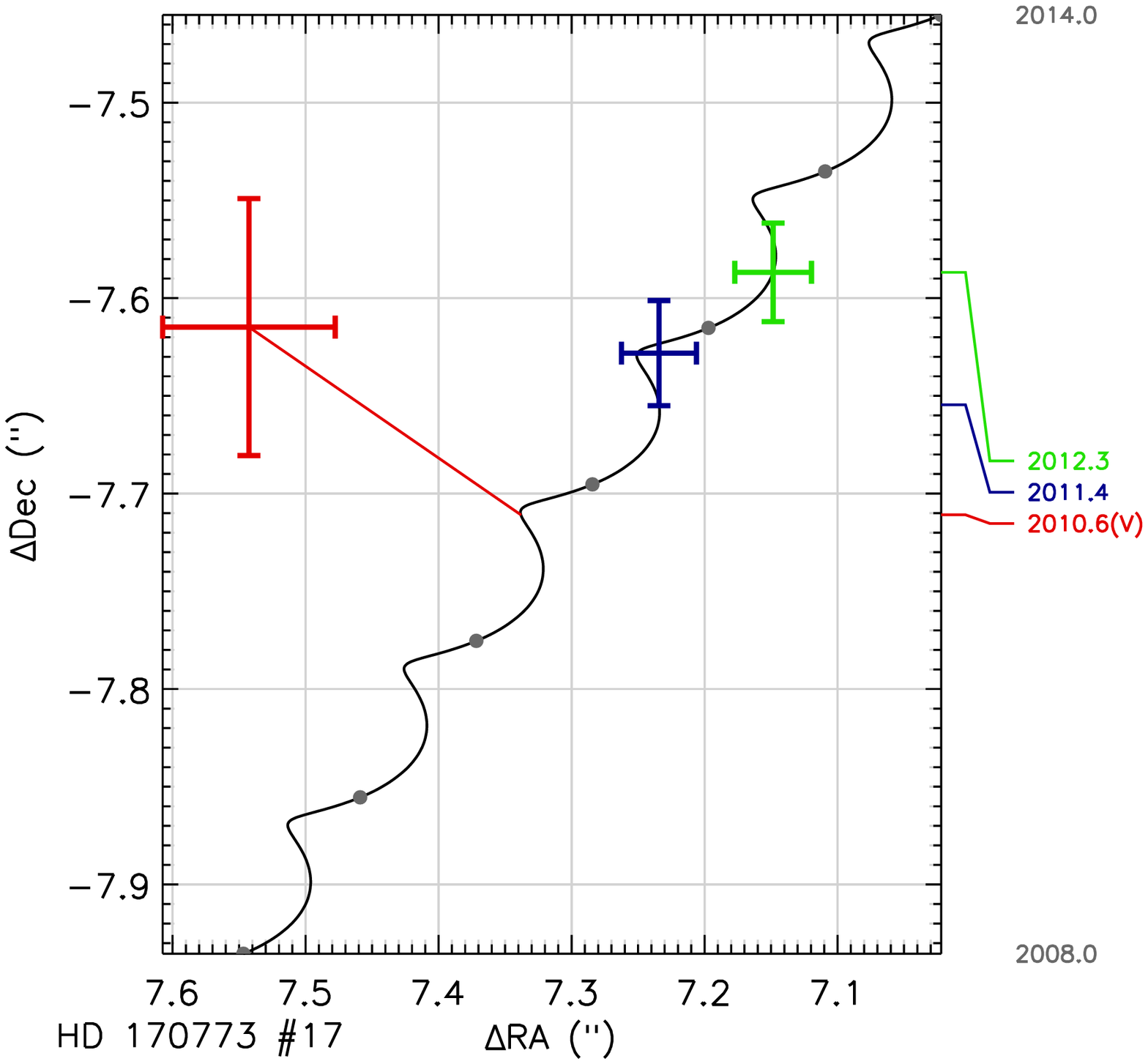}
\hskip -0.3in
\includegraphics[width=2.0in]{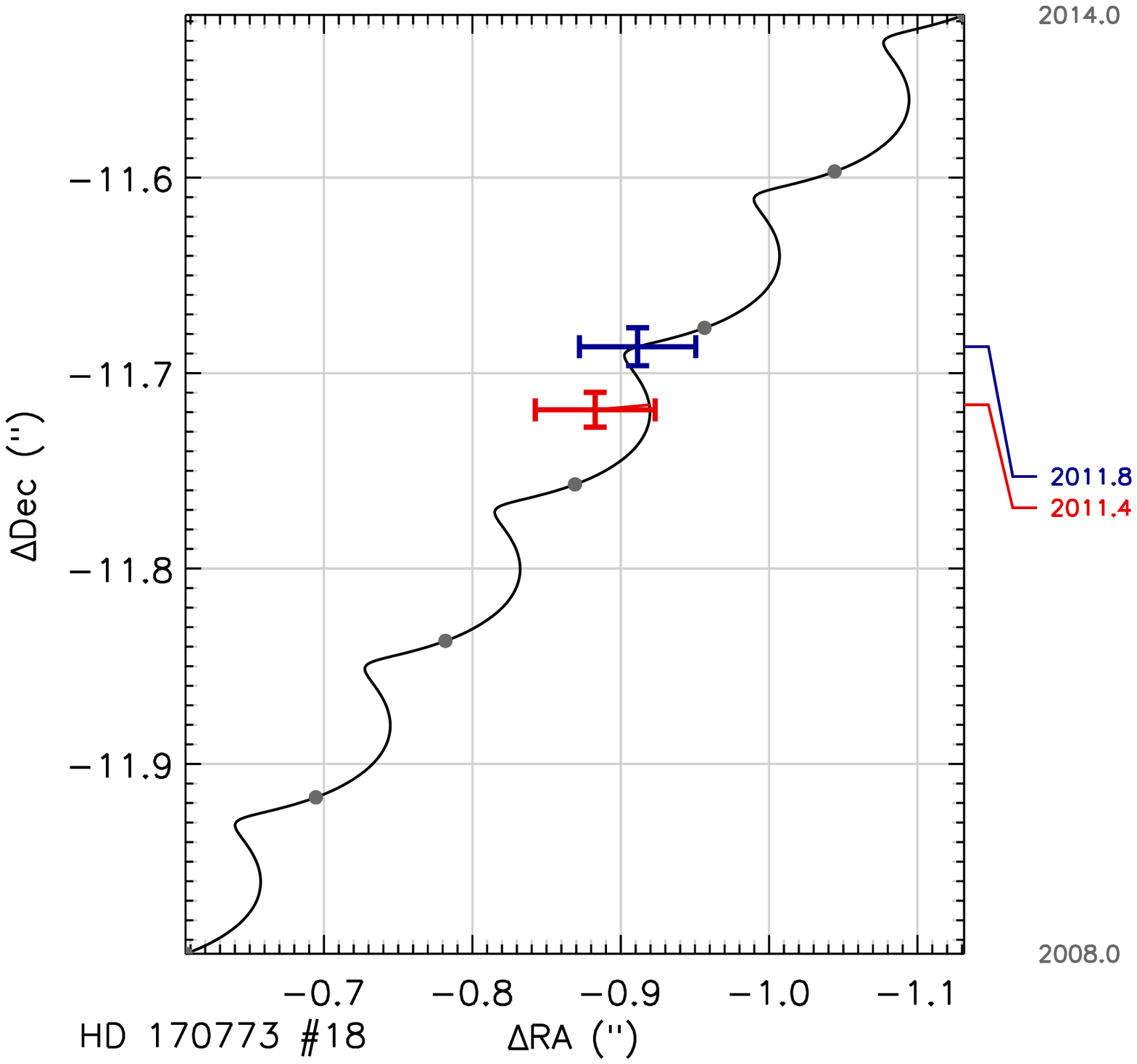}
}
\vskip -0.2in
  \caption{Same as Figure~\ref{fig:cand_motion}.}
  \label{fig:cand_motion2}
\end{figure} 

\begin{figure}
  \centerline{
    \includegraphics[width=2.0in]{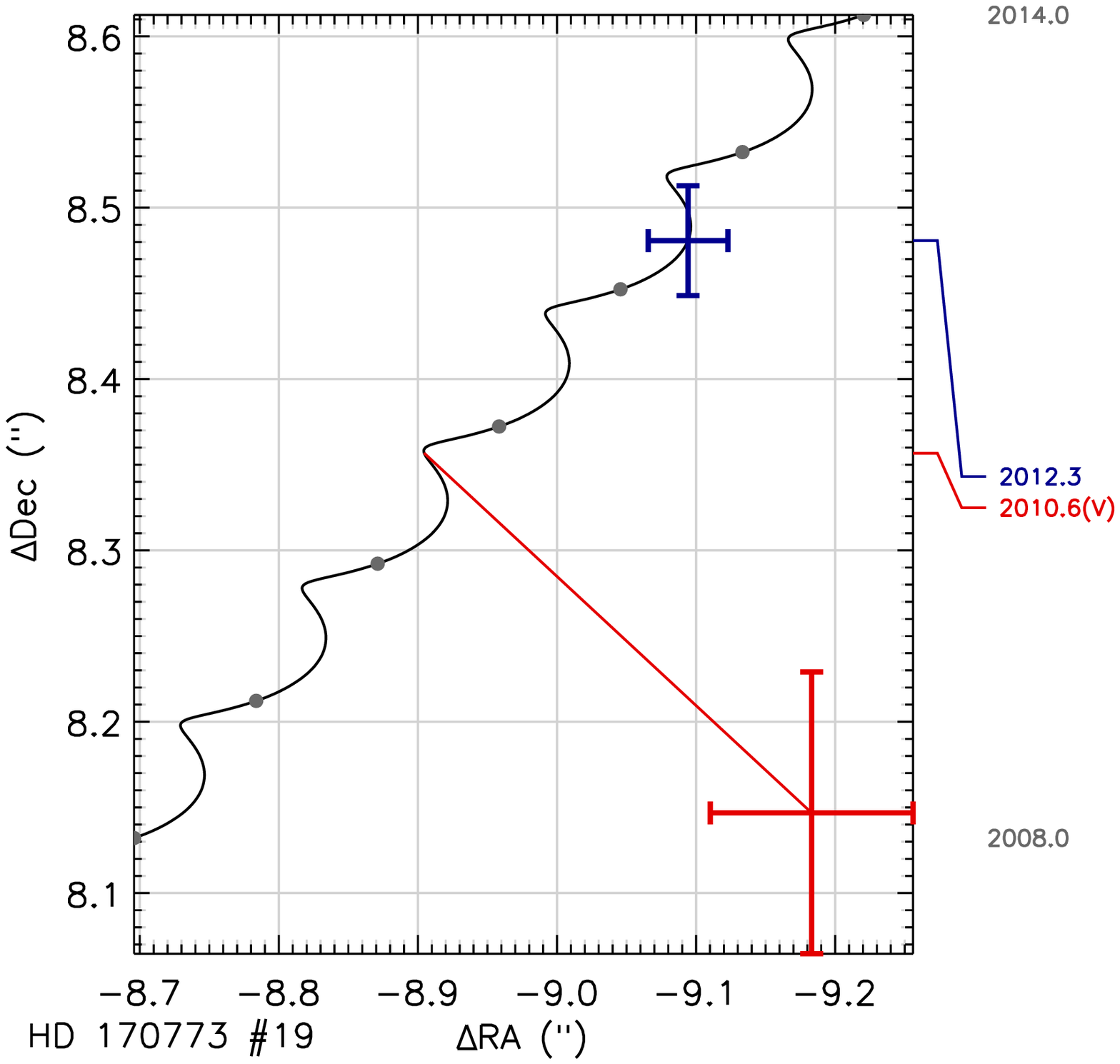}
    \hskip -0.3in
    \includegraphics[width=2in]{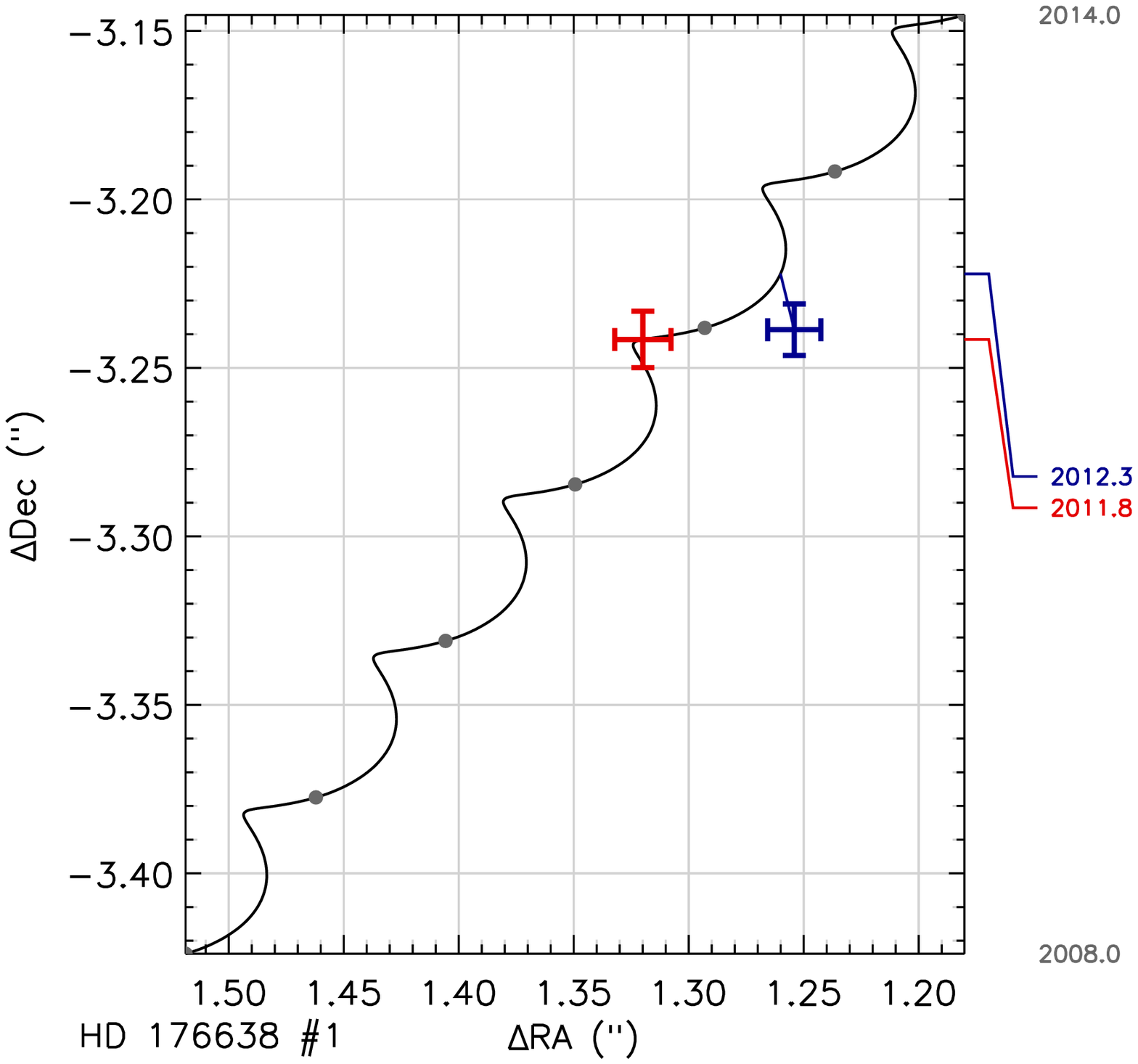}
    \hskip -0.3in
    \includegraphics[width=2in]{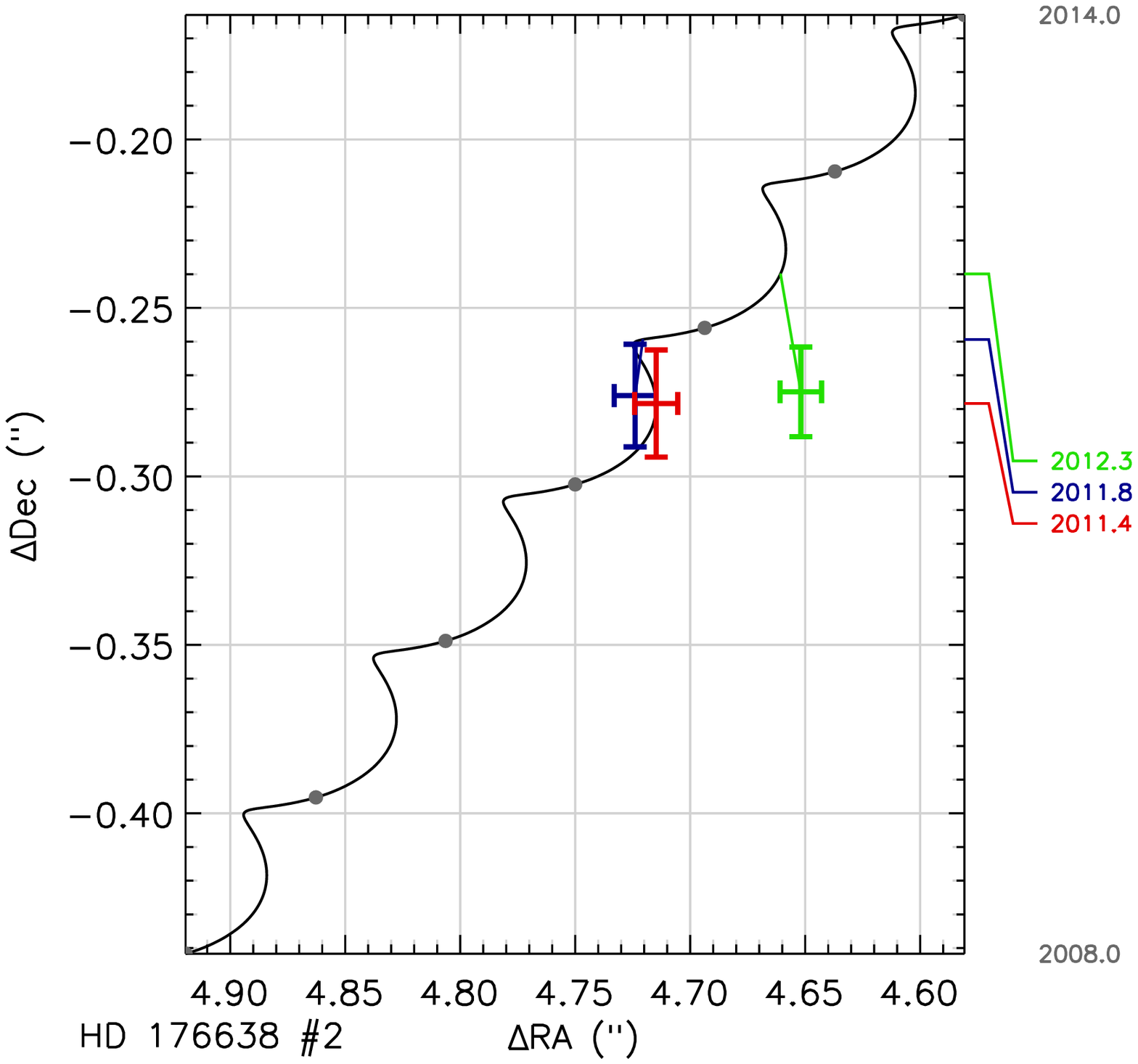}
    \hskip -0.3in
    \includegraphics[width=2in]{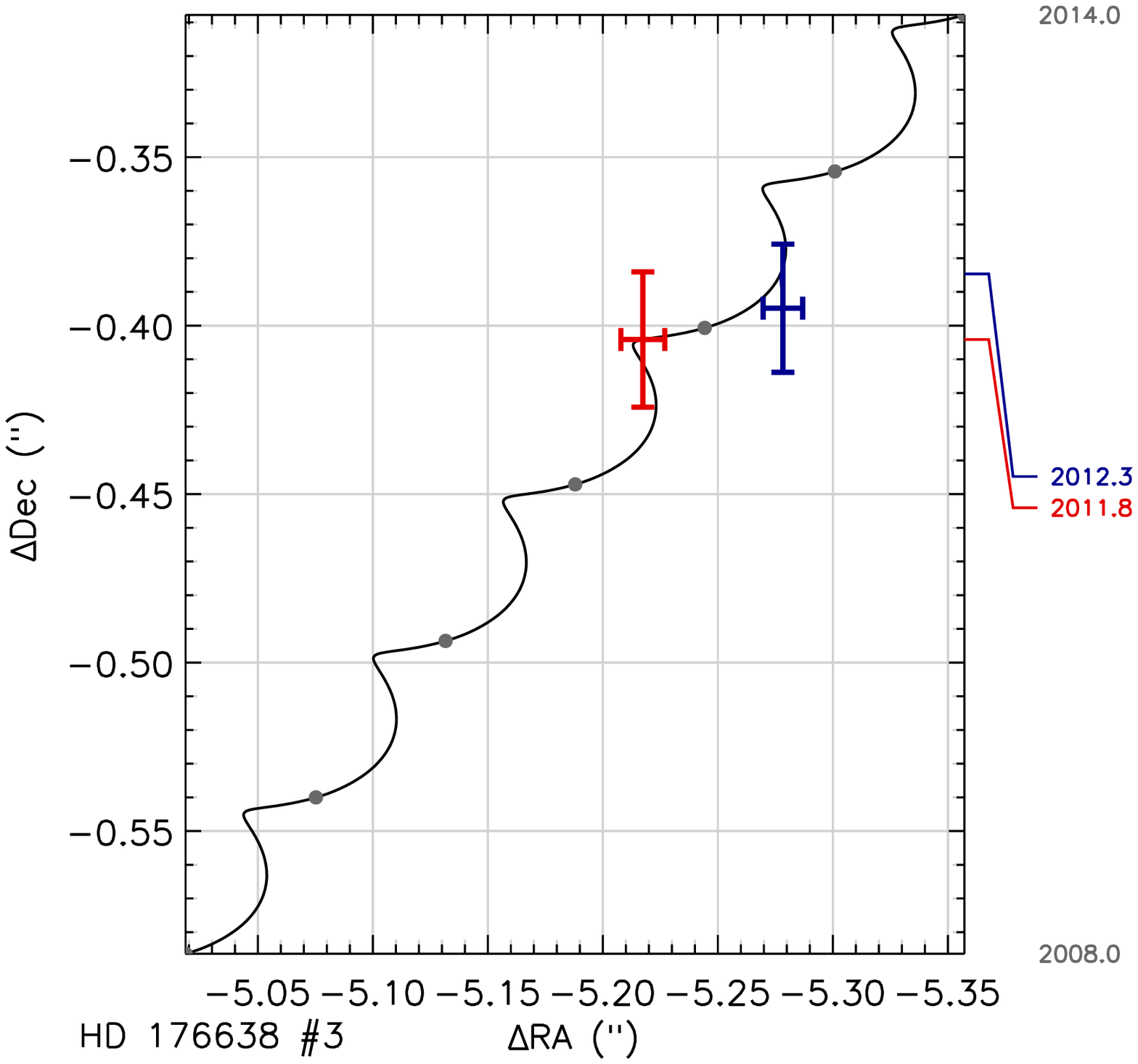}
  }
  \vskip -0.2in
  \centerline{
    \includegraphics[width=2in]{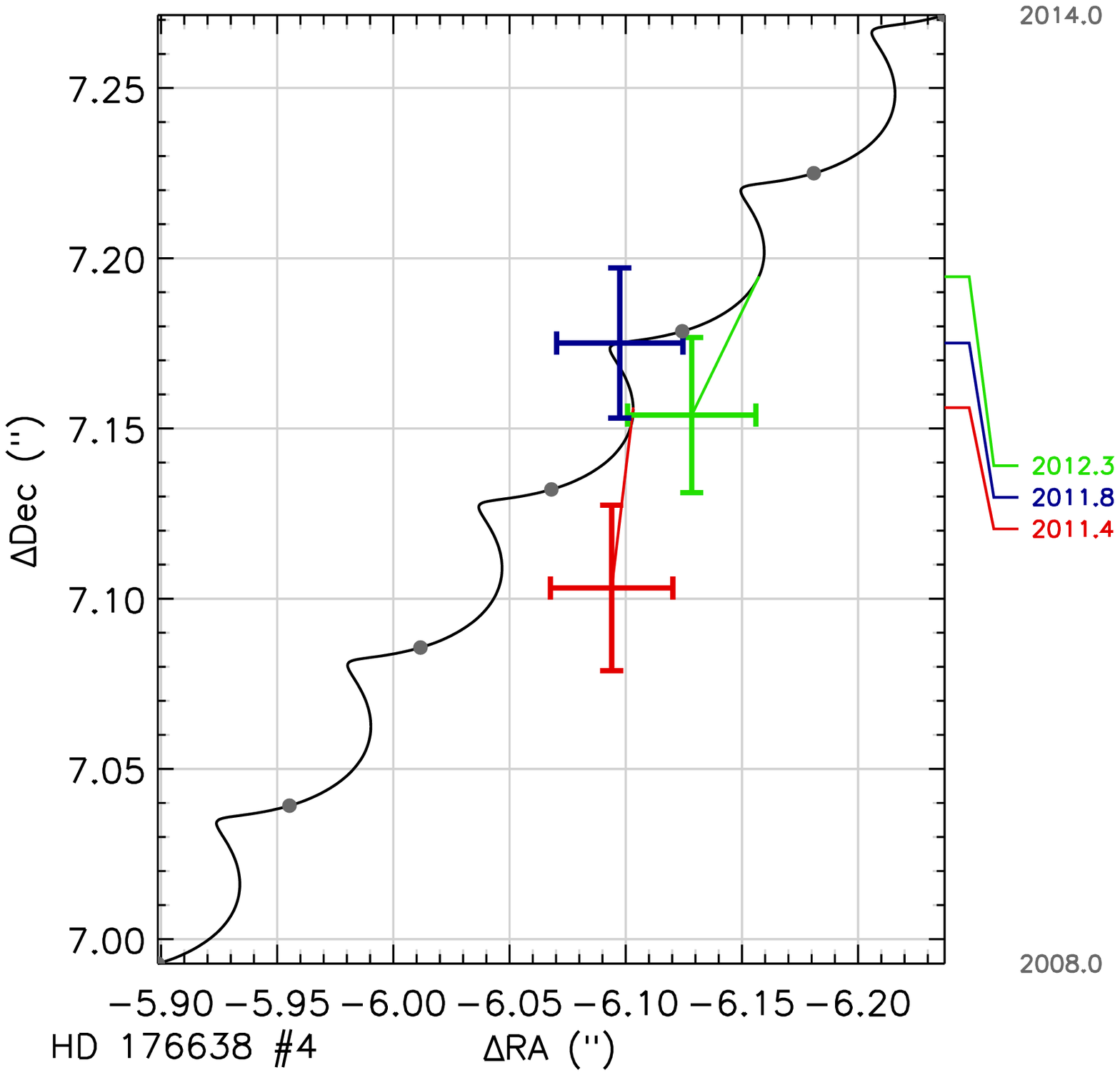}
    \hskip -0.3in
    \includegraphics[width=2in]{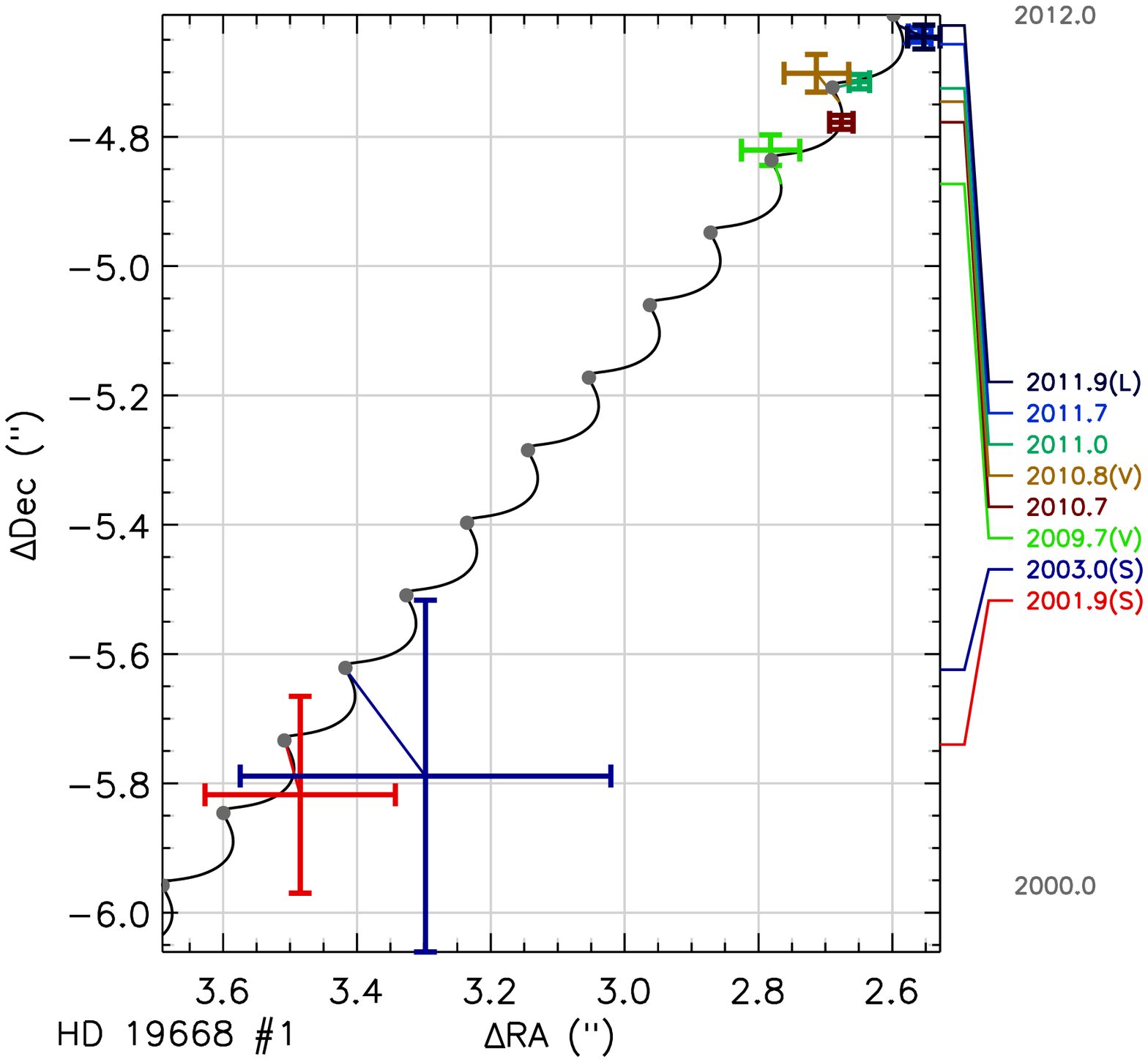}
    \hskip -0.3in
    \includegraphics[width=2in]{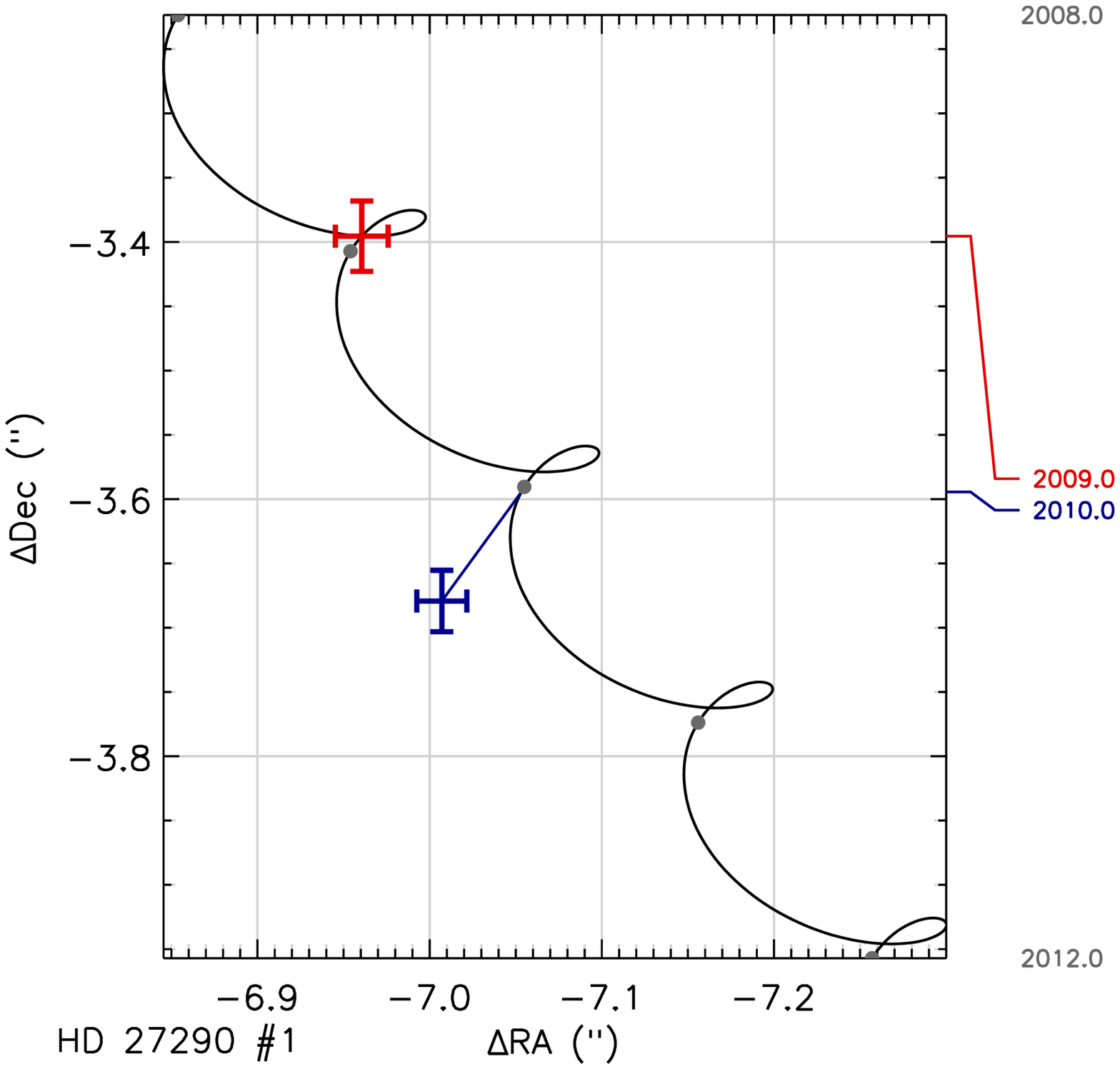}
    \hskip -0.3in
    \includegraphics[width=2in]{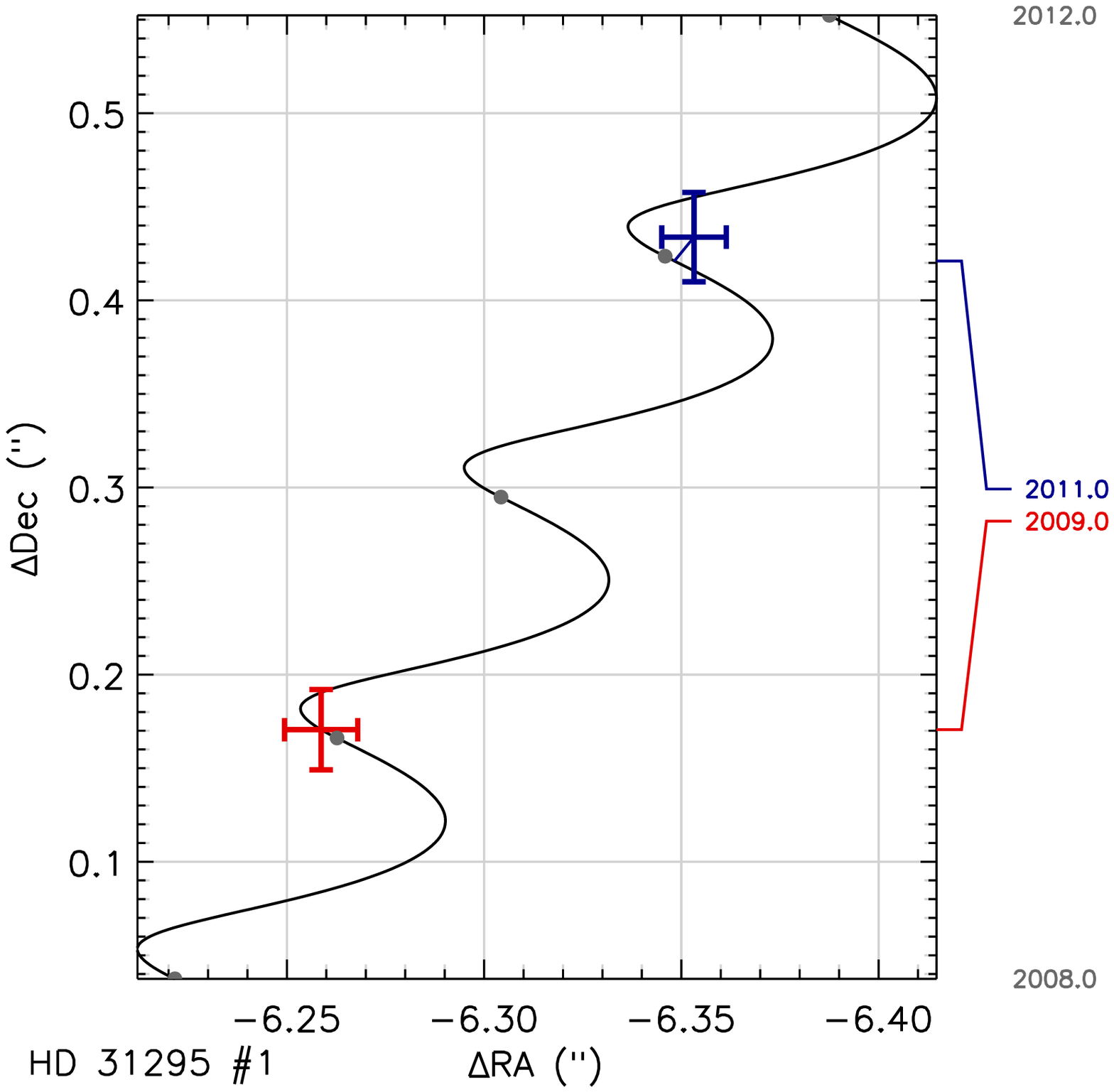}
}
\vskip -0.2in
\centerline{
  \includegraphics[width=2in]{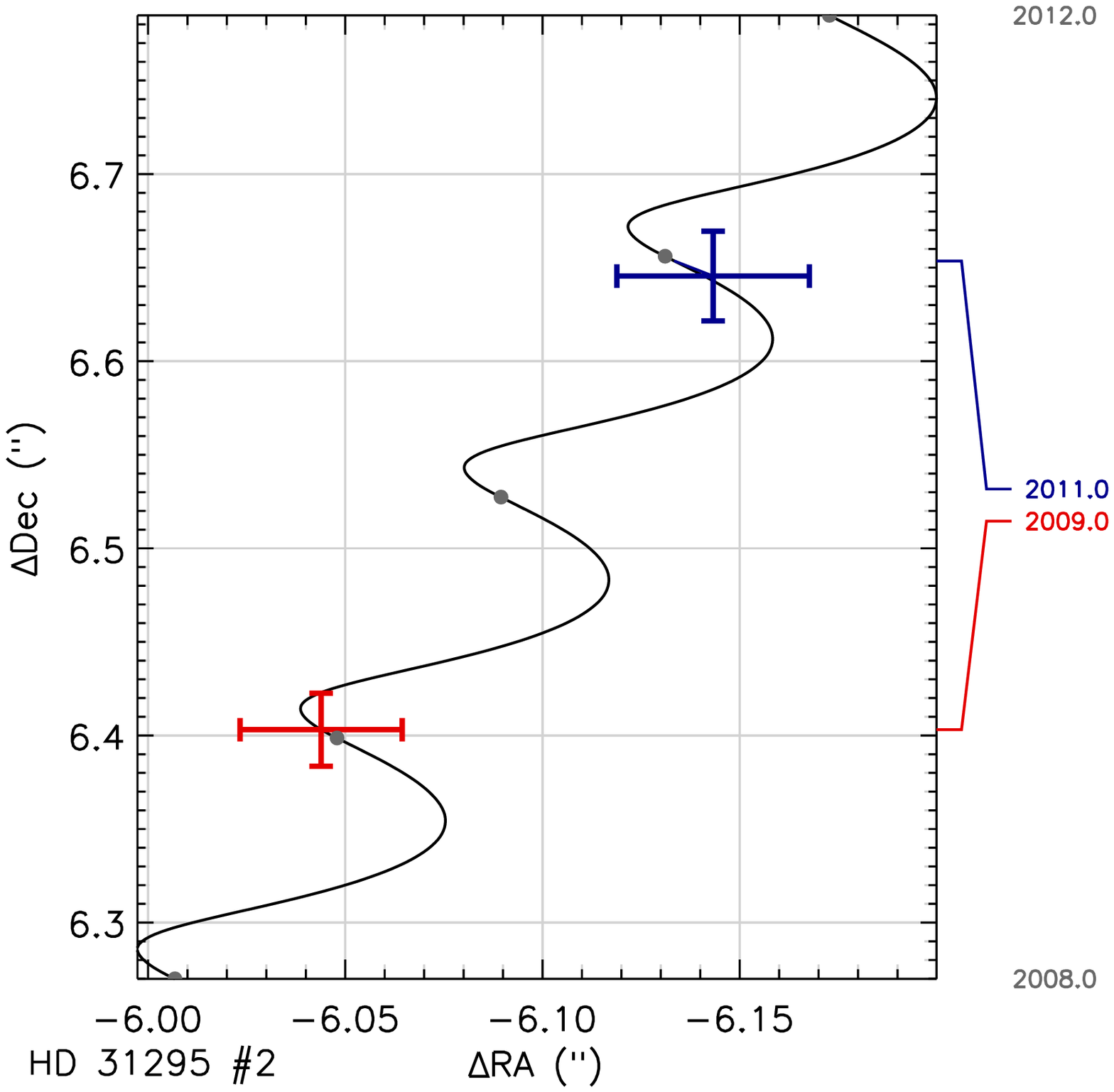}
  \hskip -0.3in
  \includegraphics[width=2in]{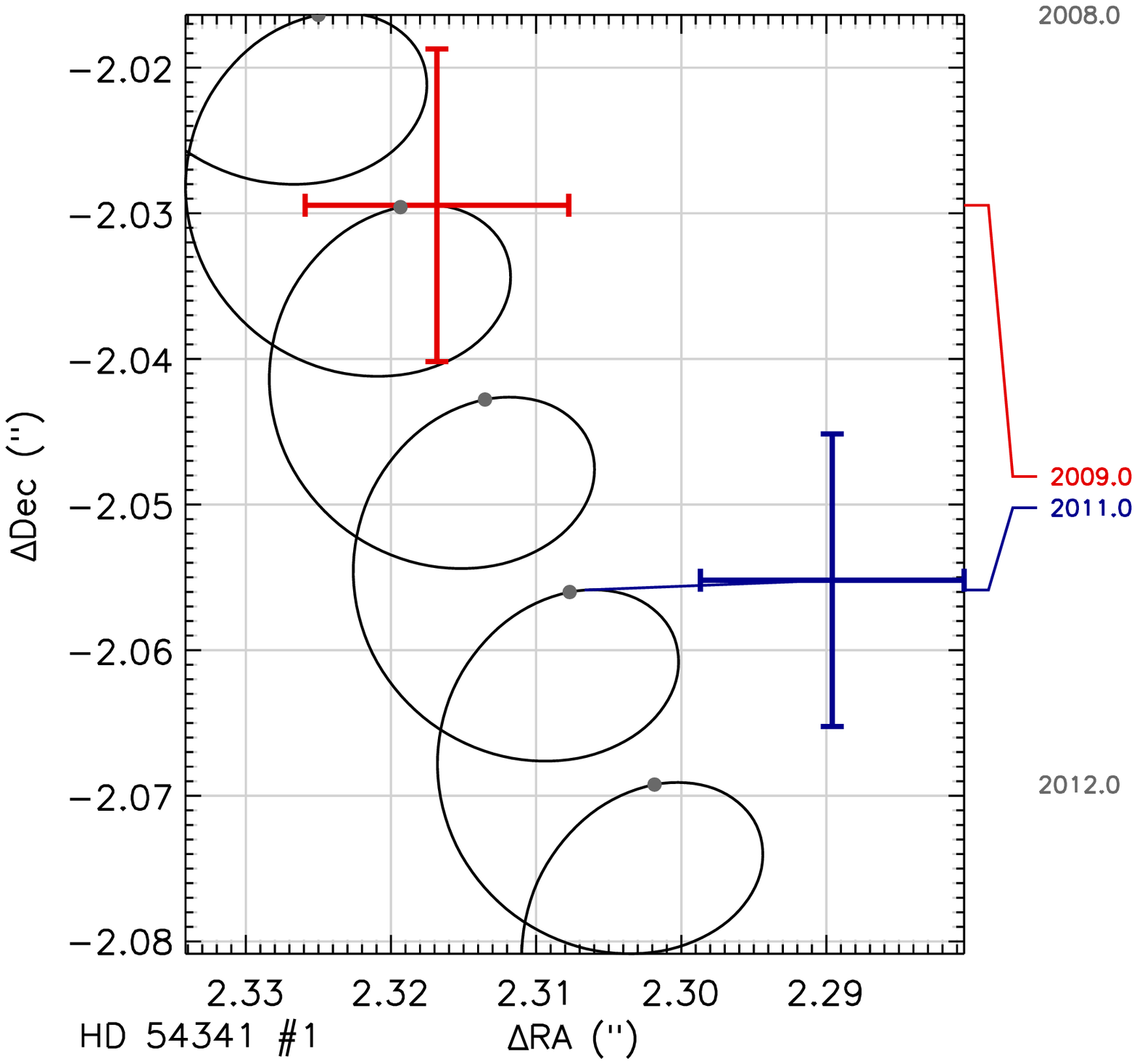}
  \hskip -0.3in
  \includegraphics[width=2in]{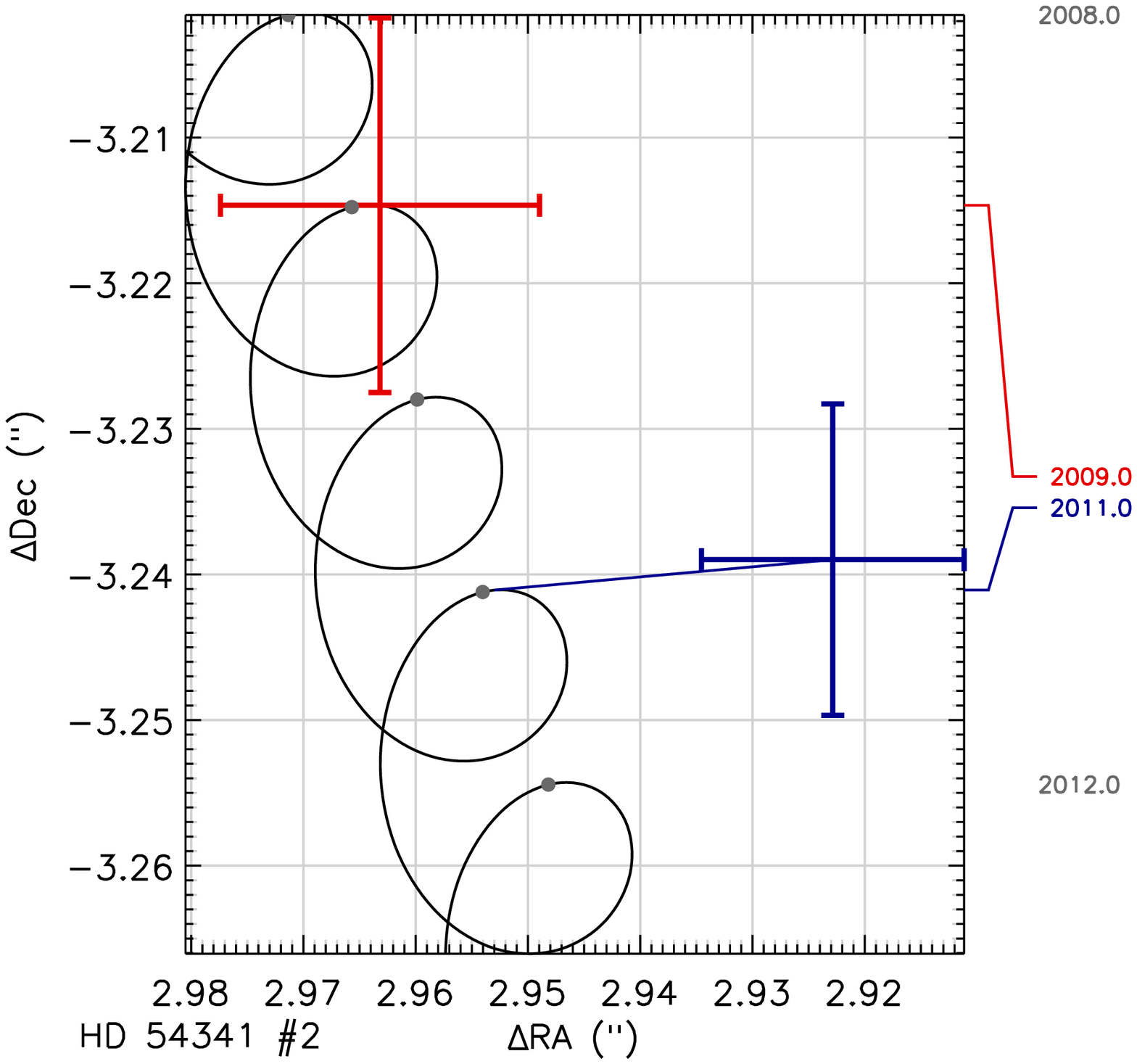}
  \hskip -0.3in
  \includegraphics[width=2in]{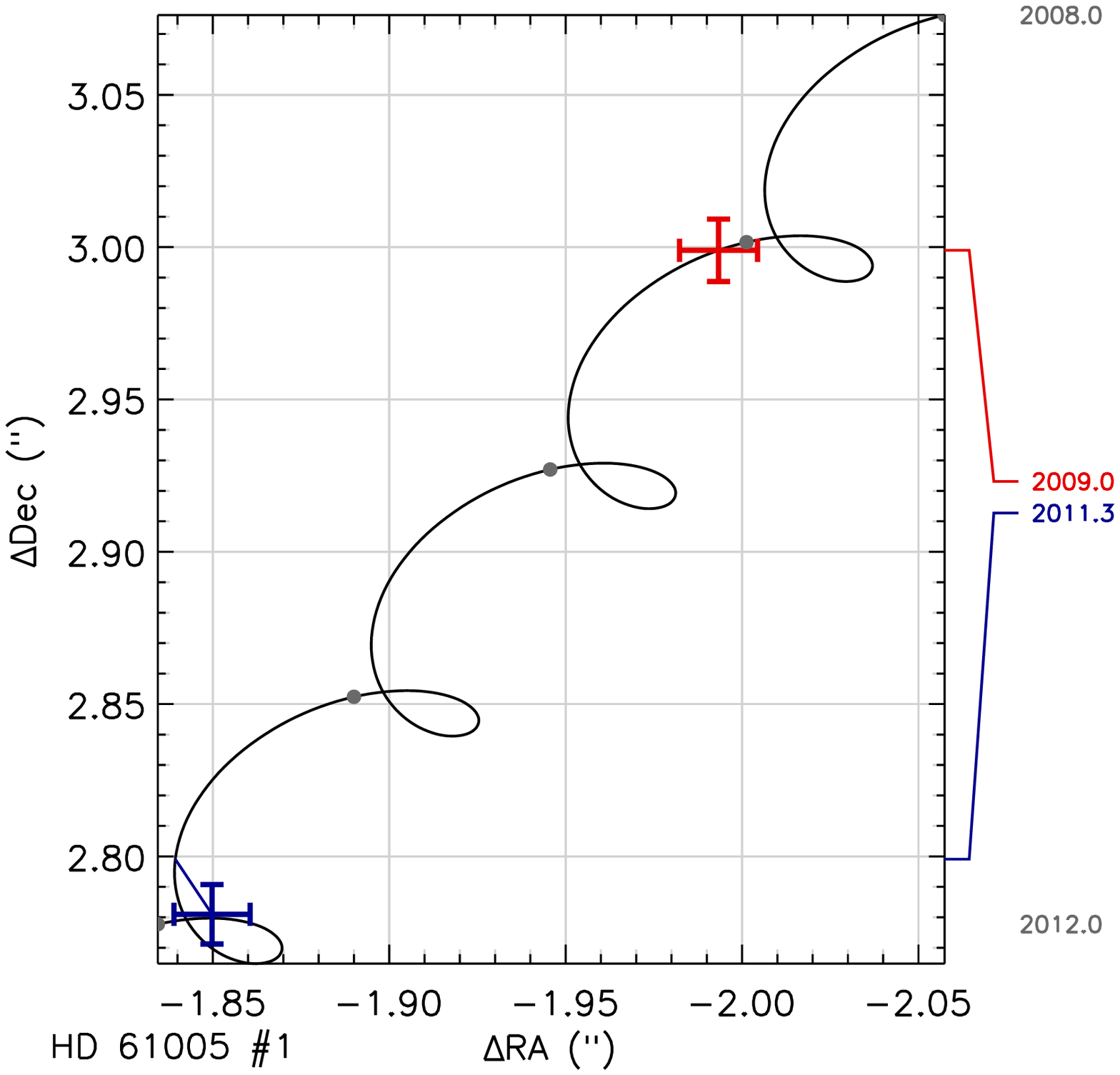}
}
\vskip -0.2in
\centerline{
  \includegraphics[width=2in]{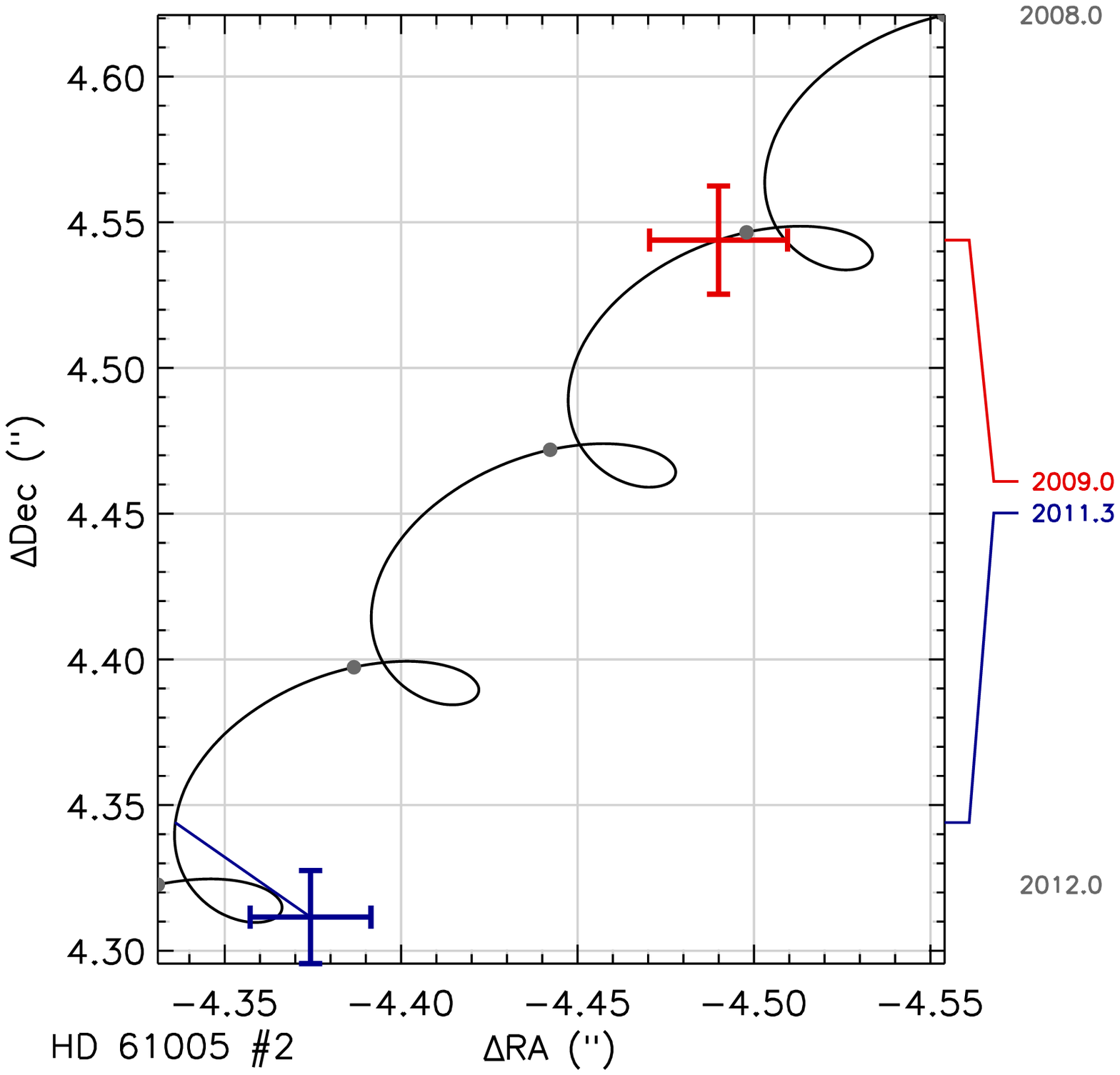}
  \hskip -0.3in
  \includegraphics[width=2in]{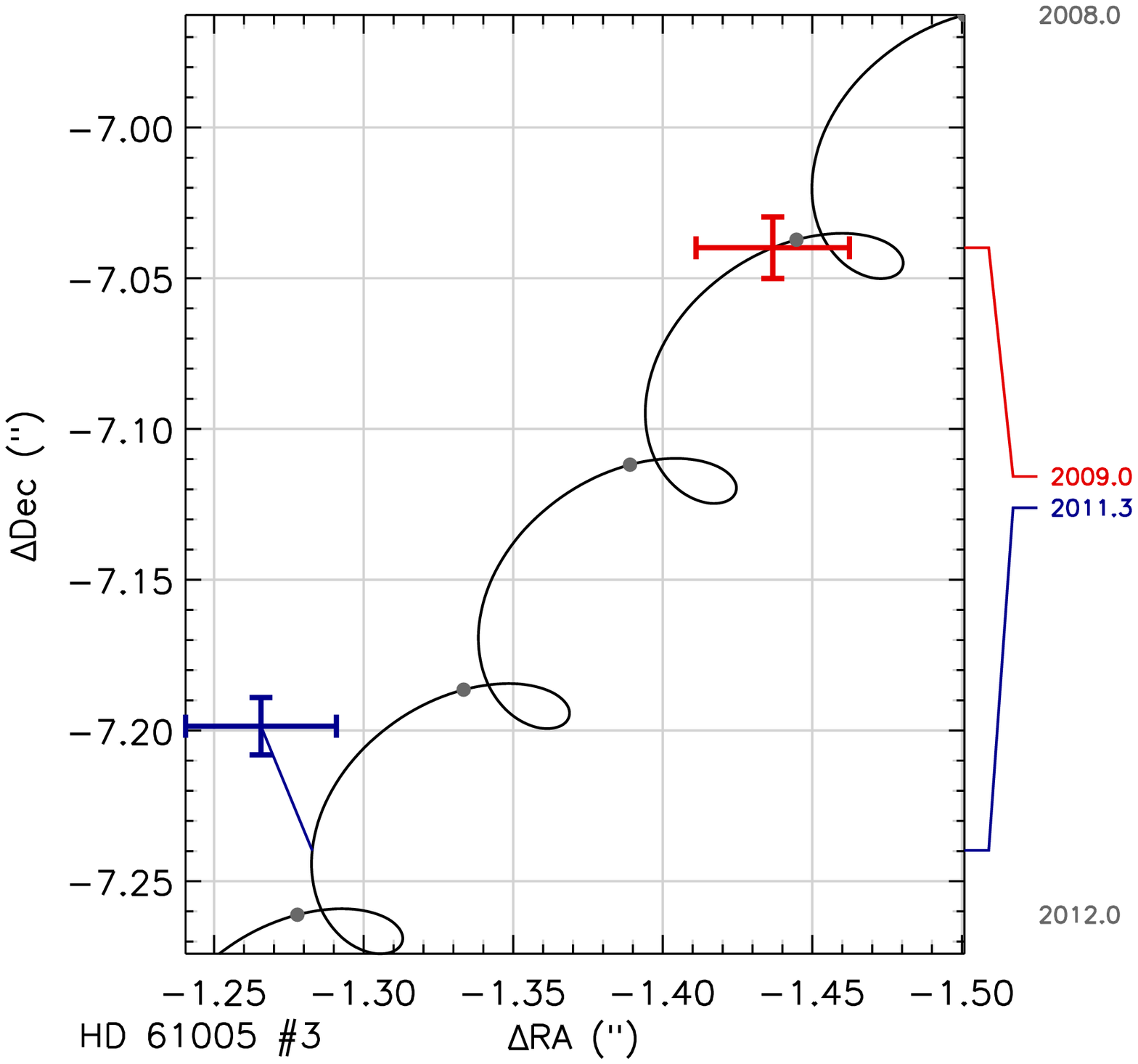}
  \hskip -0.3in
  \includegraphics[width=2in]{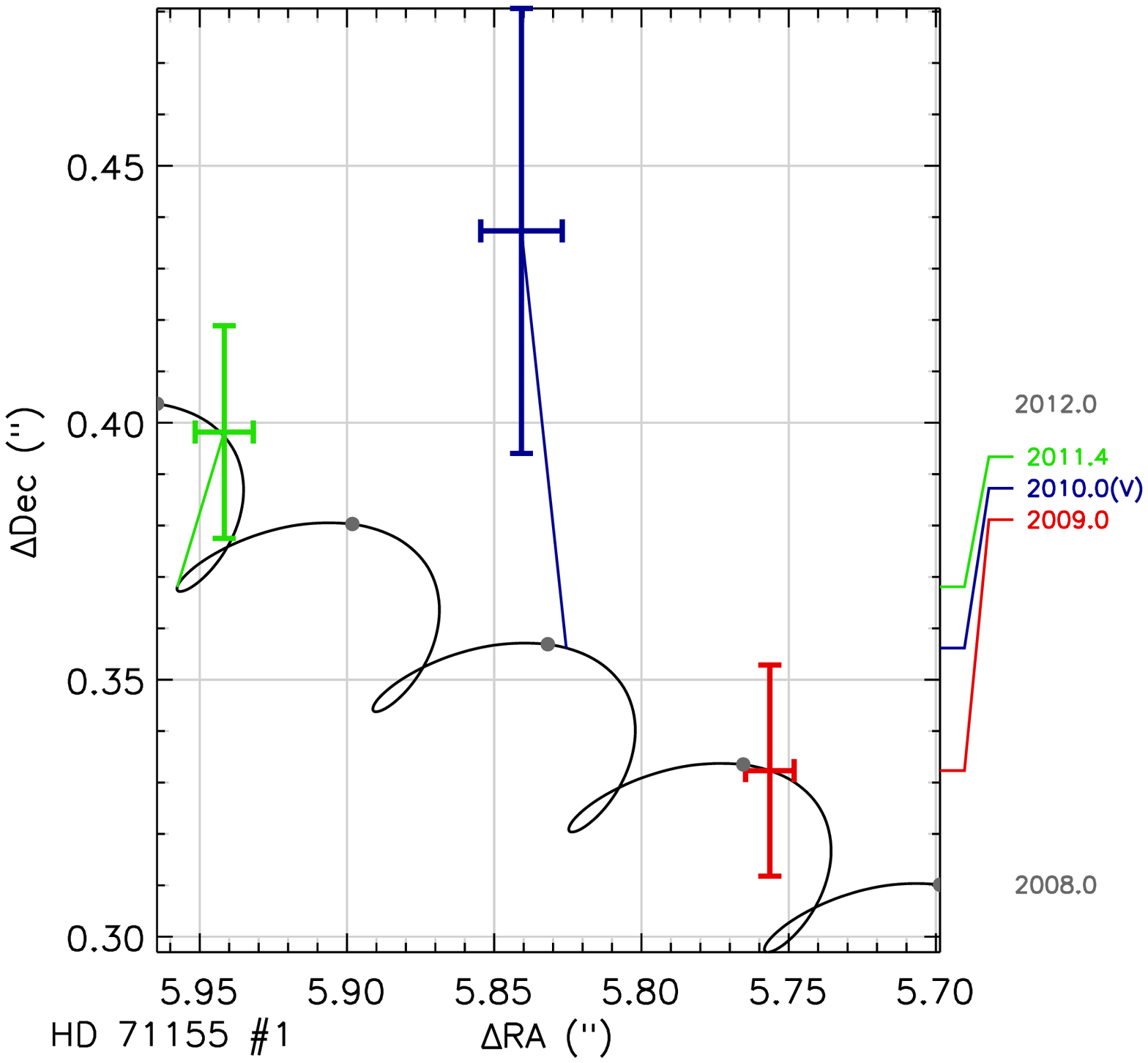}
  \hskip -0.3in
  \includegraphics[width=2in]{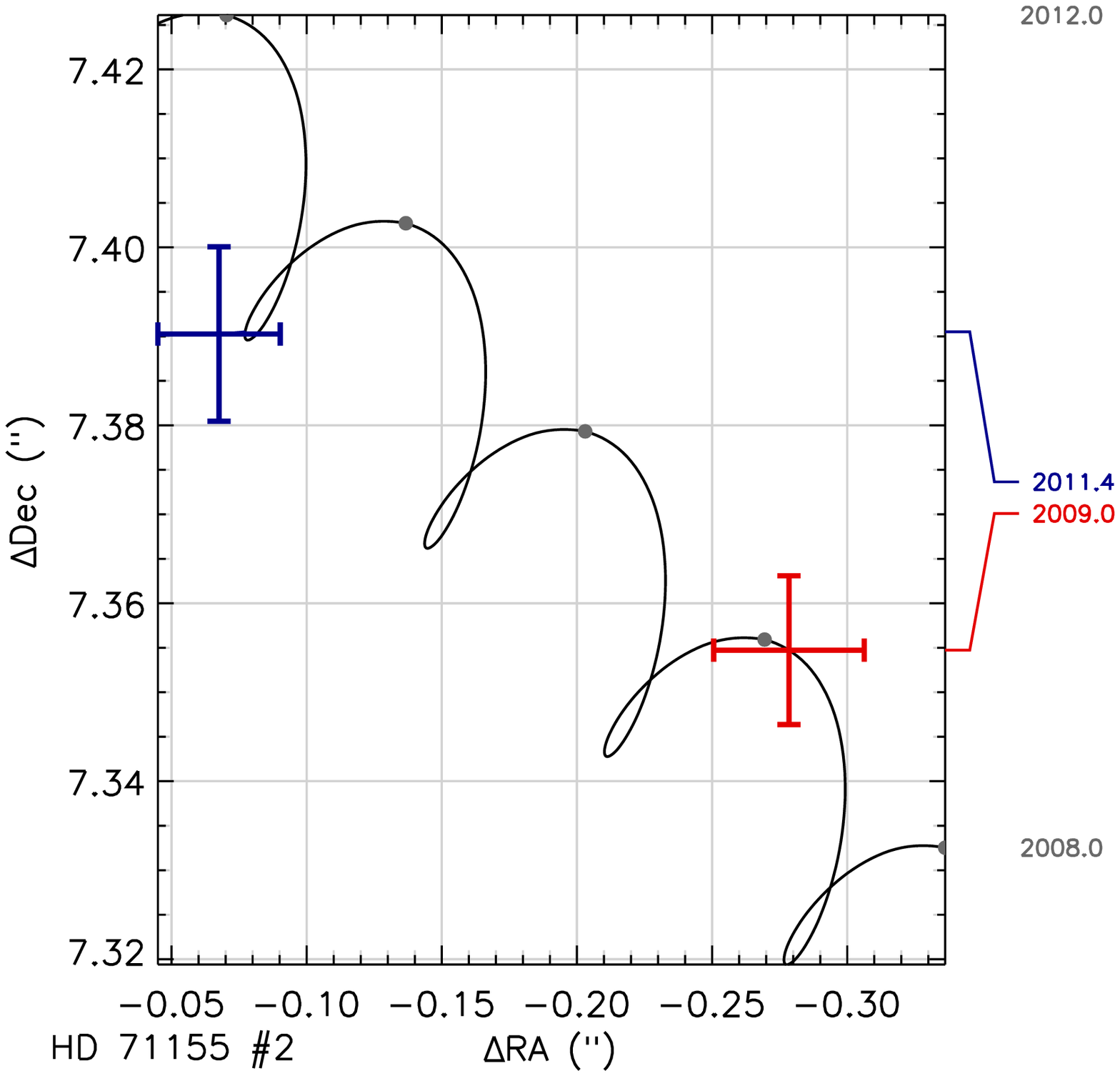}
}
\vskip -0.2in
\caption{Same as Figure~\ref{fig:cand_motion}.}
\label{fig:cand_motion3}
\end{figure}

\begin{figure}
  \centerline{
    \includegraphics[width=2in]{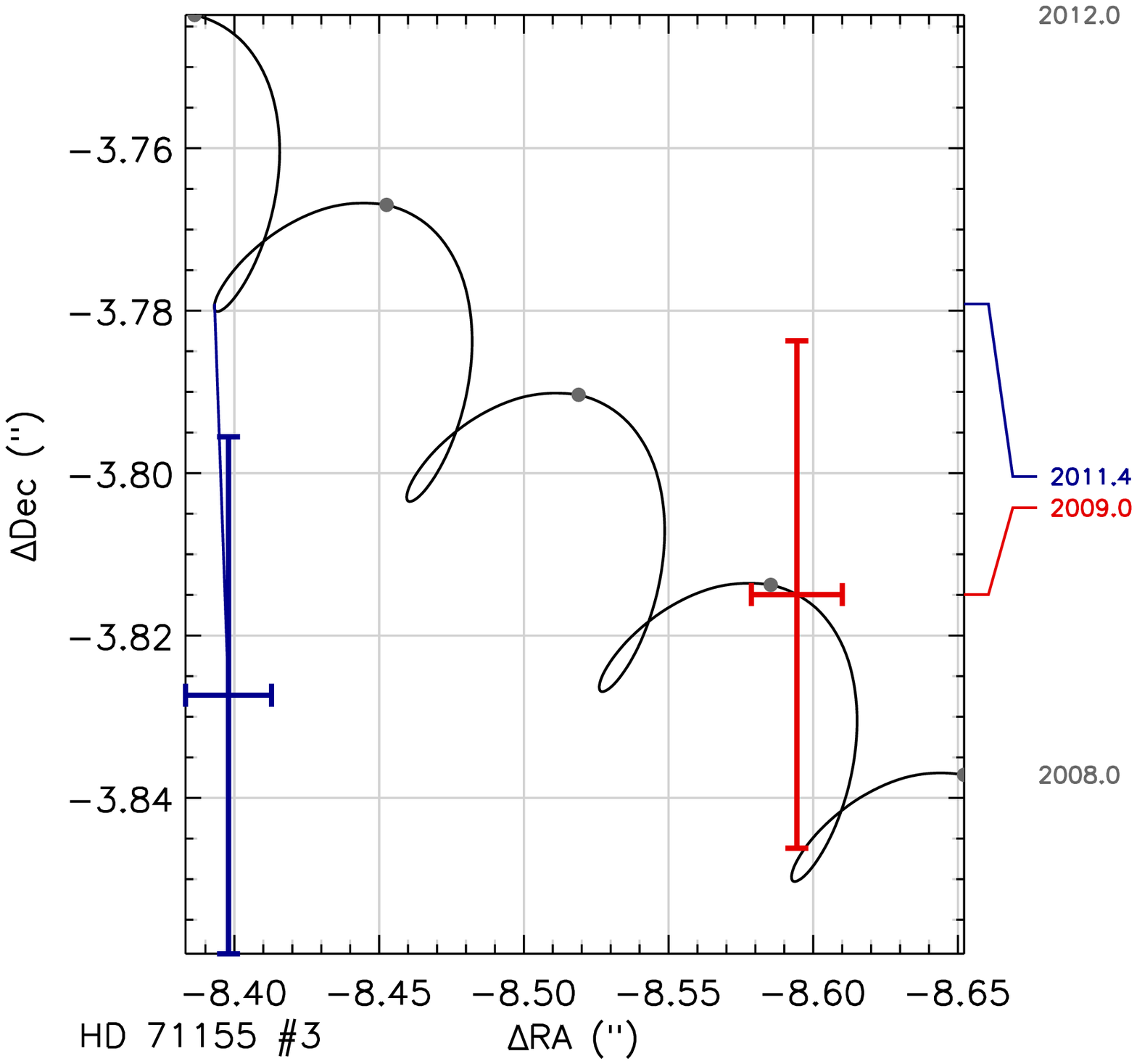}
    \hskip -0.3in
    \includegraphics[width=2in]{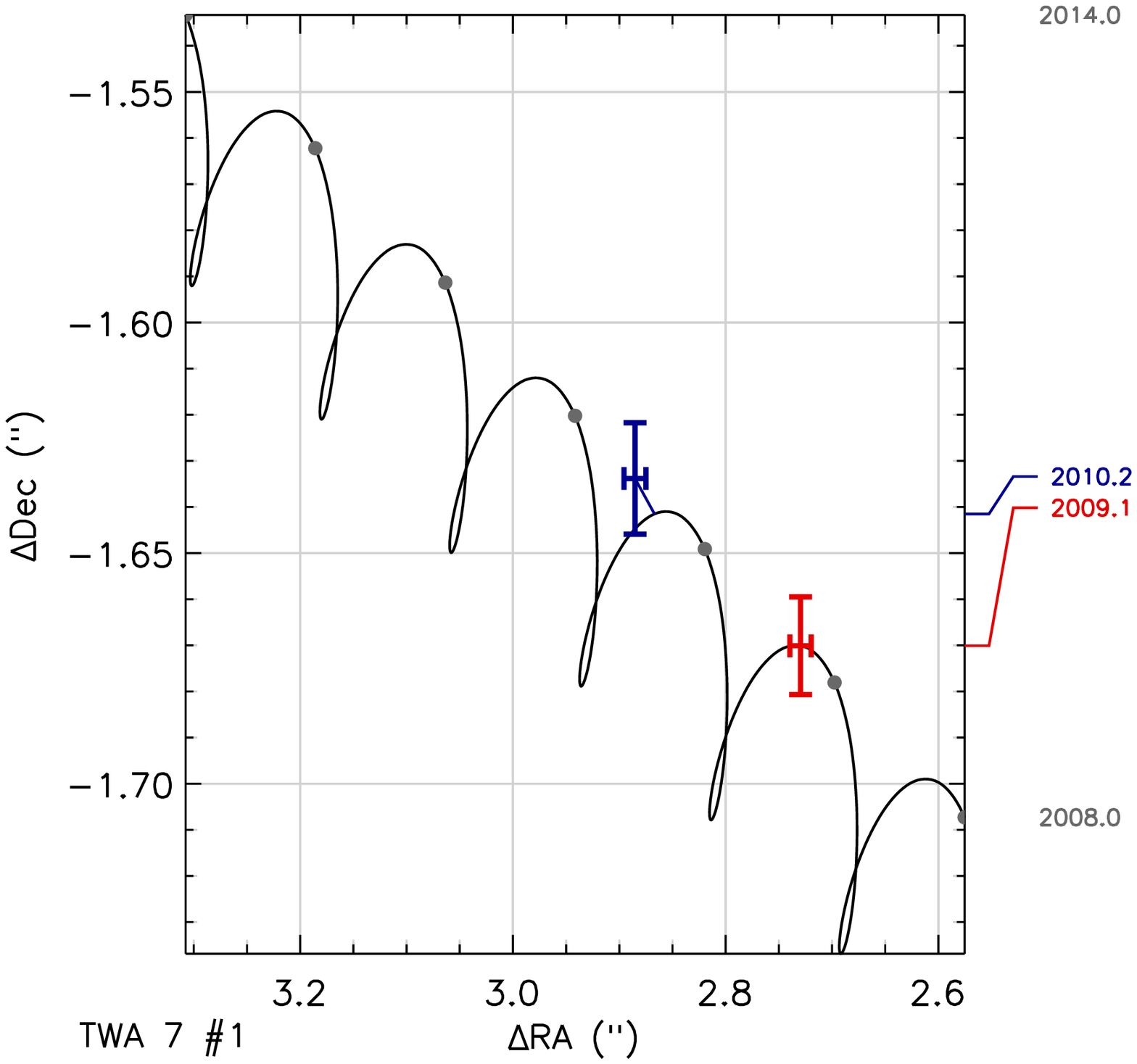}
    \hskip -0.3in
    \includegraphics[width=2in]{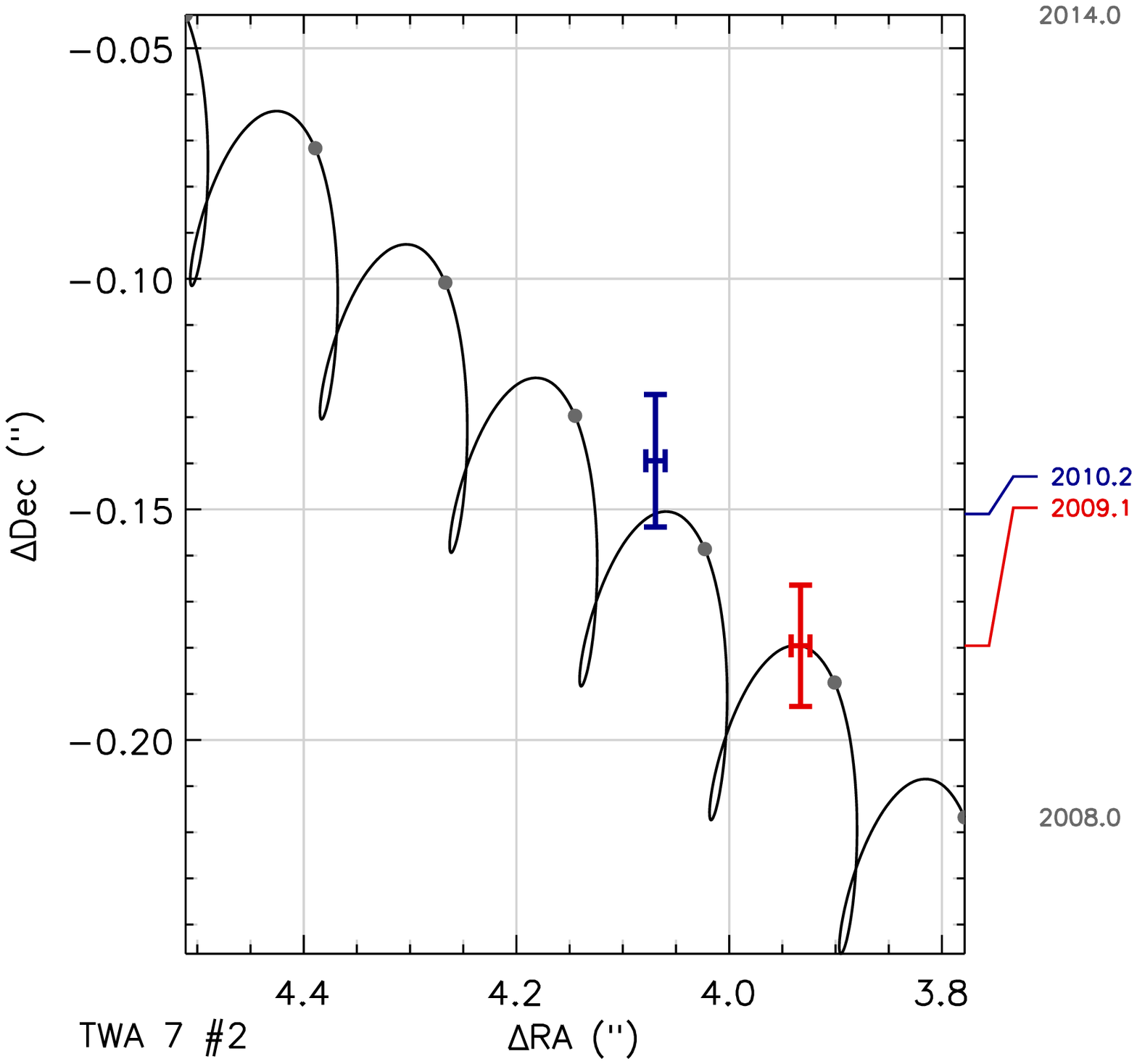}
    \hskip -0.3in
    \includegraphics[width=2in]{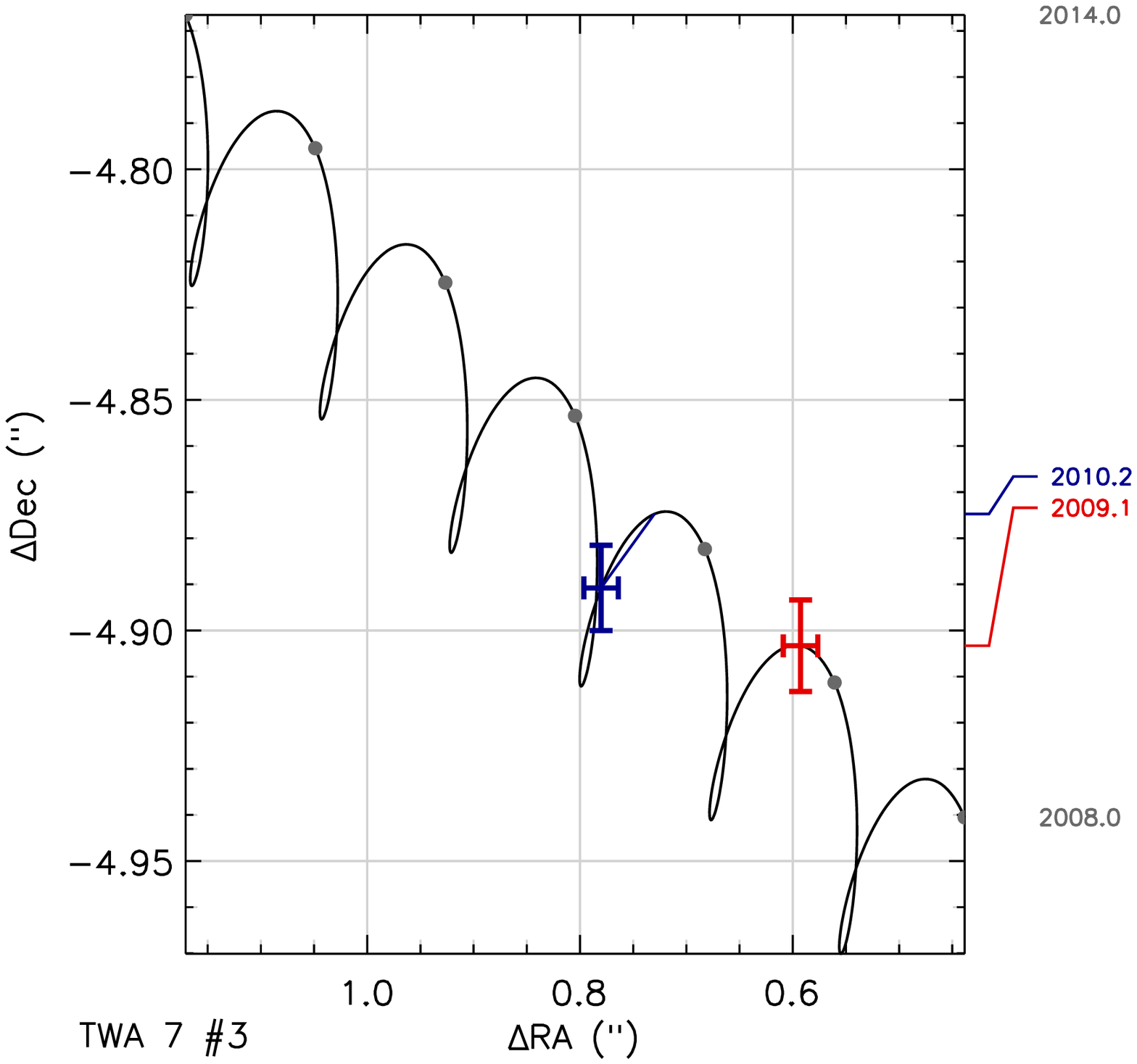}
  }
  \vskip -0.2in
  \centerline{
    \includegraphics[width=2in]{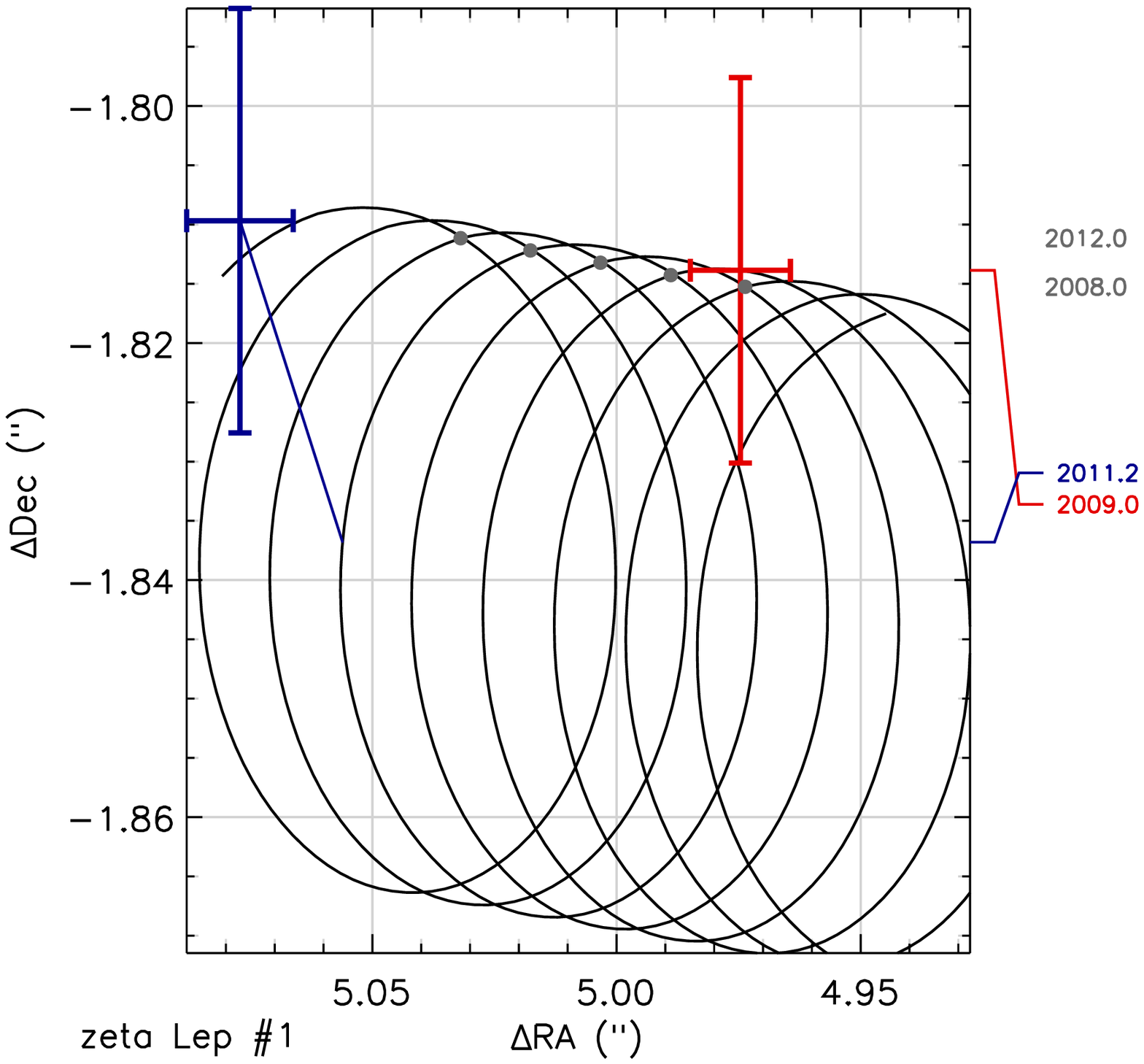}
    \hskip -0.3in
    \includegraphics[width=2in]{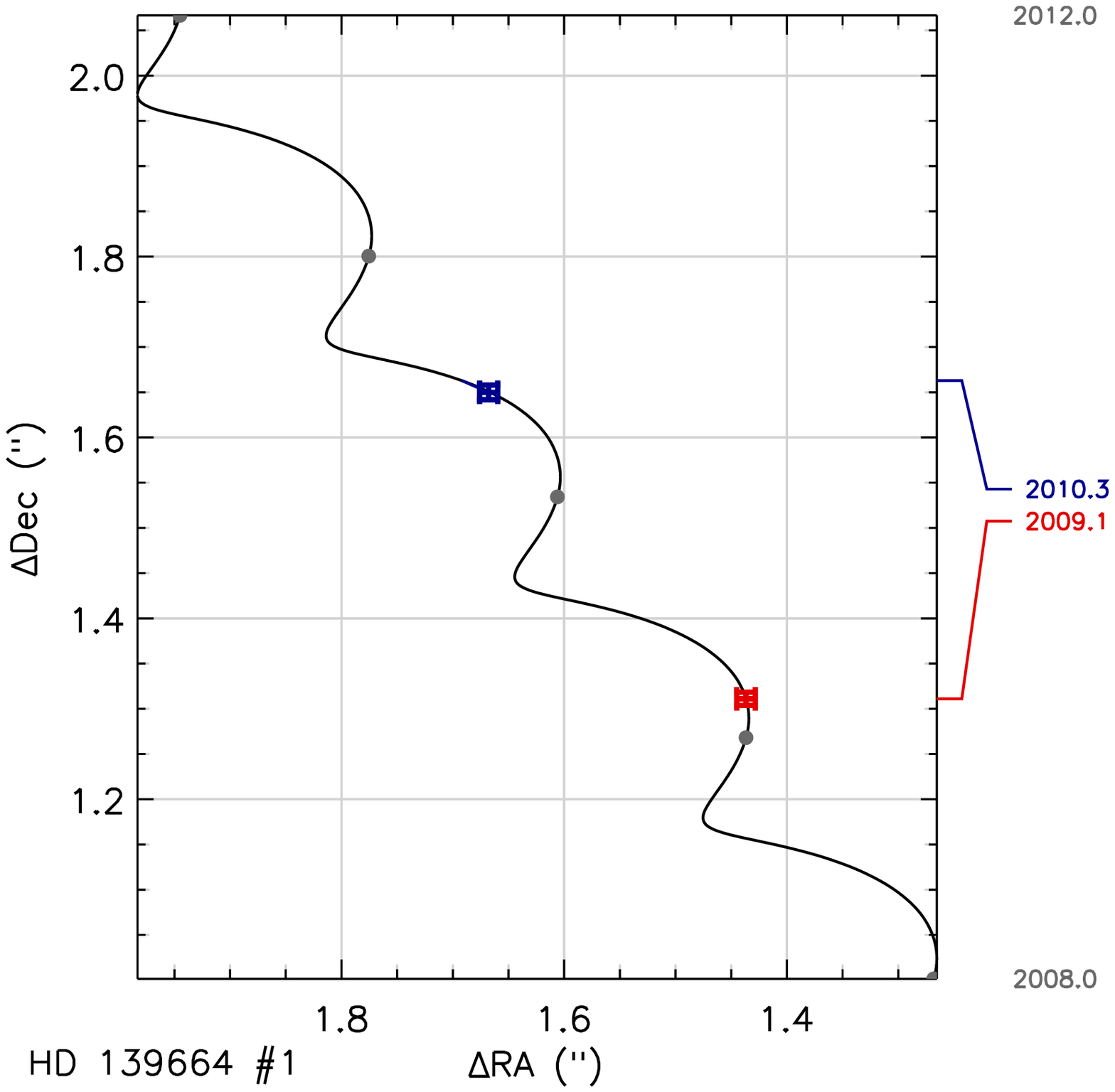}
    \hskip -0.3in
    \includegraphics[width=2in]{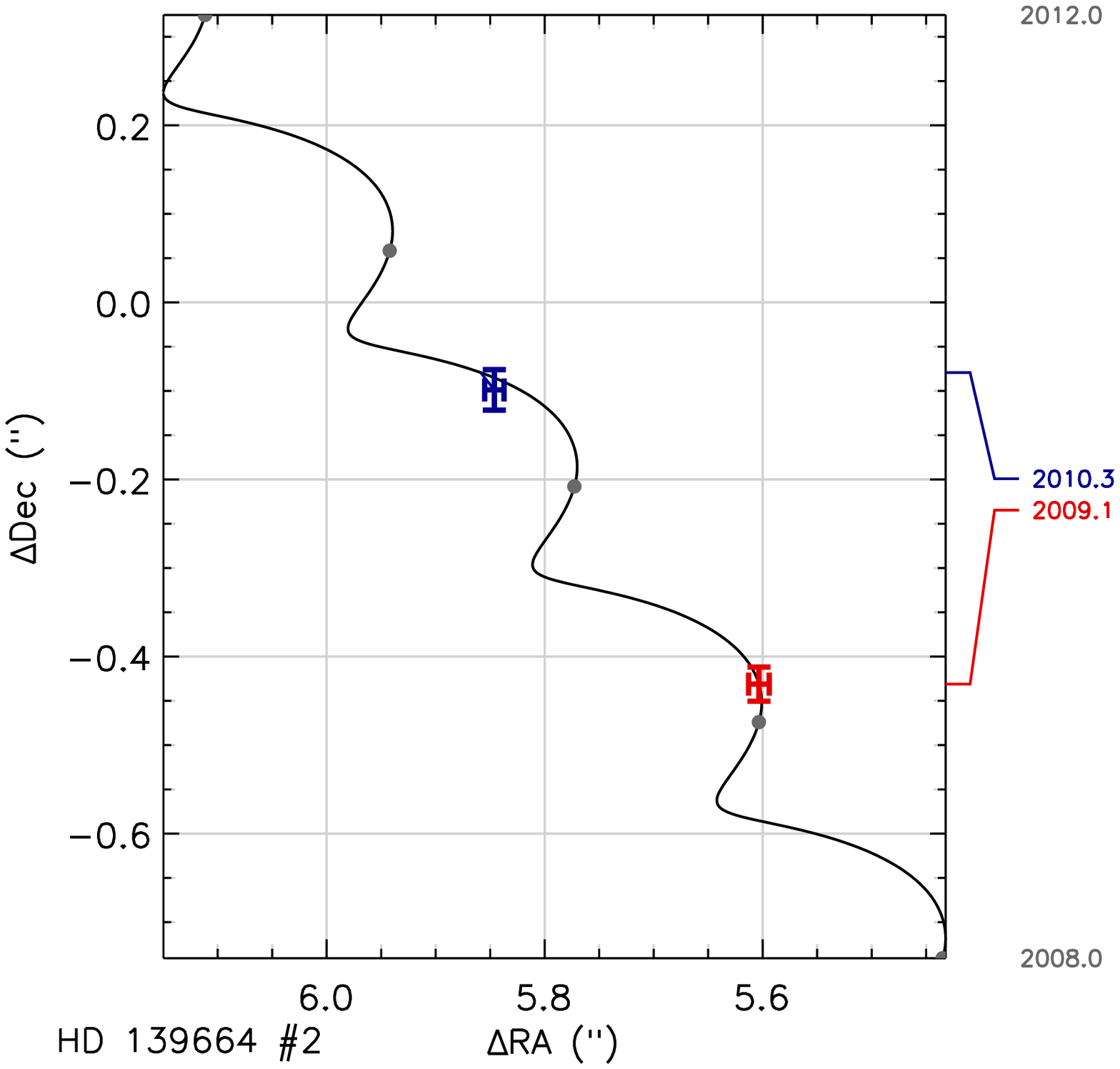}
    \hskip -0.3in
    \includegraphics[width=2in]{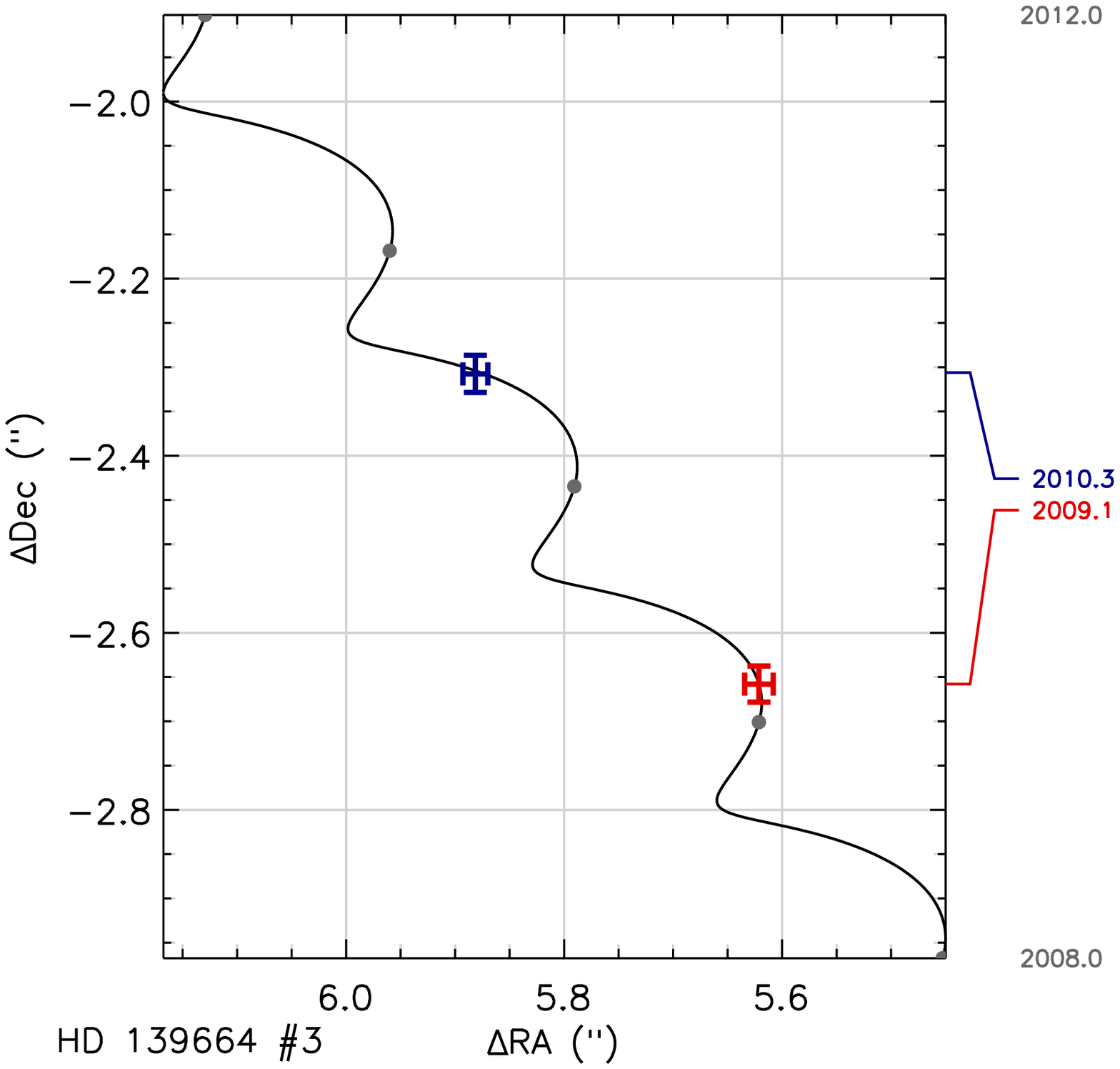}
  }
  \vskip -0.2in
  \centerline{
    \includegraphics[width=2in]{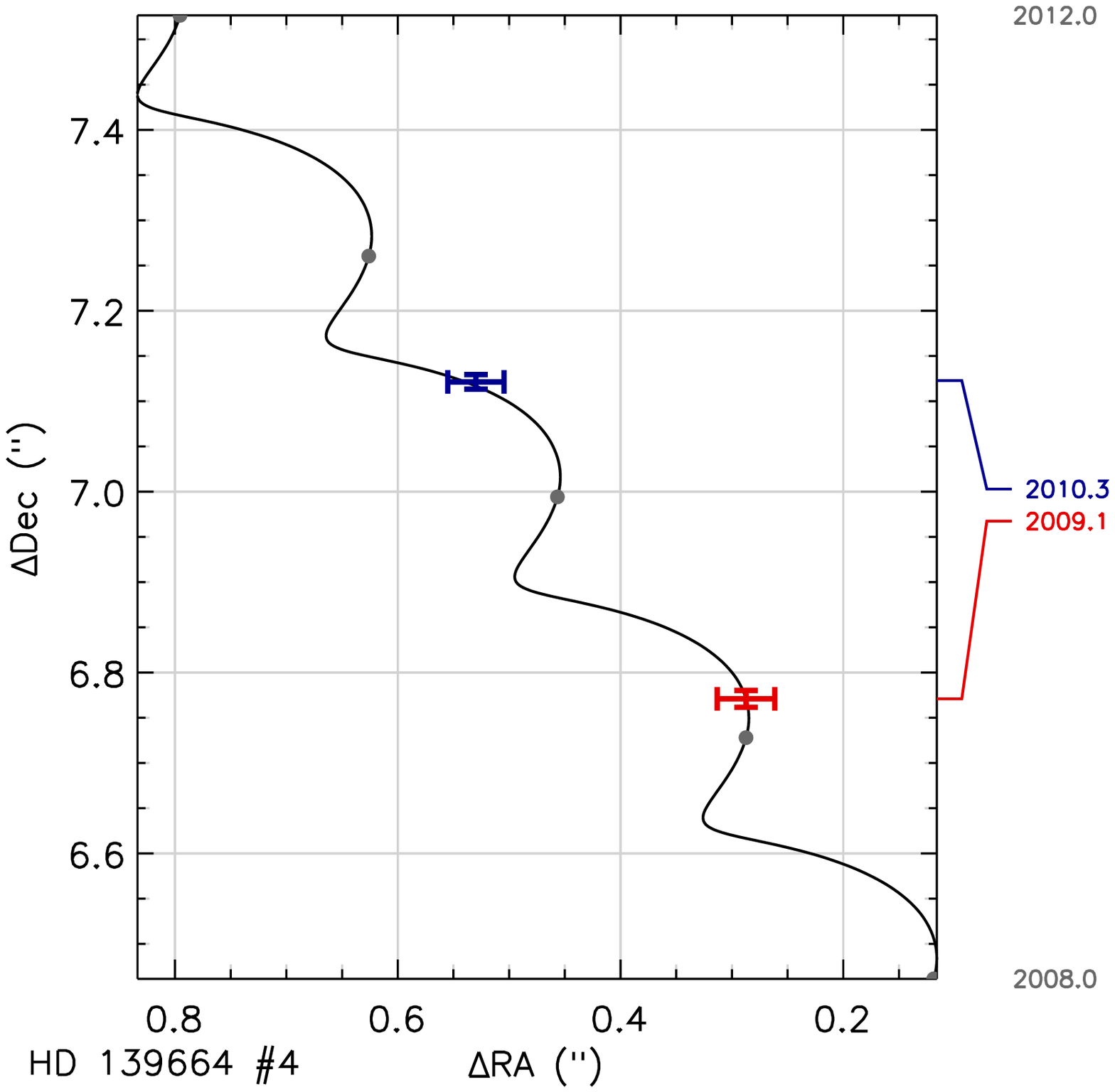}
    \hskip -0.3in
    \includegraphics[width=2in]{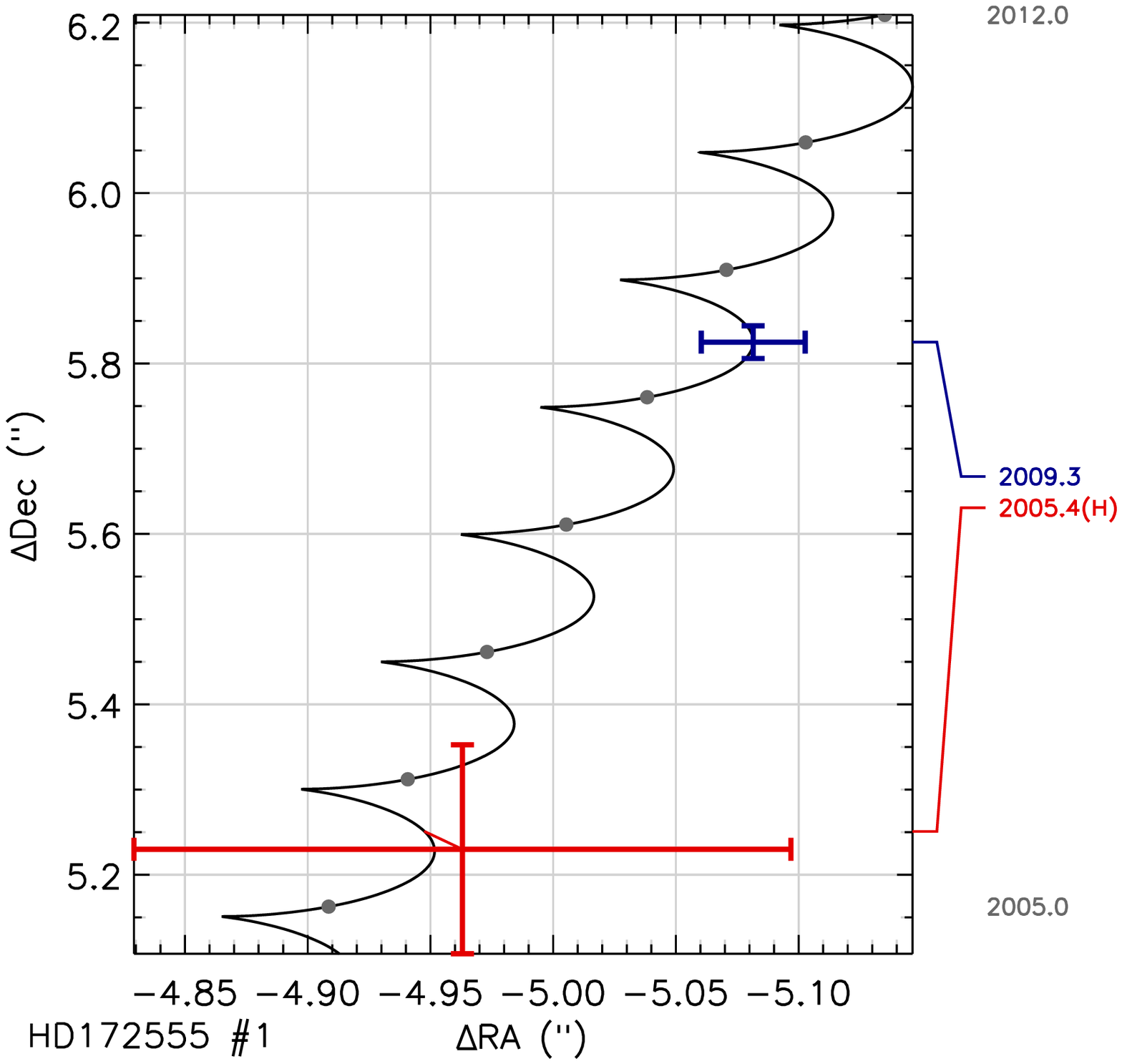}
    \hskip -0.3in
    \includegraphics[width=2in]{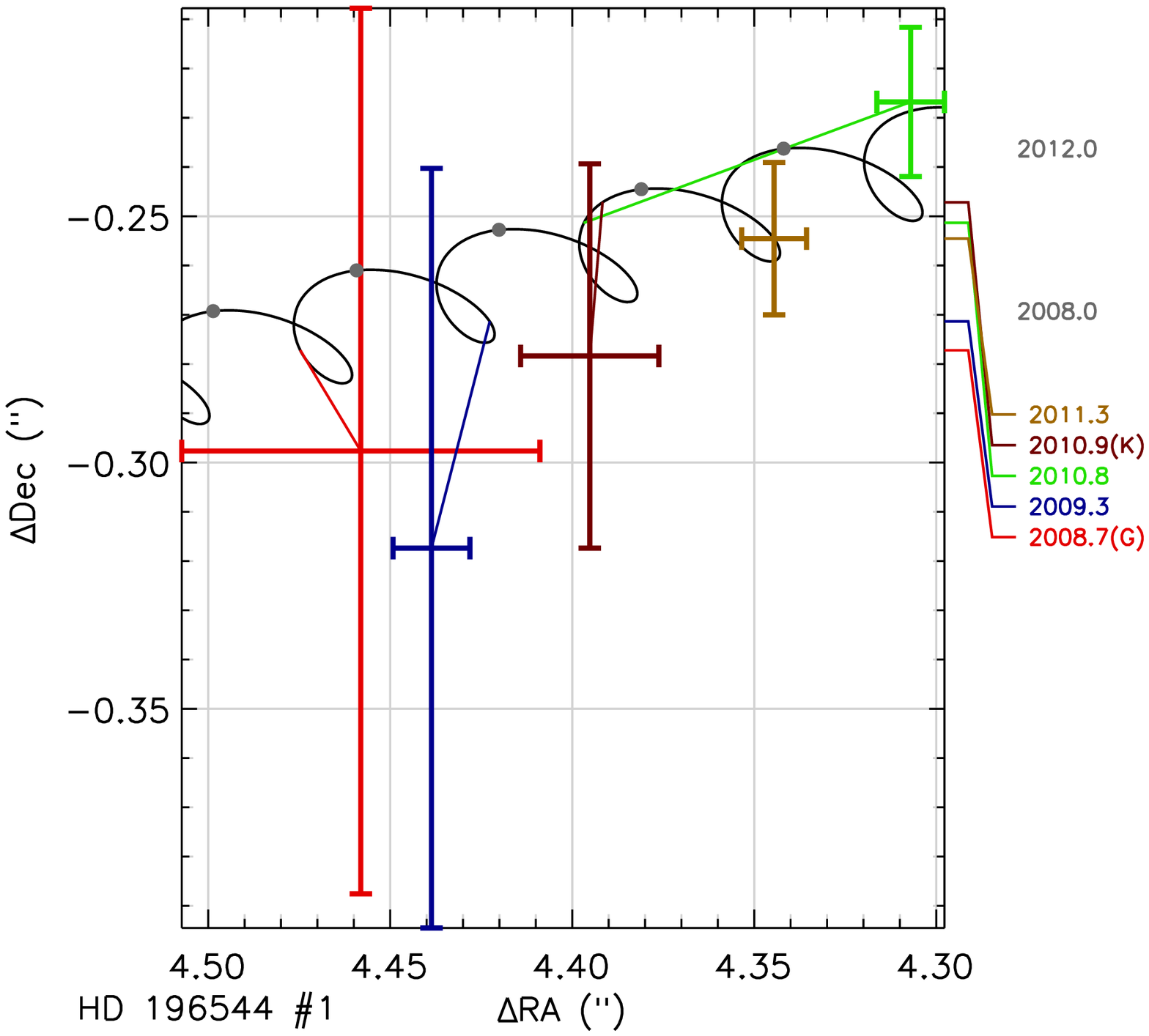}
    \hskip -0.3in
    \includegraphics[width=2in]{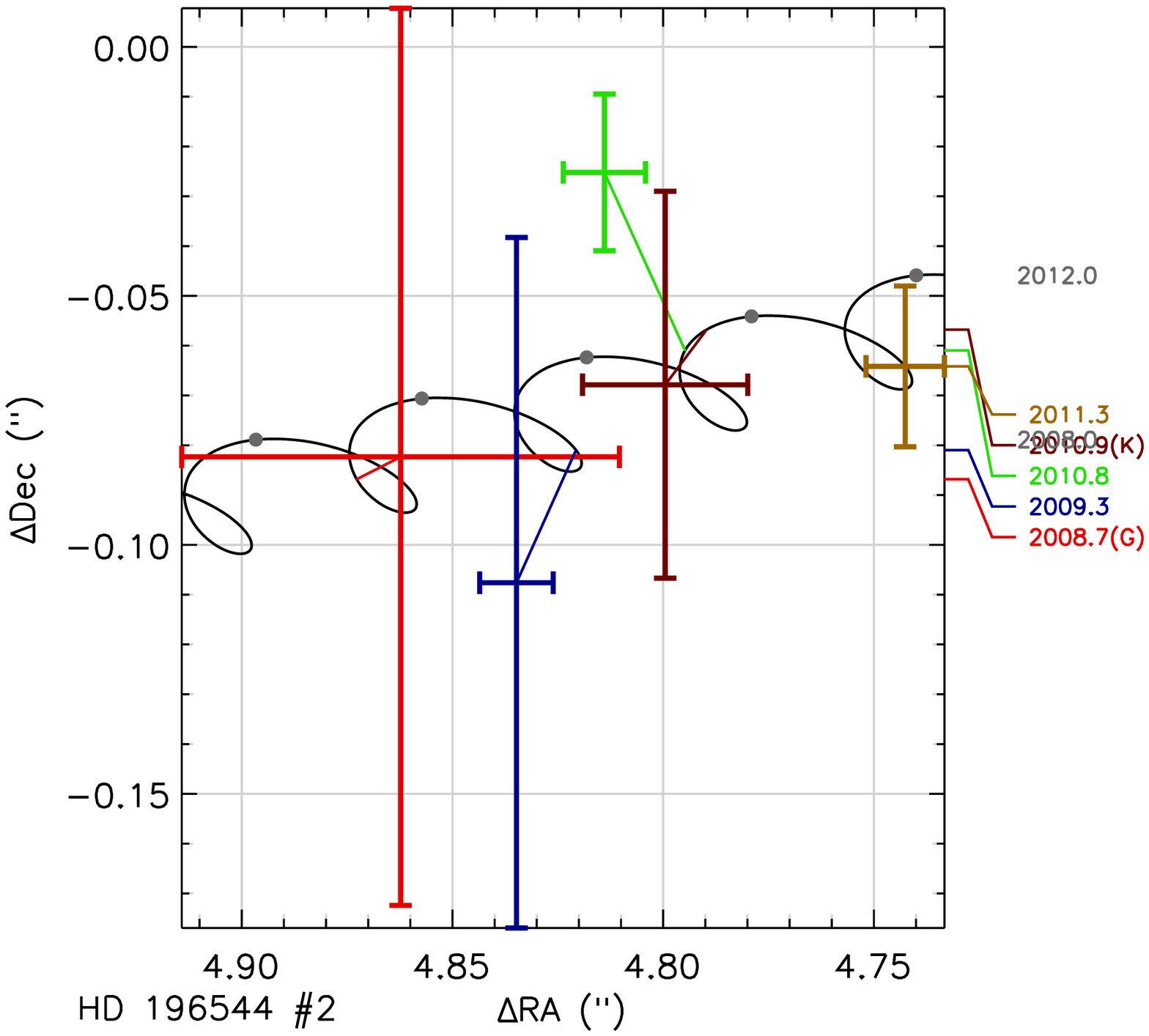}
  }
  \vskip -0.2in
  \centerline{
    \includegraphics[width=2in]{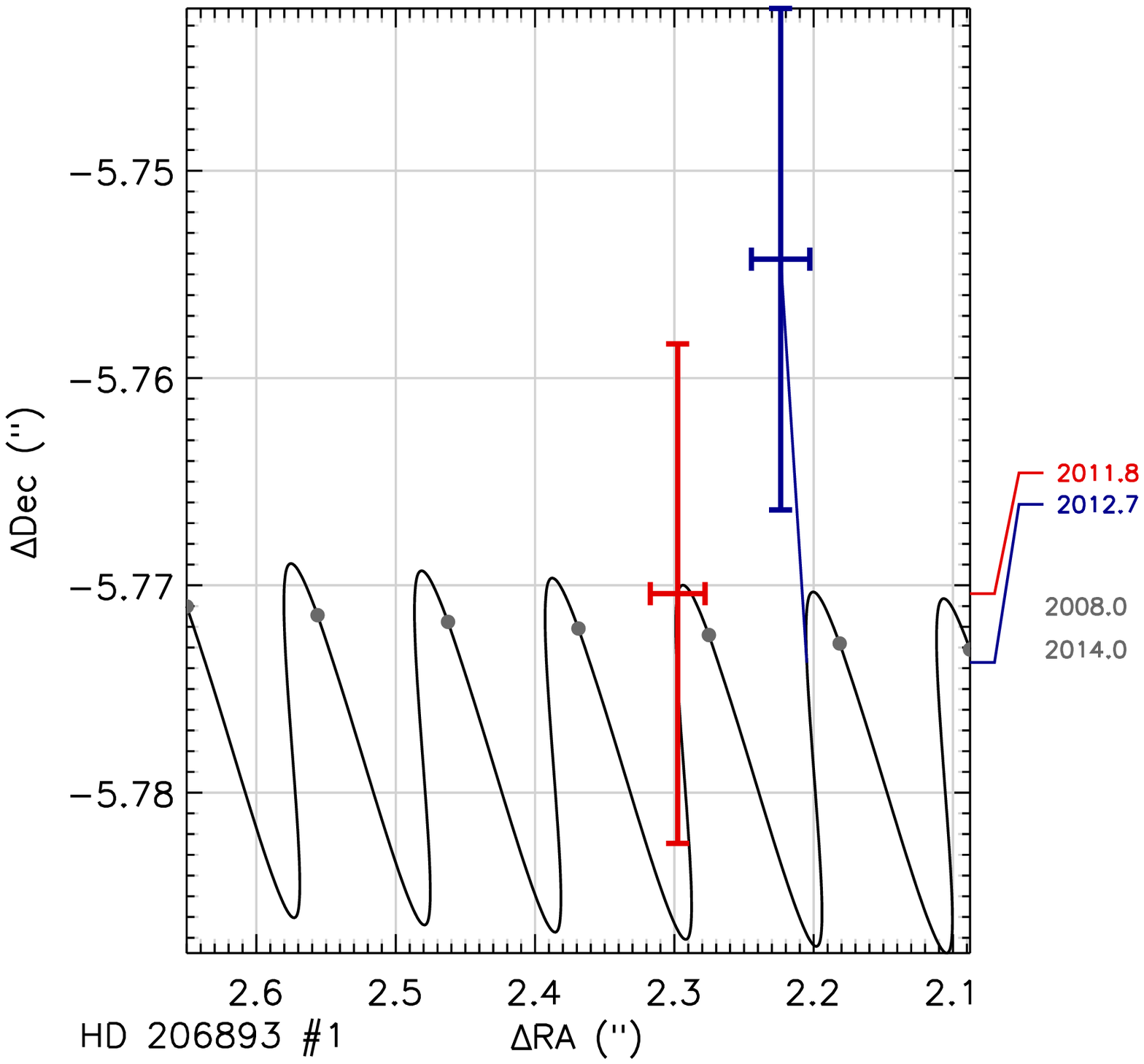}
    \hskip -0.3in
    \includegraphics[width=2in]{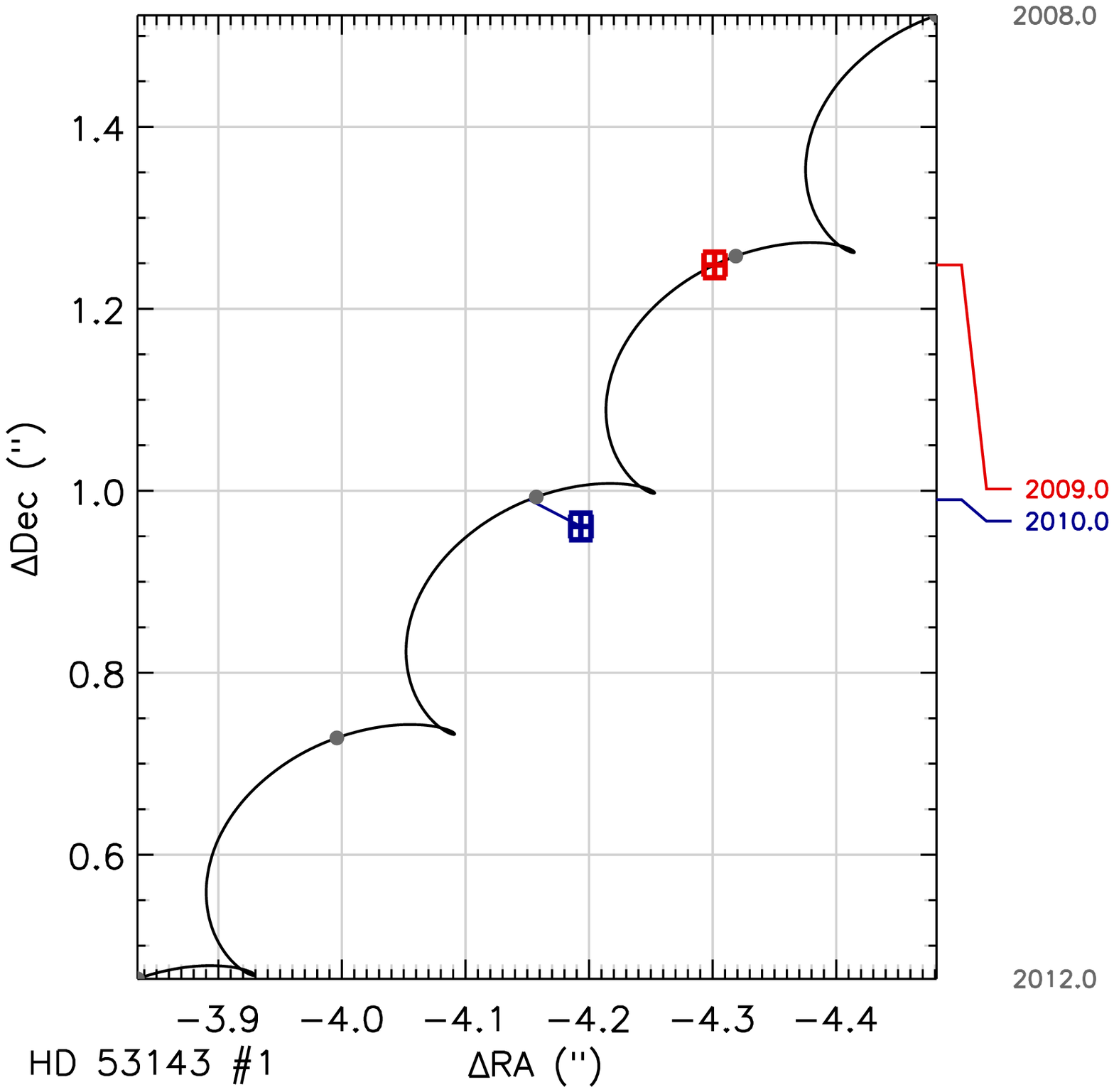}
    \hskip -0.3in
    \includegraphics[width=2in]{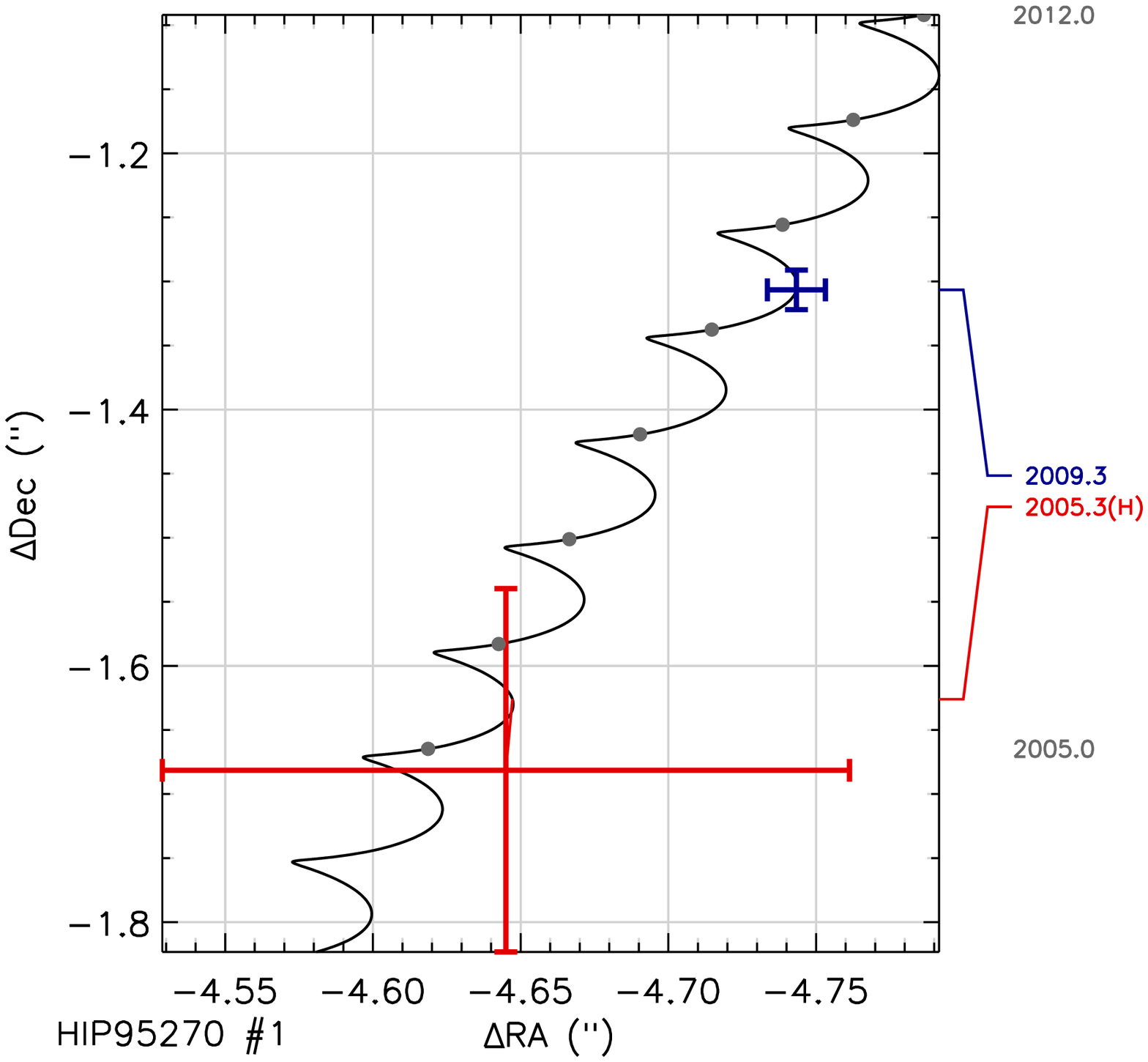}
    \hskip -0.3in
    \includegraphics[width=2in]{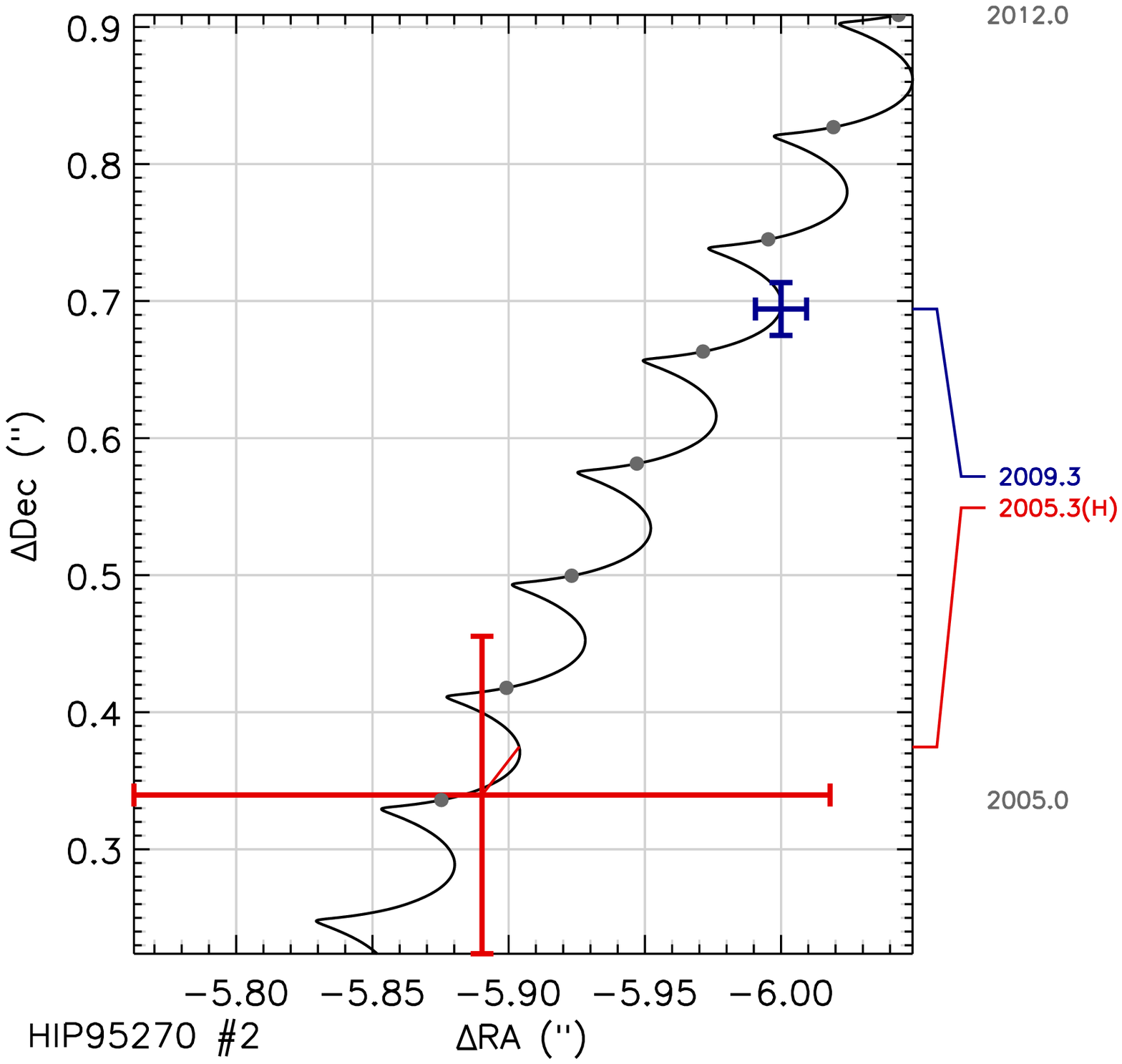}
  }
  \vskip -0.2in
  \caption{Same as Figure~\ref{fig:cand_motion}.}
  \label{fig:cand_motion4}
\end{figure} 

\begin{figure}
\centerline{
\includegraphics[width=2.0in]{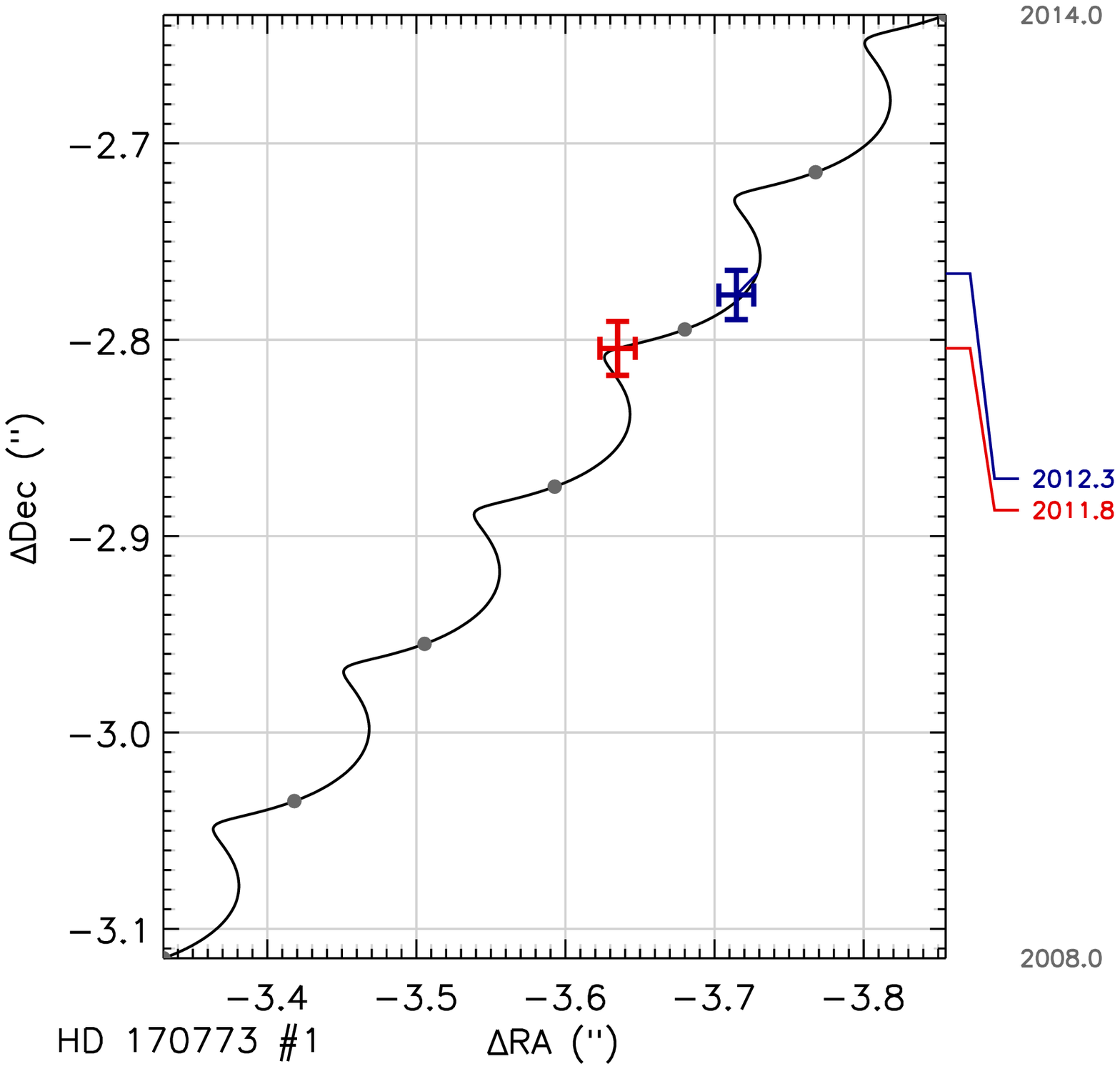}
\hskip -0.3in
\includegraphics[width=2.0in]{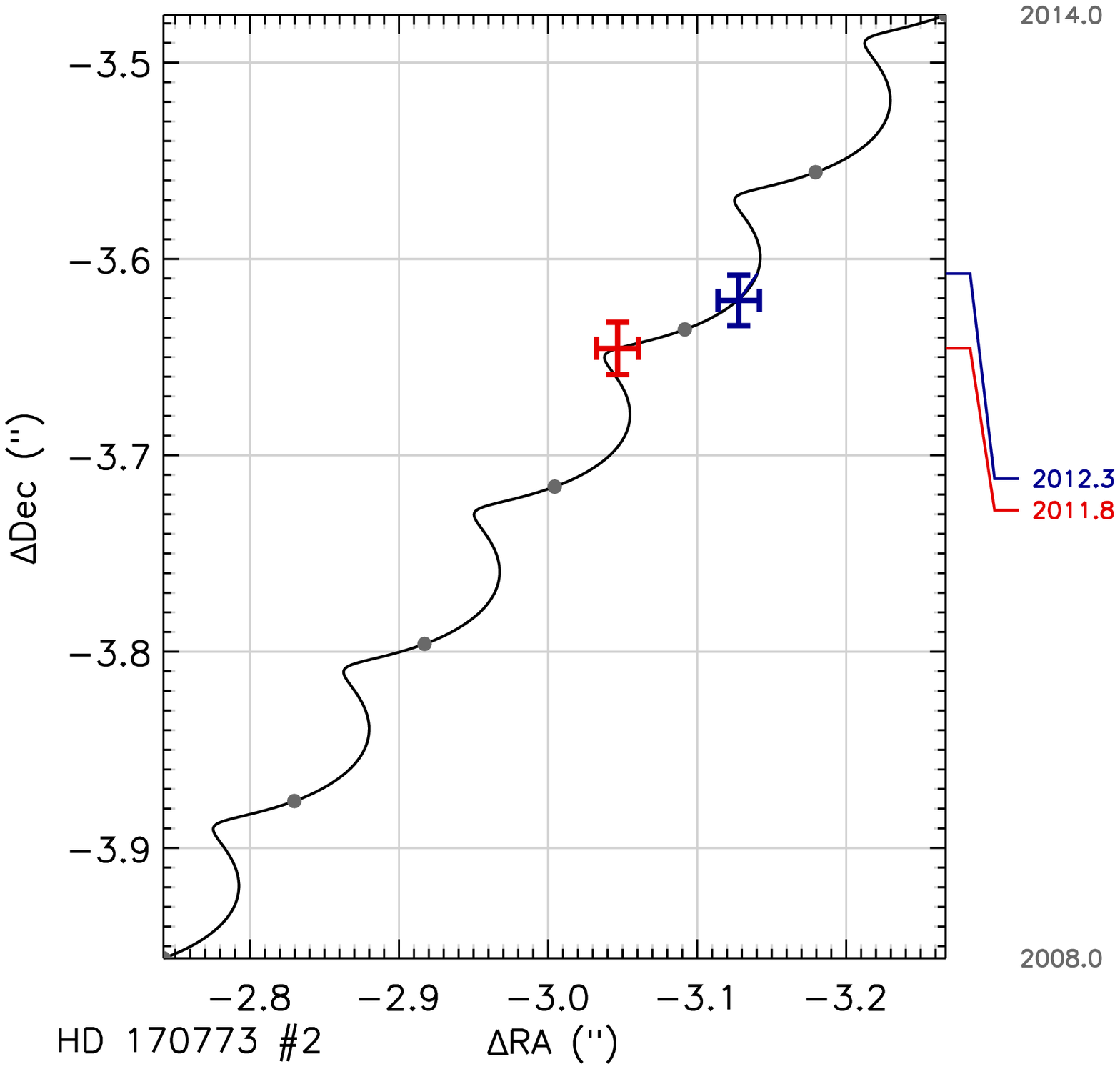}
}
\vskip -0.2in
  \caption{Same as Figure~\ref{fig:cand_motion}.}
  \label{fig:cand_motion5}
\end{figure}

\begin{figure}[H]
  \centerline{
    \hbox {
      \includegraphics[height=7cm]{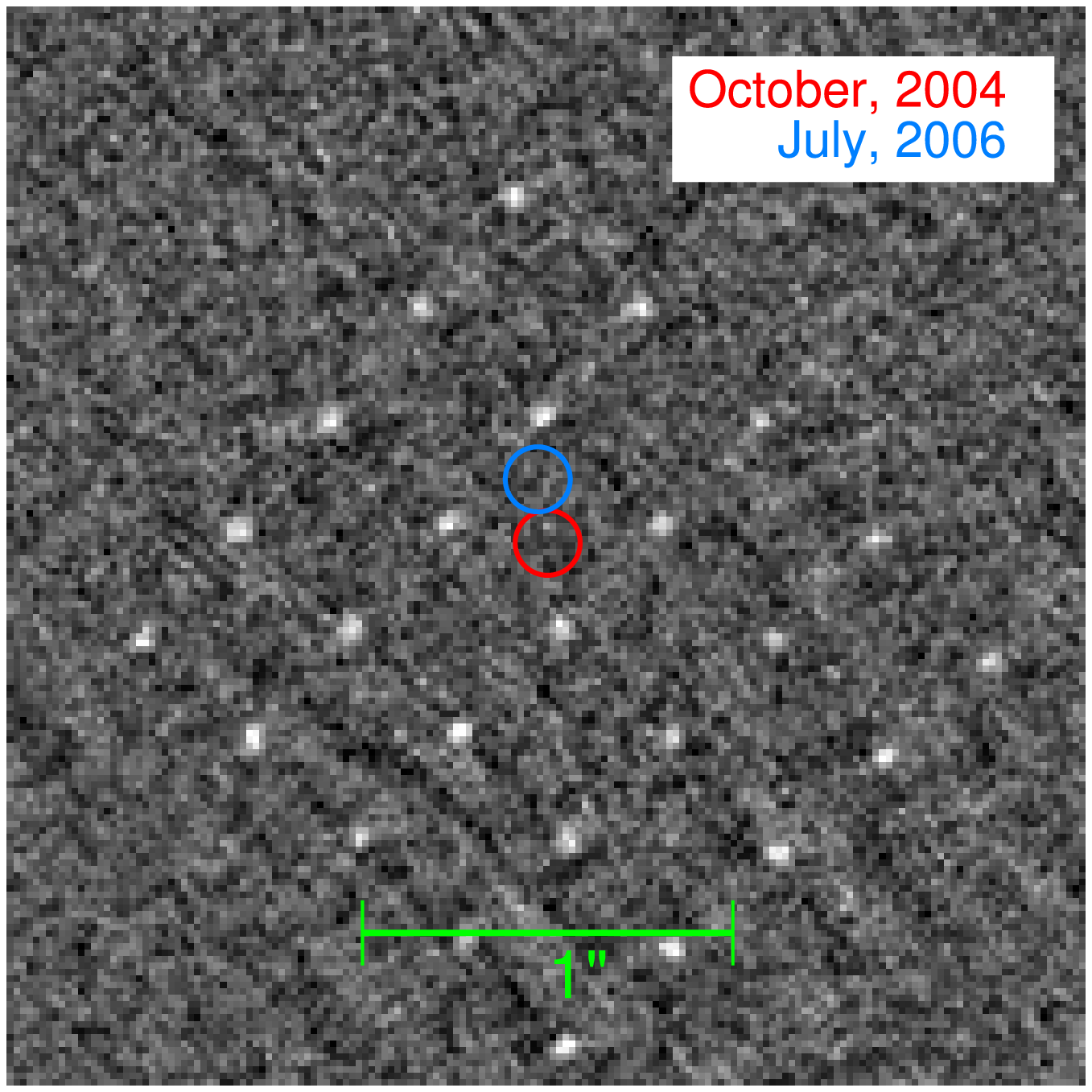}
      \includegraphics[height=7cm]{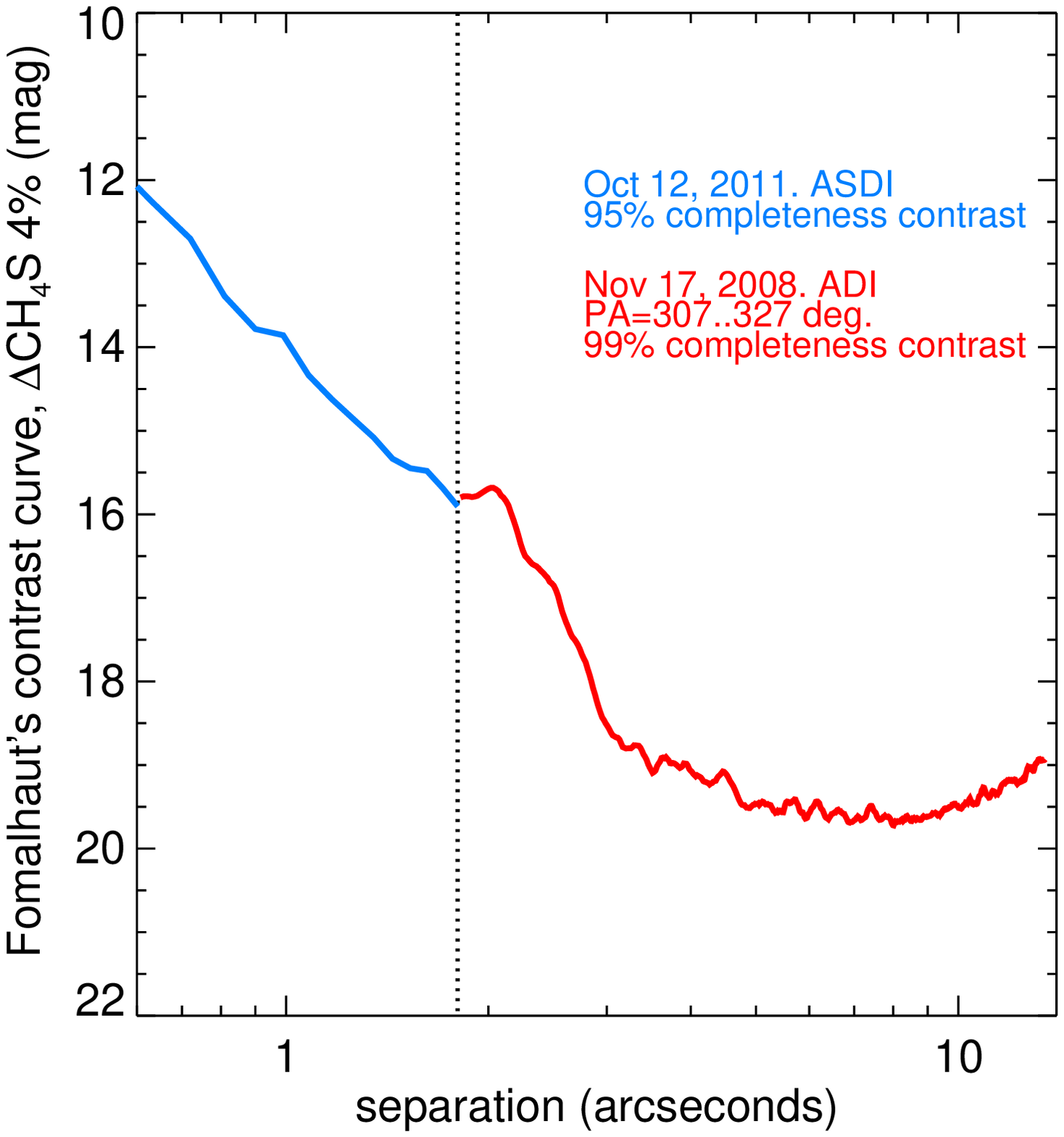}
    }}
  \caption{Left:  $CH_4S$ 4\% NICI reduced image of Fomalhaut obtained on UT 2008 November~17, showing the locations of Fomalhaut~b detections from \citet{2008Sci...322.1345K}. Also shown are simulated planets at the detection limit of the image at $20.2$~mag. Right: Contrast curves from the data sets obtained UT 2011, October 12 and UT 2008, November 17, joined at a separation of 1.8$''$. The 2008 contrast curve has been adjusted for anisoplanatism.}
  \label{fig:fomal}
\end{figure}

\begin{figure}[H]
  \centering{
    \includegraphics[width=8cm]{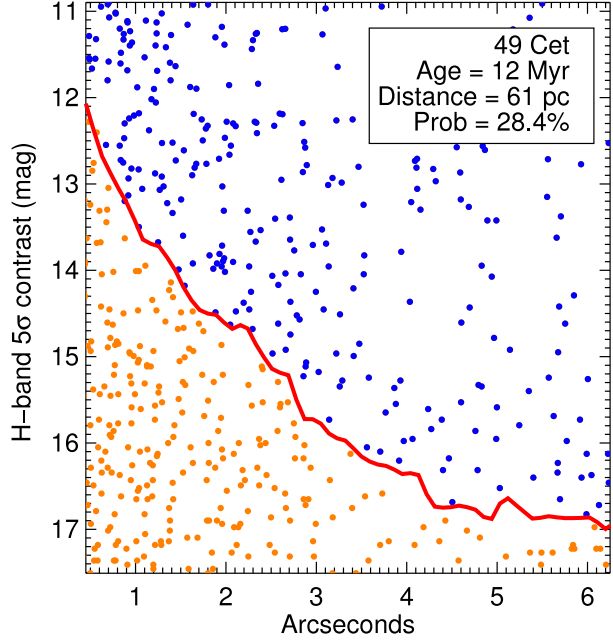}
  }  
  \caption{Simulated planets around 49 Cet. The red line is the 95\% completeness contrast we achieved for 49 Cet. 
    The blue dots represent planets which would have been detected by the Campaign observations, if they existed.
    The orange dots represent planets which would not be detected. The probability of detecting a planet, 
    assuming that exactly one planet exists around 49 Cet is 28.4\%. These are the $f_j$ values we use in our 
    Bayesian computations. The mass and SMA power-law indices, $\alpha$ and $\beta$, 
    were set to $-1.1$ and $-0.6$, respectively for the planet population model used in this particular simulation.}
  \label{fig:completeness}
\end{figure}

\begin{figure}[H]
     \hbox {
        \includegraphics[width=5cm]{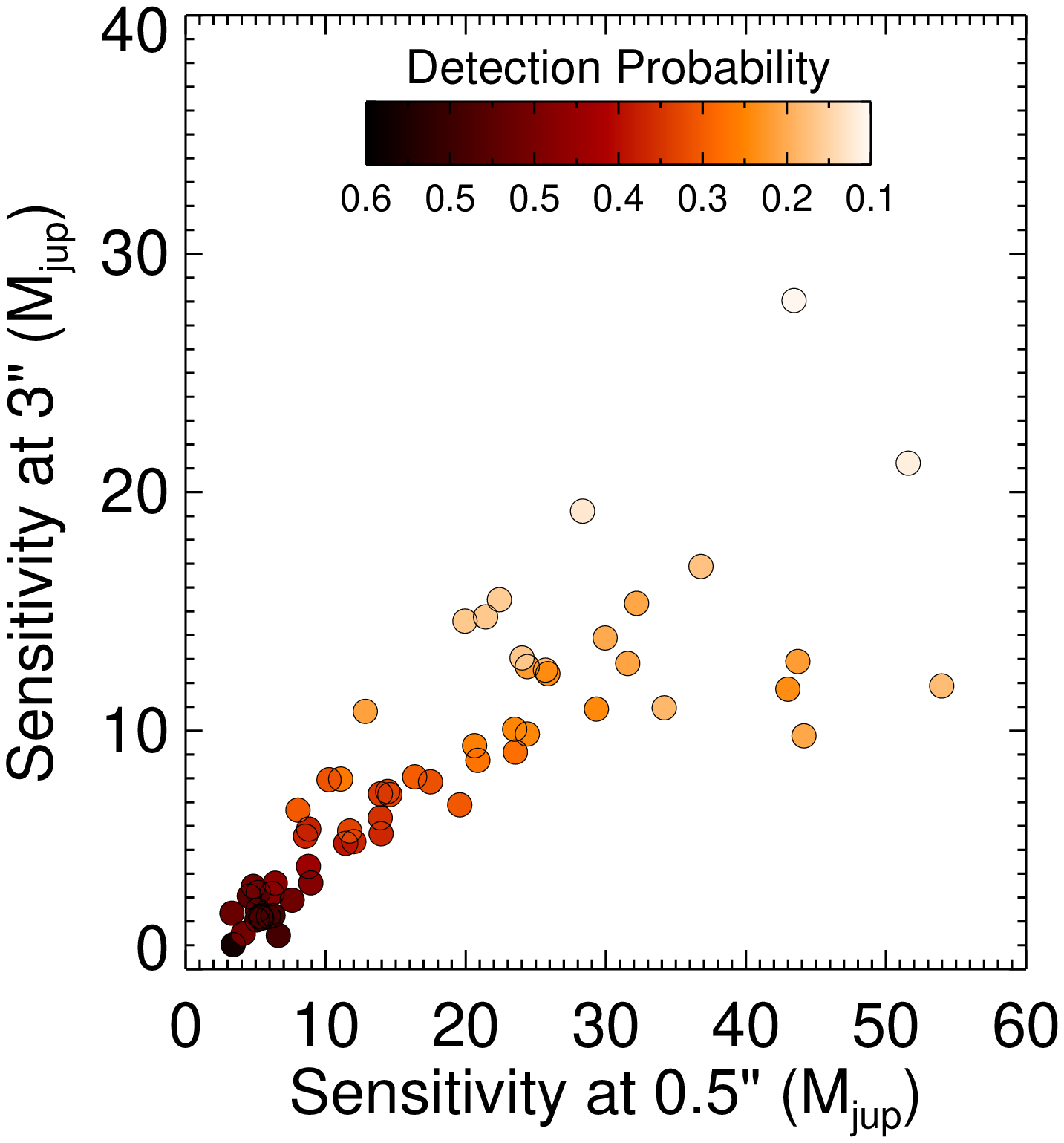}
        \includegraphics[width=5cm]{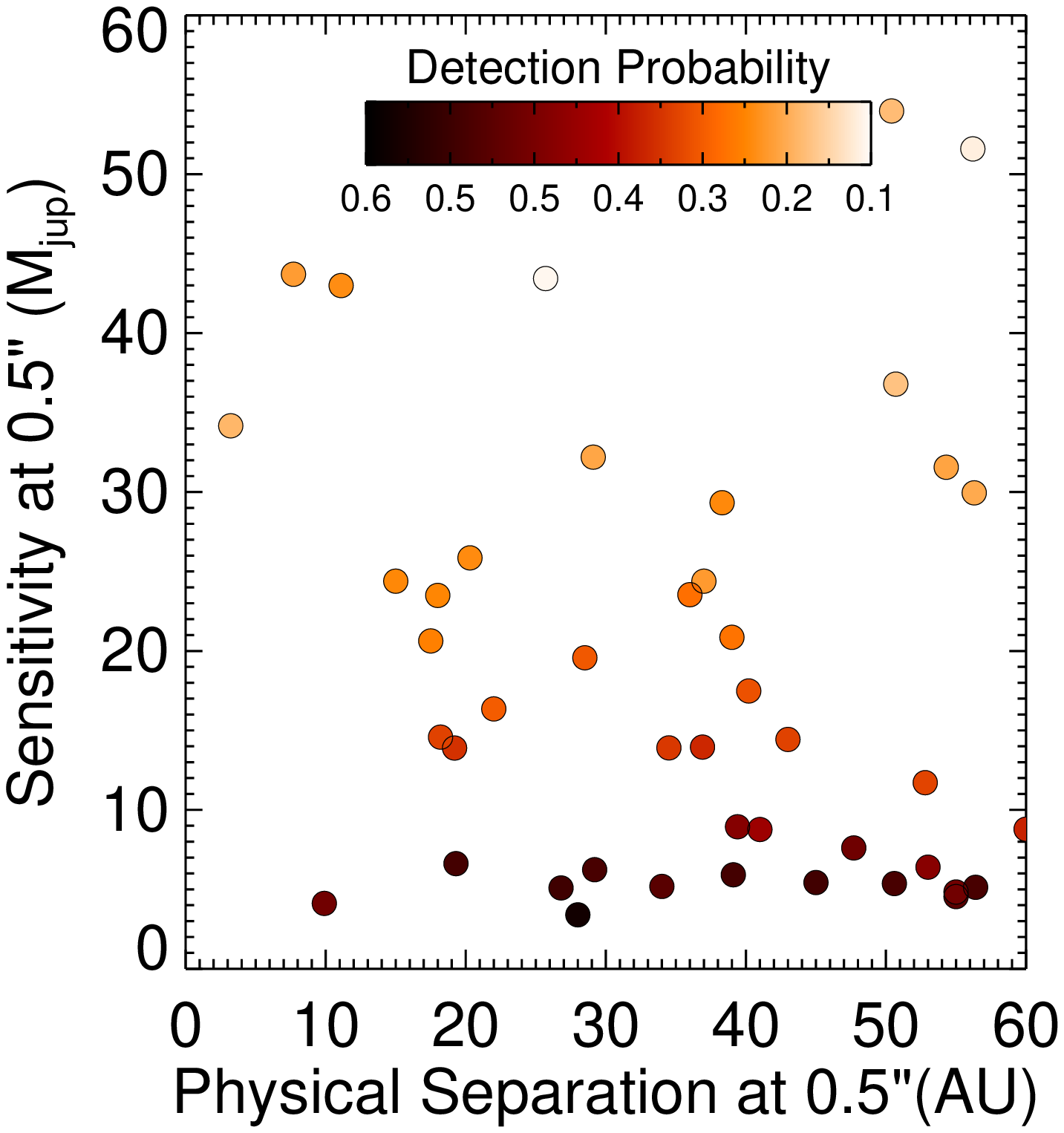}
        \includegraphics[width=5cm]{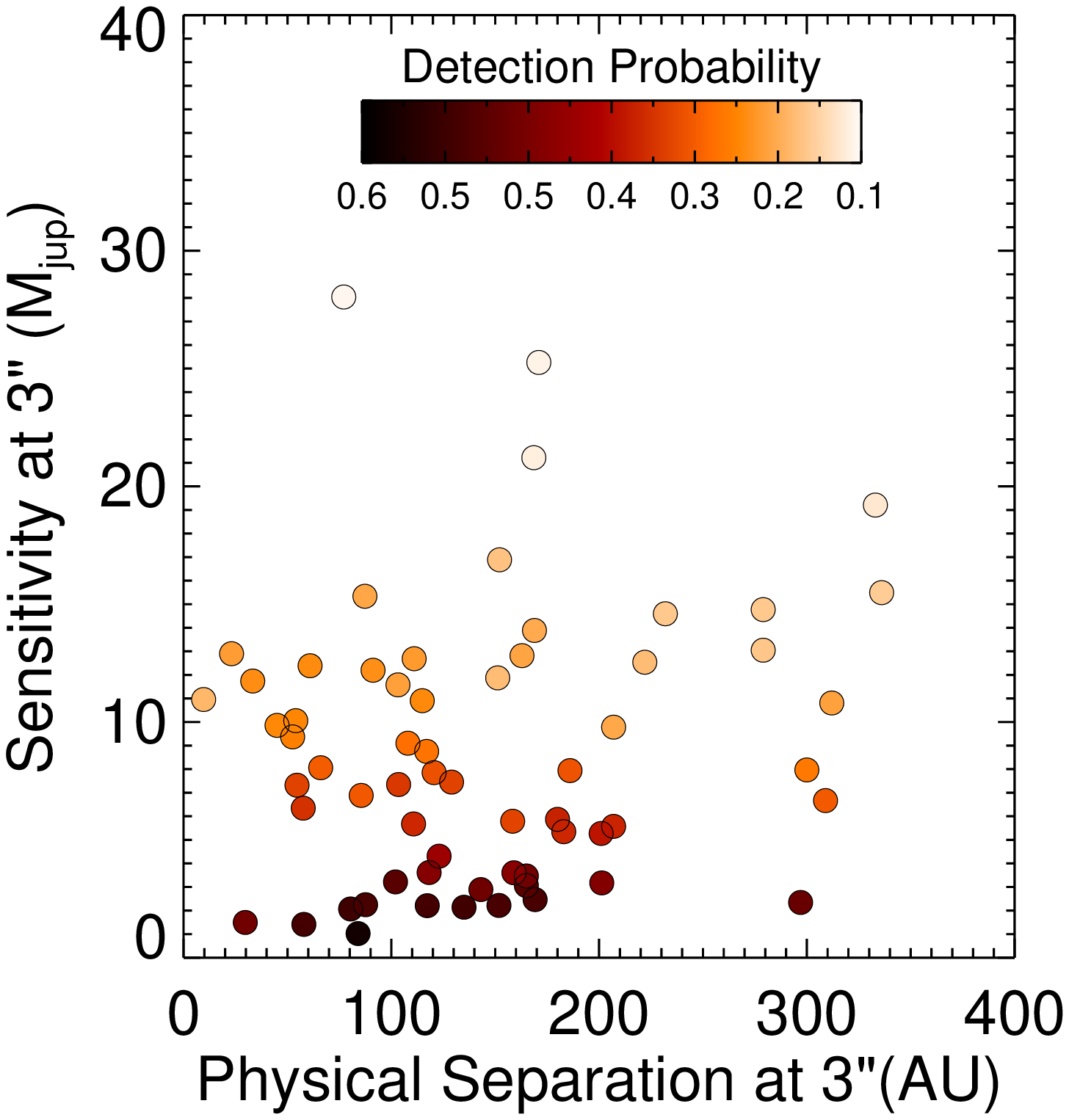}
      }  
  \caption{Comparison of the mass sensitivities reached at 0.5$''$ and 3$''$ and the detection probabilities, $f_j$.}
  \label{fig:mass_lim}
\end{figure}

\begin{figure}[H]
     \hbox {
        \includegraphics[height=8cm]{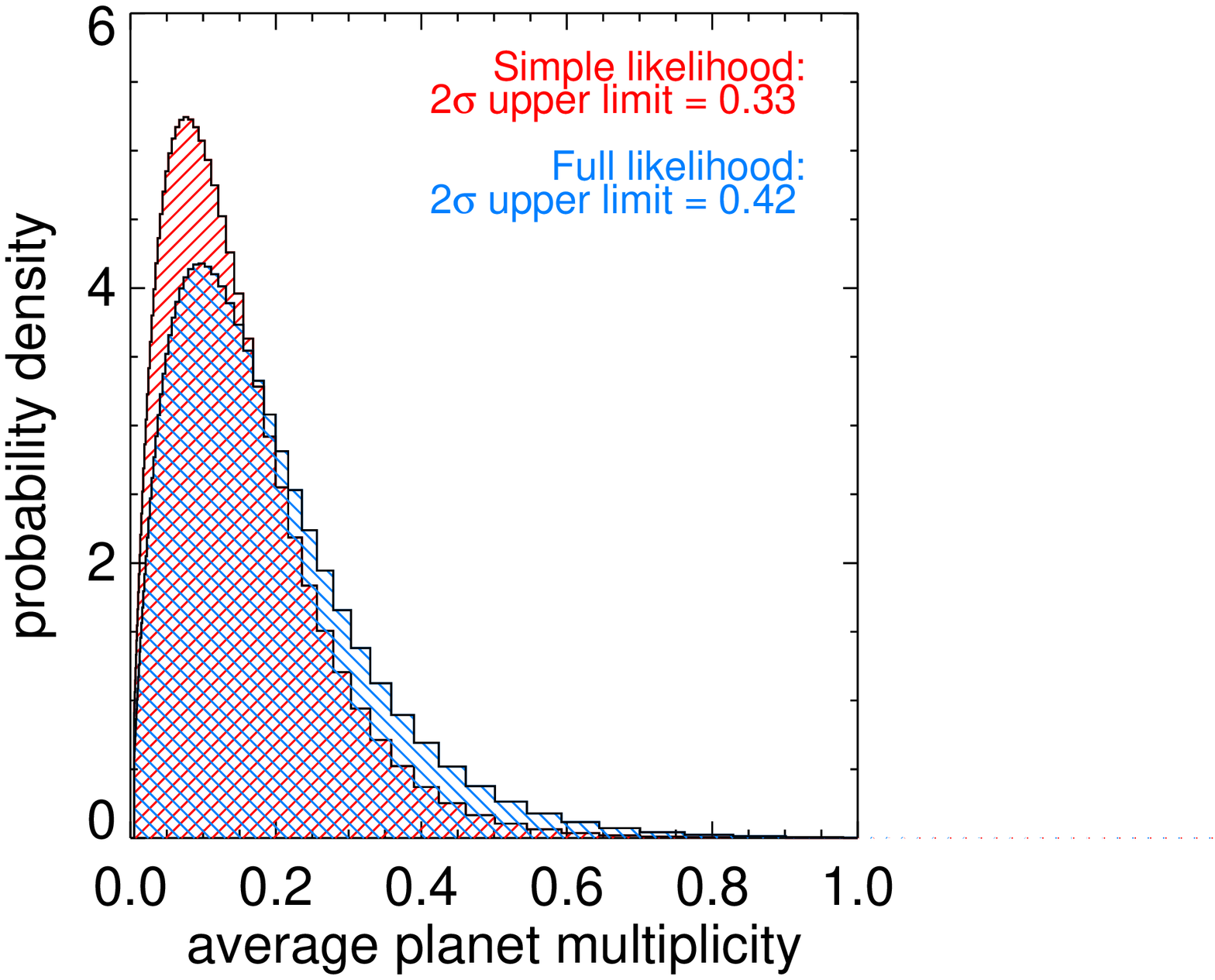}
        \includegraphics[height=8cm]{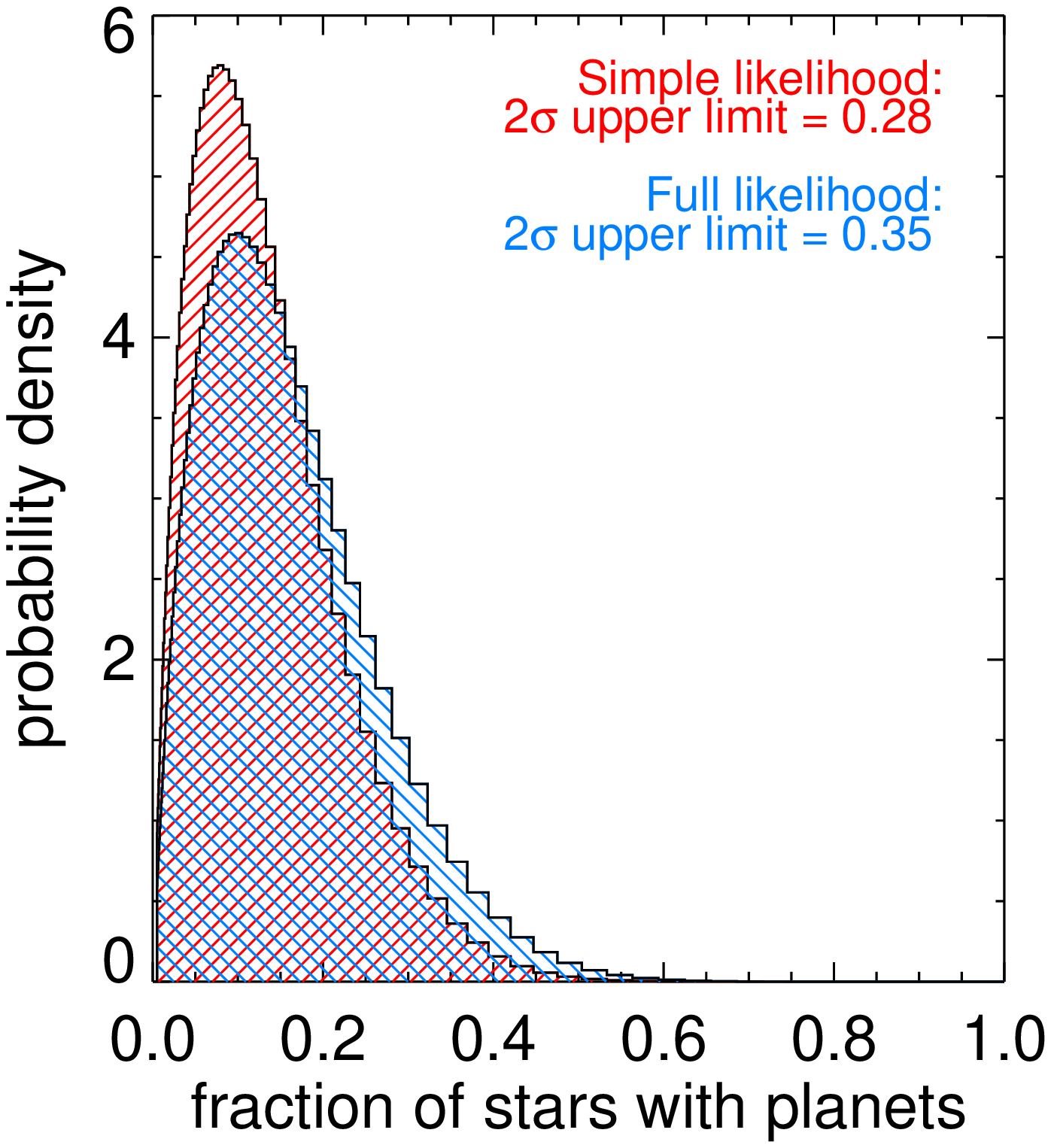}
      }  
  \caption{Left: The probability distribution for the average number
    of planets in systems with detected debris disks, $F$, when
    $\alpha = -1.31$ and $\beta = -0.61$ are set from \citet{2008PASP..120..531C}. The distribution 
    shown in red results from the simple likelihood where only the average number of planets detected is considered. The distribution in blue
    results from the full likelihood where the number of detections around a star is also considered. 
    Right: The probability distribution for the fraction of stars with planets, $F_{sp}$.}
  \label{fig:limits_fp}
\end{figure} 

\begin{figure}[H]
     \hbox {
        \includegraphics[height=8cm]{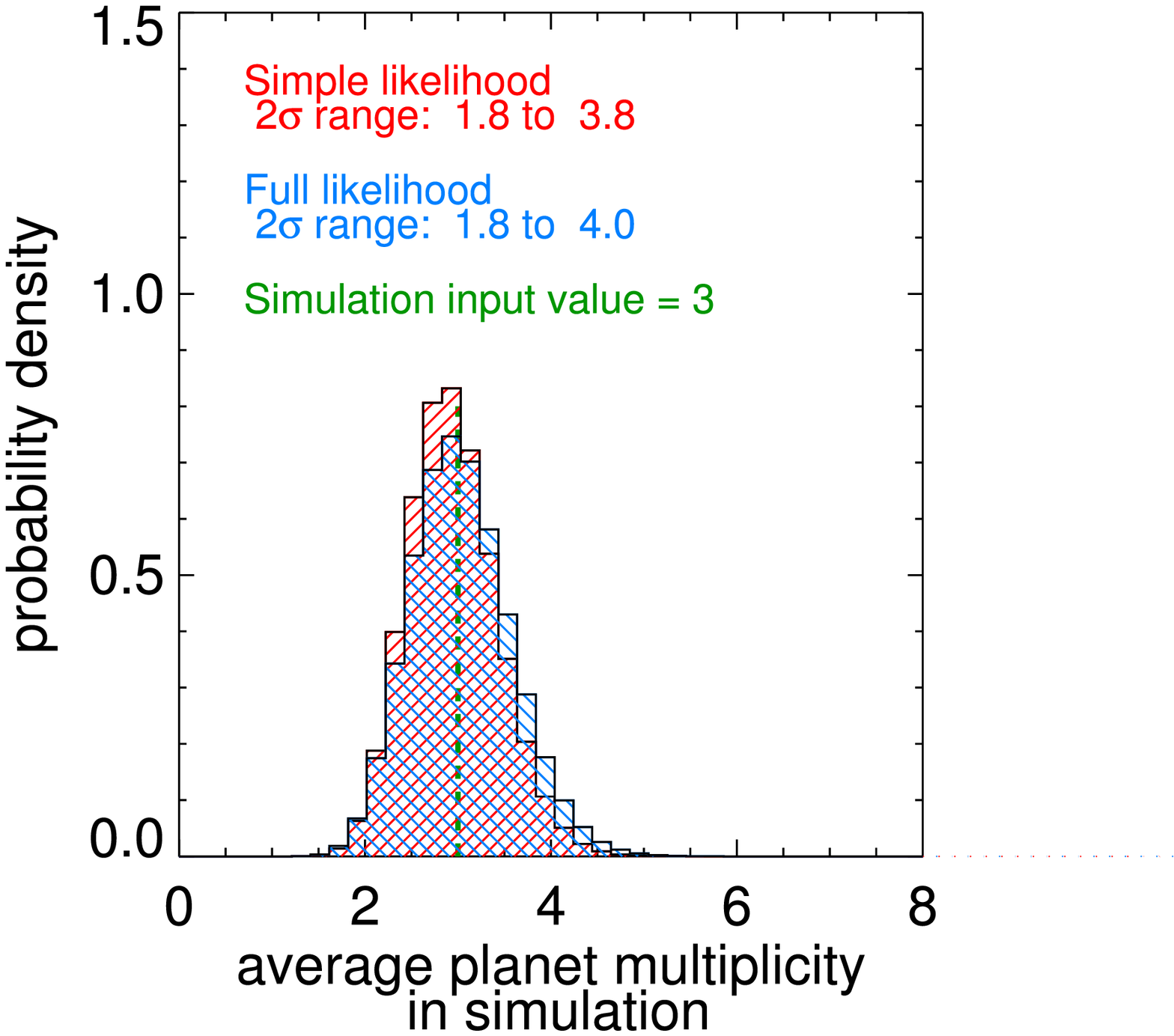}
        \includegraphics[height=8cm]{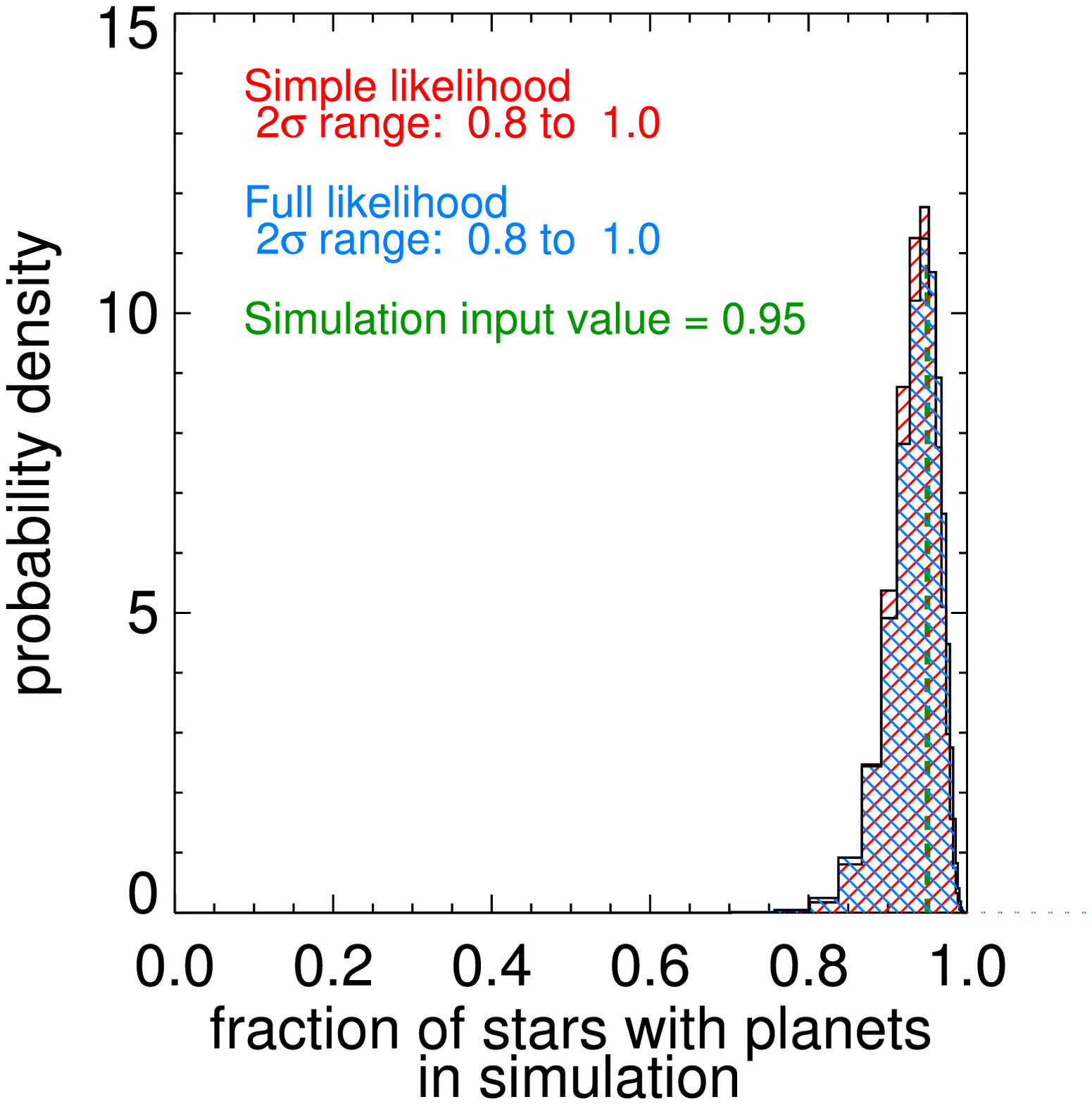}
      }  
  \caption{Probability distributions for the planet population from a
    simulation where the average planet multiplicity was 3, when
    $\alpha = -1.31$ and $\beta = -0.61$ are set from \citet{2008PASP..120..531C}.
    Left: The probability distribution obtained for the average number of
    planets ($F$)  in the simulated population. The distribution 
    shown in red results from the simple likelihood, while the distribution in blue
    results from the full likelihood. Right: The probability
    distribution for the fraction of stars with planets ($F_{sp}$) in the simulated population.
    For $F=3$, $F_{sp}$ should be 0.95. \textcolor{black}{The consistency our results 
    with the input simulated population demonstrates the robustness of
    our Bayesian method.} }
  \label{fig:fake_limits_fp}
\end{figure} 

\begin{figure}[H]
  \hbox {
    \includegraphics[width=8cm]{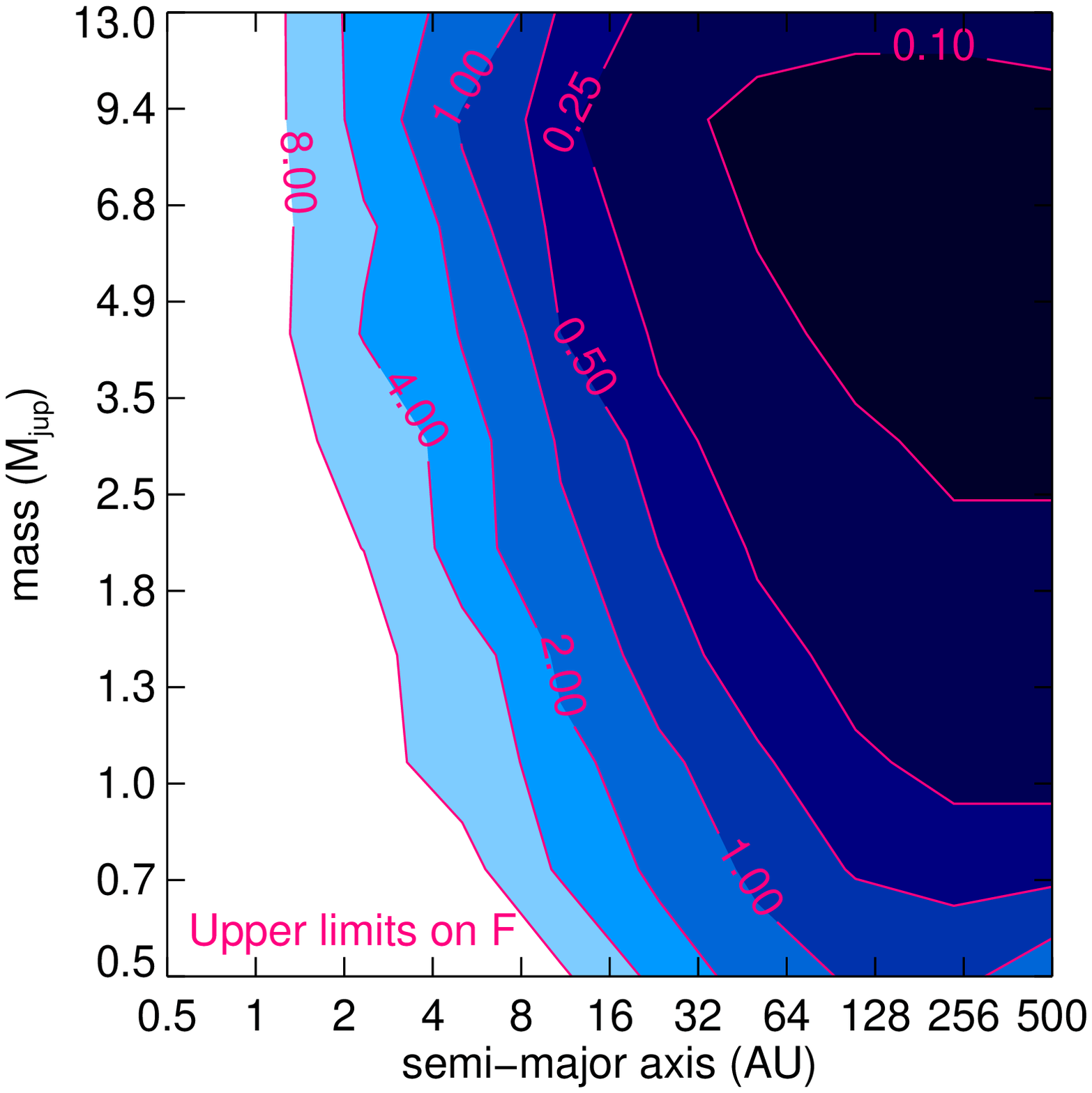}
    \includegraphics[width=8cm]{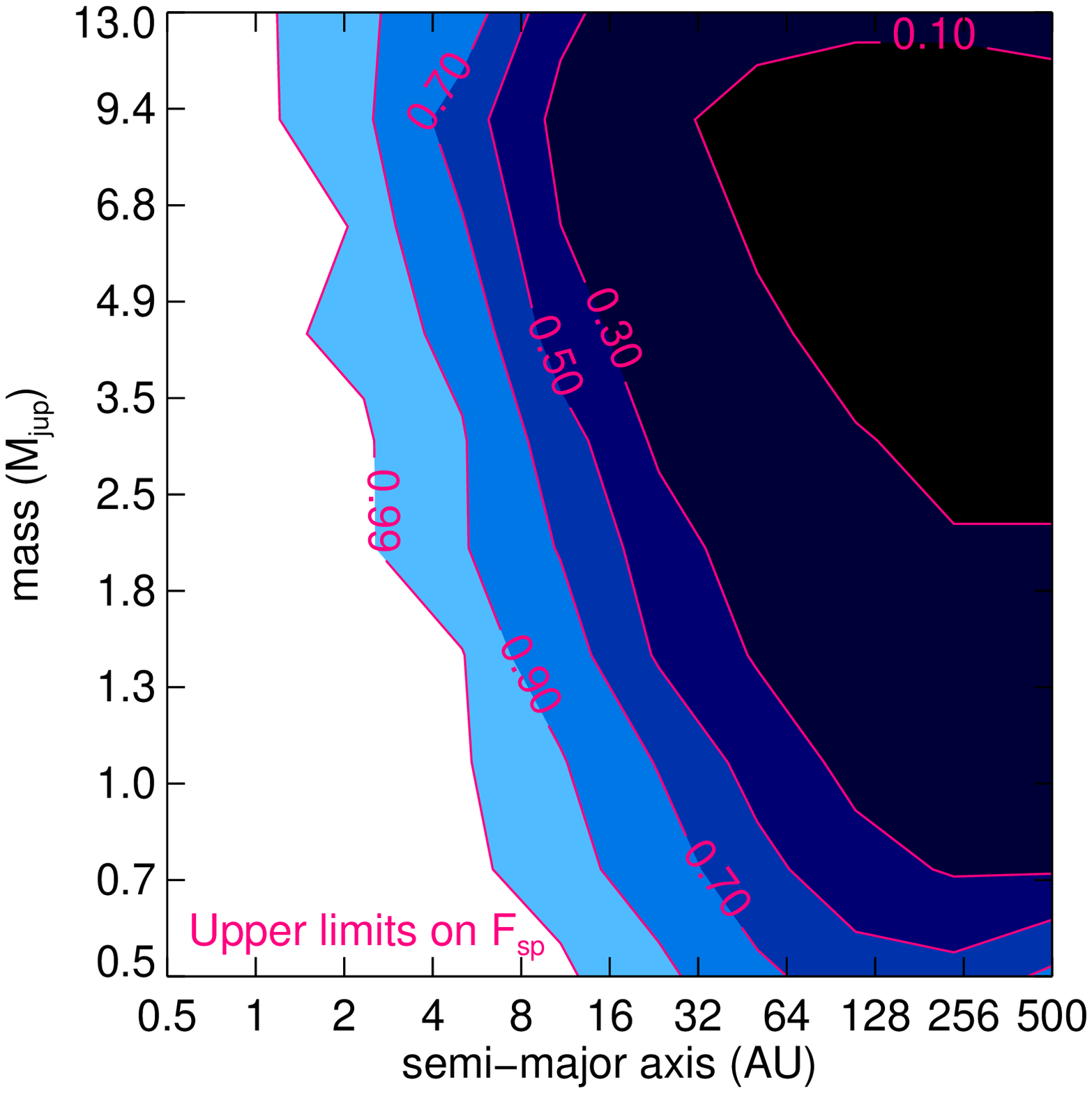}
  }  
  \caption{Left: The 2$\sigma$ upper limits on the average number of planets in systems with detected debris disks, $F$, as a function of planet mass and orbital semi-major axis.
    Right: The 2$\sigma$ upper limits on the fraction of stars with planets, $F_{sp}$, which is only valid for very small $F$ or when 
    stars are only allowed to have either one or zero planets. Again, the upper limits are given as a function of planet mass and SMA.}
  \label{fig:limits_mass_sep}
\end{figure} 

\clearpage 

\renewcommand{\baselinestretch}{1}
\begin{figure}[H]
  \vbox{
    \hbox{
      \includegraphics[height=8cm]{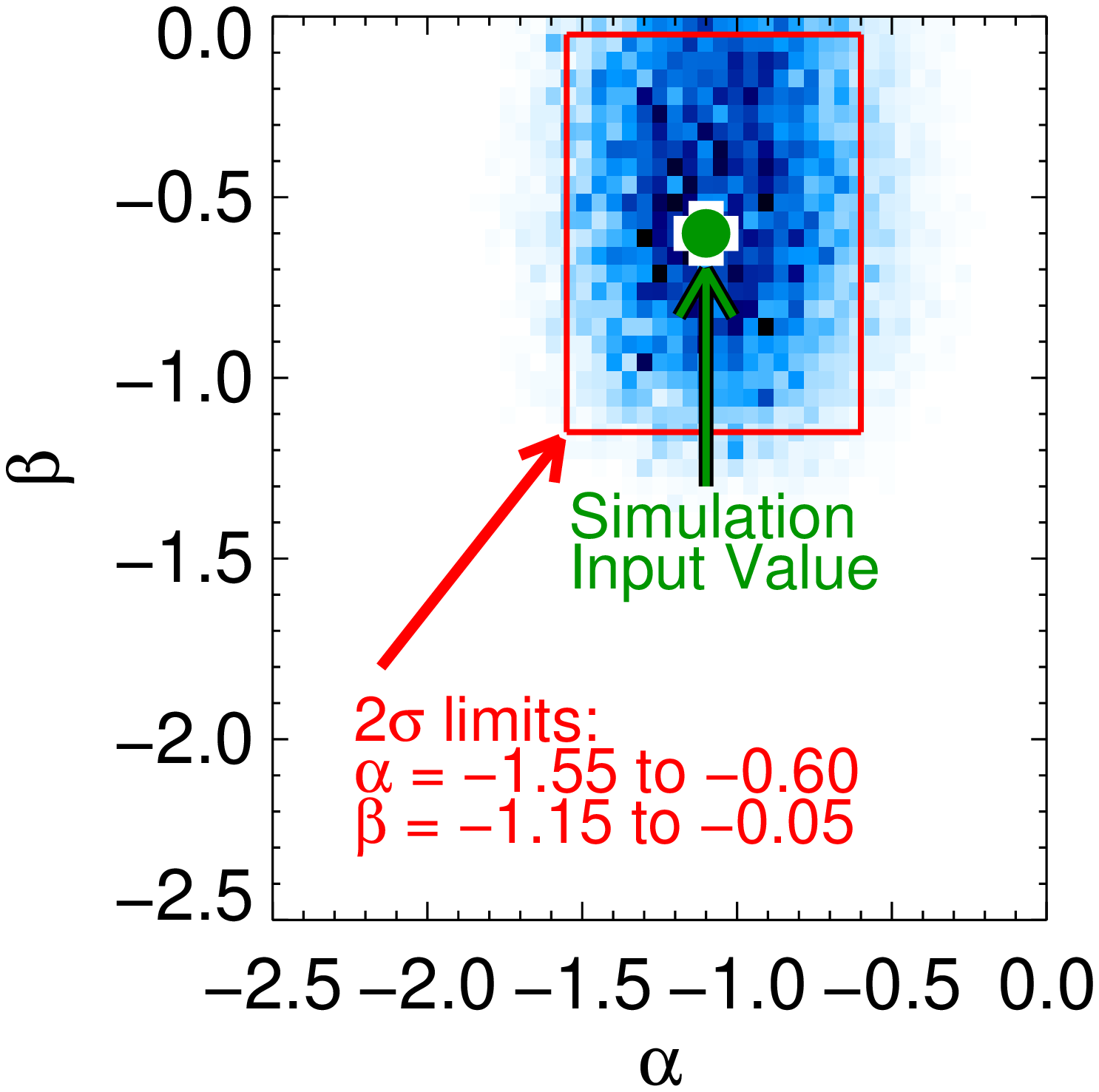}
      \includegraphics[height=8cm]{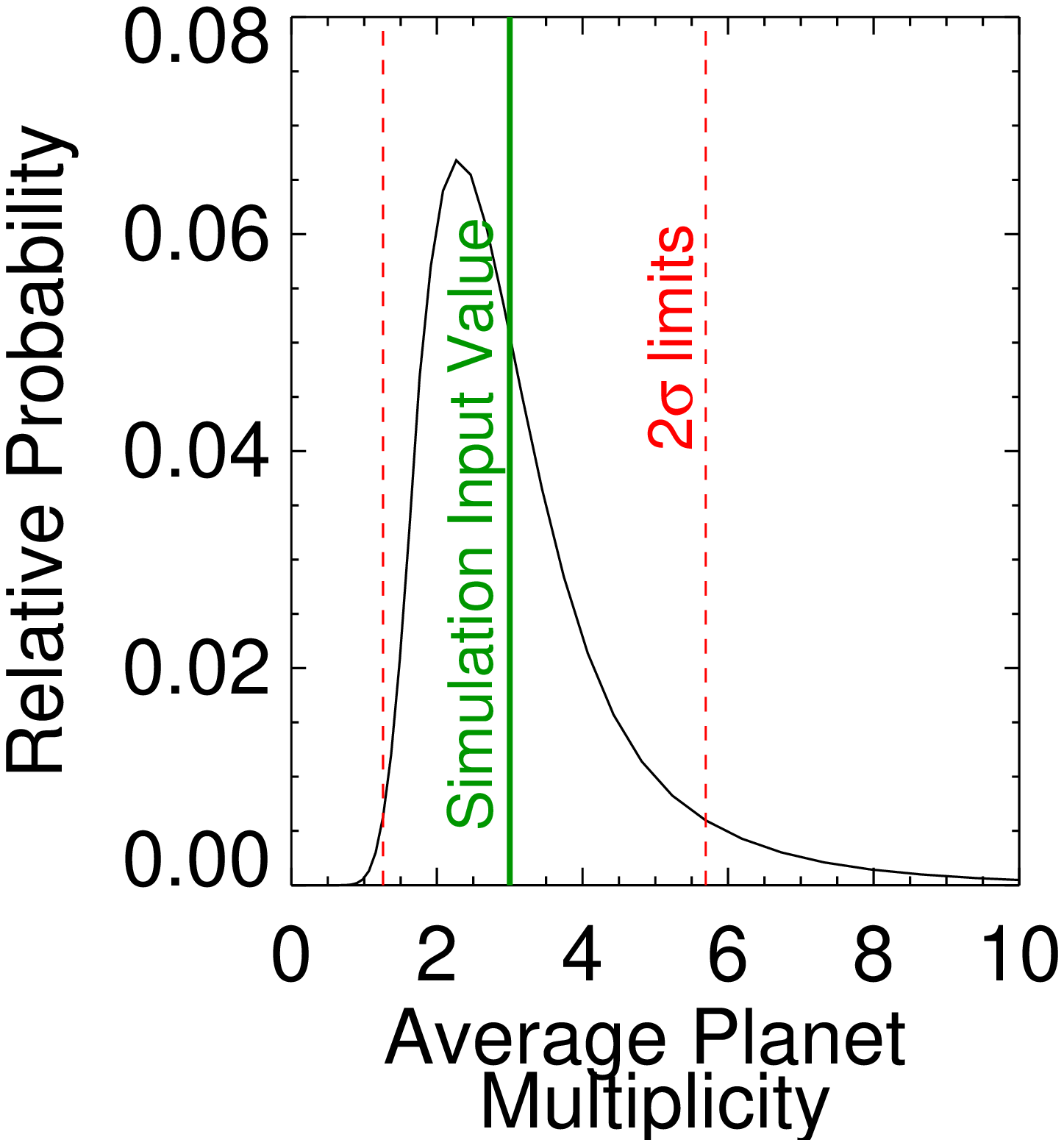}}
    \hbox{
      \includegraphics[height=8cm]{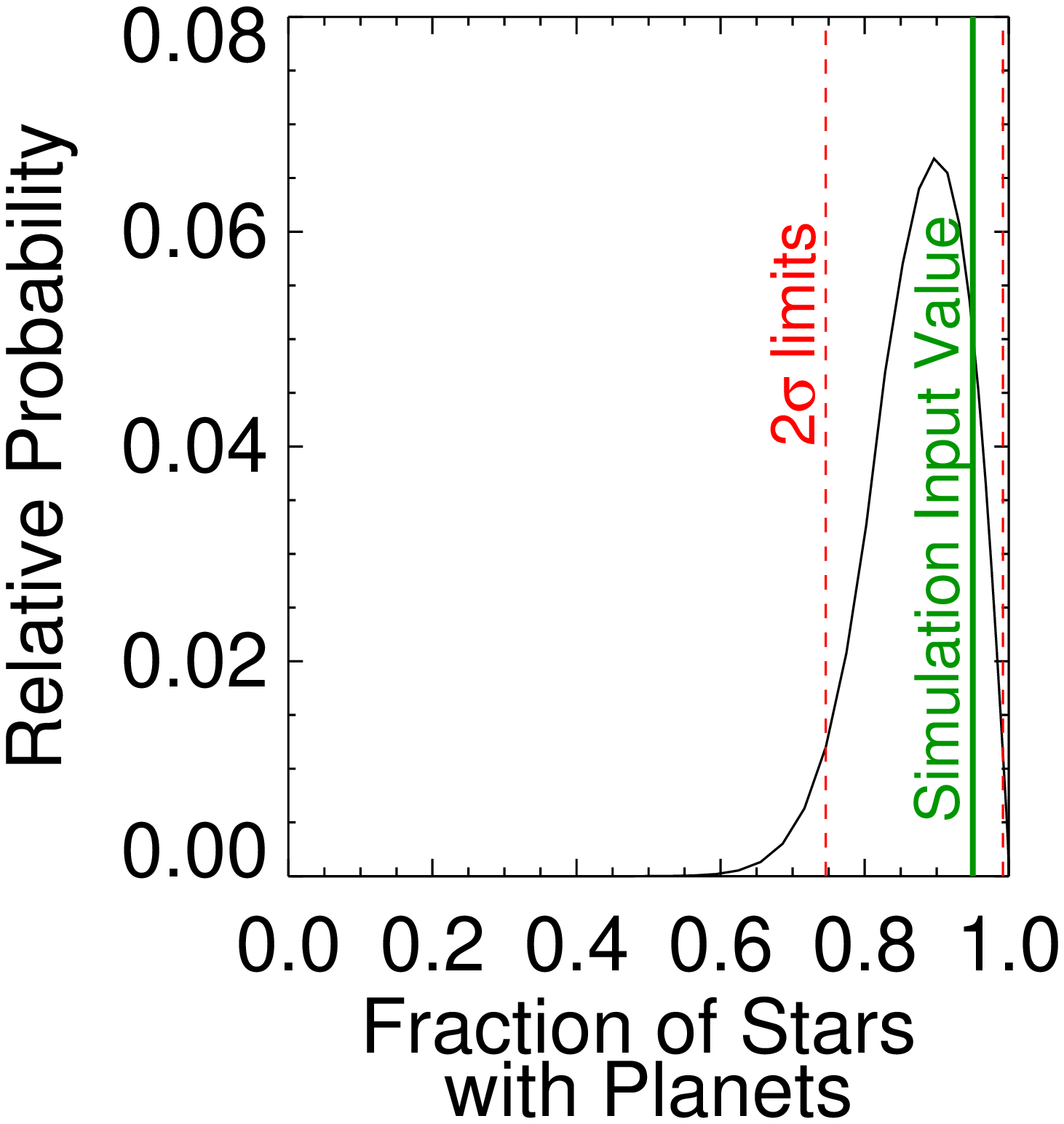}
      \includegraphics[height=8cm]{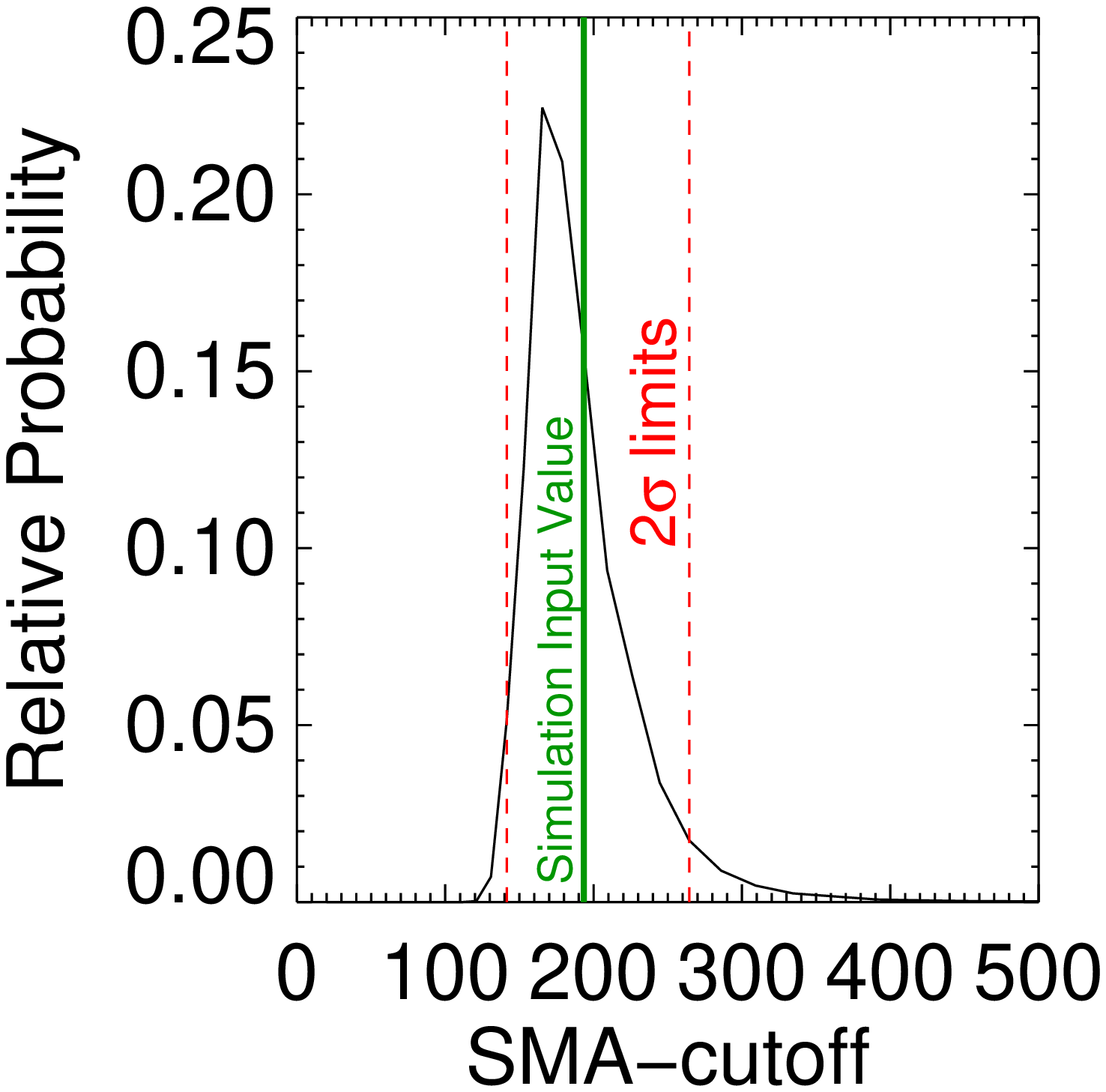}
    }}
  \caption{Constraints from a simulated survey where the input planet
    population has $\alpha=-$1.1 and $\beta=-$0.6, an average planet
    multiplicity of 3.0 and SMA cutoff of 193~AU. Darker regions
    indicate higher probability. Both astrometry and
    photometry of the model and simulated data are compared. Figures
    from left to right: the probability distributions for (1)  the
    planet mass and SMA power-law indices, $\alpha$ and $\beta$ (joint
    distribution), (2) the average planet multiplicity, (3) the
    fraction of stars with planets, and (4) the SMA cutoff beyond
    which no planets are allowed. The 2$\sigma$ constraints from the
    Bayesian analysis are indicated by the red
    lines. \textcolor{black}{The consistency of our results 
    with the input simulated population demonstrates the robustness of
    our Bayesian method.}}
  \label{fig:alpha_beta}
\end{figure} 

\clearpage 

\begin{figure}[H]
  \vbox{
    \hbox{
      \includegraphics[height=7cm]{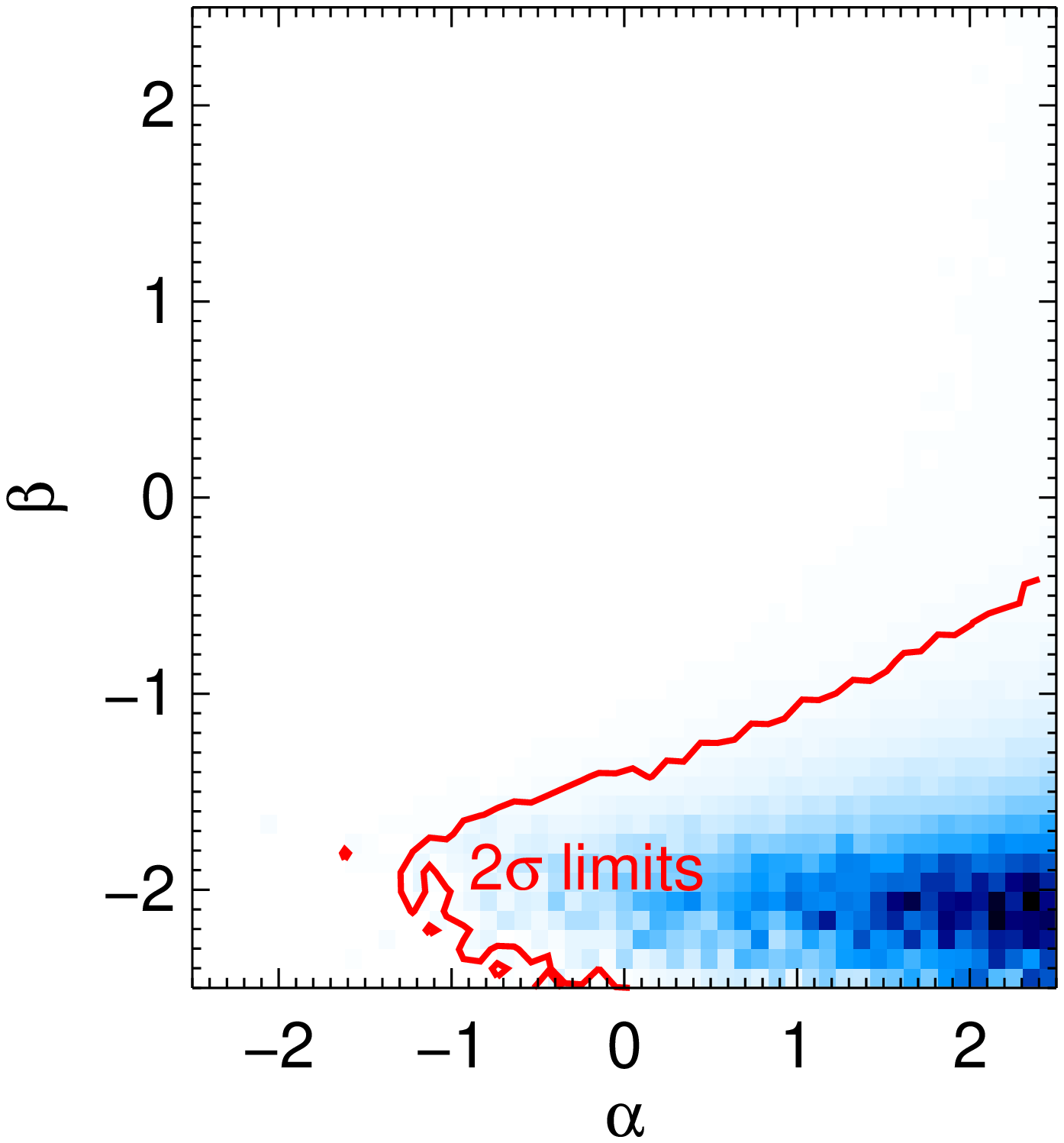}
      \includegraphics[height=7cm]{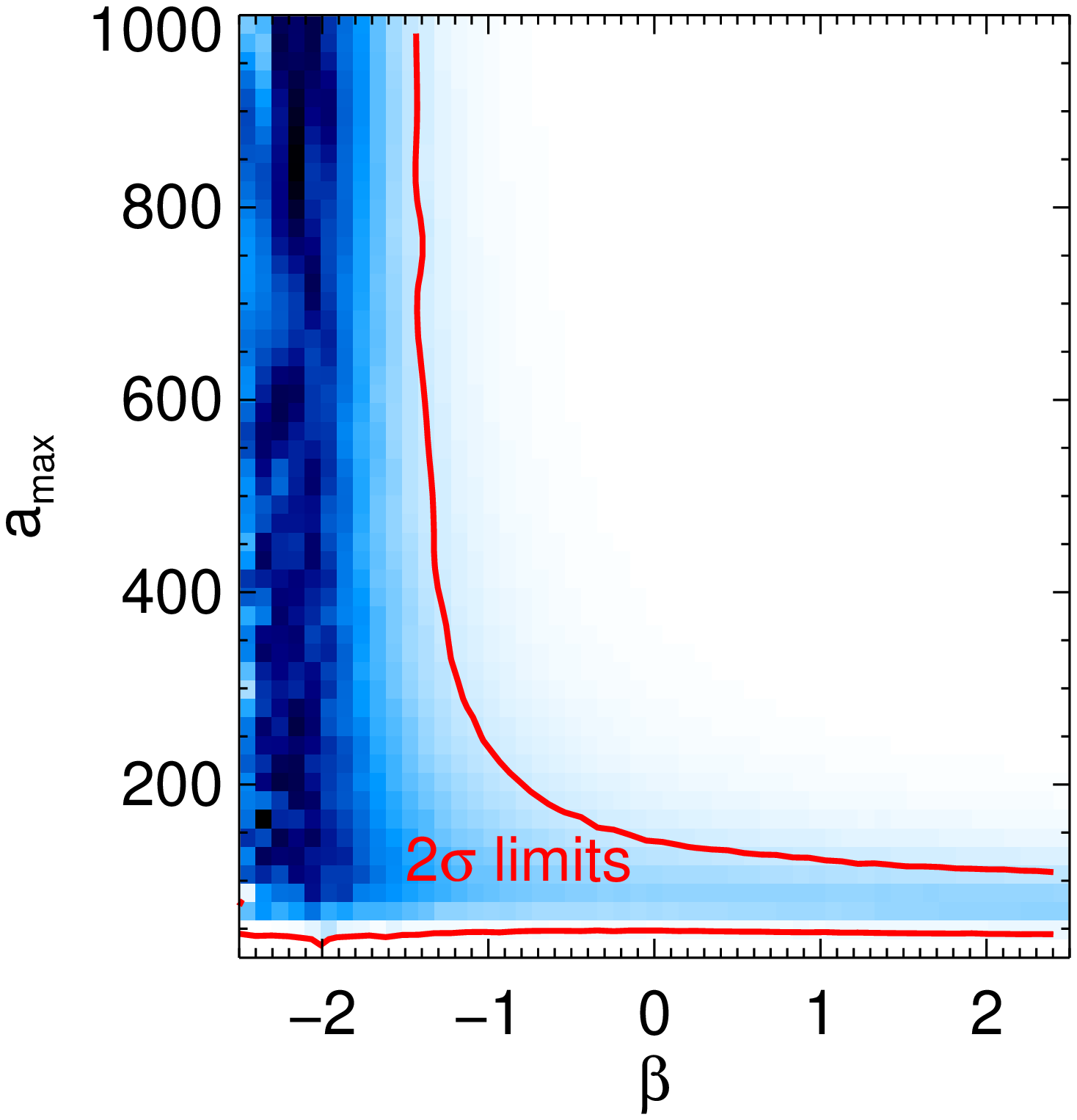}}
    \hbox{
      \includegraphics[height=7cm]{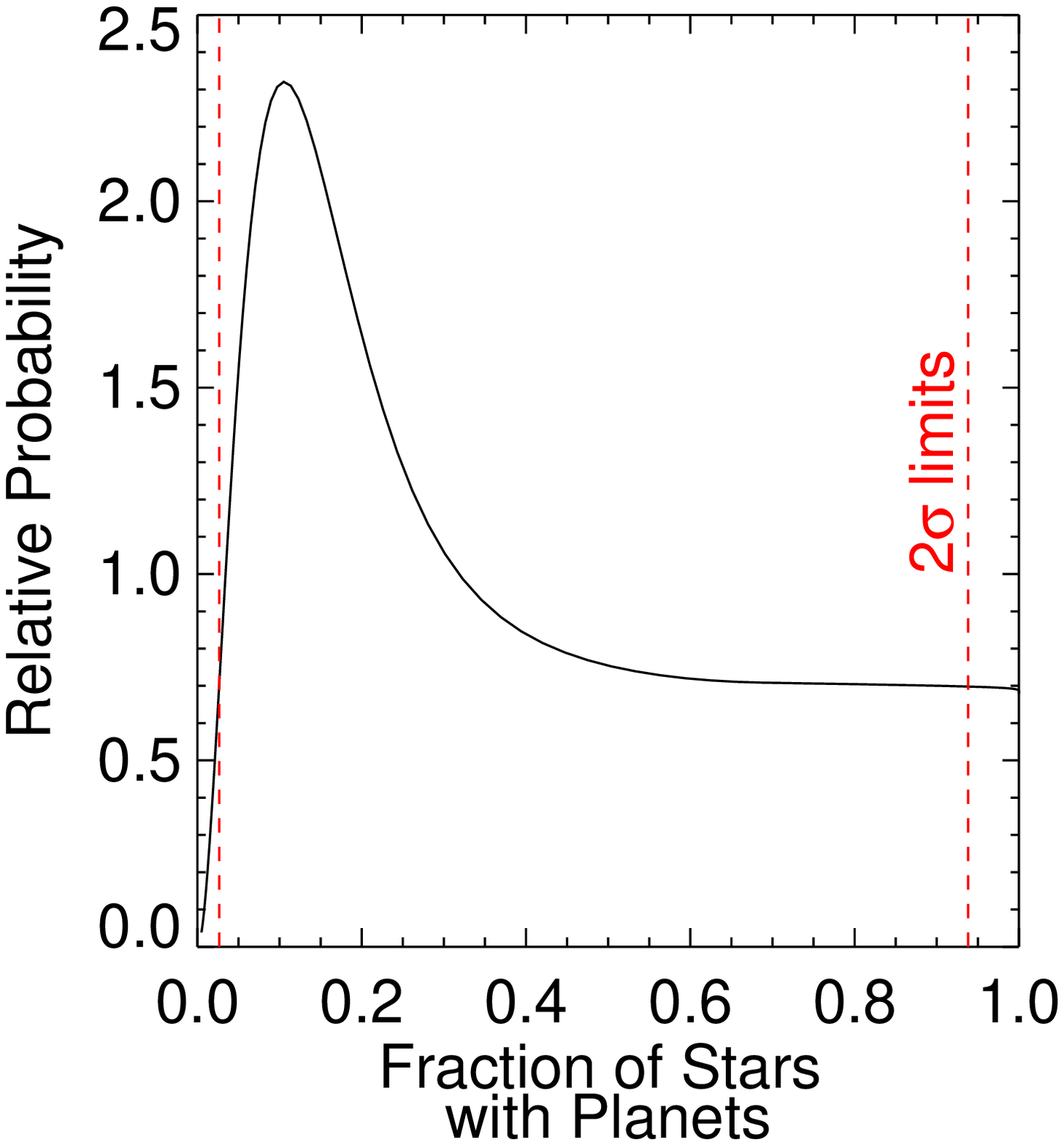}
      \includegraphics[height=7cm]{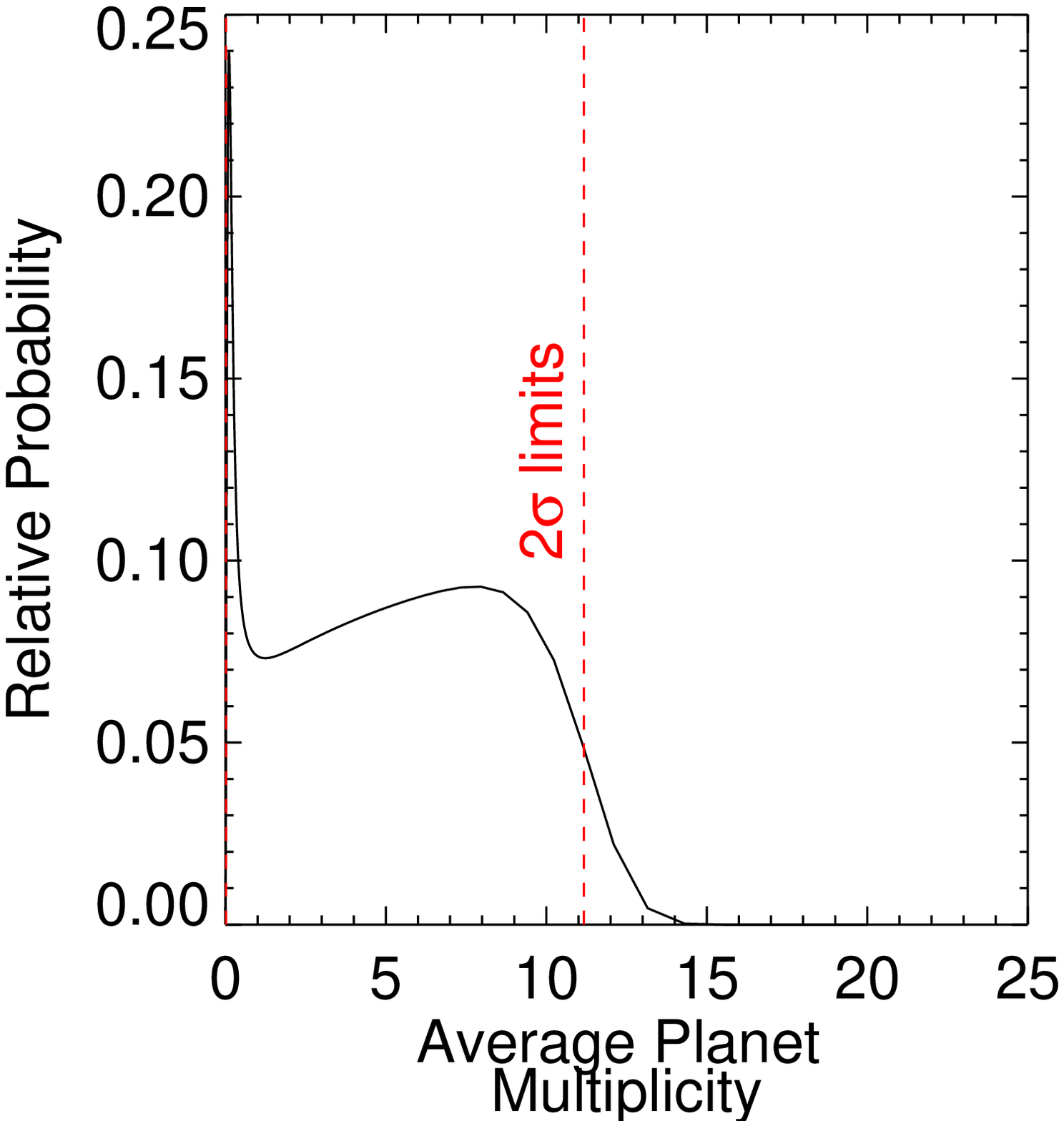}}
  }
  \caption{\textcolor{black}{ Constraints from the NICI debris disk survey combined with 
    the Vigan et al.\ 2012 debris disk survey. Darker regions
    indicate higher probability. We include the planets $\beta$ Pic~b and HR~8799~bcd in this analysis. Figures from
   left to right: the probability distributions for (1)  the planet
    mass and SMA power-law indices, $\alpha$ and $\beta$ (joint
    distribution), (2) SMA power-law and SMA cutoff  (joint
    distribution), (3) the fraction of stars with planets, and (4) the
    average multiplicity. The 2$\sigma$ constraints from the Bayesian
    analysis are indicated by the red lines. Where the constrained space
    is degenerate in two parameters, as in the top two panels, we quote 
    the constraints as follows. From the point $\beta = -0.8$ and
    $\alpha = 1.7$ on the 2$\sigma$ rejection line, we can state that 
    either $\beta < -0.8$ or $\alpha > 1.7$ (rising). With similar logic,
    we can state that either $\beta < -0.8$ or $a_{max} < 200$~AU.}}
  \label{fig:alpha_beta2}
\end{figure} 

\clearpage 

\begin{figure}[H]
  \vbox{
    \hbox{
      \includegraphics[height=7cm]{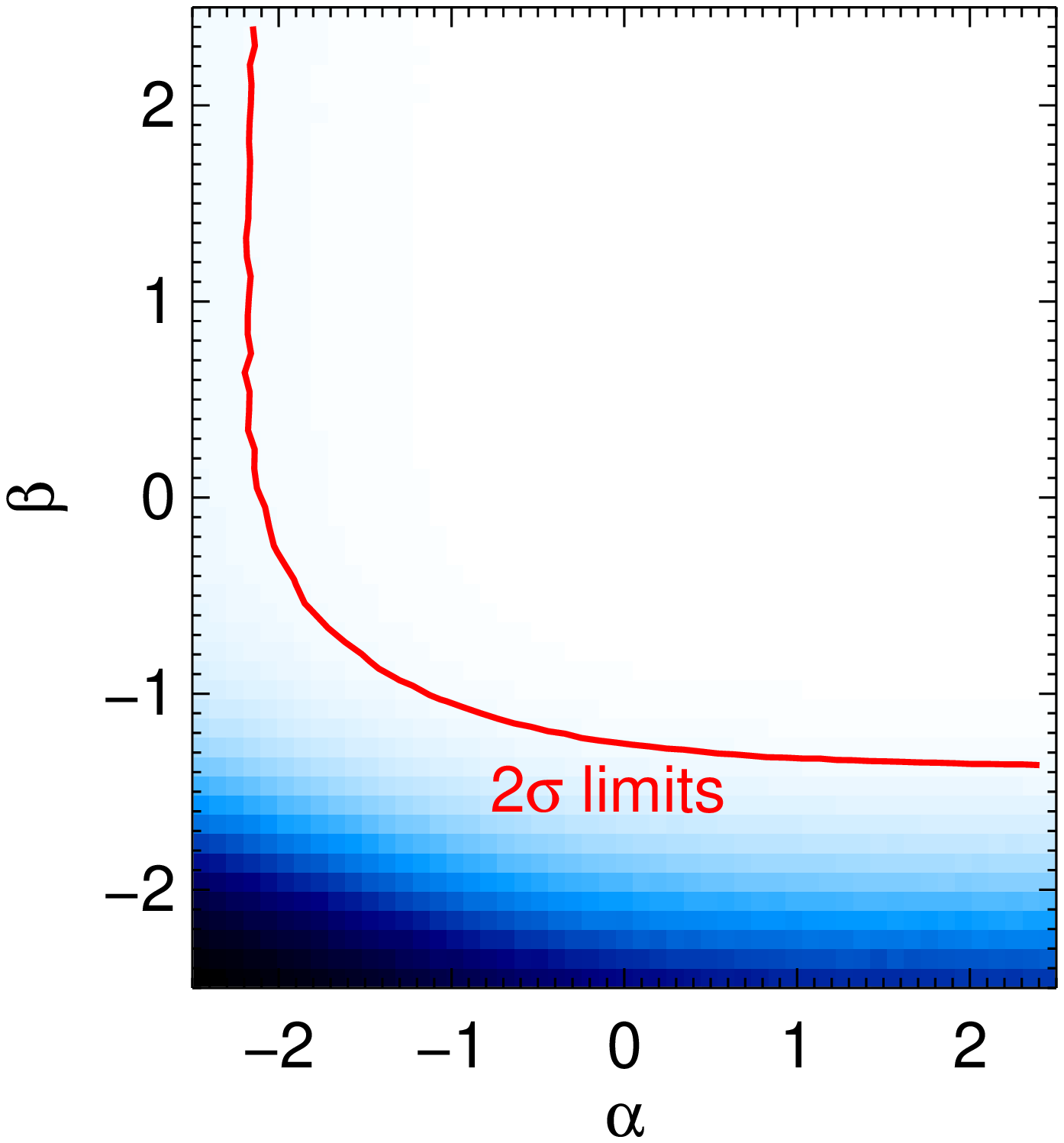}
      \includegraphics[height=7cm]{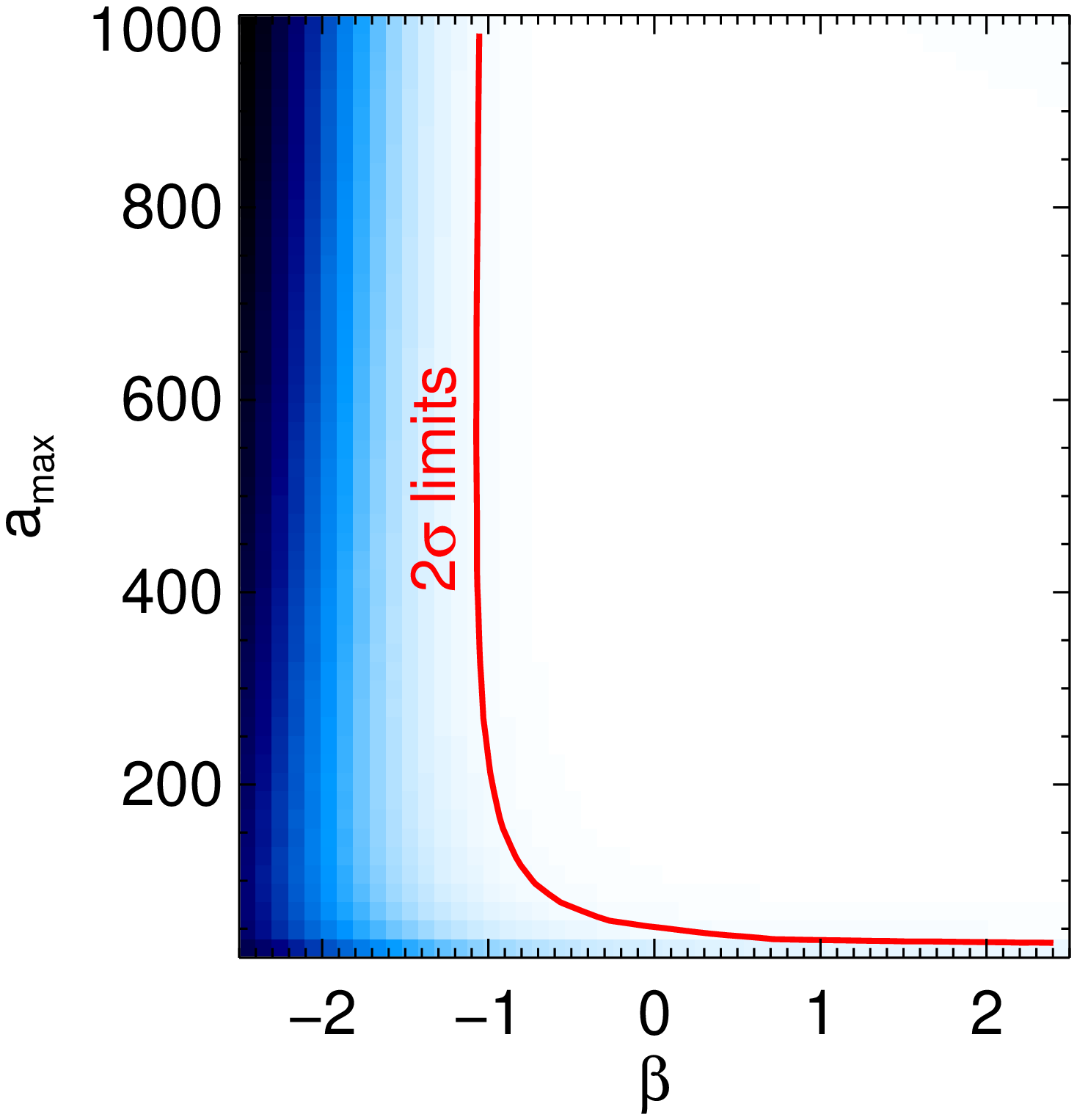}}
    \hbox{
      \includegraphics[height=7cm]{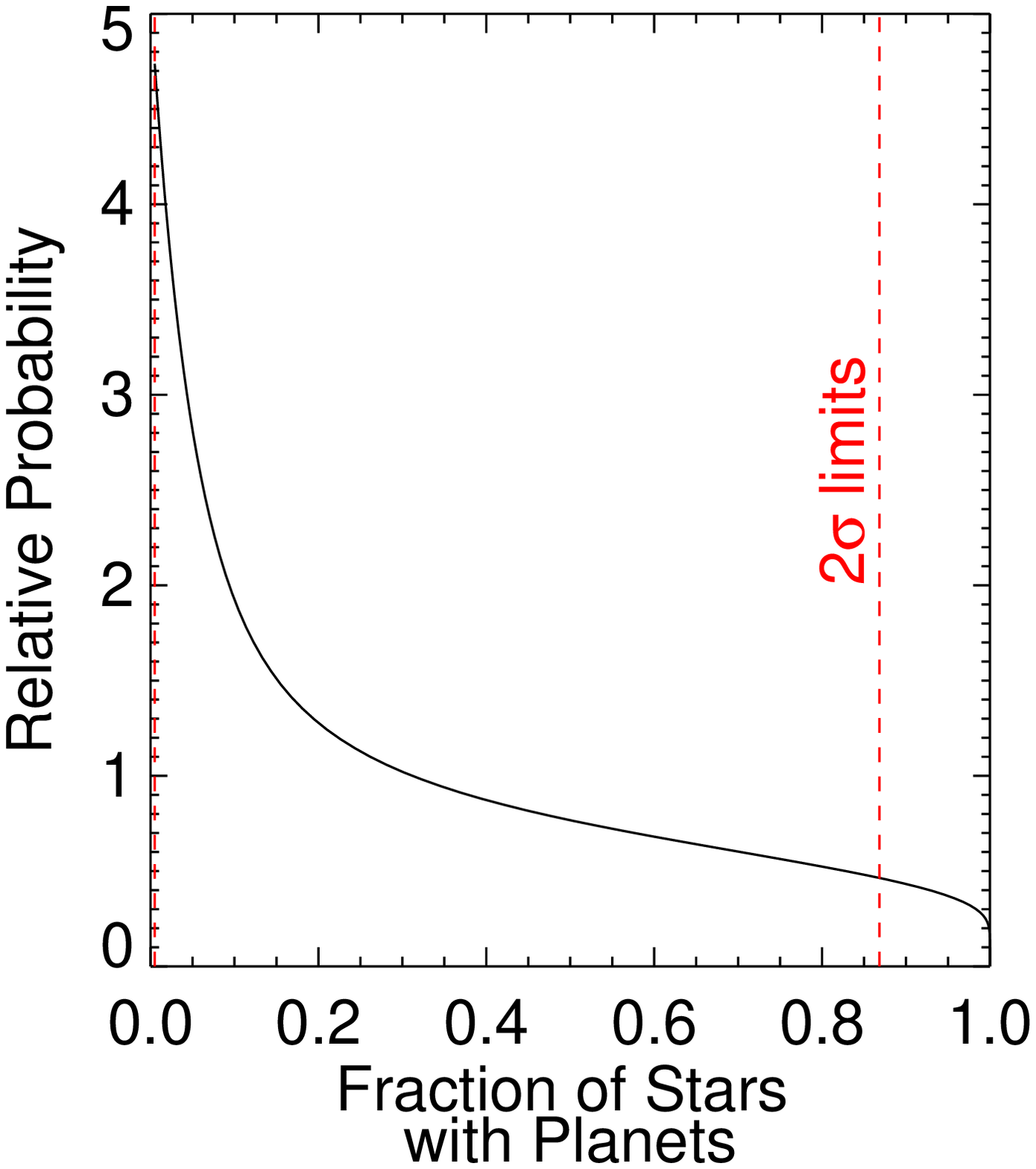}
      \includegraphics[height=7cm]{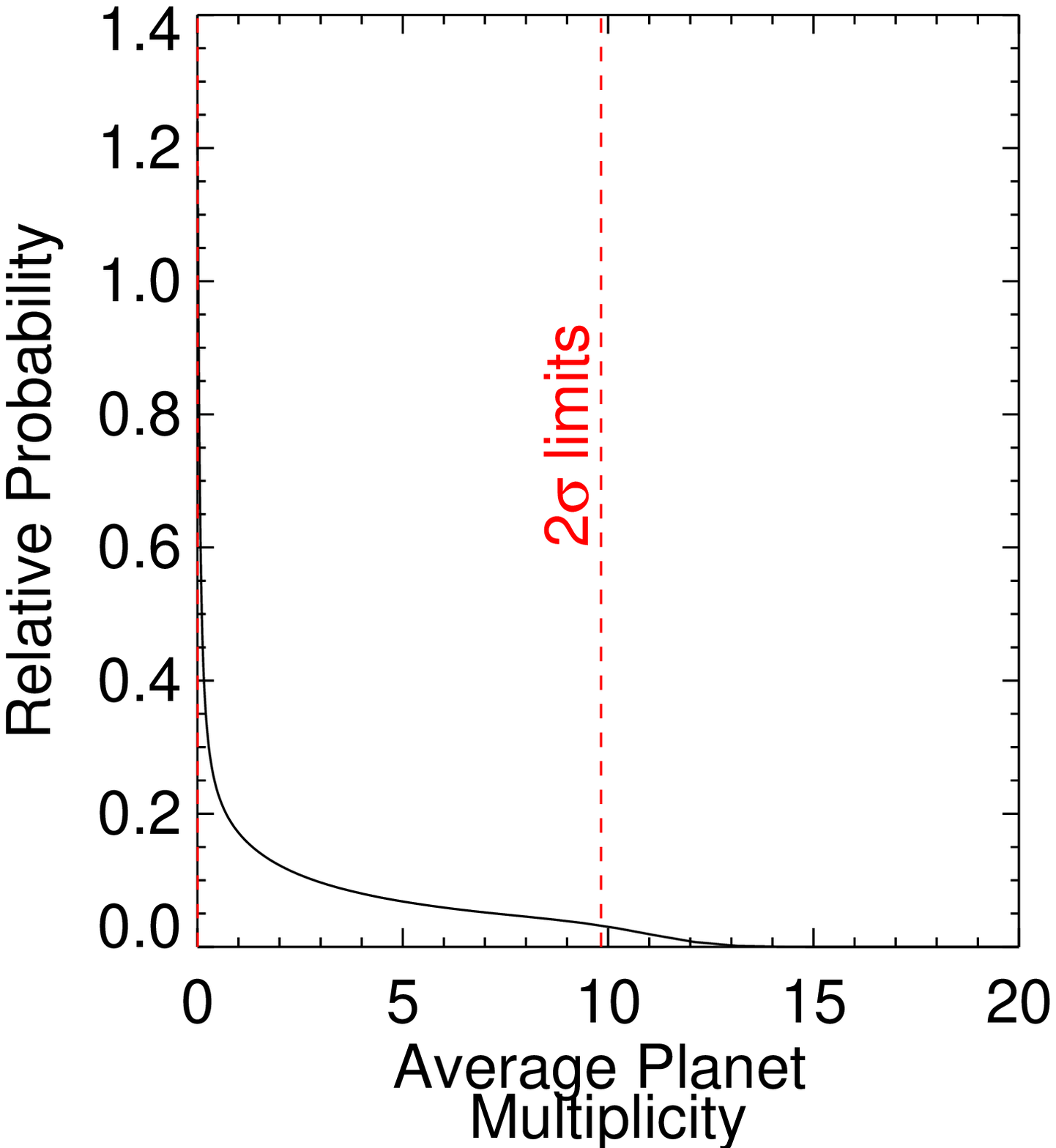}}
  } 
  \caption{\textcolor{black}{Constraints from the NICI debris disk survey combined with
    the Vigan et al.\ 2012 debris disk survey. Darker regions
    indicate higher probability. Here, we ignore 
    $\beta$ Pic~b and the HR~8799~bcd planets. Figures from
    left to right: the probability distributions for (1)  the planet
    mass and SMA power-law indices, $\alpha$ and $\beta$ (joint
    distribution), (2) SMA power-law and SMA cutoff  (joint
    distribution), (3) the fraction of stars with planets, and (4) the
    average multiplicity. The 2$\sigma$ constraints from the Bayesian
    analysis are indicated by the red lines. Where the constrained space
    is degenerate in two parameters, as in the top two panels, we quote 
    the constraints as follows. From the point $\beta = -0.8$ and
   $\alpha = -1.5$ on the 2$\sigma$ rejection line, we can state that 
   either $\beta < -0.8$ or $\alpha < -1.5$. With similar logic,
    we can state that either $\beta < -0.8$ or $a_{max} < 125$~AU.}}
   \label{fig:alpha_beta2_zd}
\end{figure} 

\clearpage

\begin{figure}[H]
  \centerline{
    \includegraphics[height=8cm]{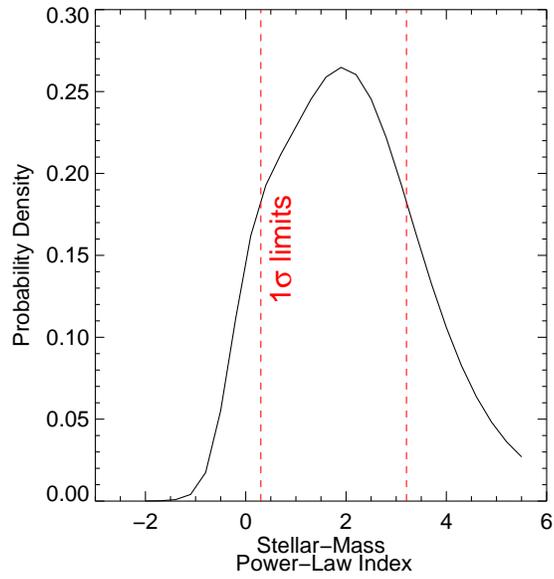}}
    \caption{\textcolor{black}{Constraints on the spectral-type power-law index from the
    augmented NICI debris disk survey, which includes $\beta$
    Pictoris~b and the HR 8799
    planets from the \cite{2012arXiv1206.4048V} survey. 
    The 1$\sigma$ limits on the index are [0.3,3.2]. Thus an A5V star is predicted to have 1.2 to 9.2 times more planets than a G2V star.}}
  \label{fig:stmass}
\end{figure}



\end{document}